\documentclass[12pt,openright,twoside,a4paper]{book}
\pdfoutput=1
\newcommand{\pdfgraphics}{\if pdf\DeclareGraphicsExtensions{.pdf,.jpg}\fi}
\usepackage{amsmath,amsfonts,amsbsy,latexsym,amssymb,amscd,amstext,graphicx,caption,slashed}
\usepackage[usenames,dvipsnames]{color}
\usepackage{dcolumn}% Align table columns on decimal point
\usepackage{bm}% bold math
\usepackage{subfigure}
\usepackage{epsfig,graphics}		
\usepackage[english]{babel}
\usepackage{multirow}
\usepackage[colorlinks=false,linkcolor=DarkBlue,citecolor=DarkGreen,urlcolor=LightBlue,bookmarksnumbered=false,bookmarksopen]{hyperref} 
\usepackage{booktabs}
\usepackage{enumerate}
\usepackage{axodraw}

\usepackage[utf8]{inputenc}
\usepackage[english]{babel}
\usepackage{makeidx}
\usepackage{latexsym} 
\usepackage{amssymb} %símbolos matemáticos más específicos
\usepackage{amsmath} %símbolos matemáticos más específicos
\usepackage{fancyhdr} %encabezados 
\usepackage[Glenn]{fncychap} 
\usepackage{graphicx}% Include figure files
\usepackage{dcolumn}% Align table columns on decimal point
\usepackage{bm}% bold math
\usepackage{amsmath,amssymb }
\usepackage{amsfonts}
\usepackage{subfigure}

\usepackage{epic}
\usepackage{eepic}
\newcommand{\beq}{\begin{equation}}
\newcommand{\eeq}{\end{equation}}
\newcommand{\bea}{\begin{eqnarray}}
\newcommand{\nn}{\nonumber}
\newcommand{\eea}{\end{eqnarray}}

\newcommand{\beqa}{\begin{eqnarray}}
\newcommand{\eeqa}{\end{eqnarray}}

\def\be{\begin{equation}}
\def\ee{\end{equation}}
\def\bea{\begin{eqnarray}}
\def\eea{\end{eqnarray}}

\def\4pig{\sfrac{4\pi G}{c^{4}}}

\def\hsp5{\hspace{5mm}}
\newcommand{\sfrac}[2]{{\textstyle{#1\over#2}}}
\def\case#1/#2{\textstyle\frac{#1}{#2}}

\usepackage{epsfig,graphics}
\usepackage{makeidx} 
\usepackage {epsfig}
\usepackage {subfigure}
\usepackage {tabularx} %para poder hacer tablas con footnotes 
\usepackage[nottoc]{tocbibind} %para incluir la bibliografía en el índice
\renewcommand{\chaptermark}[1]{%
\markboth{#1}{}}
\renewcommand{\sectionmark}[1]{%
\markright{\thesection\ #1}}
\renewcommand{\headrulewidth}{0.5pt}
\renewcommand{\footrulewidth}{0pt}
\fancypagestyle{plain}{
\fancyhead{}
\renewcommand{\headrulewidth}{0pt}}
\def\be{\begin{equation}}
\def\ee{\end{equation}}
\def\bea{\begin{eqnarray}}
\def\eea{\end{eqnarray}}

\def\e{\epsilon}

\def\bi{\begin{itemize}}
\def\ei{\end{itemize}}

\def\tr{{\rm tr\,}}
\def\Tr{{\rm Tr\,}}

\def\lsi{\raise0.3ex\hbox{$<$\kern-0.75em\raise-1.1ex\hbox{$\sim$}}}
\def\gsi{\raise0.3ex\hbox{$>$\kern-0.75em\raise-1.1ex\hbox{$\sim$}}}
\def\be{\begin{equation}}
\def\ee{\end{equation}}
\def\ba{\begin{eqnarray}}
\def\ea{\end{eqnarray}}

\definecolor{DarkGreen}{rgb}{0,0.7,0.03} 
\definecolor{DarkBlue}{rgb}{0.03,0.05,0.79}
\definecolor{LightBlue}{rgb}{0.22,0.31,0.86}
\usepackage[toc,page,titletoc]{appendix}
\usepackage{mathptmx}
\usepackage[english]{babel}
\usepackage{amssymb}
\usepackage{amsfonts}
\usepackage{amsmath}
\usepackage{eucal}
\usepackage{graphicx}
\usepackage{epsfig}
\def\erf#1{(\ref{#1})} % For references to formulas
\newcommand{\cA}{{\cal A}}  \newcommand{\cB}{{\cal B}}
  
  \newcommand{\cF}{{\cal F}}
  
\newcommand{\cI}{{\cal I}}  
  \newcommand{\cL}{{\cal L}}
\newcommand{\cO}{{\cal O}}  \newcommand{\cP}{{\cal P}}

  \newcommand{\cV}{{\cal V}}

\newcommand{\bbA}{{\mathbb A}}  \newcommand{\bbB}{{\mathbb B}}

\newcommand{\beann}{\begin{eqnarray*}}  \newcommand{\eeann}{\end{eqnarray*}}
\newcommand{\bfig}{\begin{figure}} \newcommand{\efig}{\end{figure}}
\newcommand{\bcen}{\begin{center}} \newcommand{\ecen}{\end{center}}
\newcommand{\btab}{\begin{tabular}} \newcommand{\etab}{\end{tabular}}
\def\tr{\operatorname{tr\:}}     \def\Tr{\operatorname{Tr\,}}
   
     \def\diag{\operatorname{diag}}

\newcommand{\vev}[1]{\left\langle{#1}\right\rangle}

\newcommand{\dd}{{\rm d}}

\newcommand{\mubar}{\bar{\mu}}
\newcommand{\ms}{\overline{\mathrm{MS}}}
\newtheorem{Theorem}{Theorem}[section]
\newtheorem{Lemma}{Lemma}[section]
\newtheorem{Corrolary}{Corrolary}[section]

\newcommand{\bt}{\begin{Theorem}}	\newcommand{\et}{\end{Theorem}}
\newcommand{\bl}{\begin{Lemma}}		\newcommand{\el}{\end{Lemma}}
\newcommand{\bc}{\begin{Corrolary}}	\newcommand{\ec}{\end{Corrolary}}

\begin{document}
 \thispagestyle{empty}
% \begin{tabular}{lr}
% \qquad\qquad\includegraphics[scale=1.2]{../figures/escudo_uam.pdf}\qquad
% &
% \qquad\qquad\qquad\qquad\qquad\quad\qquad\quad\includegraphics[scale=0.36]{../figures/logoift.pdf}
% \end{tabular}
%\vskip1cm
\begin{center}
\begin{tabular}{p{1cm} p{11cm}  p{2.5cm}}
 &
\large{
\begin{center}
\sc \hspace{7mm} Universidad Aut\'onoma de Madrid\\
\sc \hspace{7mm} Facultad de Ciencias\\
\sc \hspace{7mm}  Departamento de F\'isica Te\'orica
\end{center}
} &
\end{tabular}
\end{center}

\begin{picture}(0,0)
\put(0,-14){\includegraphics[height=1.8cm]{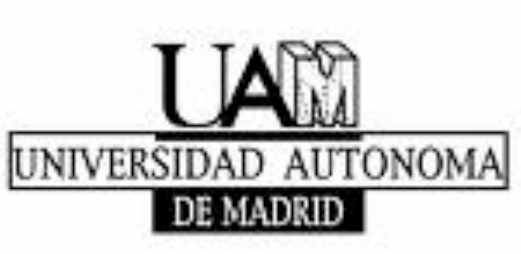}}
\end{picture}

\begin{picture}(0,0)
\put(98,-10){\includegraphics[height=2cm]{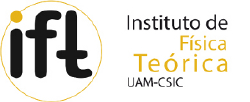}}
\end{picture}
\vskip1cm
% \begin{center}
%   {\large Universidad Aut\'onoma de Madrid}\\[.3cm]
%   {\large Facultad de Ciencias}\\[.2cm]
%   {\large Departamento de F\'isica Te\'orica}\\[2cm]
% %\vspace{1cm}
% \end{center}
\newcommand{\HRule}{\rule{\linewidth}{1mm}}
\setlength{\parindent}{1cm}
\setlength{\parskip}{1mm}
\noindent
\HRule
\begin{center}
\Huge{\bf Anomaly Induced Transport} \\ \vspace{1cm} \huge{\bf\emph{From Weak to Strong Coupling}}
\end{center}
\HRule
\vspace{1.5cm}
\begin{center}
          Memoria de Tesis Doctoral realizada por \\[0.3cm]
          {\LARGE \bf Francisco José Pe\~na Ben\'itez} \\[0.3cm]
          presentada ante el Departamento de F\'isica Te\'orica \\
          de la Universidad Aut\'onoma de Madrid \\[0.1cm]
          para la obtenci\'on del T\'itulo de Doctor en Ciencias.
\end{center}
\vspace{1cm}
% \begin{figure}[ht]
% \centering
% \begin{tabular}{lr}
% \includegraphics[scale=1.2]{../figures/escudo_uam.pdf}\qquad\qquad\qquad
% &
% \qquad\qquad\qquad\includegraphics[scale=0.36]{../figures/logoift.pdf}
% \end{tabular}
% \end{figure}
\begin{center}
          Tesis Doctoral dirigida por  \\[0.1cm]
	  {\bf Karl Landsteiner}

          del I.F.T.\\
          de la Universidad Aut\'onoma de Madrid %\\[0.1cm]        
\end{center}
\vspace{1cm}
\begin{center}
  {\large Madrid, Abril 2013.}
\end{center}
\newpage
\thispagestyle{empty}
\vspace*{1.5cm}
\newpage

% % Abstracts
% \input{abstract.tex}
 \cleardoublepage
 \cleardoublepage
 \cleardoublepage
% % %%%%%%%%%%%%%%%%%%%%%%%%%%%%%%%%%%%%%%%%%%%%%%%%%%%%%%%%%%%%%%%%%%
 \chapter*{Acknowledgements}
 \label{acknowledments}
 \addcontentsline{toc}{chapter}{Acknowledgements}
 {\it Hay muchas personas e instituciones a las que debo agradecer porque sin ellos hubiese sido casi imposible la realizaci\'on de esta tesis. Para comenzar a enumerar daré gracias a la Universidad Sim\'on Bol\'ivar por el apoyo económico dado al inicio de mis estudios doctorales, sin dicha financiaci\'on no estar\'ia aqu\'i. Karl, gracias por ser mi director de tesis, por tener tiempo para mi, por educarme, por guíarme. ¡Eres un gran tutor!. Por otro lado me gustar\'ia agradecer tambi\'en a la Comunidad de Madrid que financió la segunda etapa de mi trabajo con uno de sus  (extintos lamentablemente) contratos de formaci\'on al personal investigador. También debo dar gracias al IFT  especialmente a sus secretarias y al dept. de F\'isica te\'orica de la UAM, por este gran ambiente de trabajo. Gracias al Max Planck Insitute y en especial a la Dra. Johanna Erdmenger por acogerme en Munich durante mi estancia de investigación. Irene Amado mi hermana (cient\'ifica) mayor, Eugenio Meg\'ias quien se ha convertido ya en mi colaborador.

Ahora en un  \'ambito mas personal debo agardecer a todos los miembros de las comidas diarias en la cocina. Estas comidas son sin duda un momento de escape y relax en medio del trabajo. Javi y Pili mis mejores compañeros de despacho {\it ever} y grandes amigos. Bryan, mi hermano gracias por to asere. Por las discusiones de f\'isica y no de f\'isica, por la cañas, viajes y dem\'as gracias a Joa, Luis, Ana, Luis, Norberto, Roberto, Jaime y alguna otra persona que seguro estoy olvidando gracias a ti tambi\'en.

Para ir finalizando y no menos importante gracias a toda mi familia. A mis padres soy lo que soy ¡gracias a ustedes!. Mis abuelas maravillosas ambas. Milo\v s, no tengo palabras para expresar todo mi agrecimiento hacia ti, sin tu apoyo y paciencia nada hubiese sido tan f\'acil y agradable. Mar\'ia como no darte gracias por todo lo que hemos vivido. Emilio, Ritguey, Deivis y Luis Alfredo grandes amigos a la distancia.  Y ya para cerrar, ¿c\'omo no dar las gracias a quien me ha apoyado a muerte durante los meses de escritura y qui\'en ha aguantado mi nerviosismo predefensa de la tesis?, ¡Aingeru gracias!.}
 \cleardoublepage
 %Quotes
% \input{quote.tex}
 %\cleardoublepage
 %\chapter*{Declaration}
 %\addcontentsline{toc}{chapter}{Declaration}
%Por mis huevos
% \input{declaration.tex}
 %\cleardoublepage
% 
 \tableofcontents
 \addtocontents{toc}{\vskip-50pt}
 \pagestyle{empty} \cleardoublepage\pagenumbering{roman}
% %%%%%%%%%%%%%%%%%%%%%%%%%%%%%%%%%%%
 \chapter*{Glossary and Notation}
 \phantomsection
  \addcontentsline{toc}{chapter}{Glossary and Notation}
 \label{notation}
 \begin{itemize}
\item Capital latin letters ($M,N,P\ldots$) refers to five dimensional indices
\item Greek letters ($\mu, \nu, \rho,\ldots$) will refer to four dimensional indices
\item Lower case from the end of the alphabet latin letters ($i,j,k\ldots$) refers to three dimensional indices
\item Lower case from the beginning of the alphabet latin letters ($a,b,c\ldots$) refers to two dimensional indices
\item $g_{MN}$ is the five dimensional metric with signature $(+,-,+,+,+)$ and $g$ its determinant
\item $h_{\mu\nu}$ the induced four dimensional boundary  metric with signature $(-,+,+,+)$ and $h$ its determinant
\item The epsilon density is defined in terms of the Levi-Civita symbol as\\
$\epsilon_{MNPQR} = \sqrt{-g} \epsilon(MNPQR)$\\
$\epsilon(r0123)=1$

The Christoffel symbols,  Riemann tensor and extrinsic curvature are given by
\begin{eqnarray}
 \Gamma^M_{NP} &=& \frac 1 2 g^{MK}\left(  \partial_N g_{KP} +  \partial_P g_{KM} - \partial_K g_{NP}  \right),\\ 
 R^M\,_{NPQ} &=& \partial_P \Gamma^M_{NQ} - \partial_Q \Gamma^M_{NP} + \Gamma^M_{PK} \Gamma^K_{NQ} -   \Gamma^M_{QK} \Gamma^K_{NP} ,\\
 K_{AV} &=&   h_A^C \nabla_C n_V =  \frac 1 2 {\pounds}_n h_{AB} \,,
\end{eqnarray}
where $\pounds_n$ denotes the Lie derivative in direction of $n_A$  .
\end{itemize}
% %%%%%%%%%%%%%%%%%%%%%%%%%%%%%%%%%%%
% %%%%%%%%%%%%%%%%%%%%%%%%%%%%%%%%%%%encabezadosguais
 \pagestyle{fancy}
 \renewcommand{\chaptermark}[1]{%
 \markboth{#1}{}}
 \renewcommand{\sectionmark}[1]{%
 \markright{\thesection\ #1}}
 \fancyhf{} % borra la cabecera y pie de pagina
 \fancyhead[LE,RO]{\bfseries\thepage}
 \fancyhead[LO]{\bfseries\rightmark}
 \fancyhead[RE]{\bfseries\leftmark}
 \renewcommand{\headrulewidth}{0.5pt}
 \renewcommand{\footrulewidth}{0pt}
 \addtolength{\headheight}{0.5pt}
 \fancypagestyle{plain}{
 \fancyhead{}
 \renewcommand{\headrulewidth}{0pt}}
% %%%%%%%%%%%%%%%%%%%%%%%%%%%%%%%%%%%%%%%%%%%%%%%%%%%%%%%%%%%%%%%%%%
% % %%%%%%%%%%%%%%%%%%%%%%%%%%%%%%%%%%%%%%%%%%%%%%%%%%%%%%%%%%%%%%%%%%
   \chapter{Motivations and Introduction}
   \pagenumbering{arabic}
   \label{intro}
  The description of the high energy physics and  the interactions between elementary particles are based on gauge theories. The Standard Model and in particular QCD are examples of gauge theories. Some of the most important features of non-Abelian gauge theories and in consequence of high energy physics are not accessible through  perturbation theory. Confinement and chiral symmetry breaking are examples of phenomena which need more sophisticated techniques to be explained. 

During the seventies \cite{'tHooft:1973jz}  G. ’t Hooft realized that the perturbative series of gauge theories
can be rearranged in terms of the rank of the gauge group $N_c$ and the effective
coupling constant $\lambda_t=g_{YM}^2N _c$ called ’t Hooft coupling. In the limit $N_c\to\infty$ the series looks like an expansion summing over $2D$ surfaces, that suggested that gauge theories had an effective
description in terms of a string theory model. It wasn't until 1997 almost twenty years later that the 't Hooft ideas were realized when J. Maldacena published his very famous paper \cite{Maldacena:1997re} which revolutionized the fields of Strings and Quantum Field Theory. This paper was the starting point of the  construction of the holographic principle through the introduction of the $AdS/CFT$ correspondence which tells us that the $\mathcal{N}=4$ $SU(N_c)$ Super-Yang-Mills (SYM) theory in $4$-dimensions is dual to the type $IIB$ string theory on $AdS_5\times S_5$. This correspondence relates the string theory coupling constant $g_s$ with $1/N_c$ and the radius of the $AdS$ space with the t’ Hooft coupling. Beside the t’ Hooft’s ideas this
correspondence is also a realization of the holographic principle which says that a quantum
gravity theory should be described with the degrees of freedom living at the boundary of
the space. In this case the space of quantum gravity is the 10-dimensional space of
the string theory and the boundary is the 4-dimensional conformal boundary associated
to that space ($AdS_5\times S_5$). The degrees of freedom of the theory at the boundary are
precisely the ones of the SYM theory. In fact, at present time $AdS/CFT$ not only refers  to Maldacena's duality but a framework of many dualities realizing the holographic principle. 

The really useful fact of this duality is its weak/strong character; from the view point of the field theory the weakly coupled situation is described by the four dimensional perturbative gauge theory. But the strong coupling scenario is described by string theory in ten dimensions\footnote{In the large $N_c$ limit string theory is reduced to classical gravity}. An interesting application of the $AdS/CFT$ duality is given by asymptotically $AdS$ black holes. According to the holographic dictionary a black hole embedded in an $AdS$ space-time is dual to a thermal state in the field theory side. One of the most important and known results of holography, at least from a phenomenological point of view is the very small lower bound in the ratio of shear viscosity to entropy density in all the Holographic plasmas which are dual to an Einstein-Hilbert gravity $\eta/s \geqslant \hbar/(4\pi \kappa_B)$ \cite{Kovtun:2004de} \footnote{Recently has been shown that this bound can be violated if rotational symmetry is broken\cite{Erdmenger:2011tj}}. This result had a big impact because the measure in the experiment RHIC  of that ratio for the Quark-Gluon Plasma (QGP)  was $\sim 2.5\times \hbar/(4\pi \kappa_B)$, suggesting that the plasma is in a strongly coupled regime because the prediction for $\eta/s$ coming from weak coupling is in contrast very large.
%$$
%\frac{\eta}{s} \sim \frac{1}{\lambda^4\log{(1/\lambda^2)}}\gg 1.
%$$

The indications of the production of a quantum liquid in a strongly coupled regime at the experiments RHIC and more recently at the LHC, pushed forward $AdS/CFT$ as a very promising framework to construct phenomenological models to try to understand and predict the behaviour of the QGP.

\section{Kubo Formulae and Holographic Triangle Anomalies}

Hydrodynamics is an ancient subject. Even in its relativistic form it appeared that everything relevant to its
formulation could be found in \cite{Landau}.  Apart from stability issues that were addressed in the 1960s and 1970s  \cite{Muller:1967zza,Israel:1976tn} leading to a second order formalism there seemed little room for new discoveries. The last years witnessed however an unexpected and profound development of the formulation of relativistic hydrodynamics.  The second order contributions have been put on a much more systematic basis
applying effective field theory reasoning \cite{Baier:2007ix}. The lessons learned from applying the AdS/CFT
correspondence \cite{Maldacena:1997re} to the plasma phase of strongly coupled non-abelian gauge theories
\cite{Gubser:1996de, Witten:1998zw, Kovtun:2004de} played a major role (see \cite{CasalderreySolana:2011us} for a recent review).

The  understanding of hydrodynamics is as an effective theory applicable when the mean free path of the particles is much shorter than the characteristic length scale of the system. In this regime the system can be described with the so called constitutive relations for the energy momentum tensor and the currents (the last is present if the system is charged under some global symmetry group) plus their conservation equations.

Hydrodynamics is about a system in local thermal equilibrium which means that the intensive thermodynamical parameters pressure, temperature and chemical potential ($p$, $T$, $\mu$) are slowly varying functions through the space-time. The last comes with the implication that constitutive relations can be written as a derivative expansion in terms of the thermodynamical variables and the fluid velocity, and the so called transport coefficients\footnote{Examples of transport coefficients are electrical conductivity, shear viscosity, bulk viscosity, etc. They are intrinsic quantities associated to the system and they are determined from the microscopical theory.}.  Very generic statements such as symmetry considerations can determine the form of the constitutive relations but cannot fix the values of the transport coefficients. To read off these coefficients it is useful to consider the theory of linear response and introduce external background fields to define the so called Kubo formulae. Having a Kubo formula allow us to compute transport coefficients in terms of retarded Green's functions, a simple example of a Kubo formula is the one for the electric conductivity
$$
\sigma = \lim_{\omega\rightarrow0}\frac{i}{\omega}\langle j^xj^x\rangle(k=0),
$$
where  $\langle j^xj^x\rangle$ is the two point retarded correlator between two electric currents.
Therefore if we have a holographic description of the field theory we can use the machinery of $AdS/CFT$ to  compute transport coefficients, because of the  strong/weak nature of the duality, solving a problem of a strongly coupled field theory is reduced to a classical general relativity problem!.

During the last years a new set of transport coefficients has been discovered as a consequence of chiral anomalies. The axial anomaly of QED is responsible for two particularly interesting effects of strong magnetic fields in dense, strongly interacting matter as found in neutron stars or heavy-ion collisions. At large quark chemical potential $\mu$, chirally restored quark matter gives rise to an axial current parallel to the magnetic field \cite{Son:2004tq,Metlitski:2005pr,Newman:2005as}
\be\label{J5}
\vec J_5=\frac{eN_c}{2\pi^2}\mu \,\vec B,
\ee
which may indeed lead to observable effects in strongly magnetized neutron stars and heavy ion collisions \cite{Charbonneau:2009ax,KerenZur:2010zw}, this phenomena is known as {\it chiral separation effect} (CSE).

In the context of heavy ion collisions it was argued in \cite{Kharzeev:2007jp,Fukushima:2008xe} that the excitation of topologically non-trivial gluon field configurations in the early
non-equilibrium stages of a heavy ion collision might lead to an imbalance in the number of left- and right-handed quarks. This situation can be modelled by an axial chemical potential\footnote{As soon as thermal equilibrium is reached this imbalance is frozen and is modelled by a chiral chemical potential, at least as long as the electric field is zero}. During the collision one expects initial magnetic fields that momentarily exceed even those found in magnetars. It has been proposed by Kharzeev et al.\ \cite{Kharzeev:2004ey,Kharzeev:2007tn,Kharzeev:2007jp,Fukushima:2008xe,Kharzeev:2009pj}
that the analogous effect  \cite{Alekseev:1998ds}
\be\label{CME}
\vec J=\frac{e^2N_c}{2\pi^2}\mu_5 \,\vec B,
\ee
where $\vec J$ is the electromagnetic current and $\mu_5$ the axial chemical potential, could render observable event-by-event P and CP violations. Indeed, there is recent experimental evidence
for this {\it chiral magnetic effect} (CME) in the form of charge separation in heavy ion collisions with respect to the reaction plane \cite{Abelev:2009uh,Voloshin:2008jx} and more recentely from LHC data \cite{Abelev:2012pa}  (see however \cite{Wang:2009kd,Asakawa:2010bu}). For lattice studies of the effect, see for example \cite{Buividovich:2009wi,Buividovich:2010tn}.

On the other hand the application of the fluid/gravity correspondence \cite{Bhattacharyya:2008jc} to theories including chiral anomalies \cite{Erdmenger:2008rm,Banerjee:2008th} lead to another surprise: it was found that not only a magnetic field induces a current but that also a vortex in the fluid leads to an induced current, this effect is called {\it chiral vortical effect} (CVE).  Again it is a consequence of the presence of  chiral anomalies.  It was later realized that the chiral magnetic and vortical conductivities are almost completely fixed in the hydrodynamic framework by demanding the existence of an entropy current with positive definite divergence \cite{Son:2009tf}. That this criterion did not fix the anomalous transport coefficients completely was noted in \cite{Neiman:2010zi} and various terms depending on the temperature instead of the chemical potentials were shown to be allowed as undetermined integration constants. See also~\cite{Kalaydzhyan:2011vx} for a recent discussion of these anomaly coefficients with applications to heavy ion physics.

In the meanwhile Kubo formulae for the chiral magnetic conductivity \cite{Kharzeev:2009pj} and the chiral vortical conductivity \cite{Amado:2011zx} had been developed. Up to this point only pure gauge anomalies had been considered to be relevant since the mixed gauge-gravitational anomaly in four dimensions is of higher order in
derivatives and was thought not to be able to contribute to hydrodynamics at first order in derivatives. Therefore it came as a surprise that in the application of the Kubo formula for the  chiral vortical conductivity to a system of free chiral fermions a purely temperature dependent contribution was found. This contribution was consistent
with some the earlier found integration constants and it was shown to arise if and only if the system of chiral fermions features a mixed gauge-gravitational anomaly \cite{Landsteiner:2011cp}. In fact these contributions had been found already very early in \cite{Vilenkin:1979ui}. The connection to the presence of anomalies was however not made at that time. The gravitational anomaly contribution to the chiral vortical effect was also established in a strongly coupled AdS/CFT approach and precisely the same result as at weak coupling was found \cite{Landsteiner:2011iq}. 

The argument based on a positive definite divergence of the entropy current allows to fix the contributions from pure gauge anomalies uniquely and provides therefore a non-renormalization theorem. No such result is known thus far for the contributions of the gauge-gravitational anomaly, actually some very recent attempts to establish a non-renormalization theorems lead to the fact that the chiral vortical conductivity indeed renormalizes due to gluon fluctuations \cite{Golkar:2012kb,Hou:2012xg}. 

A gas of weakly coupled Weyl fermions in arbitrary dimensions has been studied in \cite{Loganayagam:2012pz} and confirmed that the anomalous conductivities can be obtained directly from the anomaly polynomial under substitution of the field strength with the chemical potential and the first Pontryagin density by the negative of
the temperature squared \cite{Loganayagam:2011mu} . Recently the anomalous conductivities have also been obtained in effective action approaches \cite{Jensen:2012jy,Banerjee:2012iz}. The contribution of the mixed gauge-gravitational anomaly appear on all these approaches as undetermined integrations constants. But in \cite{Jensen:2012kj} the authors argued that mixed gravitational anomaly produces a ''casimir momentum'' which fix the anomalous transport coefficients and explain why this anomaly been higher order in derivative contributes at lower orders.

In this thesis we will study these anomalous transport effects through the calculation of the anomalous conductivities via Kubo formulae and using the fluid/gravity correspondence. The advantage of the usage of Kubo formulae is that they capture all contributions stemming either from pure gauge or from
mixed gauge-gravitational anomalies. The disadvantage is that the calculations can be performed only with a particular model and only in a weak coupled or in the gravity dual of the strong coupling regime. The fluid gravity computation allow us to confirm the independence of the anomalous transport coefficients on the intensity of the external fields because Kubo formulae are valid only in the linear response limit. Also is a systematic and powerful tool to compute the influence of the anomalies into the second order transport, for which we would need three point function whether  we would want to compute it using Kubo formulae.

Along the way we will explain our point of view on some subtle issues concerning the definition of currents and
of chemical potentials when anomalies are present. These subtleties lead indeed to some ambiguous
results \cite{Yee:2009vw} and \cite{Rebhan:2009vc}. A first step to clarify these issues
was done in \cite{Gynther:2010ed} and a more general exposition of the relevant issues has appeared in \cite{Landsteiner:2011tf} and \cite{Landsteiner:2012kd}. 

The thesis tries to be self-contained and  is organized as follows. In chapter \ref{anomalies} we will briefly
summarize the relevant issues concerning anomalies. We recall how vector like symmetries can always be restored by adding suitable finite counterterms to the effective action~\cite{Bardeen:1969md}. A related but different issue is the fact that currents can be defined either as consistent or as covariant currents. The hydrodynamic constitutive relations depend on what definition of current is used. We review the notion of chemical potential in the approach of the grand canonical ensemble in the chapter \ref{chemical} and discuss subtleties in the definition of the chemical potential in the presence of an anomaly   and define our preferred scheme. Then in chapter \ref{rel_hydro} we move to the building of relativistic hydrodynamics, constitutive relations and the derivation of the Kubo formulae that allow the calculation of the anomalous transport coefficients from two point correlators of an underlying quantum field theory.

In chapter \ref{cmecse} we apply the Kubo formulae to a holographic model describing a field theory system with two $U(1)$ currents, one  conserved vector current which is associated to the electromagnetic current and an anomalous one interpreted as an axial current. We also give some arguments coming from holography in favour of our preference in the introduction of anomalous chemical potentials, this arguments are also confirmed by a three point computation in field theory in the next chapter.

In chapter \ref{freefermions} we apply the Kubo formulae to a theory of free Weyl fermions and show that two different contributions arise. They are clearly identifiable as being related to the presence of pure gauge and mixed gauge-gravitational anomalies.

In chapter \ref{strongcve} we define a holographic model that implements the mixed gauge gravitational anomaly via a mixed gauge-gravitational Chern-Simons term. We calculate the same Kubo formulae as at weak coupling, obtaining the same values for chiral axi-magnetic and chiral vortical conductivities as in the weak coupling model.

Finally in chapter \ref{fluid_grav} we apply the Fluid/Gravity correspondence to the holographic model in order to study the second order behaviour of transport coefficients due to the gauge and mixed gauge-gravitational anomaly.

We conclude this thesis with some discussions and outlook to further developments. 

% %%%%%%%%%%%%%%%%%%%%%%%%%%%%%%%%%%%%%%%%%%%%%%%%%%%%%%%%%%%%%%%%%%
% % %%%%%%%%%%%%%%%%%%%%%%%%%%%%%%%%%%%%%%%%%%%%%%%%%%%%%%%%%%%%%%%%%%
   \chapter{Quantum Anomalies}
   \label{anomalies}
  Anomalies arise by integrating over chiral fermions in the path integral. They signal a fundamental incompatibility between the symmetries present in the classical theory and the quantum theory.

Unless otherwise stated we will always think of the symmetries as global symmetries. But we still will introduce gauge fields. These gauge fields serve as classical sources coupled to the currents. As a side effect their presence promotes the global symmetry to a local gauge symmetry. It is still justified to think of it as a global symmetry as long as we do not introduce a kinetic Maxwell or Yang-Mills term in the action. 

In a theory with chiral fermions we define an effective action depending on these gauge fields by the path integral
\begin{equation}
 \label{eq:Weff}
 e^{i W_{eff}[A_\mu]/\hbar } := \int {\cal D}\Psi {\cal D}\bar \Psi e^{ i S[\psi, A_\mu]/\hbar }\,.
\end{equation}
The vector field $A_\mu(x)$ couples to a classically conserved current $J^\mu = \bar\Psi \gamma^\mu Q \Psi$. The charge operator $Q$ can be the generator of a Lie group combined with chiral projectors $\mathcal P_{\pm} = \frac{ 1}{ 2} (1\pm\gamma_5)$. General combinations are allowed although in the following we will mostly concentrate on a simple chiral $U(1)$ symmetry  for which we can take $Q=\mathcal P_+$. The fermions are minimally coupled to the gauge field and the classical action has an underling gauge symmetry 
\begin{equation}
 \label{eq:symclassical}
 \delta \Psi = -i \lambda(x) Q \Psi    ~~~~~~~~~~~~~~,~~~~~~~~~~~~ \delta A_\mu(x) = D_\mu \lambda(x) \,,
\end{equation}
with $D_\mu$ denoting the gauge covariant derivative. Assuming that the theory has a classical limit the effective action in terms of the gauge fields allows for an expansion in $\hbar$ 
\begin{equation}
 \label{eq:expansionWeff}
 W_{eff} = W_0 + \hbar W_1 + \hbar^2 W_2 + \dots
\end{equation}
We find it convenient to use the language of BRST symmetry by promoting the gauge parameter to a ghost field $c(x)$.~\footnote{A recent comprehensive review on BRST symmetry is \cite{Dragon:2012au}.}
The BRST symmetry is generated by
\begin{equation}
 \label{eq:BRST}
  s A_\mu = D_\mu c   ~~~~~~, ~~~~~ sc = -i c^2\,.
\end{equation}
It is nilpotent $s^2=0$. 
The statement that the theory has an anomaly can now be neatly formalized. Since on gauge fields the BRST symmetry acts just as the gauge symmetry,  gauge invariance translates into BRST invariance. An anomaly is present if
\begin{equation}
 \label{eq:WZanomaly}
 s W_{eff} = {\cal A}  ~~~~~~~~ \mathrm{and} ~~~~ {\cal A} \neq sB \,.
\end{equation}
Because of the nil potency of the BRST operator the anomaly has to fulfil the Wess-Zumino consistency condition 
\begin{equation}
 s{\cal A}=0\,.
\end{equation}
As indicated in (\ref{eq:WZanomaly}) this has a possible trivial solution if there exists a local functional $B[A_\mu]$ such that $sB = {\cal A}$. An anomaly is present if no such $B$ exists.  The anomaly is a quantum effect. If it is of order $\hbar^n$ and if a suitable local functional $B$ exists we could simply redefine the effective action as $\tilde{W}_{eff} = W_{eff} - B$ and the new effective action would be BRST and therefore gauge invariant. The form and even the necessity to introduce a functional $B$ might depend on the particular regularization scheme chosen. As we will explain in the case of an axial and vector symmetry a suitable $B$ can be found that always
allows to restore the vectorlike symmetry, this is the so-called Bardeen counterterm \cite{Bardeen:1969md}. The necessity to introduce the Bardeen counterterm relies however on the regularization scheme chosen. In schemes that automatically preserve vectorlike symmetries, such as dimensional regularization, the vector symmetries are
automatically preserved and no counterterm has to be added. Furthermore the Adler-Bardeen theorem guarantees that chiral anomalies appear only at order $\hbar$. Their presence can therefore by detected in one loop diagrams such as the triangle diagram of three currents.

We have introduced the gauge fields as sources for the currents
\begin{equation}
 \label{eq:sources}
 \frac{\delta}{\delta A_\mu(x)} W_{eff}[A] = \langle \mathcal J^\mu \rangle\,.
\end{equation}
For chiral fermions transforming under a general Lie group generated by $T^a$ the chiral anomaly takes the form \cite{Bertlmann:1996xk}
\begin{eqnarray}
\nonumber s W_{eff}[A] &=& -\int d^4x c^a (D_\mu \mathcal J^\mu)^a \\
\label{eq:anomalyform} &=&  -\frac{\eta}{24\pi^2}\int d^4x\, c^a
\epsilon^{\mu\nu\rho\sigma} \mathrm{tr} \left[ T^a \partial_\mu
\left( A_\nu \partial_\rho A_\sigma + \frac{ 1}{ 2} A_\nu A_\rho A_\sigma
\right)
\right]  \,.
\end{eqnarray}
Where $\eta=+1$ for left-handed fermions and $\eta=-1$ for right-handed fermions.  Differentiating with respect to the ghost field (the gauge parameter) we can derive a local form. To simplify the formulas we specialize this to the case of a single chiral $U(1)$ symmetry taking $T^a=1$
\begin{equation}
 \label{eq:localanomaly}
 \partial_\mu \mathcal J^\mu = \frac{\eta}{96\pi^2} \epsilon^{\mu\nu\rho\sigma}
F_{\mu\nu}F_{\rho\sigma}\,.
\end{equation}
This is to be understood as an operator equation. Sandwiching it between the vacuum state $\left|0\right>$ and further differentiating with respect to the gauge fields we can generate the famous triangle form of the anomaly
\begin{equation}
 \label{eq:triangleanomaly}
 \langle \partial_\mu \mathcal J^\mu(x) \mathcal J^\sigma(y) \mathcal J^\kappa(z) \rangle =
\frac{1}{12\pi^2}\epsilon^{\mu\sigma\rho\kappa} \partial^x_\mu \delta(x-y)
\partial^x_\rho \delta(x-z) \,.
\end{equation}
The one point function of the divergence of the current is non-conserved only in the background of parallel electric and magnetic fields whereas the non-conservation of the current as an operator becomes apparent in the triangle diagram even in vacuum. 

By construction this form of the anomaly fulfills the Wess-Zumino consistency condition and is therefore called the {\em consistent   anomaly}. In analogy we call the current defined by (\ref{eq:sources}) the {\em consistent current}.

For a $U(1)$ symmetry the functional differentiation with respect to the gauge field  and the BRST operator $s$ commute, 
\begin{equation}
 \label{eq:scommute}
\left[s, \frac{\delta}{\delta A_\mu(x)} \right] =0\,.
\end{equation}
An immediate consequence is that the consistent current is not BRST invariant but rather obeys
\begin{equation}
 \label{eq:constentcurrentstrafo}
s \mathcal J^\mu = \frac{1}{24\pi^2} \epsilon^{\mu\nu\rho\lambda} \partial_\nu c
F_{\rho\lambda} = - \frac{1}{24\pi^2} s K^\mu \,,
\end{equation}
where we introduced the Chern-Simons current $K^\mu = \epsilon^{\mu\nu\rho\lambda} A_\nu F_{\rho\lambda}$ with $\partial_\mu K^\mu = \frac{ 1}{ 2} \epsilon^{\mu\nu\rho\lambda} F_{\mu\nu} F_{\rho\lambda} $. 

With the help of the Chern-Simons current it is possible to define the so-called {\em covariant current} (in the case of a $U(1)$ symmetry rather the invariant current)
\begin{equation}
 \label{eq:covcurrent}
 J^\mu = \mathcal J^\mu + \frac{1}{24\pi^2} K^\mu\,.
\end{equation}
fulfilling 
\begin{equation}
 s J^\mu =0\,.
\end{equation}

The divergence of the covariant current defines the {\em covariant anomaly}
\begin{equation}
 \label{eq:covanomaly}
 \partial_\mu J^\mu = \frac{\eta}{32\pi^2} \epsilon^{\mu\nu\rho\sigma}
F_{\mu\nu}F_{\rho\sigma}\,.
\end{equation}
Notice that the Chern-Simons current cannot be obtained as the variation with respect to the gauge field of any funtional. It is therefore not possible to define an effective action whose derivation with respect to the gauge field gives the covariant current. 
 
Let us suppose now that we have one left-handed and one right-handed fermion with the corresponding left- and right-handed anomalies. Instead of the left-right basis it is more convenient to introduce a vector-axial basis by defining the vectorlike current $\mathcal J^\mu_V =\mathcal J^\mu_L +\mathcal  J^\mu_R$ and the axial current $\mathcal J_A^\mu = \mathcal J^\mu_L-\mathcal J^\mu_R$.  Let $V_\mu(x)$ be the gauge field that couples to
the vectorlike current and $A_\mu(x)$ be the gauge field coupling to the axial current. The (consistent) anomalies for the vector and axial current turn out to be 
\begin{eqnarray}
 \label{eq:vectoranomaly}
 \partial_\mu \mathcal J^\mu_V &=& \frac{1}{24\pi^2} \epsilon^{\mu\nu\rho\lambda}
F_{\mu\nu}^V F_{\rho\lambda}^A \,,\\
\label{eq:axialanomaly}
 \partial_\mu \mathcal J^\mu_A &=& \frac{1}{48\pi^2} \epsilon^{\mu\nu\rho\lambda}
(F_{\mu\nu}^V F_{\rho\lambda}^V + F_{\mu\nu}^A F_{\rho\lambda}^A) \,.
\end{eqnarray}
As long as the vectorlike current corresponds to a global symmetry nothing has gone wrong so far. If we want to identify the vectorlike current with the electromagnetic current in nature we need to couple it to a dynamical photon gauge field and now the non-conservation of the vector current is worrisome to say the least. The problem arises because implicitly we presumed a regularization scheme that treats left-handed and right-handed fermions on the same footing.  As pointed out first by Bardeen this flaw can be repaired by adding a finite counterterm (of order $\hbar$) to the effective action. This is the so-called Bardeen counterterm and has the form \begin{equation}
 \label{eq:Bardeenct}
 B_{ct} = -\frac{1}{12\pi^2} \int d^4x \,\epsilon^{\mu\nu\rho\lambda} V_\mu A_\nu F_{\rho\lambda}^V \,.
\end{equation}
Adding this counterterm to the effective action gives additional contributions of Chern-Simons form to the consistent vector and axial currents. With the particular coefficient chosen it turns out that the anomaly in the vector current is canceled whereas the axial current picks up an additional contribution such that after adding the Bardeen counterterm the anomalies are
\begin{eqnarray}
 \label{eq:vectorconserved}
 \partial_\mu J^\mu_V &=&0 \,,\\
\label{eq:bardeenaxialanomaly}
 \partial_\mu J^\mu_A &=& \frac{1}{48\pi^2} \epsilon^{\mu\nu\rho\lambda} (3
F_{\mu\nu}^V F_{\rho\lambda}^V + F_{\mu\nu}^A F_{\rho\lambda}^A) \,.
\end{eqnarray}
This definition of currents is mandatory if we want to identify the vector current with the usual electromagnetic current in nature! It is furthermore worth to point out that both currents are now invariant under the vectorlike $U(1)$  symmetry. The currents are not invariant under axial transformation, but these are anomalous anyway.

Generalizations of the covariant anomaly and the Bardeen counterterm to the non-abelian case can be found e.g. in \cite{Bertlmann:1996xk}.

There is one more anomaly that will play a major role in this thesis, the mixed gauge-gravitational anomaly \cite{Delbourgo:1972xb}.~\footnote{In $D=4k+2$ dimensions also purely gravitational anomalies can appear \cite{AlvarezGaume:1983ig}.} So far we have considered only spin one currents and have coupled them to gauge fields. Now we also want to introduce the energy-momentum tensor through its coupling to a fiducial background metric $g_{\mu\nu}$. Just as the gauge fields, the metric serves primarily as the source for insertions of the energy momentum tensor in correlation functions. Just as in the case of vector and axial currents, the mixed gauge-gravitational anomaly is the statement that it is impossible in the quantum theory to preserve at the same
time diffeomorphisms and chiral (or axial) transformations as symmetries. It is however possible to add Bardeen counterterms to shift the anomaly always in the sector of the spin one currents and preserve translational (or diffeomorphism) symmetry. If we have a set of left-handed and right-handed chiral fermions transforming under a Lie Group generated by $(T_a)_L$ and $(T_a)_R$ in the background of arbitrary gauge fields and metric, the anomaly is conveniently expressed through the non-conservation of the {\em covariants} current and energy momentum tensor as
\begin{eqnarray}
\label{eq:divTcov}
 D_\mu T^{\mu\nu} &=& F^\nu\,_\mu J^\mu + \frac{b_a}{384\pi^2}\epsilon^{\rho\sigma\alpha\beta} D_\mu\left[F_{\rho\sigma}R^{\nu\mu}\,_{\alpha\beta} \right] \,,\\
 \label{eq:gravanomaly}
 (D_\mu J^\mu)_a &=& \frac{d_{abc}}{32\pi^2} \epsilon^{\mu\nu\rho\lambda}
F^b_{\mu\nu} F^c_{\rho\lambda} + \frac{b_a}{768\pi^2}
\epsilon^{\mu\nu\rho\lambda} R^\alpha\,_{\beta \mu\nu}
R^\beta\,_{\alpha \rho\lambda}\,.
\end{eqnarray}
The purely group theoretic factors are
\begin{eqnarray}
 d_{abc} &=& \frac{ 1}{ 2} \mathrm{tr} ( T_a \{ T_b , T_c\} )_L - \frac{ 1}{ 2}
\mathrm{tr} ( T_a \{ T_b , T_c\} )_R \,, \label{eq:chiralcoeff} \\ 
 b_a &=& \mathrm{tr}(T_a)_L - \mathrm{tr}(T_a)_R\,. \label{eq:gravcoeff}
\end{eqnarray}
Chiral anomalies are completely absent if and only if $d_{abc}=0$ and $b_a$=0.

  % %%%%%%%%%%%%%%%%%%%%%%%%%%%%%%%%%%%%%%%%%%%%%%%%%%%%%%%%%%%%%%%%%%
% % %%%%%%%%%%%%%%%%%%%%%%%%%%%%%%%%%%%%%%%%%%%%%%%%%%%%%%%%%%%%%%%%%%
   \chapter{Chemical Potential For Anomalous Symmetries}
   \label{chemical}
  In statistical mechanics an equilibrium state is characterized by the grand canonical density operator $\hat\varrho$ which is constructed with the exponential of the conserved charges
\be
\label{density1}
\hat\varrho =\frac{1}{Z}e^{\beta_\mu \hat P^\mu + \sum_a\gamma_a \hat Q_a},
\ee
with $Z$ the partition function defined as $Z=\Tr e^{\beta_\mu \hat P^\mu + \sum_a\gamma_a \hat Q_a}$, $\hat P^\mu$ and $\hat Q_a$ the momentum  and charge operators.  The observables' expectation value are computed as $\langle O\rangle =\Tr [\hat\varrho O] $. The parameters $\beta_\mu$ and $\gamma_a$ are Lagrange multipliers playing the role of a generalized temperature and chemical potential, $\beta_\mu$ must be a time-like vector and can be redefined as $\beta_\mu =-\beta u^\mu$ with the normalization condition $u_\mu u^\mu=-1$ and $\gamma_a=-\beta\mu_a$. It is always possible to find a frame in which $u^\mu=(1,0,0,0)$ and the density operator recovers the usual form
\be
\label{density2}
\hat\varrho =\frac{1}{Z}e^{-\beta( \hat H + \sum_a\mu_a \hat Q_a)},
\ee
then we can interpret $u^\mu$ as a velocity and then associate to the equilibrium state a velocity field, the rest frame of the system is then the one in which the operator density looks like (\ref{density2}). 

In the context of quantum field theory is possible to find a path integral representation of the expectation value of any observable
\be
\langle \hat O\rangle = \int \mathcal{D}[\phi(x)] O[\phi(x)]e^{-S_E},
\ee
the integration must be done with the boundary conditions $ \phi(t-i\beta) =  \pm e^{ \mu \beta} \phi(t)$,\footnote{The plus sign is for bosons and minus for fermions.} where $\mu=\sum_a q_a\mu_a$, $q_a$ the charges associated to $\phi(x)$\footnote{$q^{ij}_A$ is an hermitian matrix in the space of the internal degrees of freedom labelled by the indices $i,j$.}
\be
\left[\hat Q_a,\hat \phi^i(x)\right] = -q_a^{ij}\phi^j(x)
\ee 
and $S_E$ the Euclidean action.  Often in the literature the expectation value is introduced with (anti-)periodic boundary condition, to do so we have to redefine
\be
\tilde{\phi}(x) = e^{it\mu}\phi(x),
\ee
 this redefinition has the implications that all the time derivative has to be shifted by $i\partial_0\to i\partial_0+\mu$ or equivalently the field theory Hamiltonian has to be deformed by
 \begin{equation}\label{eq:deformedH}
 H[\phi(x)] \rightarrow H[\phi(x)] - \mu Q[\phi(x)] \, .
\end{equation}

\begin{table}
\begin{center}
%
% Follow this input for your own table layout
%
\begin{tabular}{p{2cm}p{2.4cm}p{5cm}}
\hline\noalign{\smallskip}
Formalism  & Hamiltonian & Boundary condition \\
%\noalign{\smallskip}\svhline\noalign{\smallskip}
(A) &  $H-\mu Q$  & $\Psi(t_i-i\beta) = \pm  \Psi(t_i)$ \\
(B) & $H$ & $\Psi(t_i-i\beta) = \pm e^{ \beta \mu} \Psi(t_i)$\\
\noalign{\smallskip}\hline\noalign{\smallskip}
\end{tabular}
%$^a$ Table foot note (with superscript)
\caption{Two formalisms for the chemical potential}
\label{tab:formalisms}       % Give a unique label
\end{center}
\end{table}

These two formalism are completely equivalent because $\hat Q$ is the generator of a real symmetry and $[\hat H,\hat Q]=0$. In the case that concerns us $\hat Q$ represent an anomalous charge, so its commutator with the Hamiltonian is not zero\footnote{If gauge fields are present.} but a c-number $[\hat H,\hat Q]= c$. In this particular case defining the density operator as (\ref{density1}) does not make sense because $\hat Q$ is not an useful observable to label a physical state because $\dot Q \neq 0$, so in principle we do not have a first principle way to introduce the chemical potential. But still we are interested in generalize the grand canonical  partition function definition in the case of anomalous charges. To do so we will remark the physical properties of both formalism and the fact that each approach introduced before (see table: \ref{tab:formalisms}) differ by an axionic term consequence of
\be
\dot Q \propto A_1 \int d^3x\, F\wedge F + A_2 \int d^3x\, Tr[R\wedge R].
\ee 

\begin{figure}[t!]
\begin{center}
\scalebox{0.5}{\setlength{\unitlength}{4144sp}%
\begingroup\makeatletter\ifx\SetFigFont\undefined%
\gdef\SetFigFont#1#2#3#4#5{%
  \reset@font\fontsize{#1}{#2pt}%
  \fontfamily{#3}\fontseries{#4}\fontshape{#5}%
  \selectfont}%
\fi\endgroup%
\begin{picture}(8857,3265)(2416,-4607)
\thicklines
{\color[rgb]{0,0,0}\put(2701,-1861){\line( 1, 0){8550}}
\put(11251,-1861){\line( 0,-1){225}}
\put(11251,-2086){\line(-1, 0){8550}}
\put(2701,-2086){\line( 0,-1){2025}}
}%
\put(11026,-1726){\makebox(0,0)[lb]{\smash{{\SetFigFont{20}{24.0}{\familydefault}{\mddefault}{\updefault}{\color[rgb]{0,0,0}$t_f$}%
}}}}
\put(2476,-1681){\makebox(0,0)[lb]{\smash{{\SetFigFont{20}{24.0}{\familydefault}{\mddefault}{\updefault}{\color[rgb]{0,0,0}$t_i$}%
}}}}
\put(2431,-4471){\makebox(0,0)[lb]{\smash{{\SetFigFont{20}{24.0}{\familydefault}{\mddefault}{\updefault}{\color[rgb]{0,0,0}$t_i-i\beta$}%
}}}}
\end{picture}%
 }
\caption{At finite temperature field theories are defined on the
Keldysh-Schwinger contour in the complexified time plane. The initial state
at $t_i$ is specified through the boundary conditions on the fields. The
endpoint of the contour is at $t_i-i\beta$ where $\beta=1/T$.}
\label{fig:KeldyshSchwinger}       % Give a unique label
\end{center}
\end{figure}
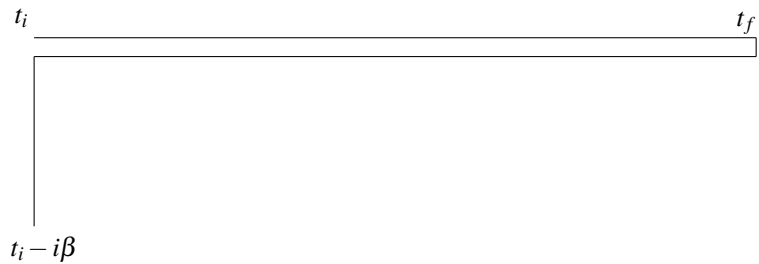

One convenient point of view on formalism (B) is the following. In a real time Keldysh-Schwinger setup we demand some initial conditions at initial (real) time $t=t_i$. These initial conditions are given by the boundary conditions in (B). From then on we do the (real) time development with the microscopic Hamiltonian $H$. In principle there
is no need for the Hamiltonian $H$ to preserve the symmetry present at times $t<t_i$. This seems an
especially suited approach to situations where the charge in question is not conserved by the real time dynamics. In the case of an anomalous symmetry we can start at $t=t_i$ with a state of certain charge. As long as we have only external gauge fields present the one-point function of the divergence of the current vanishes and the charge is conserved. This is not true on the full theory since even in vacuum the three-point correlators are sensitive to
the anomaly. For the formulation of hydrodynamics in external fields the condition that the one-point functions of the currents are conserved as long as there are no parallel electric and magnetic external fields (or a metric that has non-vanishing Pontryagin density) is sufficient.~\footnote{If dynamical gauge fields are present, such as the gluon fields in QCD even the one point function of the charge does decay over (real) time due to non-perturbative processes (instantons) or at finite temperature due to thermal sphaleron processes \cite{Moore:2010jd}. Even in this case in the limit of large number of colors these processes are suppressed and can e.g. not be seen in holographic models in the supergravity approximation.}

Let us assume now that $\hat Q$ is an anomalous charge, i.e. its 
associated current suffers from chiral anomalies.
We first consider formalism (B) and ask what happens if we do now the gauge
transformation that would bring us to formalism (A). Since the
symmetry is anomalous the action transforms as
\begin{equation}
S[A+\partial\chi] = S[A] + \int d^4x\, \chi
\epsilon^{\mu\nu\rho\lambda}\left(C_1 F_{\mu\nu}F_{\rho\lambda} + C_2
R^\alpha\,_{\beta\mu\nu} R^\beta\,_{\alpha\rho\lambda}\right) \,,
\end{equation}
with the anomaly coefficients $C_1$ and $C_2$ depending on the chiral
fermion content. It follows that formalisms (A) and (B) are physically
inequivalent now, because of the anomaly. However, we would like to
still come as close as possible to the formalism of (A) but in a form
that is physically equivalent to the formalism (B). To achieve this we
proceed by introducing a non-dynamical axion field $\Theta(x)$ and the
vertex
\begin{equation}\label{eq:Stheta}
S_\Theta[A,\Theta] = \int d^4x \, \Theta\, \epsilon^{\mu\nu\rho\lambda}\left(C_1
F_{\mu\nu}F_{\rho\lambda} + C_2 R^\alpha\,_{\beta\mu\nu}
R^\beta\,_{\alpha\rho\lambda}\right) \,.
\end{equation}
If we demand now that the ``axion'' transforms as $\Theta \rightarrow \Theta -
\chi$ under gauge transformations
we see that the action
\begin{equation}
S_{tot}[A,\Theta] = S[A] + S_\Theta[A,\Theta] 
\end{equation}
is gauge invariant. Note that this does not mean that the theory is
not anomalous now. We introduce it solely for the purpose to make
clear how the action has to be modified such that two field
configurations related by a gauge transformation are physically
equivalent. It is better to consider $\Theta$ as {\em coupling} and
not a field, i.e. we consider it a spurion field.  The gauge field
configuration that corresponds to formalism (B) is simply $A_0=0$. A
gauge transformation with $\chi=\mu t$ on the gauge invariant action
$S_{tot}$ makes clear that a physically equivalent theory is obtained
by chosing the field configuration $A_0=\mu$ and the time dependent
coupling $\Theta = -\mu t$. If we define the current through the
variation of the action with respect to the gauge field we get an
additional contribution from $S_\Theta$,
\begin{equation}
J_\Theta^\mu = 4 C_1 \epsilon^{\mu\nu\rho\lambda} \partial_\nu \Theta
F_{\rho\lambda}\,, 
\end{equation}
and evaluating this for $\Theta= -\mu t$ we get the spatial current
\begin{equation}\label{eq:Jtheta}
J_\Theta^m = 8 C_1 \mu B^m\,.
\end{equation}
We do not consider this to be the chiral magnetic effect! This is only the
contribution to the current that comes from the new coupling that we
are forced to introduce by going to formalism (A) from (B) in a
(gauge)-equivalent way. As we will see in the following
chapters the chiral magnetic and vortical effect are
on the contrary non-trivial results of dynamical one-loop calculations. 

What is the Hamiltonian now based on
the modified formalism (A)? We have to take of course the new coupling
generated by the non-zero $\Theta$. The Hamiltonian now is therefore
\begin{equation}\label{eq:Hmodified}
H - \mu \left( Q + 4 C_1 \int d^3x \epsilon^{0ijk} A_i \partial_j A_k\right)\,, 
\end{equation}
where for simplicity we have ignored the contributions from the metric terms.

For explicit computations from now on we will introduce the chemical potential through the formalism (B) by demanding twisted boundary conditions. It seems the most natural choice since the dynamics is described by the microscopic Hamiltonian $H$. The modified (A) based on the Hamiltonian (\ref{eq:Hmodified}) is however not without merits. Could be convenient in holography where it allows vanishing temporal gauge field on the black hole horizon and therefore a non-singular Euclidean black hole geometry.~\footnote{It is possible to define a generalized formalism   to make any choice for the gauge field $A_0=\nu$, so that one   recovers formalism (A) when $\nu=\mu$ and formalism (B) when $\nu=0$   as particular cases (see~\cite{Landsteiner:2012hc} for details).}

% % %%%%%%%%%%%%%%%%%%%%%%%%%%%%%%%%%%%%%%%%%%%%%%%%%%%%%%%%%%%%%%%%%%  
  \chapter{Relativistic Hydrodynamics with Anomalies}
   \label{rel_hydro}
   \subsection{Very brief introduction to thermodynamics}

The variation of the internal energy in a thermodynamical system is 
\begin{equation}
dU = -pdV +TdS +\mu_a dQ_a\, ,
\end{equation} 
where $p$, $V$, $S$, $T$, $\mu_a$ and $Q_a$ \footnote{ In this chapter we will label the number of conserved charges with the indices $a,b,\ldots$} are the pressure, volume, entropy, temperature, chemical potentials and conserve charges. The internal energy is an extensive function depending on the extensive variables $V$, $S$, $Q$, so
\begin{equation}
\label{intener} 
U(\lambda V,\lambda S, \lambda Q) = \lambda U(V,S,Q)
\end{equation}
now taking derivatives respect $\lambda$ and evaluating at $\lambda=1$ we can find that
\begin{equation}
U(V,S,Q) = -pV + TS + \mu_a Q_a,
\end{equation}
from which we can derive
\begin{equation}
Vdp = SdT + Q_a d\mu_a.
\end{equation}
But having in mind hydrodynamics is convenient to define the densities $s=S/V$, $\epsilon=U/V$ and $n_a=Q_a/V$ and rewrite the last expressions in term of the intensive variable
\begin{eqnarray}
\label{emasp}
\epsilon &=& -p + Ts + \mu_a n_a\\
dp &=& sdT + n_a d\mu_a\\
\label{diffE}
d\epsilon &=& Tds + \mu_a dn_a
\end{eqnarray}

\subsection{Relativistic hydrodynamics}
\label{subsec:hydro}

Standard thermodynamics assumes thermodynamical equilibrium, implying that the intensive parameters ($p,T,\mu$) are constant along the volume of the system, furthermore it is always possible to find a frame in which the total momentum of the system vanishes. In order to go to systems in a more interesting state, we will allow the thermodynamical parameters to vary in space and time taking our system out of equilibrium . However we will assume local thermodynamical equilibrium which means that the variables vary slowly between the points in the volume and in time, this approximation makes sense when the mean free path of the particles is much shorter than the characteristic size or length of the system $l_{mfp}\ll L$ \cite{Kovtun:2012rj}.

If we sit in a frame in which an element of fluid is at rest we will call this frame {\it fluid rest frame}. All the thermodynamical quantities defined in this frame are Lorentz invariant by construction, another important property is that local thermodyncamical equilibrium implies that in this frame in such element of fluid we will have isotropy.
The equations of motion are the (anomalous) conservation laws of the energy-momentum tensor and spin one
currents. These are supplemented by expressions for the energy-momentum tensor and the current which are organized in a derivative expansion, the so-called constitutive relations. Symmetries constrain the possible terms. Let us construct order by order in a derivative expansion all the possible terms contributing to the energy momentum tensor and current, starting with the case with no derivatives. The presence of the fluid velocity introduces a preferred direction in space-time, so we can decompose the objects in term of their longitudinal and transverse part. To do so we define the projector
\begin{equation}
\label{eq:projector}
P^\mu\,_\nu = \delta^\mu_\nu + u^\mu u_\nu\,
\end{equation} 
where $u^\mu$ is the fluid velocity satisfying the normalization condition $u^\mu u_\mu =-1$ and $P^\mu\,_\nu$ satisfy the properties $P^\mu\,_\nu u^\nu=0$, $P^\mu\,_\nu P^\nu\,_\rho=P^\mu\,_\rho$ and $P^\mu\,_\mu=3$, we also define the projector on transverse traceless tensors
\be
\Pi^{\mu\nu}\,_{\alpha\beta}= \frac{1}{2}\left(P^\mu_\alpha P^\nu_\beta + P^\nu_\alpha P^\mu_\beta -\frac{2}{3}P^{\mu\nu}P_{\alpha\beta}\right)
\ee
and the notation
\be
F^{\langle \mu \nu\rangle} = \Pi^{\mu\nu}\,_{\alpha\beta}F^{\alpha\beta}\,.
\ee
The most general decomposition of the constitutive relations is
\begin{eqnarray}
T^{\mu\nu}  &=& \mathcal{E}u^\mu u^\nu + \mathcal{P}P^{\mu\nu} + q^\alpha(P^\mu\,_\alpha  u^\nu + P^\nu\,_\alpha u^\mu) + \Pi^{\mu\nu}\,_{\alpha\beta}\tau^{\alpha\beta}  \\
J_a^\mu &=& \mathcal{N}_a u^\mu + P^\mu\,_\nu j^\nu_a .
\end{eqnarray}

At zero order there are no gauge nor diffeomorphism covariant objects we can construct, so the only contributions allowed by symmetries are
\begin{eqnarray}
T^{\mu\nu} &=& \mathcal{E}u^\mu u^\nu + \mathcal{P}P^{\mu\nu}\\
J_a^\mu &=& \mathcal{N}_a u^\mu.
\end{eqnarray}
Now we will sit in the fluid rest frame $u^\mu=(1,0,0,0)$ in order to identify $\mathcal{E}$, $\mathcal{P}$ and $\mathcal{N}_a$ with proper quantities characterizing the fluid. Then we get that the conserved quantities in matricial form look like
\begin{eqnarray}
T^\mu\,_\nu &=& \left(
\begin{array}{cccc}
-\mathcal{E} & 0 & 0 & 0\\
0					&\mathcal{P} & 0 & 0 \\
0 & 0 	& \mathcal{P} & 0 \\
0 & 0 & 0 & \mathcal{P}
\end{array}\right)\\
J_a^\mu &=& \left(\begin{array}{c}
\mathcal{N}_a\\
0\\
0\\
0
\end{array}\right)
\end{eqnarray}
now is easy to identify  the undetermined parameters, the component $T^{00}$ is the energy density, $T^{ii}$ the pressure and $J_a^0$ the charge density in the local rest frame, so we conclude that the constitutive relations at zero order in the derivative expansion are the one for an ideal fluid
\begin{eqnarray}
\label{zeroconsti1}
T_{(0)}^\mu\,_\nu &=& \epsilon u^\mu u_\nu + p P^\mu\,_\nu\\
\label{zeroconsti2}
J_{(0)a}^\mu &=& n_a u^\mu.
\end{eqnarray}

The knowledge of the constitutive relations\footnote{Including the so-called transport coefficients} and the (non) conservation laws is enough to describe the full dynamic of the fluid. In a system with diffeomorphism invariance in $3+1$ dimensions the most general conservation equations we can have are

\begin{eqnarray}
\label{eq:dT}
 D_\mu T^{\mu\nu} &=& F_a^{\mu}\,_\nu J_a^\nu +2\lambda_aD_\nu\left[\epsilon^{\rho\sigma\alpha\beta}F_{a\rho\sigma}R^{\mu\nu}\,_{\alpha\beta}\right]\,,\\
\label{eq:dJ}
 (D_\mu J^\mu)_a &=& \epsilon^{\mu\nu\rho\lambda}\left(3\kappa_{abc}F^b_{\mu\nu} F^c_{\rho\lambda} + \lambda_a R^\alpha\,_{\beta \mu\nu}R^\beta\,_{\alpha \rho\lambda}\right)\, ,
\end{eqnarray}
where we have redefined $\bar \kappa_{abc} =\frac{d_{abc}}{96\pi^2}$ and $\bar \lambda_a=\frac{b_a}{768\pi^2}$.

Now using these equation together with the zero order constitutive relations (\ref{zeroconsti1}), (\ref{zeroconsti2}) and (\ref{emasp}), (\ref{diffE}) it is straightforward to prove that for such a fluid there exists a conserved current which mimics how the local entropy varies along the fluid and that there is no entropy production per particle species
\begin{equation}
\label{zero_entro_cons}
D_\mu(su^\mu)=0, \qquad u^\mu D_\mu\left(s\,n_a^{-1}\right)=0 \, ,
\end{equation}
so we define the zero order entropy current as
\begin{equation}
S^\mu = s u^\mu.
\end{equation}
Notice that we have been able of construct the constitutive relations and the second law of thermodynamics\footnote{In this particular case the entropy current is conserved because we are working with the ideal constitutive relations in what follows when we consider dissipative process we will demand $D_\mu S^\mu \geq 0$} using symmetry arguments and the equation of motions.  The procedure at higher order in the derivatives will have the same spirit.

Before going to the higher order analysis let us remark that out of equilibrium the fluid velocity and the thermodynamical variables are not well defined quantities, the basic reason is the non existence of quantum operators to which we can associate their expectation value to such observables. So we can define many local temperatures $T(x)$ \footnote{The same happen for the rest of the variables} that differ from each other by gradients of hydro variables but coincide in their equilibrium value when the gradients vanish. This implies that the coefficients $\mathcal{E},\mathcal{N}$ and $\mathcal{P}$ have to be of the form
\begin{eqnarray}
\mathcal{E} &=& \epsilon(T,\mu) + f_\mathcal{E}(\partial T,\partial\mu,\partial u)\\
\mathcal{P} &=& p(T,\mu) + f_\mathcal{P}(\partial T,\partial\mu,\partial u)\\
\mathcal{N} &=& n(T,\mu) + f_\mathcal{N}(\partial T,\partial\mu,\partial u) \, ,
\end{eqnarray}
with $f_\mathcal{E}, f_\mathcal{P}$ and $f_\mathcal{N}$ determined by the particular definition of the fields $T(x),\mu(x)$ and $u^\mu(x)$. The choice of those fields is often called to select a {\it frame}. On the other hand the energy momentum tensor and charged current are physical quantities, so they cannot depend on the ambiguity of choosing a frame. Considering a redefinition of the type
\begin{eqnarray}
T(x) &\to & T(x) + \delta T(x)\\ 
\mu(x) &\to & \mu(x) + \delta \mu(x)\\
u^\mu(x) &\to & u^\mu(x) + \delta u^\mu(x)\, ,
\end{eqnarray}
 demanding invariance of $T^{\mu\nu}$ and $J^\mu$ under such transformations and the normalization condition it is possible to realize that for an arbitrary redefinition of hydro variables these relations have to be always satisfied
 \begin{eqnarray}
&& \delta\mathcal{E} = 0\,, \qquad \delta\mathcal{P} = 0 \, , \qquad \delta\mathcal{N} = 0 \, ,\\
 &&\delta q_\mu = -(\mathcal{E}+\mathcal{P})\delta u_\mu \,, \qquad \delta j_\mu = -\mathcal{N}\delta u_\mu\, , \\
 &&\delta \tau_{\mu\nu} = 0 \, .
\end{eqnarray}  

The tensor part of the system is frame independent consequence of these transformations but the vector and scalar parts are not. However these transformations allow us to define a frame independent vector and scalar
\begin{eqnarray}
l_a^\mu &\equiv& j_a^\mu - \frac{n_a}{\epsilon + p}q^\mu \\
f &\equiv & f_\mathcal{P} - \left(\frac{\partial p}{\partial\epsilon}\right)_n f_\mathcal{E} - \left(\frac{\partial p}{\partial n}\right)_\epsilon \cdot f_\mathcal{N}\, .
\end{eqnarray}
Without loss of generality it is always  possible to fix $f_\mathcal{E} = f_\mathcal{N} = 0$. This gauge only allows higher order corrections to the local pressure but maintain the functions $\mathcal{E}$ and $\mathcal{N}$ being the energy density and charge density respectively.

Now it is time to build up the first order constitutive relations, to do so it is useful to  decompose the derivatives of the velocity in term of transverse and longitudinal objects.
\begin{equation}
D^{\nu}u^\mu = -a^\mu u^\nu + \sigma^{\mu\nu} + \frac{1}{2}\omega^{\mu\nu} + \frac{1}{3}\theta P^{\mu\nu}\,
\end{equation}
where $a^\mu$ (acceleration), $\sigma^{\mu\nu}$ (shear tensor), $\omega^{\mu\nu}$ (vorticity tensor) and $\theta$ (expansion) are defined as
\begin{eqnarray}
\theta &=& D_\mu u^\mu = P^{\mu\nu}D_{\mu}u_{\nu} \\
a^\mu &=& u^\nu D_\nu u^\mu \equiv \mathcal{D}u^\mu\\
\sigma^{\mu\nu} &=& D^{(\mu}u^{\nu)} + u^{(\mu}a^{\nu)} - \frac{1}{3}\theta P^{\mu\nu} = D^{\langle\mu}u^{\nu\rangle}\\
\omega^{\mu\nu} &=& 2D^{[\mu}u^{\nu]} + 2u^{[\mu}a^{\nu]}  = 2P^{\mu\alpha}P^{\nu\beta}D_{[\alpha}u_{\beta]},
\end{eqnarray}
notice that it is also possible to define a vorticity pseudo vector
\begin{equation}
\omega^\mu = \frac{1}{2}\epsilon^{\mu\nu\rho\lambda}u_\nu \omega_{\rho\lambda}\, .
\end{equation}
It is straightforward to notice that the acceleration is transverse and the vorticity and shear tensors are transverse and traceless. Now using the backgroud fields $A^a_{\mu}$ and $g_{\mu\nu}$ the only objects we can construct with only one derivative are the electric and magnetic field\footnote{Notice that there are no diffeomorphism covariant quantities constructed with the metric and  containing only one derivative}
\begin{eqnarray}
F^a_{\mu\nu} &=&  \partial_{\mu} A^a_{\nu} - \partial_{\nu} A^a_{\mu} + g f^{abc}A^b_\mu A^c_\nu \\
E^a_\mu &=& F^a_{\mu\nu}u^\nu \\
B^{a\,\mu} &=& \frac{1}{2}\epsilon^{\mu\nu\rho\lambda}u_\nu F^a_{\rho\lambda}\, ,
\end{eqnarray}
finally we have to build up one derivative quantities with thermodynamical parameters, in particular we will chose the combinations $\bar{\mu}=\mu/T$ and $T$ as the independent variables. Besides we can observe that using the ideal energy and charge conservation we get that there is only one independent scalar and five independent vectors, so we choose the scalar $\theta$ and the independent vectors
\begin{equation}
P^{\mu\nu}D_\nu\bar{\mu} \quad , \quad  P^{\mu\nu}D_\nu T \quad , \quad \omega^\mu \quad , \quad E^a_\mu \quad \mathrm{and} \quad B^{a\, \mu} \, .
\end{equation}

Now we will study the ($C$, $P$) properties of each object \cite{Kharzeev:2011ds} in order to classify the transport coefficients  in terms of the anomalous and non anomalous one. Under $C$ and $P$ the metric tensor is even and the epsilon tensor even and odd respectively. The vectors $u^\mu$, $D_ \mu$, $J^\mu$, $S^\mu$ and $A^a_{\mu}$ under parity behave like vectors so they are odd, the gauge field and current are also odd under charge conjugation but the velocity, derivative and entropy current are even. From the constitutive relations $J^\mu_a =n_au^\mu+\ldots$, and $S^\mu =su^\mu+\ldots$ we conclude that $n_a$ has $(C,P) = (-,-)$ and $s$ $(+,+)$ in consequence $\mu_a$ and $T$ transform in the same way to its conjugated variables. Having this in mind we can do the following analysis, all the transport coefficients have to be a function $f(\bar{\mu},\kappa,\lambda)$\footnote{$f(\bar{\mu},\kappa,\lambda)$ is also a function of the temperature but because of its trivial behaviour under $C$ and $P$ it will be ignored in this analysis. We are also ignoring the index structure of the quantities}, each of this functions will have a definite $(C,P)$ depending on the combinations $(\bar{\mu},\kappa,\lambda,\kappa\bar{\mu},\lambda\bar{\mu})$, so the possible combinations are
\begin{eqnarray}
(+,+) \quad &:& \quad f(\bar{\mu},\kappa,\lambda) = g(\bar{\mu}^2,\kappa^2,\lambda^2,\kappa\lambda)\\
(-,+) \quad &:& \quad f(\bar{\mu},\kappa,\lambda) = \bar\mu g(\bar{\mu}^2,\kappa^2,\lambda^2,\kappa\lambda)\\
(-,-) \quad &:& \quad f(\bar{\mu},\kappa,\lambda) = \kappa g_1(\bar{\mu}^2,\kappa^2,\lambda^2,\kappa\lambda) + \lambda g_2(\bar{\mu}^2,\kappa^2,\lambda^2,\kappa\lambda)\\
(+,-) \quad &:& \quad f(\bar{\mu},\kappa,\lambda) = \bar\mu\kappa g_1(\bar{\mu}^2,\kappa^2,\lambda^2,\kappa\lambda) + \bar\mu\lambda g_2(\bar{\mu}^2,\kappa^2,\lambda^2,\kappa\lambda),
\end{eqnarray}
with this classification we can select systematically all the transport coefficients which are present if and only if the system presents anomalies. The part of the constitutive relations associated to the transport coefficients with $(C,P)=(\mp,-)$ will be split respect the ordinary part. Finally that we have classified all the contributions we can write the first order corrections to the constitutive relations as, 
\begin{eqnarray}
\label{gen_fp1}f_\mathcal{P} &=&- \zeta \theta \, ,\\
\label{gen_fp1_ano}\tilde f_\mathcal{P} &=& 0 \, ,\\
\label{gen_tau1}\tau^{\mu\nu} &=& -2\eta \sigma^{\mu\nu} \, ,\\
\label{gen_tau1_ano}\tilde \tau^{\mu\nu} &=& 0\, ,\\
\label{gen_q1} q^\alpha &=& \xi_a^1 E_a^\mu + \xi_a^2 P^{\alpha\nu}D_\nu\bar{\mu}_a + \xi^3 P^{\alpha\nu}D_\nu T  \, ,\\
\label{gen_q1_ano}\tilde q^\alpha &=& \sigma_a^{(\e,B)} B_a^\mu + \sigma^{(\e,V)} \omega^\mu \, ,\\
\label{gen_j1}j_a^\alpha &=& \Omega_{ab}^1 E_b^\mu + \Omega_{ab}^2 P^{\alpha\nu}D_\nu\bar{\mu}_b + \Omega_a^3 P^{\alpha\nu}D_\nu T  \, ,\\
\label{gen_j1_ano}\tilde j_a^\alpha &=& \sigma_{ab}^B B_b^\mu + \sigma_a^V \omega^\mu\, ,
\end{eqnarray}
we use tildes to distinguish the anomalous terms from the rest.

\begin{table}
\begin{center}
\begin{tabular}{|c|c|c|c|c|c|c|c|c|}
\hline
 & $\theta$ & $D_\mu\bar{\mu}$ & $D_\mu T$ & $a^\mu$ & $E_a^\mu$ & $B_a^\mu$ & $\omega^\mu$ & $\sigma^{\mu\nu}$ \\
 \hline
$C$ & $+$  & $ -$  & $+$ & $+$ &$ -$ & $-$ & $+$ & $+$ \\
$P$ &  $+$    & $-$ & $-$ & $-$ & $-$ & $+$ & $+$ & $+$   \\
\hline
\end{tabular}
\label{tab:first_cp}
\caption{$C$, $P$ properties of the first order scalars, vectors and tensors}
\end{center}
\end{table}

And in an frame invariant language
\begin{eqnarray}
l_a^\mu &=& -\Sigma_{ab} E_b^\mu -  \bar\Sigma_{ab} P^{\alpha\nu}D_\nu\bar{\mu}_b  + \xi_{ab}^B B_b^\mu + \xi_a^V \omega^\mu + \chi^T_a P^{\alpha\nu}D_\nu T \, ,\\
f &=& - \zeta \theta \, ,
\end{eqnarray}
with 
\begin{eqnarray}
 \Sigma_{ab}=\left(\Omega_{ab}^1 -\frac{n_a}{\e + p} \xi_b^1 \right) &,& \bar\Sigma_{ab}=\left(\Omega_{ab}^2 -\frac{n_a}{\e + p} \xi_b^2 \right) \, ,\\
 \xi_{ab}^B=\left(\sigma_{ab}^B -\frac{n_a}{\e + p} \sigma_b^{(\e,B)} \right) &,& \xi_{a}^V=\left(\sigma_{a}^V -\frac{n_a}{\e + p} \sigma^{(\e,V)} \right) \, ,\\
 \chi^T_a=\left(\Omega_a^3 -\frac{n_a}{\e + p} \xi^3 \right) \, .
\end{eqnarray}
It is possible to constrain a bit more this formulas using the second law of thermodynamics ($D_\mu S^\mu \geq 0$). Combining the (non) conservation equations up to second order corrections is possible to find a modification of the equation (\ref{zero_entro_cons})
\begin{eqnarray}\label{eq:divS}
D_\mu S^\mu &=& -\bar{\mu}_aD_\mu J^\mu_a  +\left(\frac{E^a_\mu}{T}-D_\mu\bar{\mu}_a\right)J_{(1)}^\mu\,_a - D_\beta\left(\frac{u_\alpha}{T}\right)T_{(1)}^{\alpha\beta} + D_\mu \tilde s_{(1)}^{\mu} \, ,
\end{eqnarray}
with $S^\mu$ defined as
\begin{eqnarray}
S^\mu &=& su^\mu - \bar{\mu}_a J_{(1)}^\mu\,_a - \frac{u_\alpha}{T}T_{(1)}^{\alpha\beta} + \tilde s_{(1)}^{\mu} \, ,\\
\tilde s_{(1)}^{\mu}  &=& \chi_a B_a^\mu + \tilde \chi \omega^\mu \, ,
\end{eqnarray}
This current satisfy the property $u_\mu S^\mu= - s$ which is the covariant expression to the statement that in the rest frame the zero component of $S^\mu$ is the entropy density $s$. Substituting in (\ref{eq:divS}) the equations (\ref{gen_tau1}) - (\ref{gen_j1_ano}) we can find a set of restrictions the transport coefficients must obey in order to always have a positive divergence of the entropy current. The most commons frames used are the so called {\it Landau frame} and {\it Eckart frame}. The Landau frame is defined requiring that in the rest frame of an element of fluid  the energy flux to vanish, this condition is realized in covariant form as $u_\mu T_{(n)}^{\mu\beta}=0$.\footnote{$n>0$} On the other hand the Eckart frame demand the presence of some conserved charge in the fluid and defines the velocity field as the velocity of those charges, so $ J_{(n)}^\mu=0$.

After imposing the positivity of $D_\mu S^\mu$ we get the most general form for the constitutive relations for the energy-momentum tensor and the currents in the
Landau frame are
\begin{eqnarray}\label{eq:Landauframe1}
 T^{\mu\nu} &=& \epsilon u^\mu u^\nu + ( p -\zeta\theta )P^{\mu\nu} -2\eta \sigma^{\mu \nu}  \,,\\
\label{eq:Landauframe2}
 J^\mu_a &=& n_a u^\mu + \Sigma_{ab} \left( E^\mu_b - T
P^{\mu\alpha}D_\alpha\bar\mu_b\right) +  \xi^B_{ab} B^\mu_b  + \xi^V_a
\omega^\mu \,,
\end{eqnarray}
with the dissipative transport coefficients satisfying the conditions $\eta\geq 0$, $\zeta\geq 0$,  $\Sigma_{ab}\geq 0$, $\bar\Sigma_{ab}=T\Sigma_{ab}$, $\chi^T_a =0$ and the anomalous one \cite{Son:2009tf,Neiman:2010zi}
\begin{eqnarray}
\xi^B_{ab} &=& 24\bar\kappa_{abc}\mu_c -\frac{n_a}{\epsilon+p}\left(12\bar\kappa_{abc}\mu_c\mu_d + \beta_b T^2\right)\\
\xi^V_a &=& 12\bar\kappa_{abc}\mu_b\mu_c  +\beta_a T^2 - \frac{n_a}{\epsilon+p}\left(8\bar\kappa_{bcd}\mu_b\mu_c\mu_d + 2\beta_b\mu_b T^2 + \gamma T^3\right)\\
\chi_a &=& \frac{12}{T} \bar\kappa_{abc}\mu_b\mu_c + \beta_a T\\
\tilde \chi &=& \frac{4}{T} \bar\kappa_{abc}\mu_a\mu_b\mu_c + \beta_a\mu_a T +\gamma T^2 \, ,
\end{eqnarray}
where $\beta_a$ and $\gamma$ are free numerical integration constants, notice that $\gamma \neq 0$ breaks $CPT$ \cite{Neiman:2010zi}, so a non vanishing value for that constant is allowed only for non preserving parity theories. A different story comes with the value of the constant $\beta_a$ which is completely unconstrained by this method\footnote{We will see below that $\beta_a$ is not arbitrary  but is completely fixed by the mixed gauge gravitational anomaly coefficient $\lambda_a$}. It is important to specify that these are the constitutive relations
for the covariant currents!.
 
\subsection{Linear response and Kubo formulae}
  
To compute the Kubo formulae for the anomalous transport coefficients it turns out that the Landau frame is not the most convenient one. It fixes the definition of the fluid velocity through energy transport. Transport phenomena related to the generation of an energy current are therefore not directly visible, rather they are absorbed in the definition of the fluid velocity. It is therefore more convenient to go to another frame in which we demand that the definition of the fluid velocity
is not influenced when switching on an external magnetic field or having  a vortex in the fluid. In such a frame the constitutive
relations for the system take the form
\begin{eqnarray}\label{eq:alternativeframe1}
&& T^{\mu\nu} = \epsilon u^\mu u^\nu + ( p - \zeta \theta) P^{\mu\nu} - \eta
\sigma^{\mu\nu}   + \tilde q^\mu u^\nu + \tilde q^\nu u^\mu \\
 \label{eq:alternativeframe2}
&& \tilde q^\mu = \sigma^{(\epsilon,B)} B^\mu + \sigma^{(\epsilon,V)}
\omega^\mu \\
\label{eq:alternativeframe3}
&&  J^\mu = n u^\mu + \Sigma \left( E^\mu - T
P^{\mu\alpha}D_\alpha\left(\frac{\mu}{T}\right)\right)  + \sigma^B B^\mu +
\sigma^V \omega^\mu \, .
\end{eqnarray}
In order to avoid unnecessary clutter in the equations we have specialized now to a single $U(1)$ charge. Notice that now there is a sort of ``heat'' current present in the constitutive relation for the energy-momentum tensor.

The derivation of Kubo formulae is better based on the usage of the consistent currents. Since the covariant and consistent currents are related by adding suitable Chern-Simons currents the constitutive relations for the consistent current receives additional contribution from the Chern-Simons current
\begin{equation}\label{eq:consistentconstitutive}
 \mathcal J^\mu =  J^\mu - \frac{1}{24\pi^2} K^\mu\,.
\end{equation}
If we were to introduce the chemical potential according to formalism $A$ (table: \ref{tab:formalisms}) via a deformation of the field theory Hamiltonian we would get an additional contribution to the consistent current from the Chern-Simons current. In this case it is better to go to the modified formalism $A'$ that also introduces a spurious axion field and
another contribution to the current $J_\Theta$ (\ref{eq:Jtheta}) has to be added
\begin{equation}\label{eq:thetaconstitutive}
 \mathcal J^\mu =  J^\mu - \frac{1}{24\pi^2} K^\mu + J^\mu_\Theta\,.
\end{equation}

For the derivation of the Kubo formulae it is therefore more convenient to work with formalism $(B)$ in which the chemical potential is introduced via the boundary condition shown in table: \ref{tab:formalisms}. Otherwise there arise additional contributions to the two point functions. We will briefly discuss them in the next chapter.

From the microscopic view the constitutive relations should be interpreted as the one-point functions of the operators $T^{\mu\nu}$ and $\mathcal J^\mu$ in a near equilibrium situation, i.e. gradients in the fluid velocity, the temperature or the chemical potentials are assumed to be small. From this point of view Kubo formulae can be derived. In the microscopic theory the one-point function of an operator near equilibrium is given by linear
response theory whose basic ingredient are the retarded two-point functions. If we consider a situation with the space-time slightly perturbed from Minkowski $g_{\mu\nu} = \eta_{\mu\nu} + h_{\mu\nu}$ in such a way that the only non vanishing deviation is $h_{xy}=h_{xy}(t)$ and all other sources switched off, i.e. the fluid being at rest $u^\mu=(1,0,0,0)$ and no gradients in temperature, chemical potentials or gauge fields the energy momentum tensor is simplified to 
\begin{equation}
T^{xy} \equiv \langle T^{xy} \rangle = -p h_{xy} - \eta\dot h_{xy} \,.
\end{equation}
Fourier transforming the equation $\dot h_{xy}=i\omega h_{xy}$ and
using linear response theory, the energy momentum tensor is given through the retarded two-point function by
\begin{equation}
T^{xy} = \left\langle T^{xy} T^{xy} \right\rangle h_{xy}\,.
\end{equation}
Equating the two expressions for the expectation value of the energy momentum tensor we find the Kubo formula for the shear viscosity
\begin{equation}\label{eq:shearvisc}
 \eta = \lim_{\omega\rightarrow 0} -\frac{1}{\omega} Im[\left\langle T^{xy} T^{xy} \right\rangle] \,.
\end{equation}
This has to be evaluated at zero momentum. The limit in the frequency follows because the constitutive relation are supposed to be valid only to lowest order in the derivative expansion, therefore one needs to isolate the first non-trivial term.

Now we want to find some simple special cases that allow the derivation of Kubo formulae for the anomalous conductivities. A very convenient choice is to go to the rest frame $u^\mu=(1,0,0,0)$, switch on a vector potential in the $y$-direction that depends only on the $z$ direction and at the same time a metric deformation $ds^2 =-dt^2 + h_{ty}dt dy  + d\vec x^2$ with $h_{ty}\equiv A^{g}_y(z)$. It is clear that such a gauge field introduces a background magnetic field pointing in the x direction $B_x= - \partial_z A_{y}$, analogously happens with the metric. In linearised gravity is well know that a perturbation like the introduced above behave as an abelian gauge field (see \cite{Mashhoon:2003ax}), the gravito-magnetic field will be $B^g_x = - \partial_z A^{g}_y$. To linear order in the background fields the non-vanishing components of the energy-momentum tensor and the current are
\begin{eqnarray}
 T^{0x} &=& \sigma^\epsilon_B B_x   + \sigma^\epsilon_V B^g_x\,,\\
 J^x &=&  \sigma_B B_x   + \sigma_V B^g_x\,.
\end{eqnarray}
 Notice that in the place of the vorticity field the gravito-magnetic field appear, that happens because in the rest frame the lower index velocity looks like $u_\mu=(-1,0,A^g_y,0)$, in consequence the vorticity coincide with the gravito-magnetic field, so at linear order the chiral vortical effect can be understood as a chiral gravito-magnetic effect\footnote{see~\cite{Landsteiner:2011tg}}. 

Now going to momentum space and differentiating with respect to the sources $A_y$ and $h_{ty}$ we find therefore the Kubo formulae
\cite{Fukushima:2008xe,Amado:2011zx}

\begin{equation}\label{Kformulae}
\boxed{ \begin{array}{lcl}
&&\\
  \sigma_B = \lim_{k_z\rightarrow 0}  \frac i k_z \langle J^x J^y \rangle ~~~~
&& ~~~~\sigma_V = \lim_{k_z\rightarrow 0}  \frac i k_z \langle J^x T^{ty}
\rangle \\ & \\
  \sigma^\epsilon_B = \lim_{k_z\rightarrow 0}  \frac i k_z \langle T^{tx} J^y
\rangle ~~~~
&& ~~~~\sigma^\epsilon_V = \lim_{k_z\rightarrow 0}  \frac i k_z \langle T^{tx}
T^{ty} \rangle \\
&&
 
\end{array}}
\end{equation}

All these correlators are to be taken at precisely zero frequency. As these formulas are based on linear response theory the correlators should be understood as retarded ones. They have to be evaluated however at zero frequency and therefore the order of the operators can be reversed. From this it follows that the chiral vortical conductivity coincides with the chiral magnetic conductivity for the energy flux $\sigma_V = \sigma^\epsilon_B$. These formulas are part of the key point of this thesis because we will use them to compute those transport coefficients in weakly and strongly coupled regimes.

We also want to discuss how these transport coefficients are related to the ones
in the more commonly used Landau frame.  
They are connected by a redefinition of the fluid velocity of the form 
\begin{equation}
 u^\mu \rightarrow u^\mu - \frac{1}{\epsilon+p} q^\mu \,,
\end{equation}
to go from (\ref{eq:alternativeframe1})-(\ref{eq:alternativeframe3}) to
(\ref{eq:Landauframe1})-(\ref{eq:Landauframe2}). We could also construct the frame invariant vector $l^\mu$ to directly identify the transport coefficients in such a frame.
\begin{eqnarray}
 \label{eq:cmc}\xi_{ B}  &=& \lim_{k_n\rightarrow 0} \frac{-i}{2 k_n} \sum_{k.l}
\epsilon_{nkl} \left( \left\langle \mathcal J^{k} \mathcal J^{l} \right\rangle - 
\frac{n}{\epsilon+p}\left\langle T^{0k} \mathcal J^{l} \right\rangle \right)\,,\\
\label{eq:cvc}\xi_{V} &=& \lim_{k_n\rightarrow 0} \frac{-i}{2 k_n} \sum_{k.l}
\epsilon_{nkl}\left( \left\langle \mathcal J^k T^{0l} \right\rangle - 
\frac{n}{\epsilon+p}\left\langle T^{0k} T^{0l} \right\rangle \right) \,,
\end{eqnarray}
where we have employed a slightly more covariant notation. The generalization to the non-abelian case is straightforward.

It is also worth to compare to the Kubo formulae for the dissipative transport coefficients (\ref{eq:shearvisc}). In the dissipative cases one first goes to zero momentum and then takes the zero
frequency limit. In the anomalous conductivities this is the other way around, one first goes to zero frequency and then takes the zero momentum limit. Another observation is that the dissipative transport coefficients sit in the anti-Hermitean part of the retarded correlators, i.e. the spectral function whereas the anomalous 
conductivities sit in the Hermitean part. The rate at which an external source $f_I$ does work on a system is given in terms of the spectral function of the operator $O^I$ coupling to that source as
\begin{equation}
 \frac{dW}{dt} = \frac 1 2 \omega f_I(-\omega) \rho^{IJ}(\omega) f_J(\omega) \,.
\end{equation}
The anomalous transport phenomena therefore do no work on the system, first they take place at zero frequency and second they are not contained in the spectral function $\rho = \frac{-i}{2} (G_{r}-G_r^\dagger)$.

 \subsection{Second Order expansion }
 \label{subsec:secord}

Now we want to go a step forward in the derivative expansion and to make our work easier and having applications to holography in mind, from now on we will consider the case of conformal fluids. This assumption introduces a big simplification in order to build up the second order constitutive relations because of the big symmetry restriction introduced with the conformal symmetry.
 
Some notion on conformal/Weyl covariant formalism is needed to construct the constitutive relations up to second order, for a detailed explanation see \cite{Loganayagam:2008is}.
A conformal fluid has to be invariant under the change
\begin{equation}
g_{\mu\nu}\to e^{-2\phi(x)}g_{\mu\nu},
\end{equation}
where $\phi(x)$ is an arbitrary function. We will say that a tensor is Weyl convariant with weight $w$ if transform as \begin{equation}
T^{\alpha\beta\ldots}_{\mu\nu\ldots}\to e^{w\phi(x)}T^{\alpha\beta\ldots}_{\mu\nu\ldots}.
\end{equation}

The consequences of conformal symmetry on hydrodynamics are, that the energy momentum tensor and (non)-conserved currents have to be covariant under Weyl transformations and the energy momentum has to be traceless modulo contributions from Weyl anomaly. To construct Weyl covariant quantities is necessary to introduce the Weyl connection
\begin{equation}
\mathcal{A}_\mu=u^\nu D_\mu u_\nu -\frac{1}{3}D_\nu u^\nu
\end{equation}
and a Weyl covariant derivative
\begin{eqnarray}
\nonumber\mathcal{D}_\lambda Q^{\mu\ldots}_{\nu\ldots} &=& D_\lambda Q^{\mu\ldots}_{\nu\ldots} - w \mathcal{A}_\lambda Q^{\mu\ldots}_{\nu\ldots} +\\
\nonumber &&+\left[g_{\lambda\alpha}\mathcal{A}^\mu - \delta^\mu_\lambda\mathcal{A}_\alpha - \delta^\mu_\alpha\mathcal{A}_\lambda\right]Q^{\alpha\ldots}_{\nu\ldots}+\ldots \\
&&-  \left[g_{\lambda\nu}\mathcal{A}^\alpha - \delta^\alpha_\lambda\mathcal{A}_\nu - \delta^\alpha_\nu\mathcal{A}_\lambda\right]Q^{\mu\ldots}_{\alpha\ldots}+\ldots
\end{eqnarray}

It is possible to reduce in a systematic way the number of independent sources contributing to the constitutive relations impossing Weyl covariance and the hydrodynamical equation of motions (Ward idetities). In the Refs. \cite{Kharzeev:2011ds,Erdmenger:2008rm,Banerjee:2008th} a classification in the so called Landau frame  of all the possible terms that can appear in the energy momentum tensor and U(1) current has been done up to second order. The Ward identities  in four dimensions in presence of quantum anomalies  are shown in (\ref{eq:dJ}) and (\ref{eq:dT}), the curvature part was always neglected because is fourth order in derivative and the expansion was done up to second order. But in \cite{Landsteiner:2011cp,Landsteiner:2011iq} was shown that the gravitational anomaly indeed fixed part of the transport coefficient at first order, actually in \cite{Jensen:2012kj} was understood why the derivative expansion breaks down in presence of the gravitational anomaly.

Before start writing the constitutive relations it is useful to study the Weyl weight of the hydro variables (see  table: \ref{tab:pesos}). 
\begin{table}[t]
\begin{center}
\begin{tabular}{|c|c|}
\hline
Field & weight \\
\hline
$\mu$, $T$, $u^\mu$ & 1 \\
$g_{\mu\nu}$ & -2 \\
$p$ & 4 \\
$n$, $E^\mu$, $B^\mu$  & 3 \\
\hline
\end{tabular}
\caption{Weyl weights for the chemical potential, temperature, fluid velocity, metric, pressure, charge density, electric field and magnetic field}
\label{tab:pesos}
\end{center}
\end{table}

Now with these ingredients we can write down the constitutive relations split in the equilibrium, first and second order part in the Landau Frame
\begin{eqnarray}
\label{eq:fullconstiT}T^{\mu\nu} &=& p(4u^\mu u^\nu +\eta^{\mu\nu}) + \tau_{(1)}^{\mu\nu}  + \tau_{(2)}^{\mu\nu} + \tilde\tau_{(2)}^{\mu\nu}\\
\label{eq:fullconstiJ}J^\mu &=& n u^\mu + j_{(1)}^{\mu}  + \tilde j_{(1)}^{\mu} + j_{(2)}^{\mu} + \tilde j_{(2)}^{\mu},
\end{eqnarray}
the subindex in parenthesis means the order in derivative. Weyl invariance implies the equation of state $\e = 3p$ and the vanisihing of the bulk viscosity $\zeta=0$. Remembering the contributions at first order in a Weyl covariant language
\begin{eqnarray}
\tau_{(1)}^{\mu\nu}  &=& -2\eta\sigma^{\mu\nu}\\
\tilde j_{(1)}^{\mu} &=&  \xi_V\omega^\mu + \xi_B B^\mu  \quad , 	\quad j_{(1)}^{\mu} =  -\Sigma\left(T P^{\mu\nu}\mathcal{D}_\nu\bar{\mu}-E^\mu\right),
\end{eqnarray}
with the shear and vorticity tensors rewritten as
\begin{eqnarray}
\sigma_{\mu\nu} &=& \frac{1}{2}\left( \mathcal{D}_\mu u_\nu + \mathcal{D}_\nu u_\mu \right) \\
\omega_{\mu\nu} &=&  \mathcal{D}_\mu u_\nu - \mathcal{D}_\nu u_\mu  \, .
\end{eqnarray}
 
The second order sources can be constructed using the same method as in the first order with the extra consideration of being Weyl convariant see \cite{Kharzeev:2011ds}, beside the derivative of the first order objects we have a new covariant object which is the conformal Weyl tensor of the background metric $C_{\mu\nu\rho\lambda}$. So, the full set of second order corrections are

\begin{eqnarray}
\label{eq:2consti} \tau_{(2)}^{\mu\nu}  &=& \sum_{a=1}^{a=15} \lambda_a \mathcal{T}^{(a)\mu\nu} \quad , 	\quad \tilde\tau_{(2)}^{\mu\nu}  = \sum_{a=1}^{a=8} \tilde\lambda_a  \mathcal{\tilde T}^{(a)\mu\nu},\\
 \nonumber j_{(2)}^{\mu} &=&  \sum_{a=1}^{a=10}\xi_a\mathcal{J}^{(a)\mu}\quad , 	\quad  \tilde j_{(2)}^{\mu} = \sum_{a=1}^{a=5}\tilde\xi_a\mathcal{\tilde J}^{(a)\mu},
 \end{eqnarray}
with the second order vector and tensors defined as
\be
\begin{array}{ccc}
 \mathcal{T}^{(1)\mu\nu} = u^\alpha\mathcal{D}_\alpha\sigma^{\mu\nu},  & 
  \mathcal{T}^{(2)\mu\nu} = \sigma^{\langle\mu}\,_\gamma \sigma^{\nu\rangle\gamma}, &
   \mathcal{T}^{(3)} = \sigma^{\langle\mu}\,_\gamma \omega^{\mu\rangle\gamma}, \\
  \mathcal{T}^{(4)\mu\nu} = \omega^{\langle\mu}\,_\gamma \omega^{\nu\rangle\gamma},  &  \mathcal{T}^{(5)\mu\nu} = \mathcal{D}^{\langle\mu}\mathcal{D}^{\nu\rangle} \bar{\mu}, & \mathcal{T}^{(6)\mu\nu} = \mathcal{D}^{\langle\mu}\bar{\mu}\mathcal{D}^{\nu\rangle} \bar{\mu}, \\
 \mathcal{T}^{(7)\mu\nu} = \mathcal{D}^{\langle\mu} E^{\nu\rangle},  &
 \mathcal{T}^{(8)\mu\nu} = E^{\langle\mu} \mathcal{D}^{\mu\rangle}\bar{\nu}, &
 \mathcal{T}^{(9)\mu\nu} = E^{\langle\mu} E^{\nu\rangle},  \\
 \mathcal{T}^{(10)\mu\nu} = B^{\langle\mu} B^{\nu\rangle} , &
  \mathcal{T}^{(11)\mu\nu} = \epsilon^{\gamma\delta\eta\langle\mu}u_\gamma B_\delta\sigma^{\nu\rangle}\,_\eta , & 
  \mathcal{T}^{(12)\mu\nu}=\omega^{\langle\mu} B^{\nu\rangle} ,\\
  \mathcal{T}^{(13)\mu\nu} = C^{\mu\alpha\nu\beta} P_{\alpha\beta} , &
  \mathcal{T}^{(14)\mu\nu} = \epsilon^{\mu\alpha\beta\gamma}\epsilon^{\nu\delta\eta\lambda}C_{\alpha\beta\delta\eta} u_\gamma u_\lambda , & 
  \mathcal{T}^{(15)\mu\nu}=\epsilon^{\langle\mu\gamma\delta\eta}C_{\gamma\delta}\,^{\nu\rangle\lambda}u_\eta u_\lambda,
\end{array}
\ee

\be\begin{array}{ccc}
\mathcal{\tilde T}^{(1)\mu\nu} = \mathcal{D}^{\langle\mu}\omega^{\nu\rangle} , &
\mathcal{\tilde T}^{(2)\mu\nu} = \omega^{\langle\mu} \mathcal{D}^{\nu\rangle}\bar{\mu}  ,&
\mathcal{\tilde T}^{(3)} = \epsilon^{\gamma\delta\eta\langle\mu}\sigma^{\nu\rangle}\,_\eta u_\gamma \mathcal{D}_\delta \bar{\mu},  \\ 
\mathcal{\tilde T}^{(4)\mu\nu} = \mathcal{D}^{\langle\mu} B^{\nu\rangle} , &
\mathcal{\tilde T}^{(5)\mu\nu} = B^{\langle\mu} \mathcal{D}^{\nu\rangle}\bar{\mu}  , &
\mathcal{\tilde T}^{(6)\mu\nu} = E^{\langle\mu} B^{\nu\rangle} ,\\
\mathcal{\tilde T}^{(7)\mu\nu} = \epsilon^{\gamma\delta\eta\langle\mu}\sigma^{\nu\rangle}\,_\eta u_\gamma E_\delta ,&
\mathcal{\tilde T}^{(8)\mu\nu}= \omega^{\langle\mu} E^{\nu\rangle} ,& 
\end{array}
\ee

\be
\begin{array}{ccc}
 \mathcal{J}^{(1)\mu}=\sigma^{\mu\nu}\mathcal{D}_\nu\bar{\mu} , & \mathcal{J}^{(2)\mu}=\omega^{\mu\nu}\mathcal{D}_\nu\bar{\mu} , & \mathcal{J}^{(3)\mu}=P^{\mu\nu}\mathcal{D}^\alpha\sigma_{\nu\alpha} ,\\ \mathcal{J}^{(4)\mu}=P^{\mu\nu}\mathcal{D}^\alpha\omega_{\nu\alpha} , &
\mathcal{J}^{(5)\mu}=\sigma^{\mu\nu}E_\nu, &
\mathcal{J}^{(6)\mu}=\omega^{\mu\nu}E_\nu ,\\
 \mathcal{J}^{(7)\mu}=u^\nu\mathcal{D}_\nu E^\mu ,&
 \mathcal{J}^{(8)\mu}=\epsilon^{\mu\nu\alpha\beta}u_\nu B_\alpha\mathcal{D}_\beta \bar{\mu}  , &
 \mathcal{J}^{(9)\mu}=\epsilon^{\mu\nu\alpha\beta}u_\nu E_\alpha B_\beta  ,\\
  \mathcal{J}^{(10)\mu}= \epsilon^{\mu\nu\alpha\beta}u_\nu \mathcal D_\alpha B_\beta ,& &
\end{array}
\ee
\be
\begin{array}{ccc}
\mathcal{\tilde J}^{(1)\mu} = \sigma^{\mu\nu}\omega_\nu, &
 \mathcal{\tilde J}^{(2)\mu}=\sigma^{\mu\nu}B_\nu , &
  \mathcal{\tilde J}^{(3)\mu}=\omega^{\mu\nu}B_\nu  ,\\
 \mathcal{\tilde J}^{(4)\mu} =\epsilon^{\mu\nu\alpha\beta}u_\nu E_\alpha\mathcal{D}_\beta \bar{\mu} ,& \mathcal{\tilde J}^{(5)\mu}=\epsilon^{\mu\nu\alpha\beta}u_\nu \mathcal{D}_\alpha E_\beta ,& 
\end{array}
\ee

In chapter: \ref{fluid_grav} we will compute all those second order transport coefficients using a holographic model which realize both gravitational and gauge anomalies, but we shall assume the fluid living in a flat space time, so the contribution to constitutive relations coming from the curvature tensor will be ignored.
  % % %%%%%%%%%%%%%%%%%%%%%%%%%%%%%%%%%%%%%%%%%%%%%%%%%%%%%%%%%%%%%%%%%%  
  \chapter{A Holograpic model for the chiral separation and chiral magnetic effect}
   \label{cmecse}
   
The anomalous conductivities (\ref{J5}) and (\ref{CME}) have  been studied in holographic models of QCD by introducing chemical potentials for left and right chiral quarks as boundary values for corresponding bulk gauge fields \cite{Lifschytz:2009si,Yee:2009vw}. However, it was pointed out by Ref.~\cite{Rebhan:2009vc} that in these calculations the axial anomaly was not realized in consistent form and that the corresponding electromagnetic current was not strictly conserved. Correcting the situation by means of Bardeen's counterterm \cite{Bardeen:1969md,Hill:2006ei} instead led to a vanishing result for the electromagnetic current in the holographic QCD model due to Sakai and Sugimoto \cite{Sakai:2004cn,Sakai:2005yt}\footnote{In Ref.~\cite{Gorsky:2010xu} a finite result was obtained in a bottom-up model that is nonzero only due to extra scalar fields.}, while recovering the result (\ref{J5}) for the anomalous axial conductivity. Indeed, the two anomalous conductivities (\ref{J5}) and (\ref{CME}) differ in that in the former case there is no difficulty with introducing a chemical potential for quark number, while a chemical potential for chirality refers to a chiral current that is anomalous. In \cite{Gynther:2010ed} we solved the problem of introducing chemical potential for  anomalous charges, realizing (holographically) the differences between the formalism resumed in table \ref{tab:formalisms} in the case of anomalous charges.

\subsection{Comparing formalisms and Kubo formulae computation}
 \label{subsec:contribs}
 
\begin{figure}
 \begin{center}
% %\sidecaption
% % Use the relevant command for your figure-insertion program
% % to insert the figure file.
% % For example, with the graphicx style use
 \includegraphics[scale=.45]{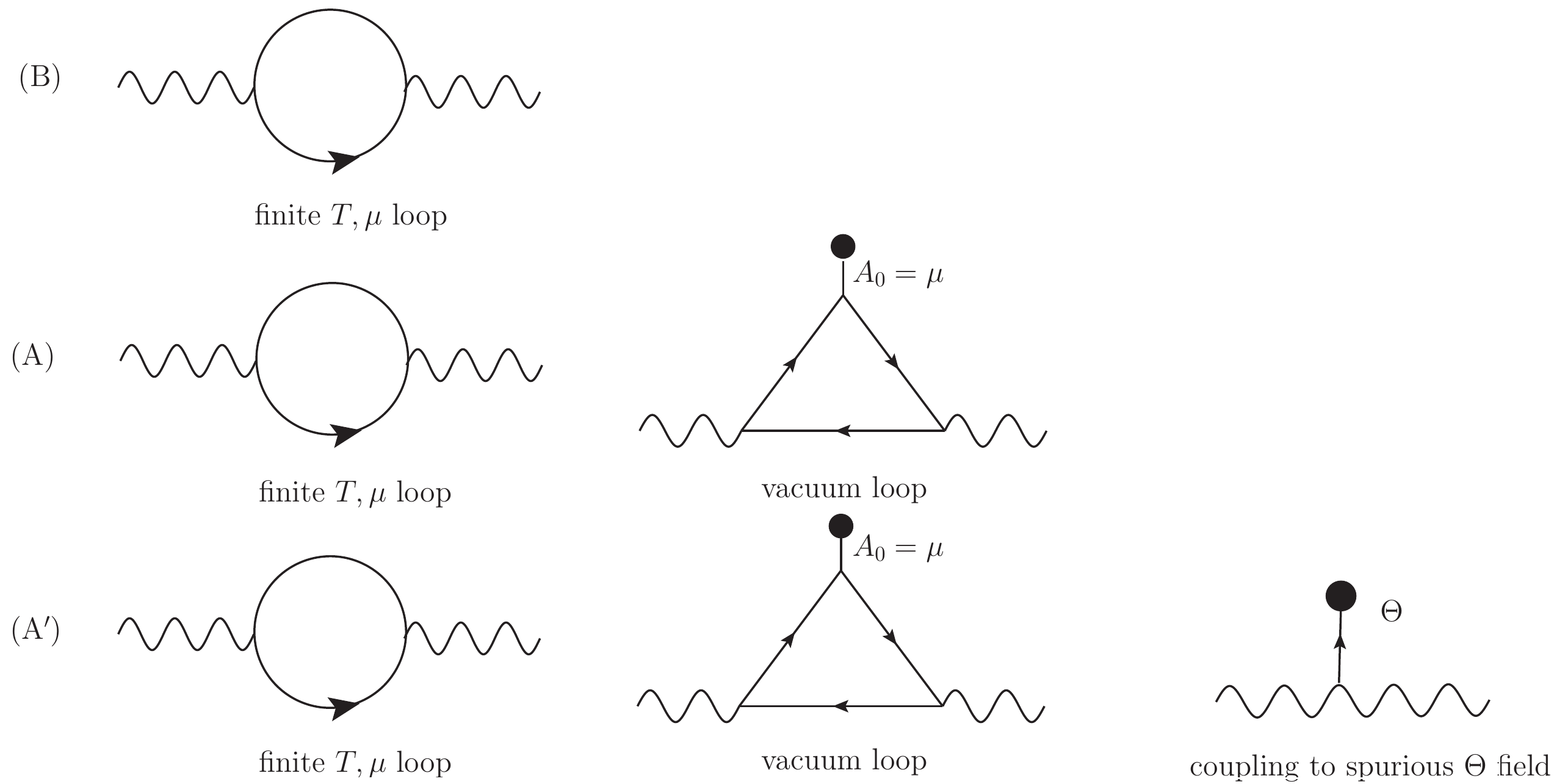}
% %
% % If no graphics program available, insert a blank space i.e. use
% %\picplace{5cm}{2cm} % Give the correct figure height and width in cm
% %
 \caption{Contributions to the Kubo formula for the chiral magnetic conductivity  in the different formalisms for the chemical potential.}
 \label{fig:contribs}       % Give a unique label
\end{center} 
 \end{figure}

 Now we want to give a detailed analysis of the different Feynman  graphs that contribute to the Kubo formulae in the different  formalisms for the chemical potentials. The simplest and most economic  formalism is certainly the one labeled (B) in which we introduce the  chemical potentials via twisted boundary conditions. The Hamiltonian
 is simply the microscopic Hamiltonian $H$. Relevant contributions  arise only at first order in the momentum and at zero frequency and in  this kinematic limit only the Kubo formulae for the chiral magnetic  conductivity is affected. In the figure (\ref{fig:contribs}) we  summarize the different contributions to the Kubo formulae in the
 three ways to introduce the chemical potential\footnote{This Feynman diagram analysis only makes sense in a weak coupled theory, but anyway gives us an useful understanding of the physics}.
 
 The first of the Feynman graphs is the same in all formalisms. It is the genuine  finite temperature and finite density one-loop contribution. This graph is  finite because the Fermi-Dirac distributions cutoff the UV momentum modes in the  loop. In the formalism $(A)$ we need to take into account that there is  also a contribution from the triangle graph with the fermions going around the  loop in vacuum.  For a non-anomalous symmetry this graph vanishes simply because on the upper  vertex of the triangle sits a field configuration that is a pure gauge. If the  symmetry
 under consideration is however anomalous the triangle diagram picks up just the  anomaly. Even pure gauge field configurations become physically distinct from  the  vacuum and therefore this diagram gives a non-trivial contribution. On the level  of the constitutive relations this contribution corresponds to the Chern-Simons
 current in (\ref{eq:consistentconstitutive}).  We consider this contribution to be unwanted. After all the anomaly would make  even a constant  value of the temporal gauge field $A_0$ observable in vacuum. An example is
 provided for a putative axial gauge field $A^5_\mu$. If present the absolute  value of  its temporal component would be observable through the axial anomaly. We can be  sure that in nature no such background field is present.   The third line $(A')$ introduces also the spurious axion field $\Theta$ the only  purpose of this field is to cancel the contribution from the triangle graph. This cancellation takes place by construction since $(A')$ is gauge equivalent  to $(B)$ in which only the first genuine finite $T,\mu$ part  contributes. It corresponds to the contribution of the current $J^\mu_\Theta$ in (\ref{eq:thetaconstitutive}).We further emphasize that these considerations are based on the usage of the consistent currents. 

In the interplay between axial and vector  currents additional contributions arise from the Bardeen
 counterterm. It turns out that the triangle or Chern-Simons current  contribution to the consistent vector current in the formalism $(A)$  cancels precisely the first one  \cite{Rebhan:2009vc,Gynther:2010ed} as we shall see below. Our take on this is that a  constant temporal component of the axial gauge field $A^5_0=\mu_5$ would  be observable in nature and can therefore be assumed to be absent. The  correct way of evaluating the Kubo formulae for the chiral magnetic
 effect is therefore the formalism $(B)$ or the gauge equivalent one  $(A')$.
 
 At this point the reader might wonder why we introduced yet another formalism $(A')$  which achieves appearently nothing but being equivalent to formalism $(B)$. At least  from the perspective of holography there is a good reason for doing so.   In holography the strong coupling duals of gauge theories  at finite temperature in the plasma phase are represented by five dimensional  asymptotically Anti- de Sitter black holes. Finite charge density translates to  charged black holes. These black holes have some non-trivial gauge flux along  the holographic direction represented by a temporal gauge field configuration  of the form $A_0(r)$ where $r$ is the fifth, holographic dimension. It is often  claimed that for consistency reasons the gauge field has to vanish on the 
 horizon of the black hole and that its value on the boundary can be identified  with the chemical potential
 \begin{equation}\label{eq:horizonzero}
  A_0(r_H) =0   ~~~~\mathrm{and}~~~ A_0(r\rightarrow \infty) = \mu\, ,
 \end{equation}
 is important to remark that the l.h.s. of (\ref{eq:horizonzero}) is just a gauge fixation and the boundary condition is the r.h.s.
  
 According to the usual holographic dictionary the gauge field values on the  boundary correspond to the sources for currents. A non-vanishing  value of the temporal component of the gauge field at the boundary is therefore
 dual to a coupling that modifies the Hamiltonian of the theory just as in  (\ref{eq:deformedH}). Thus with the boundary conditions (\ref{eq:horizonzero})  we have the holographic dual of the formalism $(A)$.  If anomalies are present they are represented in the holographic dual by  five-dimensional Chern-Simons terms of the form  $A\wedge F\wedge F$. The two  point correlator  of the (consistent) currents receives now contributions from the Chern-Simons  term that is precisely of the form of the second graph in $(A)$ in figure  \ref{fig:contribs}.  As we have argued this is an a priory unwanted contribution. We can however cure  that by introducing an additional term in the action of the form  (\ref{eq:Stheta})  living only on the boundary of the holographic space-time. In this way we can  implement the formalism $(A')$, cancel the unwanted triangle contribution  with the third graph in $(A')$ in figure \ref{fig:contribs} and maintain $A_0(r_H)=0$!
 
 The claim that the temporal component of the gauge field has to vanish at the  horizon is of course not unsubstantiated. The reasoning goes as follows. The  Euclidean section  of the black-hole space time has the topology of a disc in the $r,\tau$  directions, where $\tau$ is the Euclidean time. This is a periodic variable
with  period $\beta=1/T$  where $T$ is the (Hawking) temperature of the black hole and at the same time
the  temperature in the dual field theory. Using Stoke's law we have
 \begin{equation}
  \int_{\partial D} A_0 \,d\tau = \int_D F_{r0}\,dr\, d\tau \,,
 \end{equation}
 where $F_{r0}$ is the electric field strength in the holographic  direction and $D$ is a Disc with origin at $r=r_H$ reaching out to  some finite value of $r_f$.  If we shrink this disc to zero size,  i.e. let $r_f\rightarrow r_H$ the r.h.s. of the last equation vanishes  and so must the l.h.s. which approaches the value $ \beta A_0(r_H)$.  This implies that $A_0(r_H)=0$.  If on the other hand we assume that  $A_0(r_H)\neq 0$ then the field strength must have a delta type  singularity there in order to satisfy Stokes theorem.  Strictly  speaking the topology of the Euclidean section of the black hole is  not anymore that of a disc since now there is a puncture at the  horizon. It is therefore more appropriate to think of this as having  the topology of a cylinder.  Now if we want to implement the formalism  $(B)$ in holography we would find the boundary conditions \begin{equation}\label{eq:boundaryzero}
  A_0(r_H) = \mu 
 \end{equation}
and the gauge fixation $A_0(r\rightarrow \infty) = 0$ and precisely such a singularity at the horizon would arise. In addition we would need to impose twisted boundary conditions around the Euclidean time  $\tau$ for the fields  just as in (table: \ref{tab:formalisms}). Now the presence of the singularity seems to be a  good thing: if the space time would still be smooth at the horizon it would be
 impossible to demand these twisted boundary conditions since the circle in $\tau$ shrinks  to zero size there. If this is however a singular point of the geometry we can  not  really shrink the circle to zero size. The topology being rather a cylinder than  a disc allows now for the presence of the twisted boundary conditions. 
 
 It is also important to note that in all formalisms the potential difference  between the boundary and the horizon is given by $\mu$. This has a very nice  intuitive  interpretation. If we bring a unit test charge from the boundary to the horizon  we need the energy $\Delta E = \mu$. In the dual field theory this is just the  energy cost  of adding one unit of charge to the thermalized system and coincides with the  elementary definition of the chemical potential. 
  \begin{figure}
\begin{center}
 \scalebox{0.4}{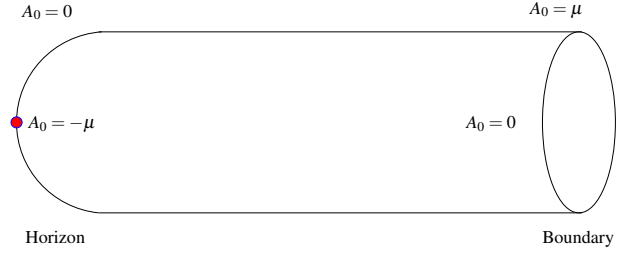 }
 \caption{A sketch of the Euclidean black hole topology. A singularity at the  horizon arises if we do not choose the temporal component of the gauge field to  vanish there. On the other hand allowing the singularity to be present changes the topology to  the one of a cylinder and this in turn allows twisted boundary conditions.}
 \label{fig:EuclideanBH}      %  Give a unique label
\end{center}
 \end{figure}
  In this chapter we will consider the ``boundary condition''
 \begin{equation}\label{eq:boundarymixed}
  A_0(r_H) -  A_0(r\rightarrow \infty) = \mu
 \end{equation}
with non gauge fixation for that component of the gauge field in order to illustrate the above discussion.

It is important to distinguish between thermodynamic state variables such as chemical potentials
and background gauge fields (as also pointed out by  Ref.\ \cite{Rubakov:2010qi}).  Recall that the holographic dictionary instructs us to construct a functional of boundary fields and that n-point functions are obtained by functional differentiation with respect to the boundary fields. For a gauge field the expansion close to the boundary takes the form
$$
A_\mu(x,r) = A_\mu^{(0)}(x) + \frac{ A^{(2)}(x)}{r^2} + \dots\,.
$$
 The leading term in this expansion is the source for the current $J^\mu$. The subleading term is often identified with the one-point function of the current. This is, however, not true in general. As has been pointed out in Ref.\ \cite{Rebhan:2009vc}, in the presence of a bulk Chern-Simons term, the current receives also contributions from the Chern-Simons term and $A_\mu^{(2)}(x)$ can, in general, not be identified with the {\it vev}  of the current. On the other hand a constant value of $A_0^{(0)}$ is often identified with a chemical  potential. This is, however, slightly misleading since the holographic realization of the chemical potential is given by the potential difference between the boundary and the horizon and only in a gauge in which the $A_0$ vanishes at the horizon such an identification can be made. Even in this case we have to keep in mind that the boundary value of the gauge field is the source of the current whereas the potential difference between horizon and boundary is the chemical potential.  

\section{The (holographic) Model}

We will consider the simplest possible holographic model for one quark flavor in a chirally restored deconfined phase.\footnote{The even simpler model considered in Ref.~\cite{Rubakov:2010qi} is instead closer to a single quark flavor in a chirally broken phase where right and left chiralities are living on the two boundaries of a single brane.}  It consists of taking two gauge fields corresponding to the two chiralities for each quark flavor in a five dimensional AdS black hole background. 

The action is given by two Maxwell actions for left and right gauge fields plus separate Chern Simons terms corresponding to separate  anomalies for left and right chiral quarks. The Chern-Simons terms are however not unique but can be modified by adding total derivatives. A total derivative which enforces invariance under vector gauge transformations $\delta V_M = \partial_M \lambda_{(V)}$ corresponds to the so-called Bardeen counterterm \cite{Bardeen:1969md,Hill:2006ei},  leading to the action
{\small \begin{equation}
\label{eq:action}
S = \int d^5x\sqrt{-g} \left( -\frac{1}{4 g_V^2} F^{(V)}_{MN} F_{(V)}^{MN}-\frac {1}{4 g_A^2} F^{(A)}_{MN} F_{(A)}^{MN}  +\frac{\kappa}{2}
\epsilon^{MNPQR} A_M ( F^{(A)}_{NP} F^{(A)}_{QR} + 3 F^{(V)}_{NP} F^{(V)}_{QR} ) \right)
\end{equation}}
Since the Chern-Simons term depends  explicitly on the gauge potential $A_M$ the action is gauge invariant under $\delta A_M = \partial_M \lambda_{(A)}$ only up to a boundary term. This non-invariance is the holographic implementation of the axial $U(1)$ anomaly, when identifying the gauge fields as holographic sources for the currents of global $U(1)$ symmetries in the dual field theory.  A rigorous string-theoretical realization of such a setup is provided for example by the Sakai-Sugimoto model \cite{Sakai:2004cn,Sakai:2005yt}. As usually done in the latter, we neglect the backreaction of the bulk gauge fields on the black hole geometry.

In order to compute the field equations and the boundary action, from which we shall obtain the two- and three-point functions of various currents, we expand around fixed background gauge fields and to second order in fluctuations. The gauge fields are written as
\begin{eqnarray}
A_M = A^{(0)}_M + a_M\,,\qquad V_M = V^{(0)}_M + v_M
\end{eqnarray}
where the $A^{(0)}_M$ and $V^{(0)}_M$ are the background fields and the lower case letters are the fluctuations.

After a little algebra we find to first order in the fluctuations
\begin{eqnarray}\label{eq:firstorderaction}
\delta S^{(1)}_{\mathrm{bulk}+\partial} &=&  \int_{\mathrm{bulk}} dr d^4x \sqrt{-g} \left\lbrace  a_M\left[ \frac{1}{g_A^2}\nabla_N \cF_{(A)}^{NM} + \frac{3\kappa}{2} \epsilon^{MNPQR}(  \cF_{NP}^{(A)} 
\cF_{QR}^{(A)} + \cF_{NP}^{(V)} \cF_{NP}^{(V)})\right] +\right. \nonumber \\
&&   v_M\left.\left[ \frac{1}{g_V^2}\nabla_N \cF_{(V)}^{NM} + 3\kappa \epsilon^{MNPQR}(\cF_{NP}^{(A)} 
\cF_{QR}^{(V)})\right] \right\rbrace+\\
&&  \int_{\partial} d^4x \left[a_\mu \left( \frac{1}{g_A^2}\sqrt{-g}\cF_{(A)}^{\mu r} +2 \kappa  \epsilon^{\mu\nu\rho\lambda} A_\nu \cF^{(A)}_{\rho\lambda}\right)
 \right.\nonumber \\
&&\left. + v_\mu\left( \frac{1}{g_V^2}\sqrt{-g}\cF_{(V)}^{\mu r} + 6 \kappa \epsilon^{\mu\nu\rho\lambda} A_\nu \cF^{(V)}_{\rho\lambda}\right)\right]\nonumber
\, ,
\end{eqnarray}
where calligraphic strength tensors refer to the background ones. From the bulk term we get the equations of motion and from the boundary terms we can read the expressions for the non renormalized consistent currents,
\begin{eqnarray}\label{eq:currents}
\mathcal J^\mu &=&  \left[\frac{1}{g_V^2}\sqrt{-g}\cF_{(V)}^{\mu r} + 6 \kappa \epsilon^{\mu\nu\rho\lambda} A_\nu \cF^{(V)}_{\rho\lambda}\right]_\partial \,,\\
\mathcal J_5^\mu &=&  \left[\frac{1}{g_A^2}\sqrt{-g}\cF_{(A)}^{\mu r} +2 \kappa  \epsilon^{\mu\nu\rho\lambda} A_\nu \cF^{(A)}_{\rho\lambda}\right]_\partial \,.
\end{eqnarray}
On-shell {\it they} obey
\begin{eqnarray}\label{eq:anomalies}
\partial_\mu\mathcal  J^\mu &=& 0\,,\nonumber\\
\partial_\mu \mathcal J_5^\mu &=& - \frac{\kappa}2 \epsilon^{\mu\nu\rho\lambda} \left(3 F_{\mu\nu}^{(V)} F_{\rho\lambda}^{(V)} + F_{\mu\nu}^{(A)} F_{\rho\lambda}^{(A)}\right)\,.
\end{eqnarray}
As expected, the vector like current is exactly conserved. Comparing with the standard result from the one loop triangle calculation we find $\kappa = -\frac{N_c}{24 \pi^2}$ for a dual 
strongly coupled $SU(N_c)$ gauge theory for a  massless Dirac fermion in the fundamental representation. 

We emphasize that only by demanding an exact conservation law for the vector current we can consistently couple it to an (external) electromagnetic field. This leaves no ambiguity in the definitions of the above currents as the
ones obtained by varying the action with respect to the gauge fields and which obey (\ref{eq:anomalies}). In particular, we have to keep the contributions from the Chern-Simons terms in the action, which are occasionally ignored in holographic calculations.

The second order term in the expansion of the action is
\begin{eqnarray}
S^{(2)}_{\mathrm{bulk}+\partial} &=& \int_{\mathrm{bulk}} \left\lbrace  a_M\left[ \frac{1}{2g_A^2} \nabla_N f_{(A)}^{NM} + \frac{3\kappa }{2} \epsilon^{MNPQR}(\cF_{NP}^{(A)} f_{QR}^{(A)} +  f_{NP}^{(V)} \cF_{QR}^{(V)} )\right] +\right. \nonumber \\
&&   v_M\left.\left[ \frac{1}{2g_V^2} \nabla_N f_{(V)}^{NM} + \frac{3\kappa}{2} \epsilon^{MNPQR}(f_{NP}^{(A)} \cF_{QR}^{(V)} + f_{NP}^{(V)} \cF_{QR}^{(A)})\right] \right\rbrace+\\
&& \int_{\partial}  \left[\frac{\sqrt{-g}}{2}(\frac{1}{g_A^2}a_\mu f_{(A)}^{\mu r} + \frac{1}{g_V^2}v_\mu f_{(V)}^{\mu r}) +  \kappa \epsilon^{\mu\nu\rho\lambda} 
( \cA_\nu a_\mu f^{(A)}_{\rho\lambda} + 3 v_\mu \cA_\nu f^{(V)}_{\rho\lambda} + 3 v_\mu a_\nu \cF^{(V)}_{\rho\lambda} ) \right]\,,\nonumber
 \end{eqnarray}
where $f_{MN}$ is the field strength of the fluctuations. Again the action is already in the form of bulk equations of motion plus boundary term.

As gravitational background we take the planar AdS Schwarzschild metric 
\begin{equation}
ds^2 = -f(r) dt^2 + \frac{dr^2}{f(r)} + \frac{r^2}{L^2} (dx^2+dy^2+dz^2)\,.
\end{equation}
with $f= \frac{r^2}{L^2}-\frac{r_+^4}{r^2}$. The temperature is given in terms of the horizon by $r_+ = L^2\pi T$. We rescale the $r$ coordinate such that the horizon lies at $r=1$ and we also will set the AdS scale $L=1$. Furthermore we also rescale time and space coordinates accordingly. To recover the physical values of frequency and momentum we thus have to do replace 
$(\omega, k) \rightarrow (\omega/(\pi T) , k/(\pi T) )$.

The background gauge fields are
\begin{eqnarray}\label{eq:gaugebg}
A^{(0)}_0(r) &=& \Phi(r) = \alpha-\frac{\beta}{r^2}\,,\\
V^{(0)}_0(r) &=& \Psi(r) = \nu-\frac{\gamma}{r^2}\,.
\end{eqnarray}
As we said before we will introduce the chemical potential as the difference of energy in the system with a unit of charge at the boundary and a unit of charge at the horizon. the integration constants $\beta$ and $\gamma$ are thus fixed to 
\begin{eqnarray}\label{eq:chempot}
\beta &=&  \mu_5\\
\gamma &=&  \mu
\end{eqnarray}
where $\mu$ is the chemical potential of the vector symmetry and $\mu_5$ the chemical potential of the axial $U(1)$. 
The constants $\alpha$ and $\nu$ we take to be arbitrary and we will eventually consider them as sources for insertions of the operators $J^0$ and $J_5^0$ at zero momentum. Due to our choice of coordinates the physical value of the chemical potentials is recovered by $\mu \rightarrow \pi T \mu$.

We can now compute the charges present in the system from the zero components of the currents   (\ref{eq:currents}) 
\begin{eqnarray}
J^0 &=& \frac{2\gamma}{g_V^2} \\
J^0_5 &=& \frac{2\beta}{g_A^2}
\end{eqnarray}
It is important to realize that without a Chern-Simons term the action for a gauge field in the bulk depends only on the field strengths and is therefore independent of constant boundary values of the gauge field. The action does, of course, depend on the physically measurable difference of the potential between the horizon and the boundary.  For our particular model, the choice of  the Chern-Simons term results, however, also in an explicit dependence on the integration constant $\alpha$. It is crucial to keep in mind that $\alpha$ is a priori unrelated to the chiral chemical potential but plays the role of the source for the operator $J_5^0$ at zero momentum.

For the fluctuations we choose the gauge $a_r=0$. We take the fluctuations to be of plane wave form with frequency $\omega$ and momentum $k$ in $x$-direction. The relevant polarizations are then the $y$- and $z$-components, i.e. the transverse gauge field fluctuations. The equations of motion are
\begin{eqnarray}
v_a'' + (\frac{f'}{f}+\frac{1}{r}) v_a' +\frac{(\omega^2r^2-fk^2)}{f^2r^2} v_a + \frac{12i \kappa g_V^2 k}{f r} \epsilon_{ab} ( \Phi' v_b + \Psi' a_b) &=& 0\,, \\
a_a'' + (\frac{f'}{f}+\frac{1}{r}) a_a' +\frac{(\omega^2r^2-fk^2)}{f^2r^2} a_a + \frac{12i \kappa g_A^2 k}{f r} \epsilon_{ab} ( \Phi' a_b + \Psi' v_b) &=& 0\,. \label{eq:probe_axial}
\end{eqnarray}
Prime denotes differentiation with respect to the radial coordinate $r$. The two-dimensional epsilon symbol is $\epsilon_{yz} =1$.

There is also a longitudinal sector of gauge field equations. They receive no contribution from the Chern-Simons term and so are uninteresting for our purposes.

The boundary action in Fourier space in the relevant transversal sector is
\begin{eqnarray}
S^{(2)} = \int_{\partial} dk \left[ -\frac{r f}{2}  (\frac{1}{g_A^2}a_{-k}^b (a_k^b)' + \frac{1}{g_V^2}v_{-k}^b (v_k^b)')- 2 i k  \kappa\epsilon_{bc}\alpha \left( a^b_{-k} a^c_{k}  + 3 v^b_{-k} v^c_{k}\right)\right]
\end{eqnarray}
As anticipated, the second order boundary action depends on the boundary value of the axial gauge field but not on the boundary value of the vector gauge field.

From this we can compute the holographic Green function. The way to do this is to compute four linearly independent solutions that fulfill infalling boundary conditions on the horizon \cite{Son:2002sd,Herzog:2002pc}. At the AdS boundary we require that the first solution asymptotes to the vector $(v_y, v_z, a_y, a_z) = 
(1,0,0,0)$, the second solution to the vector $(0,1,0,0)$ and so on. We can therefore build up a matrix of solution $F_k\,^I\,_J(r)$ where each column corresponds to one of these solutions \cite{Kaminski:2009dh}. Given a set of boundary fields $a_i^{(0)}(k)$, $v_i^{(0)}(k)$ which we collectively arrange in the vector $\varphi^{I,(0)}(k)$, the bulk solution corresponding to these boundary fields is 
\begin{equation}
\label{eq:f}
\varphi^I(k,r) = F_k^I\,_J \varphi^{J,(0)}(k)
\end{equation}
$F$ is the (matrix valued) bulk-to-boundary propagator for the system of coupled differential equations. 

The holographic Green function is then given by
\begin{eqnarray}
\label{eq:GR}
G_{IJ} =  - 2\lim_{r\rightarrow \infty} (\bbA_{IL} (F_k\,^L\,_J)' + \bbB_{IJ} )
\end{eqnarray}
The matrices $\bbA$ and $\bbB$ can be read off from the boundary action as
\begin{equation}
\bbA = -\frac 1 2 r f  \left( 
\begin{array}{cc}
\frac{1}{g_V^2} &  0\\ 
0& \frac{1}{g_A^2}\\ 
\end{array}  
\right)~~~,~~~~~\bbB = -2 i\kappa k \alpha  \left( 
\begin{array}{cc}
3 \epsilon_{ij} &  0\\ 
0&  \epsilon_{ij}\\ 
\end{array}  
\right)
\end{equation}
(notice that $F$ becomes the unit matrix at the boundary).

We are interested here only in the zero frequency limit and to first order in an expansion in the momentum $k$.\footnote{In this approximation the on shell action does not need to be renormalized} In this limit the differential equations can be solved explicitly. To this order the matrix bulk-to-boundary propagator is
\begin{equation}
F = \left(
\begin{array}{cccc}
1 &  -g_A^2\mu_5 g(r) & 0 &  -g_V^2\mu g(r) \\
 g_A^2\mu_5 g(r) & 1&   g_V^2\mu g(r)&0 \\
0 & -g_V^2\mu g(r) & 1 &  -g_A^2\mu_5 g(r) \\
 g_V^2\mu g(r) & 0 &  g_A^2\mu_5 g(r)&1 \\
\end{array}
\right)
\end{equation}
where $g(r) =  6 i k \kappa \log(1+1/r^2) $.
We find then the holographic current two-point functions in presence of the background boundary gauge fields $A_0=\nu$ and $A^5_0=\alpha$
\begin{eqnarray}\label{eq:correlators}
\langle \mathcal J^a\mathcal J^b \rangle &=& -12 i \kappa k (\mu_5 - \alpha) \epsilon_{ab}\\
\langle\mathcal J_5^a\mathcal J^b \rangle &=& -12 i \kappa k \mu \epsilon_{ab}\\
\langle\mathcal J_5^a\mathcal J^b_5 \rangle &=& -4 i \kappa k (3\mu_5 - \alpha) \epsilon_{ab}
\end{eqnarray}
Although $\mu,\mu_5$ and the boundary gauge field value $\alpha$ enter in very similar ways in this result, we need to remember their completely different physical meaning. The chemical potentials $\mu$ and $\mu_5$ are gauge invariant physical state variables whereas $\alpha$ is the source for insertions of $J_5^0(0)$. Had we chosen the ``gauge'' $\alpha=\mu_5$ we would have concluded (erroneously) that the two-point correlator of electric currents vanishes. We see now that with $\mu_5$ introduced separately from $\alpha$ that this not so.
We simply have obtained expressions for the correlators in the physical state described by $\mu$ and $\mu_5$ in the external background fields $\alpha$ and $\nu$. Due to the gauge invariance of the action under vector gauge transformations the constant mode of the source $\nu$ does not appear. The physical difference between the chemical potentials and the gauge field values is clear now. The susceptibilities of the two-point functions obtained by differentiating with respect to the chemical potentials are different from the three-point functions obtained by differentiating with respect to the gauge field values. Finally, we remark that the temperature dependence drops out due to the opposite scaling of $k$ and $\mu$, $\mu_5$.

To compute the anomalous conductivities we therefore have to evaluate the two-point function  for vanishing background fields $\nu=\alpha=0$. We obtain, in complete agreement with the well-known weak coupling results,
\begin{eqnarray}\label{eq:conductivities}
\mathcal J^i=e^2 \sigma_{\rm CME} B^i,\qquad
\sigma_{\rm CME} &=& \lim_{k\to0}\frac{i\epsilon_{ab}}{2k}\langle\mathcal J^a\mathcal J^b \rangle|_{\nu=\alpha=0}
=\frac{N_c %N_f
}{2\pi^2} \mu_5, \\
\mathcal J^i_5=e \sigma_{\mathrm{axial}} B^i,\qquad\;
\sigma_{\mathrm{CSE}} &=& \lim_{k\to0}\frac{i\epsilon_{ab}}{2k}\langle\mathcal J_5^a\mathcal J^b \rangle|_{\nu=\alpha=0}
=\frac{N_c %N_f
}{2\pi^2} \mu,\\ \label{eq:conductivities3}
\mathcal J^i_5=\sigma_{55} B_5^i,\qquad\quad
\sigma_{55} &=& \lim_{k\to0}\frac{i\epsilon_{ab}}{2k}\langle\mathcal J_5^a\mathcal J_5^b \rangle|_{\nu=\alpha=0}
=\frac{N_c %N_f
}{2\pi^2} \mu_5.
\end{eqnarray}
We are tempted to call all $\sigma$'s conductivities. This is, however, a slight misuse of language in the case of $\sigma_{55}$. Formally $\sigma_{55}$ measures the response due to the presence of an axial magnetic field $\vec{B}_5 = \nabla \times \vec{A_5}$. Since such fields do not exist in nature, we cannot measure $\sigma_{55}$ in the same way as $\sigma_{CME}$ and  $\sigma_{CSE}$.

Since the two-point functions (\ref{eq:correlators}) still depend on the external source $\alpha$ we can also obtain the three
point functions in a particular kinematic regime. Differentiating with respect to $\alpha$ (and $\nu$) we find the three
point functions

\begin{eqnarray}\label{eq:threepts1}
\langle\mathcal J^a(k)\mathcal J^b(-k)\mathcal J^0(0)\rangle &=& 0, \\ \label{eq:threepts2}
\langle\mathcal J_5^a(k)\mathcal J^b(-k)\mathcal J^0(0)\rangle &=& 0, \\ \label{eq:threepts3}
\langle\mathcal J_5^a(k)\mathcal J_5^b(-k)\mathcal J^0(0)\rangle &=& 0, \\ \label{eq:threepts4}
\langle\mathcal J^a(k)\mathcal J^b(-k)\mathcal J^0_5(0)\rangle &=& -i k \frac{N_c}{2\pi^2} \epsilon_{ab}, \label{eq:jjj5}\\ \label{eq:threepts5} 
\langle\mathcal J_5^a(k)\mathcal J^b(-k)\mathcal J_5^0(0)\rangle &=& 0, \\ \label{eq:threepts6}
\langle\mathcal J_5^a(k)\mathcal J_5^b(-k)\mathcal J_5^0(0)\rangle &=& -i k \frac 1 3  \frac{N_c }{2\pi^2} \epsilon_{ab}.
\end{eqnarray}
Note the independence on chemical potentials and temperature. Therefore, these expression hold also in
vacuum.

Although the anomaly is conventionally expressed through the divergence of the axial current we can also see it from these three-point functions containing the axial current at zero momentum. The zero component of the current at zero momentum is nothing but the total charge $Q$ and $Q_5$. Since all currents are neutral they should commute with the charges and if we are not in a situation of spontaneous symmetry breaking the vacuum should be annihilated by the charge. From this it follows that insertions of $Q$ into correlation functions of currents should annihilate them. And this is indeed what an insertion of the electric charge $Q$ does.  Insertion of $Q_5$ however does result in a non trivial three-point correlator and therefore expresses the non-conservation of axial charge! 

Equations (\ref{eq:threepts4}) and (\ref{eq:threepts6}) show the sensitivity of the theory to a constant temporal component of the axial gauge field even at zero temperature and chemical potentials. If the axial $U(1)$  symmetry was exactly conserved, such a constant field value would be a gauge degree of freedom and the theory would be insensitive to it. Since this symmetry is, however, anomalous, it couples to currents through these threepoint functions. The correlators (\ref{eq:threepts4}) and (\ref{eq:threepts6}) can therefore be understood as expressing the anomaly in the axial $U(1)$ symmetry.

In the next chapter we will check these results in vacuum at weak coupling by calculating the triangle diagram in the relevant kinematic regimes, this consistency check will come as a confirmation that our intuition comparing formalism (table \ref{tab:formalisms}) is the right one and that we have to compute expectation values at finite temperature with anomalous charges either in formalism (B) or formalism (A').

%  % %%%%%%%%%%%%%%%%%%%%%%%%%%%%%%%%%%%%%%%%%%%%%%%%%%%%%%%%%%%%%%%%%%
  \chapter{Gas of Free Fermions}
   \label{freefermions}
e   An important property of the two- and three-point functions we just calculated is that they are independent of temperature. The three-point functions are furthermore independent of the chemical potentials. Therefore, the results for the three-point function should coincide with correlation functions in vacuum. So in this chapter we will start computing the three point functions (\ref{eq:threepts1})-(\ref{eq:threepts6}) in vacuum and then we will move to a finite temperature and chemical potential situation to compute two point functions and use the Kubo formulae (\ref{Kformulae}) to extract the anomalous transport coefficients.

\section{Three point functions at weak coupling}

 At weak coupling all the three-point functions can be obtained from a single 1-loop Feynman integral. We only need to evaluate the diagram with two vector currents and one axial current. The diagram with three vector currents vanishes identically  (due to C-parity) and the one with three axial currents can be reduced to the one with only one axial current by anticommuting $\gamma_5$ matrices (when a regularization is applied that permits this). Similarly, it can be seen that the diagram with two axial vector currents can be reduced to the one with none, which vanishes.

When computing the three-point function, it is crucial to check the resulting anomalies. Gauge invariant regulators, like dimensional regularization, should yield the correct anomaly, such that the vector currents are identically conserved. On the other hand, for example cutoff regularization breaks gauge invariance and further finite renormalizations may be needed in order to restore gauge invariance. In the following, we apply both dimensional and cutoff regularizations to compute the three-point function and show that they give consistent results with each other and with Eqs.~(\ref{eq:threepts1})-(\ref{eq:threepts6}).
 
\begin{figure}
\begin{center}
\includegraphics[scale=1]{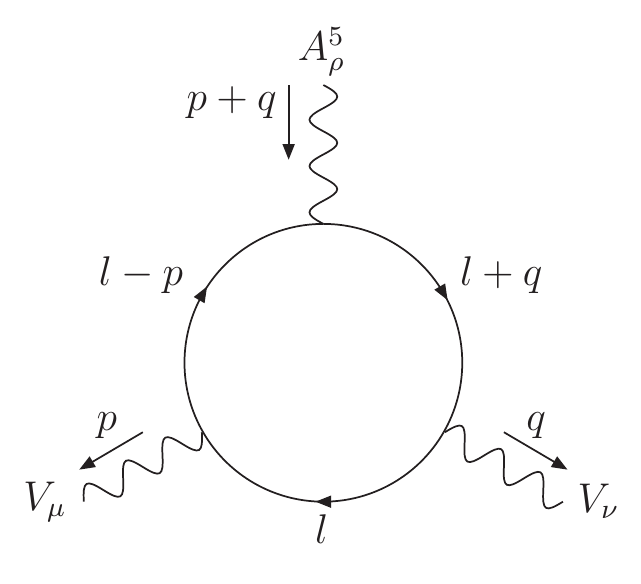}
\caption{The triangle diagram.}
\label{fig:triangle}
\end{center}
\end{figure}

\subsection{Triangle diagram with one axial current}
 
The triangle diagram, shown in Fig.~\ref{fig:triangle}, with one axial current and two vector currents is given by
\begin{eqnarray}\label{eq:triangle}
\Gamma^{\mu\nu\rho}(p,q) &=& (-1) (i e)^2 (i g) (i)^3 \int \frac{d^d l}{(2\pi)^d} \mathrm{tr}\left( 
\gamma_5 
 \frac{\slashed{l}-\slashed{p}}{(l-p)^2}
\gamma^\mu
\frac{ \slashed{l}}{l^2} \gamma^\nu \frac{\slashed{l}+\slashed{q}}{(l+q)^2} \gamma^\rho  \right) \nonumber\\
 & &+(\mu \leftrightarrow \nu , p \leftrightarrow q).
\end{eqnarray}
The factors are a $(-1)$ from the fermion loop, the couplings to vector and axial gauge fields and $i$ for
each fermion propagator. We will simply set the electric and axial couplings $e$ and $g$ to one. Evaluation of the integral with dimensional and cutoff regularizations is presented in some detail in Appendix \ref{app:triangle}. 

The anomalies of the various currents coupled to the triangle diagram are obtained by contracting the three-point function above by the corresponding momenta. Applying dimensional regularization, we get immediately
\begin{eqnarray}
 p_\mu \Gamma^{\mu\nu\rho}_\mathrm{DR}(p,q) &=&0,\\
 q_\nu \Gamma^{\mu\nu\rho}_\mathrm{DR}(p,q) &=&0,\\
 (p+q)_\rho \Gamma^{\mu\nu\rho}_\mathrm{DR}(p,q)&=&  \frac{i}{2\pi^2} p_\alpha q_\beta \epsilon^{\alpha\beta\mu\nu},
 \end{eqnarray}
 yielding the correct Adler-Bell-Jackiw anomaly. In terms of cutoff regularization, we however find
 \begin{eqnarray}
 p_\mu \Gamma^{\mu\nu\rho}_\mathrm{CO}(p,q) &=&-\frac{i}{6\pi^2}p_\alpha q_\beta\epsilon^{\alpha\beta\nu\rho},\\
 q_\nu \Gamma^{\mu\nu\rho}_\mathrm{CO}(p,q) &=&\frac{i}{6\pi^2}p_\alpha q_\beta\epsilon^{\alpha\beta\mu\rho},\\
 (p+q)_\rho \Gamma^{\mu\nu\rho}_\mathrm{CO}(p,q)&=&  \frac{i}{6\pi^2} p_\alpha q_\beta \epsilon^{\alpha\beta\mu\nu}.
 \end{eqnarray}
 In order to cancel the anomalies in the vector current, we must perform an additional finite renormalization by adding the Bardeen counterterm,
 \begin{equation}
 \Gamma^{\mathrm{c.t.}} = c \int\, d^4x\, \epsilon^{\mu\nu\rho\lambda} \,V_\mu\, \,A^5_\nu \,F^V_{\rho\lambda},
 \end{equation}
 where $F^V_{\rho\lambda} = \partial_\rho V_\lambda -  \partial_\lambda  V_\rho$. This vertex brings an additional contribution to the three-point function, and the full result reads
\begin{equation}
\Gamma^{\mu\nu\rho} =  \Gamma_{\mathrm{CO}}^{\mu\nu\rho}(p,q) + 2 i c (p_\lambda - q_\lambda ) \epsilon^{\lambda\mu\nu\rho}.
\end{equation}
Choosing the coefficient $c$ of the Bardeen counterterm appropriately, $c=\frac{1}{12 \pi^2}$, we find the anomaly equations
\begin{eqnarray}
 p_\mu \Gamma^{\mu\nu\rho}(p,q) &=&0,\\
 q_\nu \Gamma^{\mu\nu\rho}(p,q) &=&0,\\
 (p+q)_\rho \Gamma^{\mu\nu\rho}(p,q)&=& \frac{i}{2\pi^2} p_\alpha q_\beta \epsilon^{\alpha\beta\mu\nu},
 \end{eqnarray}
in full agreement with the the result from dimensional regularization and the conservation of the vector current.

 We next want to evaluate the triangle diagram in the special kinematic regimes 
 of Eqs.~(\ref{eq:threepts1})-(\ref{eq:threepts6}). Taking $q=-p$, corresponding to the three-point function in Eq.~(\ref{eq:threepts4}), only the integrands $A$ and $B$ in Eqs.~(\ref{eq:coeffs_begin})-(\ref{eq:coeffs_end}) contribute  
 and take the values $1/2$ and $-1/2$ in dimensional regularization and $1/6$ and $-1/6$ in cutoff, respectively. The three-point function is then 
\begin{equation}
\Gamma^{\mu\nu 0}(p,-p)  = \frac{i}{2\pi^2} \epsilon^{\alpha\mu\nu 0} p_\alpha \,,
\end{equation}
in agreement with Eq.~(\ref{eq:threepts4}). Note that with cutoff regularization, $\frac 1 3$ of this result comes from the loop diagram and $\frac 2 3$ comes from the counterterm.

Let us next take $p=0$, {\it i.e.} we put zero momentum on one of the vector currents. The corresponding loop integral vanishes in dimensional regularization, while the loop contribution in cutoff regularization is precisely cancelled by the contribution from the counterterm,
\begin{equation}
\Gamma^{0\nu\rho}(0,-q)=0\,.
\end{equation}
This result is in agreement with Eq.~(\ref{eq:threepts2})

\subsection{Triangle diagram with three axial currents}

From the same one loop integral we can also compute the correlator of three axial currents\footnote{However, as this requires commuting the $\gamma_5$ with the rest of the $\gamma$ matrices, only cutoff regularization can be applied.}. Since we can anticommute the $\gamma_5$
and use $\gamma_5^2=-1$, we can reduce this diagram to (\ref{eq:triangle}). The Bardeen counterterm, however, does not contribute this time, and we therefore find
\begin{equation}
\Gamma_5^{\mu\nu 0} = \frac{1}{3} \frac{i}{2 \pi^2} \epsilon^{\alpha \mu\nu 0} p_\alpha  \,,
\end{equation}
just as in Eq.~(\ref{eq:threepts6}). The factor $\frac 1 3$ is fixed by demanding Bose symmetry on the external legs. 

All other current three-point functions can be related to the triangle with three vector currents which is known to vanish. Therefore we have indeed reproduced the holographic results in  Eqs.~(\ref{eq:threepts1})-(\ref{eq:threepts6})! which is a non trivial check that we have introduced in a right way the anomalous chemical potential.

\section{Thermal two point functions}

In the present section we will use the Kubo formulae deduced in chapter \ref{rel_hydro}  to compute anomalous transport in a system of free chiral fermions. The anomalous magnetic conductivity has been derived and applied in \cite{Kharzeev:2009pj} whereas the one for the anomalous vortical conductivity has been established first in \cite{Amado:2011zx}. They are\footnote{Notice that we are using  in this chapter the capital letters $A,B,C\ldots$ to label the number of conserved currents. }
\begin{eqnarray}
\sigma^\cB_{AB} &=& \lim_{k_n\rightarrow 0} \sum_{ij}\epsilon_{ijn} \frac{-i}{2k_n} \langle J_A^i J_B^j \rangle|_{\omega=0} \, ,\label{eq:sigmaB}\\
\sigma^\cV_{A}  &=& \lim_{k_n\rightarrow 0} \sum_{ij}\epsilon_{ijn} \frac{-i}{2k_n} \langle J_A^i T^{0j} \rangle|_{\omega=0} \, ,
 \label{eq:sigmaV} 
 \end{eqnarray}
where $J_A^\mu$ are the (anomalous) currents and $T^{\mu\nu}$ is the energy momentum tensor.

We will now evaluate the Kubo formulae (\ref{eq:sigmaB}), (\ref{eq:sigmaV})  
for a theory of $N$ free right-handed fermions $\Psi^f$ transforming under a global symmetry group $G$ generated by matrices $(T_A)^f\,_g$.  We denote the generators in the Cartan subalgebra by $H_A$. Chemical
potentials $\mu_A$ can be switched on only in the Cartan subalgebra. Furthermore the presence of the chemical potentials breaks the group $G$ to a subgroup $\hat G$. Only the currents that lie in the unbroken subgroup are conserved (up to anomalies) and participate in the hydrodynamics. The chemical potential for the fermion $\Psi^f$
is given by $\mu^f= \sum_A q_A^f  \mu_A$, where we write the Cartan generator $H_A = q_A^f\delta^f\,_g$ in terms of its eigenvalues, the charges $q_A^f$. The unbroken symmetry group $\hat G$ is generated by those
matrices $T_A^f\,_g$ fulfilling 
\begin{equation}\label{eq:unbroken}
 T_A^f\,_g \mu^g = \mu^f T_A^f\,_g\,.
\end{equation}
There is no summation over indices in the last expression. From now on we will assume that all currents $\vec{J}_A$ lie in directions indicated in  (\ref{eq:unbroken}). We define the chemical potential through boundary conditions on the fermion fields around the thermal circle using formalism (B),
\be \Psi^f(\tau) = - e^{\beta \mu^f} \Psi^f(\tau-\beta),
\ee
with $\beta=1/T$. Therefore the eigenvalues of $\partial_\tau$ are $i\tilde\omega_n+\mu^f$ for the fermion species $f$ with $\tilde\omega_n=\pi T(2n+1)$ the fermionic Matsubara frequencies \cite{Landsman:1986uw}.  A convenient way of expressing the currents is in terms of Dirac fermions and writing
\begin{eqnarray}
J^i_A &=& \sum_{f,g=1}^N T_A^g\,_f \bar\Psi_g \gamma^i \cP_+ \Psi^f \,, \label{eq:JA}\\
T^{0i} &=&  \frac i 2 \sum_{f=1}^N\bar\Psi_f  ( \gamma^0  \partial^i + \gamma^i \partial^0  ) \cP_+\Psi^f\,, \label{eq:JE}
\end{eqnarray}
where we used the chiral projector $\cP_\pm = \frac 1 2 (1\pm\gamma_5)$.
The fermion propagator is
\begin{eqnarray}
S(q)^f\,_g &=&  \frac{\delta^f\,_g}{2} \sum_{t=\pm} \Delta_t(i\tilde\omega^f,\vec{q}) \cP_+ \gamma_\mu \hat q^\mu_t \,,\\
\Delta_t( i\tilde\omega^f, q) &=& \frac{1}{i\tilde\omega^f - t E_q}\,,
\end{eqnarray}
with  $i\tilde\omega^f = i\tilde\omega_n + \mu^f$, $\hat q_t^\mu = (1, t \hat q)$, $\hat{q} = \frac{\vec{q}}{E_q}$ and $E_q=|\vec q |$. 
We can easily include left-handed fermions as well. They contribute in all our calculations in the same way as the right handed ones up to a relative minus sign.

\subsection{\label{sec:kubo1}Evaluation of Kubo formulas}

We will address in detail the computation of the vortical conductivities Eq.~(\ref{eq:sigmaV}) and sketch only the calculation of the magnetic conductivities since the latter one is a trivial extension of the calculation of the chiral magnetic conductivity in \cite{Kharzeev:2009pj}.

\subsubsection{\label{sec:vortical-conductivity}Vortical conductivity}

\begin{figure}[tbp]
\begin{center}
\includegraphics[scale=1]{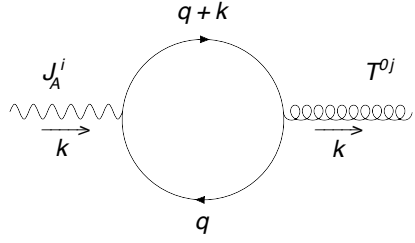}
\caption{1 loop diagram contributing to the vortical conductivity Eq.~(\ref{eq:sigmaV}).}
\label{fig:1loop}
\end{center}

\end{figure}

 The vortical conductivity is defined from the retarded correlation
 function of the current $J^i_A(x)$~(\ref{eq:JA}), and the
 energy momentum tensor or energy current  $T^{0j}(x^\prime)$~(\ref{eq:JE}), i.e.
\begin{equation}
G_A^\cV(x-x^\prime) = \frac{1}{2} \epsilon_{ijn}\,i \, \theta(t-t^\prime)  \,\langle [J^i_{A}(x),T^{0j}(x^\prime)] \rangle \,.
\end{equation}
Going to Fourier space, one can evaluate this quantity as
\be
G_A^\cV(k)= \frac{1}{4}\sum_{f=1}^N T_A^f\,_f \frac{1}{\beta}\sum_{\tilde\omega^f} \int\frac{d^3q}{(2\pi)^3} \epsilon_{ijn} \tr \Bigg[ S^f\,_f(q) \gamma^i  S^f\,_f(q+k) \left(  \gamma^0 q^j + \gamma^j i\tilde\omega^f \right)  \Bigg] \,, \label{eq:Gav}
\ee
which corresponds to the one loop diagram of Fig.~\ref{fig:1loop}. 
The vertex of the two quarks with the graviton is $\sim \delta^f\,_g$, and therefore we find only contributions from the diagonal part of the group~$\hat G$. Our metric is $g_{\mu\nu} = \diag(1,-1,-1,-1)$. We can split $G_A^\cV$ into two contributions, i.e.
\begin{equation}
G_A^\cV(k) = G_{A,(0j)}^\cV(k) + G_{A,(j0)}^\cV(k) \,, \label{eq:Gav2}
\end{equation}
which correspond to the terms $\gamma^0 q^j$  and $\gamma^j i\tilde\omega^f$ in Eq.~(\ref{eq:Gav}) respectively. We will focus first on the computation of $G_{A,(0j)}^\cV$. The integrand of Eq.~(\ref{eq:Gav}) for $G_{A,(0j)}^\cV$ can be written as
\be\label{eq:integrand0j}
 \cI_{A,(0j)}^\cV = \frac{1}{4}q^j\sum_{t,u=\pm} \epsilon_{ijn} \tr[\gamma_\mu \gamma^i \gamma_\nu \gamma^0 \cP_+ ]  \Delta_t(i\tilde\omega^f,\vec{q}) \Delta_u(i\tilde\omega^f + i\omega_n,\vec{q}+\vec{k}) {\hat q}^\mu_t {\widehat{(q+k)}}^\nu_u \,.
\ee
From a computation of the Dirac trace in Eq.~(\ref{eq:integrand0j}) one has two contributions
\begin{eqnarray}
&&\epsilon_{ijn} \tr[\gamma_\mu \gamma^i \gamma_\nu \gamma^0] a^\mu b^\nu = 4 \epsilon_{ijn} ( a^i b^0 + a^0 b^i ) \,, \label{eq:trace1} \\
&&\epsilon_{ijn} \tr[\gamma_\mu \gamma^i \gamma_\nu \gamma^0 \gamma_5] a^\mu b^\nu = 4i(a_j b_n - a_n b_j) \,.  \label{eq:trace2}
\end{eqnarray}
Using Eqs.~(\ref{eq:integrand0j}), (\ref{eq:trace1}) and (\ref{eq:trace2}) one can express $G_{A,(0j)}^\cV(k)$ as
\begin{eqnarray}
G_{A,(0j)}^\cV(k) &=& \frac{1}{8} \sum_{f=1}^N  T_A^f\,_f \frac{1}{\beta}\sum_{\tilde\omega^f} \int \frac{d^3q}{(2\pi)^3}   q^j \sum_{t,u=\pm} \Bigg[  \epsilon_{ijn} \Bigg( t\frac{q^i}{E_q} 
+ u \frac{k^i + q^i}{E_{q+k}}\Bigg) + \nonumber \\ 
&+& i\frac{tu}{E_q E_{q+k}} \left( q_j k_n - q_n k_j  \right) \Bigg]   \Delta_t(i\tilde\omega^f,\vec{k}) \Delta_u(i\tilde\omega^f+i\omega_n,\vec{q}+\vec{k})  \,. \label{eq:Gav0j2}
\end{eqnarray}
At this point one can make a few simplifications. Note that due to the antisymmetric tensor $\epsilon_{ijn}$, the two terms proportional to $q^i$ inside the bracket in Eq.~(\ref{eq:Gav0j2}) vanish. Regarding the term  $\epsilon_{ijn} q^j k^i$, it leads to a contribution $\sim \epsilon_{ijn} k^j k^i$ after integration in $d^3q$, which is zero. Then the only term which remains is the one not involving $\epsilon_{ijn}$. We can now perform the sum over fermionic Matsubara frequencies. One has
\be
\frac{1}{\beta}\sum_{\tilde\omega^f} \Delta_t(i\tilde\omega^f,\vec{q}) \Delta_u(i\tilde\omega^f+i\omega_n,\vec{q}+\vec{p}) = \frac{t n(E_q - t\mu^f) - u n(E_{q+k}-u \mu^f) + \frac{1}{2}(u-t)}{i\omega_n + t E_q - u E_{q+k}} \,, \label{eq:Matsubara1}
\ee
where $n(x) = 1/(e^{\beta x} + 1)$ is the Fermi-Dirac distribution function. In Eq.~(\ref{eq:Matsubara1}) we have considered that $\omega_n = 2\pi T n$ is a bosonic Matsubara frequency. This result is also obtained in Ref.~\cite{Kharzeev:2009pj}. After doing the analytic continuation, which amounts to replacing $i\omega_n$ by $k_0 + i\epsilon$ in Eq.~(\ref{eq:Matsubara1}), one gets
\begin{eqnarray}
\nonumber G_{A,(0j)}^\cV(k) &=& -\frac{i}{8} \sum_{f=1}^N T_A^f\,_f \int \frac{d^3q}{(2\pi)^3}  \frac{ \vec{q}^2 k_n - (\vec{q}\cdot\vec{k}) q_n }{E_q E_{q+k}}   \\
\label{eq:Gav0j3} &&\quad\times \sum_{t,u=\pm}  \frac{u n(E_q - t\mu^f) - t n(E_{q+k}-u \mu^f) + \frac{1}{2}(t-u)}{k_0 + i\epsilon + t E_q - u E_{q+k}}\,. 
\end{eqnarray} 
The term proportional to~$\sim \frac{1}{2}(t-u)$ corresponds to the vacuum contribution, and it is ultraviolet divergent. By removing this term the finite temperature and chemical potential behavior is not affected, and the result becomes ultraviolet finite because the Fermi-Dirac distribution function exponentially suppresses high momenta. By making both the change of variable $\vec{q} \to -\vec{q} -\vec{k}$ and the interchange $u \to -t$ and $t \to -u$ in the part of the integrand involving the term $-tn(E_{q+k}-u\mu^f)$, one can express the vacuum subtracted contribution of Eq.~(\ref{eq:Gav0j3}) as
\begin{eqnarray}
\widehat{G}_{A,(0j)}^\cV(k) &=& \frac{i}{8} k_n \sum_{f=1}^N T_A^f\,_f \int \frac{d^3 q}{(2\pi)^3} \frac{1}{E_q E_{q+k}} \left( \vec{q}^2 - \frac{(\vec{q}\cdot\vec{k})^2}{\vec{k}^2}\right)  \nonumber \\
&&\times\sum_{t,u=\pm} u \frac{n(E_q - \mu^f) + n(E_q+ \mu^f)}{k_0 + i\epsilon + t E_q + u E_{q+k}}\,. \label{eq:Gav0j4}
\end{eqnarray}
where we have used that $n(E_q - t\mu^f) + n(E_q+ t\mu^f) = n(E_q - \mu^f) + n(E_q+ \mu^f)$ since $t=\pm 1$. 
The result has to be proportional to $k_n$, so to reach this expression we have replaced $q_n$ by $(\vec{q}\cdot \vec{k})k_n/\vec{k}^2$ in Eq.~(\ref{eq:Gav0j3}). At this point one can perform the sum over $u$ 
by using $\sum_{u = \pm} u/(a_1+u a_2) = -2a_2/(a_1^2-a_2^2)$, 
and the integration over angles. 
by considering $\vec{q} \cdot \vec{k} = E_q E_k x $ and $E_{q+k}^2 = E_q^2+E_k^2 + 2 E_q E_k x$, where $x := \cos(\theta)$ and $\theta$ is the angle between $\vec{q}$ and $\vec{k}$. 
Then one gets the final result
\begin{eqnarray}
\widehat{G}_{A,(0j)}^\cV(k) &=& \frac{i}{16\pi^2}\frac{k_n}{k^2}(k^2 - k_0^2) \sum_{f=1}^N T_A^f\,_f \int_0^\infty dq \, q \, f^\cV(q)\Bigg[ 1 + \frac{1}{8qk} \sum_{t=\pm} \left[k_0^2-k^2+4q(q+t k_0)\right] \nonumber \\
&& \times\log \left( \frac{\Omega_t^2 - (q+k)^2}{\Omega_t^2 - (q-k)^2}\right) \Bigg] \,,
\label{eq:Gav0j5} 
\end{eqnarray} 
where~$\Omega_t = k_0 + i\epsilon + t E_q\,,$ and
\begin{equation}
f^\cV(q) = n(E_q-\mu^f) + n(E_q + \mu^f) \,.   \label{eq:fV}
\end{equation}
The steps to compute $G_{A,(j0)}^\cV$ in Eq.~(\ref{eq:Gav2}) are similar. In this case the Dirac trace leads to a different tensor structure, in  which the only contribution comes from the trace involving~$\gamma_5$, i.e.
\begin{equation}
\epsilon_{ijn} \tr[\gamma_\mu \gamma^i \gamma_\nu \gamma^j \gamma_5] a^\mu b^\nu =8i ( a_n b_0 - a_0 b_n ) \,. \label{eq:trace3}
\end{equation}
The sum over fermionic Matsubara frequencies involves an extra~$i\tilde\omega^f$. Following the same procedure as explained above, the vacuum subtracted contribution writes ,~i.e.
\begin{eqnarray}\nonumber
&&\frac{1}{\beta}\sum_{\tilde\omega^f} i{\tilde\omega^f} \Delta_t(i\tilde\omega^f,\vec{q}) \Delta_u(i\tilde\omega^f+i\omega_n,\vec{q}+\vec{k}) =    \frac{1}{i\omega_n + t E_q - u E_{q+k}} \Bigg[ E_q n(E_q - t\mu^f)   \\
\label{eq:Matsubara2} &&- (E_{q+k} - ui\omega_n) n(E_{q+k} -u \mu^f) -\frac{1}{2}\left( E_q - E_{q+k} + ui\omega_n \right) \Bigg] \,. 
\end{eqnarray}
The last term inside the bracket in the r.h.s. of Eq.~(\ref{eq:Matsubara2}) corresponds to the vacuum contribution which we choose to remove, as it leads to an ultraviolet divergent contribution after integration in $d^3q$. Making similar steps as for $\widehat{G}_{A,(0j)}^\cV$, one finds the equation analogous to Eq.~(\ref{eq:Gav0j4}), which writes
\begin{eqnarray}
\widehat{G}_{A,(j0)}^\cV(k) &=& \frac{i}{4} \sum_{f=1}^N T_A^f\,_f \int \frac{d^3 q}{(2\pi)^3} \sum_{t,u=\pm} \left( t \frac{q_n}{E_q} + u \frac{q_n + k_n}{E_{q+k}} \right) \nonumber \\
&&\qquad \times \frac{E_q \left[ n(E_q - \mu^f) + n(E_q+ \mu^f) \right] + t k_0 n(E_q + t \mu^f)}{k_0 + i\epsilon + t E_q + u E_{q+k}}\,. \label{eq:Gavj04}
\end{eqnarray}
After performing the sum over $u$ and integrating over angles, one gets the final result
\begin{eqnarray}
\widehat{G}_{A,(j0)}^\cV(k) &=& -\frac{i}{32\pi^2} \frac{k_n}{k^3}\sum_{f=1}^N T_A^f\,_f \int_0^\infty dq \sum_{t=\pm} f^\cV_t(q,k_0) \label{eq:Gavj05} \\
&&\times \Bigg[  4 t q k k_0  -\left(k^2-k_0^2\right) (2q + t k_0) \log\left( \frac{\Omega_t^2  - (q+k)^2}{\Omega_t^2 - (q-k)^2 } \right) \Bigg] \,, \nonumber
\end{eqnarray}
where
\begin{equation}
f^\cV_t(q,k_0) = q f^\cV(q) + t k_0 n(E_q + t\mu^f)   \,. \label{eq:fVt}
\end{equation}
The result for $\hat{G}_A^\cV(k)$ writes as a sum of Eqs.~(\ref{eq:Gav0j5}) and (\ref{eq:Gavj05}), according to Eq.~(\ref{eq:Gav2}). From these expressions one can compute the zero frequency, zero momentum, limit. Since
\begin{equation}
\lim_{k\to 0} \lim_{k_0 \to 0} \sum_{t=\pm}\log \left( \frac{\Omega_t^2 - (q+k)^2}{\Omega_t^2 - (q-k)^2}\right) = \frac{2k}{q} \,, \label{eq:limlog}
\end{equation}
and 
\begin{equation}
\lim_{k\to 0} \lim_{k_0 \to 0}  \sum_{t=\pm} \left[k_0^2-k^2+4q(q + t k_0)\right] \log \left( \frac{\Omega_t^2 - (q+k)^2}{\Omega_t^2 - (q-k)^2}\right)  = 8qk \,, \label{eq:limAlog}
\end{equation}
the relevant integrals are
\begin{equation}
\int_0^\infty dq \, q \, f^\cV(q) = \int_0^\infty dq \, f^\cV_t(q,k_0 = 0) = \frac{(\mu^f)^2}{2} + \frac{\pi^2}{6} T^2 \,.
\end{equation}
Finally it follows from Eqs.~(\ref{eq:Gav0j5}) and (\ref{eq:Gavj05}) that the zero frequency, zero momentum, vortical conductivity writes
\begin{eqnarray}
\sigma_A^\cV &=& \frac{1}{8\pi^2} \sum_{f=1}^N T_A^f\,_f \Bigg[ ( \mu^f)^2 + \frac{\pi^2}{3} T^2 \Bigg] \label{eq:sigmaV2} \\
&=& \frac{1}{16\pi^2}  \Bigg[ \sum_{B,C} \tr\left(T_A \{ H_B , H_C \}\right) \mu_B \, \mu_C  + \frac{2\pi^2}{3} T^2 \tr \left( T_A \right) \Bigg]  \,. \nonumber
\end{eqnarray}
Both $\widehat{G}_{A,(0j)}^\cV$ and $\widehat{G}_{A,(j0)}^\cV$ lead to the same contribution in~$\sigma_A^\cV$.  More interesting is the term $\sim T^2$ which is proportional to the gravitational anomaly \cite{Delbourgo:1972xb,Eguchi:1976db,AlvarezGaume:1983ig} (see chapter \ref{anomalies}). Left handed fermions contribute in the same way but with a relative minus sign. 

If instead of having taken the zero momentum limit at zero frequency, one took the zero frequency limit at zero momentum, the result would be $1/3$ of the result quoted in Eq.~(\ref{eq:sigmaV2}). The same factor appears in the magnetic conductivity when one interchanges the two limits~\cite{Kharzeev:2009pj}.

% \subsection{\label{sec:magnetic-energy}Magnetic Energy conductivity}
% 
% $\sigma_B^{\epsilon,\cB}$ computes in the same way as the vortical conductivity in Sec.~\ref{sec:vortical-conductivity}. From an evaluation of the corresponding Feynman diagram one finds that the result is the same as Eq.~(\ref{eq:Gav}). Then one concludes that
% \begin{equation}
% \sigma_B^{\epsilon,\cB} = \sigma_A^\cV \,.
% \end{equation}

\subsubsection{\label{sec:magnetic-conductivity}Magnetic conductivity}

The magnetic conductivity in the case of a vector and an axial $U(1)$ symmetry was computed at weak coupling in~\cite{Kharzeev:2009pj}. Following the same method, we have computed it for the unbroken (non-abelian) symmetry group~$\hat G$.
The relevant Green function is 
\be
G^\cB_{AB} = \frac 1 2 \sum_{f,g} T_A^g\,_f T_B^g\,_f \frac 1 \beta  \sum_{\tilde\omega^f} \int\frac{d^3q}{(2\pi)^3} \epsilon_{ijn}  \tr \Bigg[ S^f\,_f(q) \gamma^i S^f\,_f(q+k) \gamma^j    \Bigg] \,. \label{eq:Gabb}
\ee
The evaluation of this expression is exactly as in \cite{Kharzeev:2009pj} so we skip the details. 
 The result is 
\begin{equation}
\sigma_{AB}^\cB =  \frac{1}{4\pi^2} \sum_{f,g=1}^N T_A^f\,_g T_B^g\,_f \, \mu^f = \frac{1}{8\pi^2} \sum_C \tr\left( T_A \{ T_B , H_C \} \right) \, \mu_C \,. \label{eq:sigmaB1}
\end{equation}
In the second equality of Eq.~(\ref{eq:sigmaB1}) we have made use of Eq.~(\ref{eq:unbroken}). No contribution proportional to the gravitational anomaly coefficient
is found in this case.

% \subsection{\label{sec:vortical-energy}Vortical Energy conductivity}
% 
% This quantity follows from the correlation function of two energy momentum tensors, cf. Eq.~(\ref{eq:sigmagV}). In the zero frequency, zero momentum, limit there are three contributions out of the eight possible terms. One of these terms involves a sum over fermionic Matsubara frequencies of the form
% \begin{eqnarray}
% &&\frac{1}{\beta}\sum_{\tilde\omega^f} (i{\tilde\omega^f})^2 \Delta_t(i\tilde\omega^f,\vec{q}) \Delta_u(i\tilde\omega^f+i\omega_n,\vec{q}+\vec{k}) \label{eq:Matsubara3}   \\
% &&=  {\cal F}(i\omega_n,E_q,E_{q+k},t,u) + \frac{1}{i\omega_n + t E_q - u E_{q+k}} \nonumber \\
% &&\times\Bigg[ t E_q^2  n(E_q - t\mu^f) - u (E_{q+k} - ui\omega_n)^2 n(E_{q+k} -u \mu^f)  \Bigg]  \,, \nonumber
% \end{eqnarray}
% where $\cal F$ corresponds to the ultraviolet divergent vacuum contribution. The result writes
% \begin{eqnarray}
% \sigma^{\epsilon,\cV} &=& \frac{1}{16\pi^2} \sum_{f=1}^N \bigg[ (\mu^f)^3 + \pi^2 T^2 \mu^f \bigg] \nonumber \\
% &=& \frac{1}{32\pi^2}  \Bigg[ \sum_{A,B,C} \tr\left(H_A \{ H_B , H_C \}\right) \mu_A \mu_B \, \mu_C  \nonumber \\
% &&\qquad\qquad  + 2\pi^2 T^2 \sum_A \tr\left(H_A\right) \mu_A \Bigg] \nonumber \,.  \label{eq:sigmagV1}
% \end{eqnarray}

It is also interesting to specialize our results to the case of one vector and one axial current with chemical potentials $\mu_R = \mu+\mu_5$, $\mu_L= \mu -\mu_5$, charges $q^R_{V,A}=(1,1)$ and $q^L_{V,A}=(1,-1)$ for one right-handed and one left-handed fermion. We find (for a vector magnetic field)
\begin{eqnarray}
\sigma^\cB_{VV} = \frac{\mu_5}{2\pi^2}  &,&  \sigma^\cB_{AV} = \frac{\mu}{2\pi^2} \,, \nonumber\\
\sigma^\cV_{V} = \frac{\mu\mu_5}{2\pi^2} &,& \sigma^\cV_{A} = \frac{\mu^2+\mu_5^2}{4\pi^2}+ \frac{T^2}{12}   \,.
\end{eqnarray}
Here $\sigma_{VV}^\cB$ is the chiral magnetic conductivity \cite{Kharzeev:2009pj},  $\sigma_{AV}^\cB$ describes the generation of an axial current due to a vector magnetic
field \cite{Son:2004tq}, $\sigma^\cV_{V}$ is the vector vortical conductivity in which the contributions of the gravitational anomaly cancel between right- and left-handed fermions. Finally
$\sigma_{A}^\cV$ is the axial vortical conductivity and it is this one that is sensitive to the presence of a gravitational anomaly.

% % %%%%%%%%%%%%%%%%%%%%%%%%%%%%%%%%%%%%%%%%%%%%%%%%%%%%%%%%%%%%%%%%%%
   \chapter{Holographic gravitational anomaly and Transport}
   \label{strongcve}
  In the previous chapter the general Kubo formulae (\ref{Kformulae}) were evaluated for a theory of free chiral fermions. The results showed a somewhat surprising appearance of the anomaly coefficient $b_A$ for the gravitational anomaly. More precisely the chiral vortical conductivity for the symmetry generated by $T_A$ was found to have two contributions, one depending only on the chemical potentials and proportional to the axial anomaly coefficient $d_{ABC}$ and a second one with a characteristic $T^2$ temperature dependence proportional to the gravitational anomaly coefficient $b_A$.

The usage of Kubo formulae has here a clear advantage, it fixes all integration constants automatically. In this way it was possible in the previous chapter\footnote{See \cite{Landsteiner:2011cp}} to show that the coefficient in front of the $T^2$ term in the chiral vortical conductivity is essentially given by the gravitational anomaly coefficient $b_A$. The disadvantage of Kubo formulae is of course that we have to calculate the potentially complicated correlations functions of a quantum field theory. They are easy to evaluate only in certain limits, such as the weak coupling limit considered in \cite{Landsteiner:2011cp}. In principle the results obtained in this limit can suffer renormalization due to the model dependent interactions \cite{Golkar:2012kb,Hou:2012xg}.  The gauge-gravity correspondence \cite{Maldacena:1997re,Gubser:1998bc,Witten:1998qj,Aharony:1999ti} makes also the strong coupling limit easily accessible. 

We would like to understand the effects anomalies have on the transport properties of relativistic fluids.  Anomalies are very robust features of quantum field theories and do not depend on the details of the interactions.  Therefore a rather general model that implements the correct anomaly structure in the gauge-gravity setup is
sufficient for our purpose even without specifying in detail to which gauge theory it corresponds to.  Our approach will therefore be a ``bottom up'' approach in which we simply add appropriate Chern-Simons terms that reproduce the relevant anomalies to the Einstein-Maxwell theory in five dimensions with negative cosmological constant.~\footnote{Very successful holographic bottom up approaches to QCD have been studied recently, either to describe non-perturbative phenomenology in the vacuum, see e.g.~\cite{Andreev:2006ct,Galow:2009kw}, or the strongly coupled plasma~\cite{Gursoy:2008za,Megias:2010ku,Veschgini:2010ws}.}

We will introduce a model that allows for a holographic implementation of the mixed gauge-gravitational anomaly via a mixed gauge-gravitational Chern-Simons term of the form
\begin{equation}
S_{CS} = \int d^5x \, \sqrt{-g} \epsilon^{MNPQR} A_M R^A\,_{BNP} R^B\,_{AQR} \,.
\end{equation}

Gravity in four dimensions  augmented by a similar term with a scalar field instead of a vector field has attracted
much interest recently \cite{Jackiw:2003pm} (see also the review \cite{Alexander:2009tp}). A four dimensional
holographic model with such a term has been shown to give rise to Hall viscosity in \cite{Saremi:2011ab}. The 
quasinormal modes of this four dimensional model have been studied in \cite{Delsate:2011qp}.

\section{Holographic Model}
\label{sec:holo_model}

In this section  we will define our model given the action
\bea
\nonumber
S &=& \frac{1}{16\pi G} \int d^5x \sqrt{-g} \left[ R - 2 \Lambda -   \frac 1 4 F_{MN} F^{MN} \right.\\ &&\left.+ \epsilon^{MNPQR} A_M   \left( \frac\kappa 3 F_{NP} F_{QR} + \lambda R^A\,_{BNP} R^B\,_{AQR}   \right) \right] + S_{GH} + S_{CSK} \,,\\ 
  S_{GH} &=& \frac{1}{8\pi G} \int_\partial d^4x \sqrt{-h} \, K \,,\\ S_{CSK} &=& - \frac{\lambda}{2\pi G}
\int_\partial d^4x \sqrt{-h} \,  \epsilon^{MNPQR} n_M A_N
K_{PL} D_Q K_R^L \,,
 \eea
we define an outward pointing normal vector $n_A \propto g^{AB} \frac{\partial \bar r}{\partial x^B}$ to the holographic boundary of an asymptotically AdS space with unit norm $n_A n^A =1$ so that bulk metric can be decomposed as 
\begin{equation}\label{eq:ucemetric}
g_{AB} = h_{AB} + n_A n_B\, ,
\end{equation}
where $S_{GH}$ is the usual Gibbons-Hawking boundary term and $D_A$ is the induced covariant derivative on the four dimensional hypersurface such that $D_A h_{BC}=0$. The second boundary term $S_{CSK}$ is needed if we want the model to reproduce the gravitational anomaly at general hypersurface.

In general a foliation with timelike surfaces defined through $\bar r(x) = C$ can be written as
\begin{equation}
 ds^2 = (N^2 + N_\mu N^\mu) d\bar r^2 + 2N_\mu dx^\mu d\bar r+ h_{\mu\nu}dx^\mu dx^\nu\,.
\end{equation}

  To study the behavior of our model under the relevant gauge and diffeomorphism gauge symmetries we note that the action is diffeomorphism invariant. The Chern Simons terms are well formed volume forms and as such are diffeomorphism invariant. They do depend however explicitly on the gauge connection $A_M$.  Under gauge transformations $\delta A_M = \nabla_M \xi$ they are therefore invariant only up to a boundary term. We have
\begin{eqnarray}
 \delta S &=& \frac{1}{16\pi G} \int_\partial d^4x \sqrt{-h} \,\xi  \epsilon^{MNPQR} \left( \frac{\kappa}{3}n_M F_{NP}F_{QR} +\lambda n_MR^A\,_{BNP} R^{B}\,_{AQR}\right) +\,\nonumber\\
& &- \frac{\lambda}{4\pi G} \int_\partial d^4x \sqrt{-h} \,n_M \epsilon^{MNPQR}D_N \xi K_{PL}D_Q K^L_R \,.
\end{eqnarray}
This is easiest evaluated in Gaussian normal coordinates (see next section) where the metric takes the form $ds^2 = d\bar r^2 + h_{\mu\nu} dx^\mu dx^\nu$. All the terms depending on the extrinsic curvature cancel thanks to the contributions from $S_{CSK}$! The gauge variation of the action depends only on the intrinsic four dimensional curvature of the boundary and is given by
\begin{equation}
 \delta S = \frac{1}{16 \pi G} \int_\partial d^4 x \sqrt{-h} \epsilon^{\mu\nu\rho\lambda}\left( \frac{\kappa}{3} \hat{F}_{\mu\nu} \hat{F}_{\rho\lambda} + \lambda \hat R^{\alpha}\,_{\beta\mu\nu} \hat R^{\beta}\,_{\alpha\rho\lambda}\right) \,.
\end{equation}

This has to be interpreted as the anomalous variation of the effective quantum action of the dual field theory.
The anomaly is therefore in the form of the consistent anomaly. Since we are dealing only with a single $U(1)$ 
symmetry the (gauge) anomaly is automatically expressed in terms of the field strength. We use this to fix $\kappa$ to the anomaly coefficient for a single chiral fermion transforming under a $U(1)_L$ symmetry. To do so we compare with (\ref{eq:anomalyform}), simply set $T_A=1$  which fixes the anomaly coefficient $d_{ABC} = \frac 1 2 \mathrm{Tr}(T_A\{T_B,T_C\}) = 1$ and therefore
\begin{equation}
\label{kappafixed}
-\frac{\kappa}{48 \pi G} = \frac{1}{96 \pi^2}\, ,
\end{equation}
similarly we can fix $\lambda$ and find
\begin{equation}
\label{lambdafixed}
-\frac{\lambda}{16\pi G} = \frac{1}{768 \pi^2} \,.
\end{equation}
As a side remark we note that the gravitational anomaly could in principle also be shifted into the diffeomorphism sector. This can be done
by adding an additional (Bardeen like) boundary counterterm to the action 
\begin{equation}\label{eq:gravBardeen}
 S_{ct} = \int_\partial d^4x \, \sqrt{-h} A_\mu\, I^\mu\,,
\end{equation}
with $I^\mu=\epsilon^{\mu\nu\rho\lambda}( \hat\Gamma^\alpha_{\nu \beta}\partial_\rho \hat\Gamma^\beta_{\lambda \alpha} + \frac 2 3 \hat\Gamma^\delta_{\nu \alpha}\hat\Gamma^\alpha_{\rho \beta}\hat\Gamma^\beta_{\lambda\delta})$ fulfilling $D_\mu I^\mu = \frac 1 4 \epsilon^{\mu\nu\rho\lambda} \hat R^{\alpha}\,_{\beta\mu\nu} \hat R^{\beta}\,_{\alpha\rho\lambda}$. Since this term depends explicitly on the four dimensional Christoffel connection it breaks diffeomorphism invariance. 

The bulk equations of motion are
\begin{eqnarray}\label{eqgrav}
 G_{MN} + \Lambda g_{MN} &=& \frac 1 2 F_{ML} F_N\,^L - \frac 1 8 F^2 g_{MN} + 2 \lambda \epsilon_{LPQR(M} \nabla_B\left( F^{PL} R^B\,_{N)}\,^{QR} \right) \,, \label{eq:Gbulk}\\\label{eqgauge}
\nabla_NF^{NM} &=& - \epsilon^{MNPQR} \left( \kappa F_{NP} F_{QR} + \lambda  R^A\,_{BNP} R^B\,_{AQR}\right) \,,  \label{eq:Abulk}
\end{eqnarray}
and they are gauge and diffeomorphism covariant. We note that keeping all boundary terms in the variations that lead to the bulk equations of motion we end up with boundary terms that contain derivatives of the metric variation normal to the boundary. We will discuss this issue in more detail in the next section where we write down the Gauss-Codazzi decomposition of the action.

\section{Holographic Renormalization}
\label{sec:holo_renorm}

In order to go through the steps of the holographic renormalization program within the Hamiltonian approach \cite{Martelli:2002sp,Papadimitriou:2004ap}, first of all we establish some notations.  Without loss of generality
we choose a gauge with vanishing shift vector $N_\mu=0$, lapse $N=1$ and $A_r=0$.  In this gauge the bulk metric can be written as
\begin{equation}
 ds^2 = d\bar r^2 + h_{\mu\nu} dx^\mu dx^\nu \,.
\end{equation}
The non vanishing Christoffel symbols are
\begin{eqnarray}
-\Gamma^{\bar r}_{\mu\nu} &=&  K_{\mu\nu} = \frac 1 2 \dot{h}_{\mu\nu} \, ,\\
\Gamma^\mu_{\nu \bar r} &=& K^\mu_\nu \,,  
\end{eqnarray}
and $\hat\Gamma^\mu_{\nu\rho}$ are four dimensional Christoffel symbols computed with $h_{\mu\nu}$. Dot denotes differentiation respect $\bar r$. All other components of the extrinsic curvature vanish, i.e. $K_{\bar r\bar r}=K_{\bar r\mu} =0$. Another useful table of formulas is
\begin{eqnarray}
\dot{\hat\Gamma}^\lambda_{\,\mu\nu} \,\,\,&=& D_\mu K^\lambda_\nu + D_\nu K^\lambda_\mu - D^\lambda K_{\mu\nu} \,, \\
R^{\bar r}\,_{\mu \bar r\nu} &=& -\dot{K}_{\mu\nu} + K_{\mu\lambda}K^\lambda_\nu \,,\\
R^\mu\,_{\bar r\nu \bar r} &=& -\dot{K}^\mu_{\nu} - K^\mu_{\lambda}K^\lambda_\nu \,,\\
R^{\bar r}\,_{\mu\nu\rho} &=& D_\rho K_{\mu\nu} - D_\nu K_{\mu\rho} \,,\\
R^\lambda\,_{\mu \bar r\nu} &=& D_\mu K^\lambda_\nu- D^\lambda K_{\nu\mu} \,,\\
R^\mu\,_{\nu\rho\lambda} &=& \hat{R}^\mu\,_{\nu\rho\lambda} - K^\mu_\rho K_{\nu\lambda} +  K^\mu_\lambda K_{\nu\rho}\,.
\end{eqnarray}
Note that indices are now raised and lowered with $h_{\mu\nu}$, e.g. $K=h^{\mu\nu} K_{\mu\nu}$, and intrinsic four dimensional curvature quantities are denoted with a hat, so $\hat{R}^\mu\,_{\nu\rho\lambda}$ is the intrinsic four dimensional Riemann tensor on the $\bar r(x)=C$ surface. Finally the Ricci scalar is
\begin{equation}
 R = \hat{R} - 2 \dot{K} - K^2 - K_{\mu\nu}K^{\mu\nu} \,.
\end{equation}

Now we can calculate the off shell action. It is useful to divide it up in three terms. The first one is the usual gravitational bulk and gauge terms with the usual Gibbons-Hawking term and the other two the gauge Chern Simons and the Mixed gauge-gravitational Chern Simons.
{\small\begin{eqnarray}
\label{eq:Sb1} 
S^0 &=& \frac{1}{16 \pi G} \int  d^5x\,\sqrt{-h} \left[ \hat{R} - 2 \Lambda + K^2 -
K_{\mu\nu}K^{\mu\nu} - \frac 1 2 E_\mu E^\mu - \frac 1 4 \hat{F}_{\mu\nu}\hat{F}^{\mu\nu} \right] \, ,\\
\label{eq:Sb2}
S^{1}_{CS} &=& -\frac{\kappa}{12\pi G} \int d^5x\,\sqrt{-h}  \epsilon^{\mu\nu\rho\lambda} A_\mu E_\nu \hat{F}_{\rho\lambda}  \,,\\
\label{eq:Sb3}
S^2_{CS} &=& -\frac{8 \lambda}{16\pi G} \int d^5x \sqrt{-h}  \epsilon^{\mu\nu\rho\lambda} \bigg[ 
A_\mu \hat{R}^\alpha\,_{\beta\rho\lambda} D_\alpha K^\beta_\nu + E_\mu K_{\nu \alpha}D_\rho K^\alpha_\lambda + \frac 1 2 \hat{F}_{\mu\rho} K_{\nu\alpha}\dot{K}^\alpha_\lambda \bigg] \, . 
\end{eqnarray}}
We have used implicitly here the gauge $A_{\bar r}=0$ and denoted $\dot{A}_\mu= E_\mu$. The purely four dimensional field strength is denoted with a hat. 

Of particular concern is the last term in $S^2_{CS}$ which contains explicitly the normal derivative of the extrinsic curvature $\dot K_{\mu\nu}$. For this reason the field equations will be generically of third
order in $r$-derivatives and that means that we can not define a well-posed Dirichlet problem by fixing the $h_{\mu\nu}$ and $K_{\mu\nu}$ alone but generically we would need to fix also $\dot K_{\mu\nu}$.
Having applications to holography in mind we will however impose the boundary condition that the metric has an asymptotically AdS expansion of the form
\begin{equation}
h_{\mu\nu} = e^{2 \bar r} \left( g^{(0)}_{\mu\nu} + e^{-2\bar r} g^{(2)}_{\mu\nu} + e^{-4\bar r} (g^{(4)}_{\mu\nu}
  + 2\bar r \tilde{g}^{(4)}_{\mu\nu} )+ \cdots\right) \,.
\end{equation}
Using the on-shell expansion of $K_{\mu\nu}$ obtained in the appendix \ref{sec:app_holo_renorm}
we can show that the last term in the action does not contribute in the limit $r\rightarrow\infty$. Therefore the boundary action depends only on the boundary metric $h_{\mu\nu}$ but not on the derivative $\dot h_{\mu\nu}$.  This is important because otherwise the dual theory would have additional operators that are sourced by the derivative. Similar issues have arisen before in the holographic theory of purely gravitational anomalies of two dimensional field theories \cite{Saremi:2011ab,Yee:2011yn,Kraus:2005zm}.  Alternatively one could restrict the field space to configurations with vanishing gauge field strength on the boundary. Then the last term in $S^2_{CS}$ is absent. We note that the simple form of the higher derivative terms arises only if we include $S_{CSK}$ in the action. An analogous term in four dimensional Chern-Simons gravity has been considered before in \cite{Grumiller:2008ie}.

The renormalization procedure follows from an expansion of the four dimensional quantities in eigenfunctions of the dilatation operator
\begin{equation}
\delta_D = 2 \int d^4x\, h_{\mu\nu} \frac{\delta}{\delta h_{\mu\nu}} \,. 
\end{equation}
We explain in much details the renormalization in appendix~\ref{sec:app_holo_renorm}. The result one gets for the counterterm coming from the regularization of the boundary action is
\begin{eqnarray}
S_{ct} &=& - \frac{(d-1)}{8\pi G} \int_{\partial_*} d^4x \sqrt{-h} \bigg[ -\frac{2\Lambda}{d(d-1)}  + \frac{1}{(d-2)}P  \nonumber  \\
&&\qquad\qquad- \frac{1}{4(d-1)} \left( P^\mu_\nu P^\nu_\mu - P^2 -  \frac{1}{4} \hat{F}_{(0)}\,_{\mu\nu} \hat{F}_{(0)}\,^{\mu\nu} \right)\log e^{-2\bar r_*} \bigg] \,, \label{eq:Sct}
\end{eqnarray}
where $d=4$, $\bar r_*$ the UV cutoff and
\begin{equation}
P = \frac{\hat{R}}{2(d-1)} \,, \qquad  P^\mu_\nu = \frac{1}{(d-2)} \left[ \hat{R}^\mu_\nu - P \delta^\mu_\nu \right] \,. \label{eq:PPij} 
\end{equation}
As a remarkable fact there is no contribution in the counterterm coming from the gauge-gravitational Chern-Simons term. This has also been derived in \cite{Clark:2010fs} in a similar model that does however not contain $S_{CSK}$.

\section{Currents and Ward identities}
\label{sec:holo_higher_der}

As we discussed before the action is third order in $r$ derivatives, so in order to get the correct one point functions we have to take into account this fact and include the assumption that the bulk space is asymptotically anti-de Sitter. Asymptotically $AdS$ is enough to get a well defined boundary value problem just in terms of the field  theory sources. Let us analyze now what this implies for a general Lagrange density.

\subsection{The holographic dictionary with higher derivatives}

Let us assume a general renormalized Lagrangian for an arbitrary set  of fields that we will call $\phi$  after the four dimensional ADM decomposition\footnote{For simplicity we will omit internal indices of the field $\phi$}, 
$$
S = \int  \,d^4x\, dr\,\cL ( \phi,  \dot\phi, D_\mu \phi,D_\mu\dot\phi,\ddot{\phi} ),
 $$
where dot indicates derivative with respect to the radial coordinate. A general variation of the action leads now to
\be
\delta S = \int_B \,d^4x\, dr\,\left[ \frac{\partial \cL}{\partial \phi} \delta \phi + \frac{\partial \cL}{\partial \dot\phi} \delta \dot\phi+ 
\frac{\partial \cL}{\partial (D_\mu\phi)} \delta(D_\mu
\phi) 
+ \frac{\partial \cL}{\partial (D_\mu\dot\phi)} \delta(D_\mu\dot
\phi)
+ \frac{\partial \cL}{\partial \ddot\phi} \delta \ddot\phi \right] \,.
\ee
Through a series of partial integrations we can bring this into the following form,
\begin{equation}
\delta S = \int\, d^4x\,dr\, \textrm{E.O.M.} \, \delta\phi + \int_{\partial_* } \,d^4x\, \left[ \left( \frac{\partial \cL}{\partial \dot\phi} -  D_\mu \left( \frac{\partial \cL}{\partial       (D_\mu\dot\phi)} \right) - \left( \frac{\partial
      \cL}{\partial \ddot\phi} \right)^. \right) \delta\phi  +   \frac{\partial       \cL}{\partial \ddot\phi}\delta\dot{\phi}  \right]\,.
\end{equation}

The bulk terms are the equations of motion. For a generic boundary, the form of the variation shows that Dirichlet boundary conditions can not be imposed.  Vanishing of the action rather imposes a relation between $\delta\phi$ and $\delta\dot\phi$.

If we have applications of holography in mind, there is however another way of dealing with the boundary term. We suppose now that we are working in an asymptotically anti-de Sitter space. The field $\phi$ has therefore a boundary expansion 
$$
\phi = e^{(\Delta-4)r} \phi^{(0)} + \mathrm{subleading}\,,
$$
here $\Delta$ is the dimension (conformal weight) of the operator that is sourced by $\phi^{(0)}$. Since this is a generic property of holography in AdS spaces, we can relate the derivative of the variation to the variation itself,
$$
\delta\dot\phi = (\Delta-4) e^{(\Delta-4)r}\delta \phi^{(0)} + \mathrm{subleading}\,.
$$
Using that, the consistent operator $\cO_\phi$ is defined as the variation of the on-shell action with respect to the source $\phi^{(0)}$. We find therefore 
\be
\label{curr}
\sqrt{-h^{(0)}}\cO_\phi = \lim_{r\rightarrow r_*} e^{(\Delta-4)r} \left[ 
\frac{\partial \cL}{\partial \dot\phi} -  D_\mu \left( \frac{\partial \cL}{\partial
      (D_\mu\dot\phi)} \right) - \frac{d}{dr}\left( \frac{\partial
      \cL}{\partial \ddot\phi} \right)
+(\Delta-4) \left( \frac{\partial
      \cL}{\partial \ddot\phi}\right) 
\right] \,.
\ee

 From this we can compute the bare consistents U(1) current and the energy-momentum tensor, and the result is
\begin{eqnarray}
\label{eq:Jc} 16\pi G \mathcal J^\mu &=&  -\lim_{r\rightarrow r_*}\frac{\sqrt{-h}}{\sqrt{-g^{(0)}}}\left[ F^{r\mu} + \frac{4}{3}\kappa\epsilon^{\mu\nu\rho\lambda} A_\nu \hat F_{\rho\lambda} \right]\,, \\
\nonumber 8\pi G T^\mu _{(c)\nu} &=& \lim_{r\rightarrow r_*}\frac{\sqrt{-h}}{\sqrt{-g^{(0)}}}\left[ K^\mu_\nu - K \gamma^\mu_{\nu} + 4\lambda \epsilon^{(\mu \alpha\beta\rho} \left( \frac{1}{2}\hat{F}_{\alpha\beta}\hat{R}_{\nu)\rho} + \nabla_\delta(A_{\alpha} \hat{R}^\delta\,_{\nu)\beta\rho}) \right) \right]  \,.\\
 \label{eq:Tc} &&
\end{eqnarray}

Now taking the divergence of these expressions and using Codazzi form of the equations of motion shown in appendix \ref{app:codazziEq}, we get the anomalous charge conservation and the energy-momentum conservation relations respectively, 
\begin{eqnarray}
 D_\mu \mathcal J^\mu &=& -\frac{1}{16\pi G} \epsilon^{\mu\nu\rho\lambda} \left( \frac\kappa 3\hat F_{\mu\nu}\hat F_{\rho\lambda}+ \lambda \hat R^\alpha\,_{\beta\mu\nu}\hat R^\beta\,_{\alpha\rho\lambda} \right) \,,\\
\label{ener_conser} D_\mu T_{(c)}^{\mu\nu} &=&  \hat F^{\nu\mu}\mathcal J_\mu + A^\nu D_\mu \mathcal J^\mu \, .
\end{eqnarray}
These are precisely the consistent Ward identities for a theory invariant under diffeomorfisms with a mixed gauge gravitational anomaly plus a pure gauge anomaly. 

\subsection{Covariant form of the current and energy-momentum tensor}

We have computed the currents as the derivative of the field theory
quantum action, and the anomaly is therefore in the form of the
consistent anomaly. Since we are dealing only with a single $U(1)$
symmetry, the (gauge) anomaly is automatically expressed in terms of
the field strength. However it is always possible to add a
Chern-Simons current and to redefine the charge current $J^\mu
\rightarrow J^\mu + c \epsilon^{\mu\nu\rho\lambda}A_\nu
F_{\rho\lambda}$, and the energy-momentum tensor $T^{\mu\nu}
\rightarrow T^{\mu\nu} + c' \epsilon^{\alpha(\mu \rho\lambda}D_\beta
\left( A_\alpha R^{\beta \nu)}\,_{\rho\lambda}\right)$. These redefined
quantities can not be expressed as the variation of a local functional
of the fields with respect to the gauge and metric fields
respectively. In particular the so-called covariant form of the
anomaly differs precisely in such a redefinition of the
current.~\footnote{Note that the  approaches used in~\cite{Son:2009tf,Neiman:2010zi} and in subsequent works,  typically make use of the covariant form of the anomaly.} 

Adding such a terms to the consistent current and energy-momentum tensor~(\ref{eq:Jc})-(\ref{eq:Tc}), we can write the covariant expressions for these quantities which are the ones we will use to construct the hydrodynamical constitutive relations in the fluid/gravity approach,
\begin{eqnarray}
16\pi G J^\mu &=&  -\frac{\sqrt{-h}}{\sqrt{-g^{(0)}}} F^{r\mu}  \,, \label{eq:Jcov}\\
 8\pi G T^{\mu\nu} &=& \frac{\sqrt{-h}}{\sqrt{-g^{(0)}}}\left[ K^{\mu\nu} - K h^{\mu\nu} + 2\lambda \epsilon^{(\mu \alpha\beta\rho} \hat{F}_{\alpha\beta}\hat{R}^{\nu)}_\rho  \right]  \,. \label{eq:Tcov}
\end{eqnarray}

Of course these one point functions either in their consistent or covariant form  have to be renormalized in order to make sense, so is necessary to include the contributions coming from the counterterm (\ref{eq:Sct}) and then take the limit $r_*\to\infty$.

\section{Kubo formulae, anomalies and chiral vortical conductivity}
\label{sec:kubo2}
We are now going to evaluate the Kubo formulas for anomalous transport in our holographic model. We will do that in the same way as we did it in chapter \ref{cmecse}. Since we are interested in the linear response limit, we split the metric and gauge field into a background part and a linear perturbation,\footnote{Do not confuse the background metric here $g^{(0)}_{MN}$ with the boundary metric used in the asymptotic expansion above }
\bea
g_{MN} &=& g^{(0)}_{MN} + \epsilon \, h_{MN} \,,\\ 
A_{M} &=& A^{(0)}_{M} + \epsilon \, a_{M} \, .
\eea
Inserting these fluctuations-background fields in the action and expanding up to second order in $\epsilon$ we can read the second order action which is needed to get the desired propagators. 

The system of equations (\ref{eqgrav})-(\ref{eqgauge}) admit the following exact background AdS Reissner-Nordstr\"om black-brane solution
 \bea 
 \dd s^2&=& \frac{ r^2}{L^2}\left(-f( r) \dd t^2 +\dd \vec{x}^2\right)+\frac{L^2}{ r^2 f( r)} \dd  r^2\,,\nn\\ A^{(0)}&=&\phi( r)\dd t = \left(\beta-\frac{\mu \,   r_{+}^2}{ r^2}\right)\dd t\,, 
 \eea
 where the horizon of the black hole is located at $ r= r_{+}$, the cosmological constant is $\Lambda = -6/L^2$ and the blackening factor of the metric is
  \be
   f( r)=1-\frac{m L^2}{   r^4}+\frac{q^2 L^2}{ r^6}\,. 
   \ee
The parameters $M$ and $Q$ of the RN black hole are related to the chemical potential $\mu$ and the horizon $ r_H$ by 
\be m=\frac{ r_{\rm     H}^4}{L^2}+\frac{q^2}{ r_{+}^2}\quad,\quad q=\frac{\mu\,
   r_{+}^2}{\sqrt{3}}\,.  
 \ee 
The Hawking temperature is given in terms of these black hole parameters as 
\be T=\frac{ r_{\rm
    H}^2}{4\pi\, L^2} f( r_{+})' = \frac{ \left(2\, 
  r_{+}^2\, m - 3\, q^2 \right)}{2 \pi \, r_{+}^5} \,.
\ee
The pressure of the gauge theory is $P=\frac{m}{16\pi GL^3}$ and its energy density is $\epsilon=3P$ due to the underlying conformal symmetry .

Without loss of generality we consider perturbations of momentum $k$ in the $y$-direction at zero frequency. To study the effect of anomalies  we just  turned on the shear sector (transverse momentum fluctuations) $a_a$ and $h^a_{\,t}$, where $a,b\ldots=x,z$.\footnote{Since we are in the zero frequency case the fields $h_y^a$ completely decouple of the system and take a constant value, see appendix \ref{app:sheareq}. } For convenience we redefine new parameters and radial coordinate
\be
\bar \lambda=\frac{4\mu \lambda L}{ r_H^2}\quad;\qquad \bar \kappa=\frac{4\mu \kappa L^3}{ r_H^2}\quad;\qquad a=\frac{\mu^2L^2}{3 r_{+}^2}\quad;\qquad u=\frac{ r_+^2}{r^2}\,.
\ee
Now the horizon sits at $u=1$ and the AdS boundary at $u=0$. Finally we can write the system of differential equations for the shear sector, that consists on four second order equations. Since we are interested in computing correlators at hydrodynamics regime, we will solve the system up to first order in $k$.  The reduced system can be written as

\bea
\label{htx1} \nonumber 0&=&  h^{a''}_t(u) - \frac{h^{a'}_t(u)}{u} 
- 3a u B'_a(u) +i\bar\lambda k \epsilon_{ab}\left[\left(24a u^3-6(1-f(u))\right)\frac {B_b(u)}{u}\right.\\
    &&\left.+(9a u^3-6(1-f(u)))B'_b(u)+2 u (uh^{b'}_t ( u ) )' \right] \,, \\ 
\nonumber\label{bx1} 0&=& B_a'' ( u )+\frac { f' ( u ) }{f(u)}B_a'( u )-  \frac{ h^{a'}_t ( u
 )}{f(u)} \\
 &&+ik\epsilon_{ab}\left(\ \frac{3}{u f(u)} \bar\lambda \left( \frac{2}{a} (f(u)-1)+ 3u^3 \right) h_t^{b'}(u)+\bar \kappa \frac{B_b(u)}{f(u)} \right) \, ,
\eea
with the gauge field redefined as $B_a=a_a/\mu$, do not confuse this $B$-field with the magnetic field!. The complete system of equations depending on frequency and momentum is showed in appendix~\ref{app:sheareq}. This system consists of six dynamical equations and two constraints.

In order to get solutions at first order in momentum we expand the fields in the dimensionless momentum $p=k/4\pi T$ such as 
\bea
h^a_t(u) &=& h^{(0)a}_t(u)+p \,h^{(1)a}_t(u)
\,,\\ 
B_a(u) &=& B^{(0)}_a(u)+p \,B^{(1)}_a(u)\,.  
\eea
The relevant physical boundary conditions on fields are: $h^a_t(0)=\tilde H^a$, $B_a(0)=\tilde B_a$; where the `tilde' parameters are the sources of the boundary operators.  The second condition compatible with the ingoing one at the horizon is regularity for the gauge field and vanishing for the metric
fluctuation (see Appendix \ref{ap:fluiddep} for a discussion on boundary conditions and frame selection in the field theory side) \cite{Amado:2011zx}.

After solving the system perturbatively (see appendix~\ref{ap:anSol} for solutions), we can go back to the formula (\ref{eq:GR}) and compute the corresponding holographic Green's functions. If we consider the vector of fields to be 
\be
\Phi_k^{\top} (u) = \Big{(} B_x(u) ,\, h^x_{\,t}(u)  ,\, B_z(u) ,\, h^z_{\,t}(u) \Big{)} \,,
\ee
the $\mathbb A$ and $\mathbb B$ matrices for that setup take the following form

\be
\mathbb A=\frac{ r_{+}^4}{16\pi G L^5} \,{\rm Diag}\left( -3a f,\, \frac{1}{u} ,\, -3a f,\, \frac{1}{u} \right) \,,
\ee

\bea
\hspace{-0.8cm}\mathbb B_{AdS+\partial}=
\frac{ r_{+}^4}{16\pi G L^5}
\left(
\begin{array}{cccc}
0 & -3a  & \frac{4 \kappa i k \mu^2 \phi L^5}{3 r_{+}^4} & 0  \\
0 & -\frac{3}{u^2} &  0 & 0  \\
\frac{-4 \kappa i k \mu^2 \phi L^5}{3 r_{+}^4} & 0 & 0 & -3a  \\
0 & 0 & 0 & -\frac{3}{u^2}  \\
\end{array}
\right)\,,
\eea

\bea
\mathbb B_{CT}=
\frac{ r_{+}^4}{16\pi G L^5}
\left(
\begin{array}{cccc}
\vspace{0.15cm}
0&  0 & 0 & 0 \\
\vspace{0.15cm}
0 & \frac{3}{u^2 \sqrt{f\,}} & 0 & 0  \\
\vspace{0.15cm}
0 & 0  & 0 & 0\\
\vspace{0.15cm}
0 & 0  & 0 & \frac{3}{u^2 \sqrt{f\,}}  \\
\end{array}
\right)\,,
\hspace{-0.8cm}\,
\eea
where $\mathbb B=\mathbb B_{AdS+\partial}+\mathbb B_{CT}$. Notice that there is no contribution to the matrices coming from the Chern-Simons gravity part, the corresponding contributions vanish at the boundary. These matrices and the perturbative solutions are the ingredients to compute the matrix of propagators. Undoing the vector field redefinition introduced in (\ref{htx1}) and (\ref{bx1}) the non-vanishing retarded correlation functions at zero frequency are then
\bea
\label{eq:giti}G_{x,tx} &=& G_{z,tz} =\frac{\sqrt{3}\, q}{4 \pi\, G\, L^3  }\,, \\
\label{eq:gxz}G_{x,z} &=& - G_{z,x} =\frac{i\, \sqrt{3}\, k\, q\,  \kappa}{2 \pi\, G \,  r_{+}^2}+\frac{i\, k\, \beta \, \kappa}{6\pi\, G  }\,,\\
G_{x,tz} &=&G_{tx,z} = -G_{z,tx}=-G_{tz,x}=\frac{3\, i\, k\, q^2\,  \kappa }{4 \pi\,G\,   r_{+}^4}+\frac{2ik\lambda \pi T^2}{G} \,,\\
G_{tx,tx} &=& G_{tz,tz}=\frac{m}{16\pi\, G\, L^3 }\,,\\
\label{eq:gtiti}G_{tx,tz} &=& -G_{tz,tx}=+\frac{i\, \sqrt{3}\, k\, q^3\, \kappa}{2\pi\, G\, r_{+}^6} + \frac{4\pi i\sqrt{3}  k q T^2 \lambda}{G \, r_{+}^2}\,.
\eea

Using the Kubo formulae \ref{Kformulae}  and setting the deformation parameter $\beta$ to zero we recover the conductivities  
\begin{eqnarray}
\label{eq:sigb}\sigma_B &=& -\frac{\sqrt{3}\,  q \, \kappa}{2 \pi   \,G\,  r_{+}^2} =  \frac{ \mu}{4 \pi^2}\,,\\
\label{eq:sigv}\sigma_V &=&\sigma_B^\epsilon=-\frac{3\, q^2\,  \kappa }{4 \pi\,G\,  r_{+}^4}-\frac{2\lambda \pi T^2}{G}= \frac{\mu^2}{8\pi^2} +\frac{T^2}{24} \,, \\
\label{eq:sigve}\sigma_V^\epsilon &=&-\frac{ \sqrt{3}\, q^3\,   \kappa}{2\pi\, G\,  r_{+}^6} - \frac{4\pi \sqrt{3}   q T^2   \lambda}{G \, r_{+}^2} =\frac{\mu^3}{12 \pi^2}+\frac{\mu   T^2}{12}\,. 
\end{eqnarray}
The first expression is in perfect agreement with the literature and the second one shows the extra $T^2$ term predicted in \cite{Landsteiner:2011cp} and shown in the previous chapter. In fact the numerical coefficients coincide precisely with the ones obtained in weak coupling. We also point out that the $T^3$ term that appears as undetermined integration constant in the hydrodynamic considerations in \cite{Neiman:2011mj} should make its appearance in $\sigma_V^\epsilon$. We do not find  any such term which is consistent with the argument that this term is absent due to CPT invariance.

It is also interesting to write down the vortical and magnetic conductivity as they appear  in the Landau frame,
\begin{eqnarray}
\xi_B &=& -\frac{\sqrt{3} q ( mL^2+3  r_{+}^4)\kappa}{8\pi G m L^2  r_{+}^2}+\frac{\sqrt{3}q\lambda\pi T^2}{Gm} \nonumber \\
\label{eq:xib}&=& \frac{1}{4\pi^2}\, \left( \mu -  \frac{1}{2}\frac{n(\mu^2+\frac{\pi^2 T^2}{3})}{\epsilon+P}\right)\,,\\
\nonumber\xi_V &=&-\frac{3 q^2  \kappa }{4\pi G  mL^2 }-\frac{2\pi\lambda T^2( r_{+}^6-2L^2q^2)}{ G m L^2  r_{+}^2}\\
\label{eq:xiv}&=&  \frac{\mu^2}{8\pi^2}\, \left( 1 - \frac{2}{3}
  \frac{n \mu}{\epsilon+P}\right)+\frac{T^2}{24}\left(1-\frac{2 n\mu}{\epsilon+P}\right)\,.
\end{eqnarray}

Finally let us also note that the shear viscosity is not modified by the presence of the gravitational anomaly. We know that $\eta\propto\lim_{w\to 0}\frac{1}{w}<T^{xy}T^{xy}>_{k=0}$, so we should solve the system at $k=0$ for the fluctuations $h^i_y$  but the  anomalous coefficients always appear with a momentum $k$ as we can see in (\ref{eq_Hys}), therefore if we switch off the momentum, the system looks precisely as the theory without anomalies. In \cite{Bonora:2011mf} it has been shown that the black hole entropy doesn't depend on the extra mixed Chern-Simons term, therefore the shear viscosity entropy ratio remain the same in this model\footnote{For a four dimensional
holographic model with gravitational Chern-Simons term and a scalar field this has also been shown in \cite{Delsate:2011qp}.}.

\subsection{Frequency dependence}

In order to study the frequency dependence of the chiral conductivities, we can use Eq. (\ref{Kformulae})  to define 
\bea
\sigma_B(\omega)  &=& \lim_{k_m\to0}\frac{-i}{k_m} \epsilon_{mij} \vev{ J^i J^{j}}\,,\\
\sigma_V(\omega) &=&\lim_{k_m\to0}\frac{-i}{k_m} \epsilon_{mij} \vev{ J^i T^{tj}}\,.
\eea

It is important to notice that these, and not the $\xi_V$ and $\xi_B$, are the relevant conductivities at finite frequency in the laboratory rest frame. The latter correspond to the conductivities measured in the local rest frame of the fluid, where one subtracts the contribution to the current due to the energy flux generated when we put the system in a background magnetic or vorticity field.

To study that dependence holographically, we have to resort to numerics. The nature of the system allows us to integrate from the horizon out to the boundary, so we should fix boundary conditions at the first one, even though we would like to be free to fix the AdS boundary values of the fields, hence the operator sources. Imposing infalling boundary conditions, the fluctuations can be written as
\bea
h^c_t(u) &=& (1-u)^{-i w+1}\,H^c_t(u)\,,\\
h^c_x(u) &=& (1-u)^{-i w}\,H^c_x(u)\,,\\
B^c(u) &=& (1-u)^{-i w}\,b^c(u)\,,
\eea
where $w=\omega/4\pi T$. As we saw, the remaining gauge symmetry acting on the shear channel implies that $h^c_t$ and $h^c_x$ are not independent. So if we fix the horizon value of the $\{b^c,H^c_t\}$ fields, the constraints \erf{constraints} fixes
\begin{eqnarray}
\nonumber H^a_y(1)&=&-\frac{3ia( 1+384 (-2+a)^2 (-2+3 a) p^2 \lambda ^2 ) }{(2-a)p-768a (2-a)^3  p^3 \lambda ^2}b^a(1) -\frac{ (i+w)}{(2-a)p}H^a_t(1) +\\
&&+ 24 \sqrt{3a} \lambda\epsilon_{ab}\frac{(2-5 a) }{1-768a (2-a)^2 p^2 \lambda ^2}b^b(1)\,.
\end{eqnarray}

In order to find a maximal set of linearly independent solutions, we can construct four of them using linearly independent combinations of these horizon free parameters. In this way we construct the following independent horizon valued vectors
\bea
\begin{array}{cc}
\left(\begin{array}{c}
1\\
0\\
-\frac{3ia( 1+384 (-2+a)^2 (-2+3 a) p^2 \lambda ^2 ) }{(2-a)p-768a (2-a)^3  p^3 \lambda ^2}\\
0\\
0\\
-\frac{24 \sqrt{3} (2-5 a) \sqrt{a} \lambda }{1-768 (-2+a)^2 a p^2 \lambda ^2}
\end{array}\right) ,&
\left(\begin{array}{c}
 0\\
 1\\
 -\frac{i+w}{(2-a)p}\\
 0\\
 0\\
 0
 \end{array}\right), \\
 &\\
\left(\begin{array}{c}
0\\
  \frac{24 \sqrt{3} (2-5 a) \sqrt{a} \lambda }{1-768 (-2+a)^2 a p^2 \lambda ^2}\\
 0\\
 1\\
0\\
 \frac{3ia( 1+384 (-2+a)^2 (-2+3 a) p^2 \lambda ^2 ) }{(2-a)p-768a (2-a)^3  p^3 \lambda ^2}\\
 \end{array}\right), &
\left(\begin{array}{c}
 0\\
 0\\
 0\\
 0\\
 1\\
 -\frac{i+w}{(2-a)p} 
 \end{array}\right).
 \end{array}
 \eea
The remaining two are given by pure gauge solutions arising from gauge transformations of the trivial one. We choose them to be
\bea
\Phi(u)=\begin{array}{cccc}
\left(\begin{array}{c}
0\\
w\\
-p\\
0\\
0\\
0
\end{array}\right) \,,&
\left(\begin{array}{c}
 0\\
 0\\
 0\\
 0\\
 w\\
 -p
 \end{array}\right)\,.
 \end{array}
 \eea
 Using the corresponding solutions we construct the $F$ matrix of \erf{eq:f} in this way:
 \be
 F^I_J(u) = H^I_M(u) H^{-1M}_J(0)\,,
 \ee
where $H^I_J(u)=(\Phi^I(u))_J$, for the numerical computation we used the numerical values $\kappa=1$ and $\lambda=1/24$ since we know that for a single chiral fermion the ratio $\lambda/\kappa=1/24$ (see (\ref{kappafixed}) and (\ref{lambdafixed})). 

\begin{figure}[t!]
\begin{center}
\includegraphics[scale=0.58]{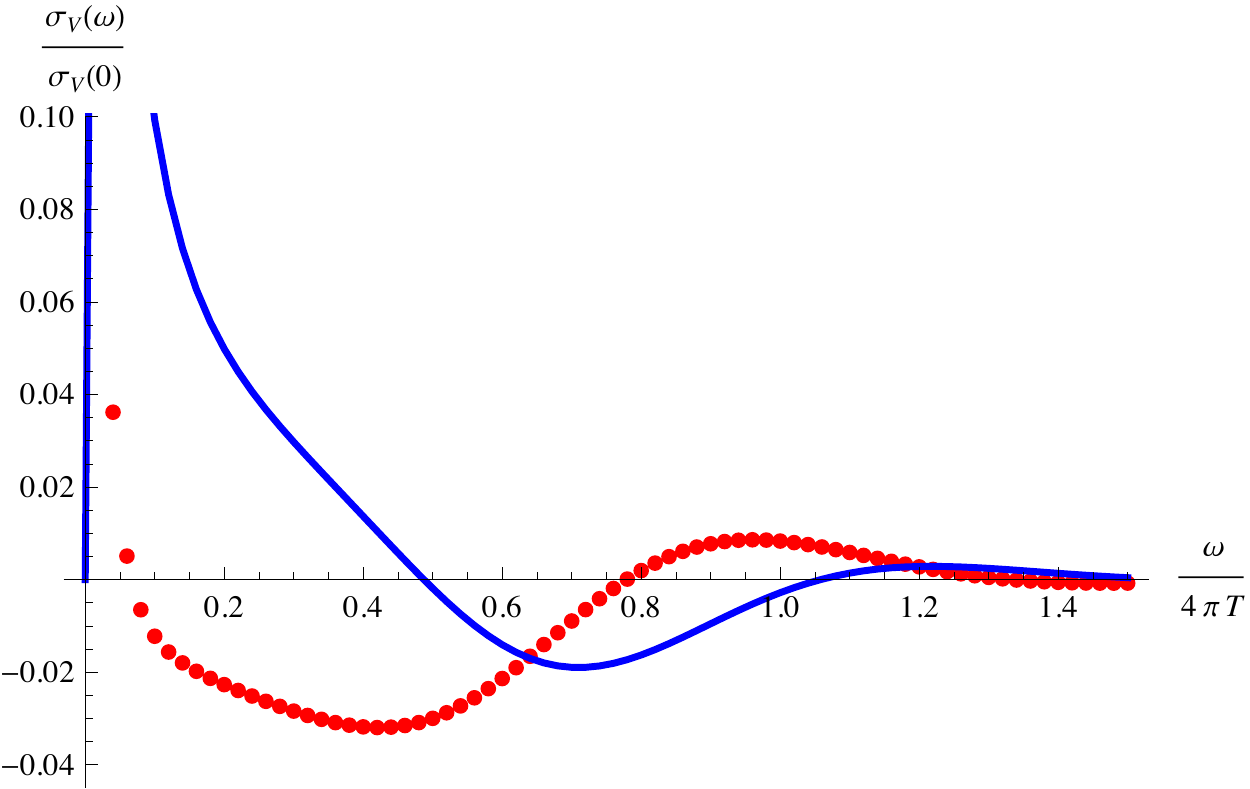}\hfill \includegraphics[scale=0.58]{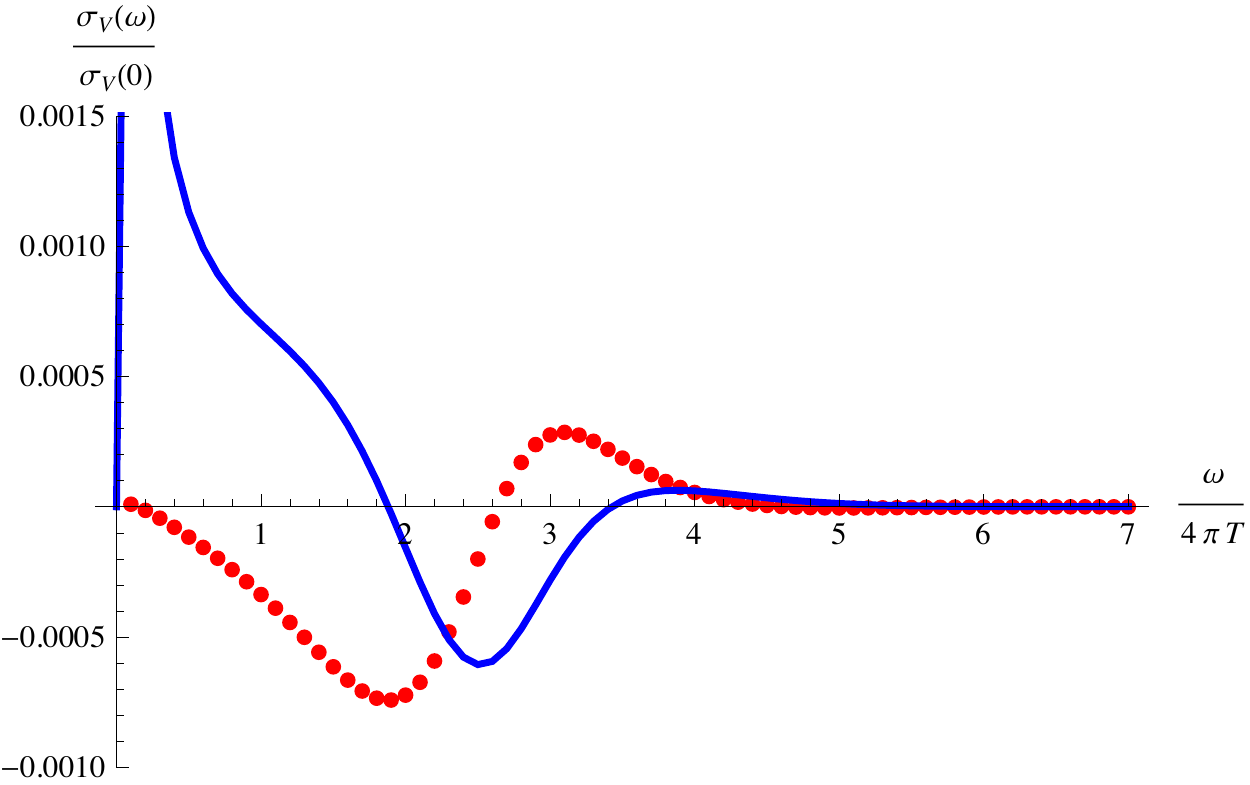} \\
\vspace{0.23cm}
\includegraphics[scale=0.58]{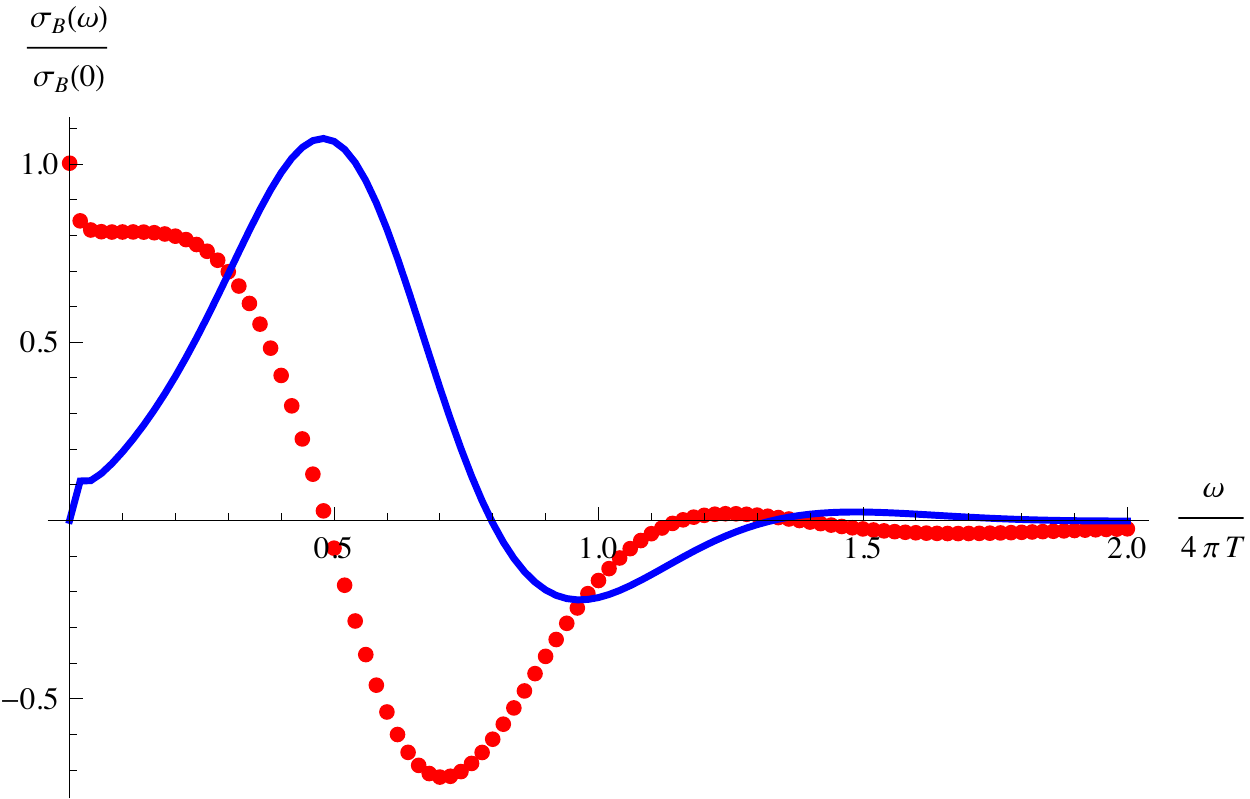} \hfill\includegraphics[scale=0.58]{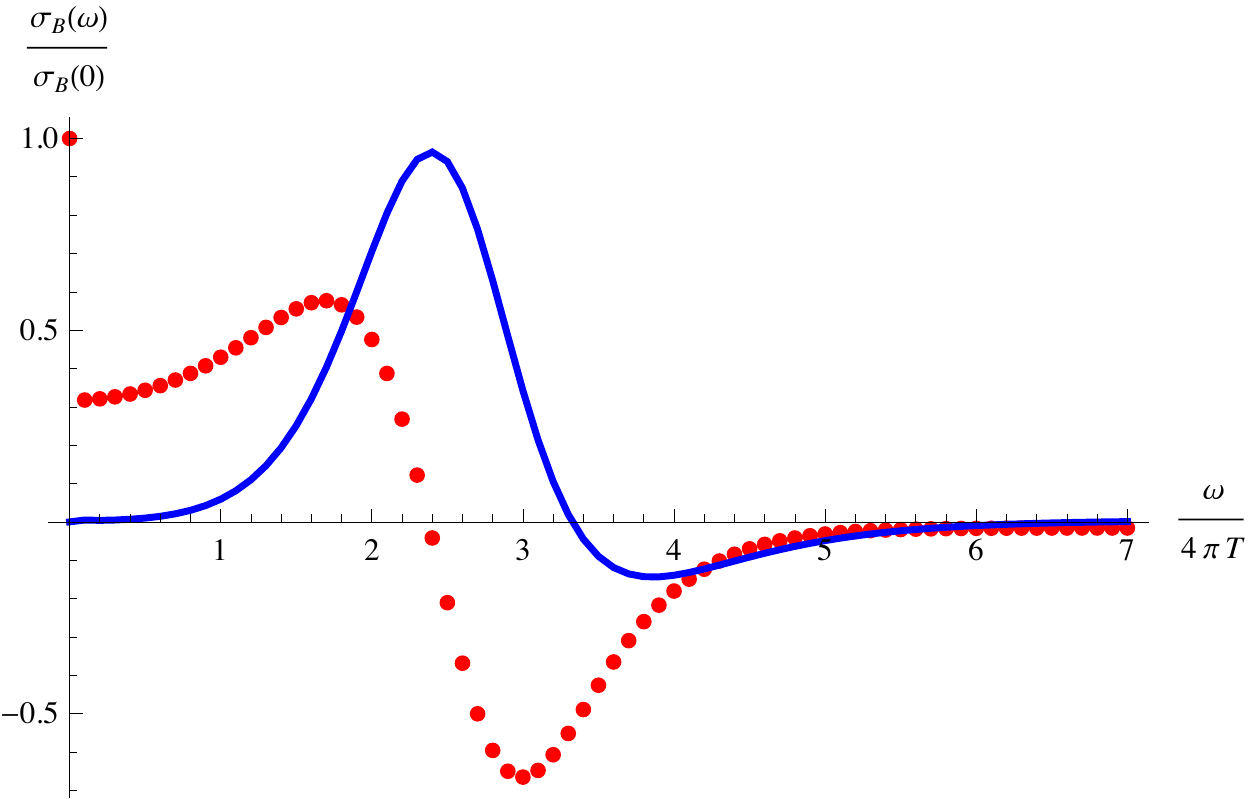} 
\end{center}
\caption{\label{plots1}Chiral vortical (up) and magnetic (bottom) conductivities as function of the frequency at $\tau=36.5$ (left) and $\tau=0.24$ (right). Red doted points represent real part and thick blue line the imaginary conductivity.}
\end{figure}

\begin{figure}[t!]
\begin{center}
\includegraphics[scale=0.5]{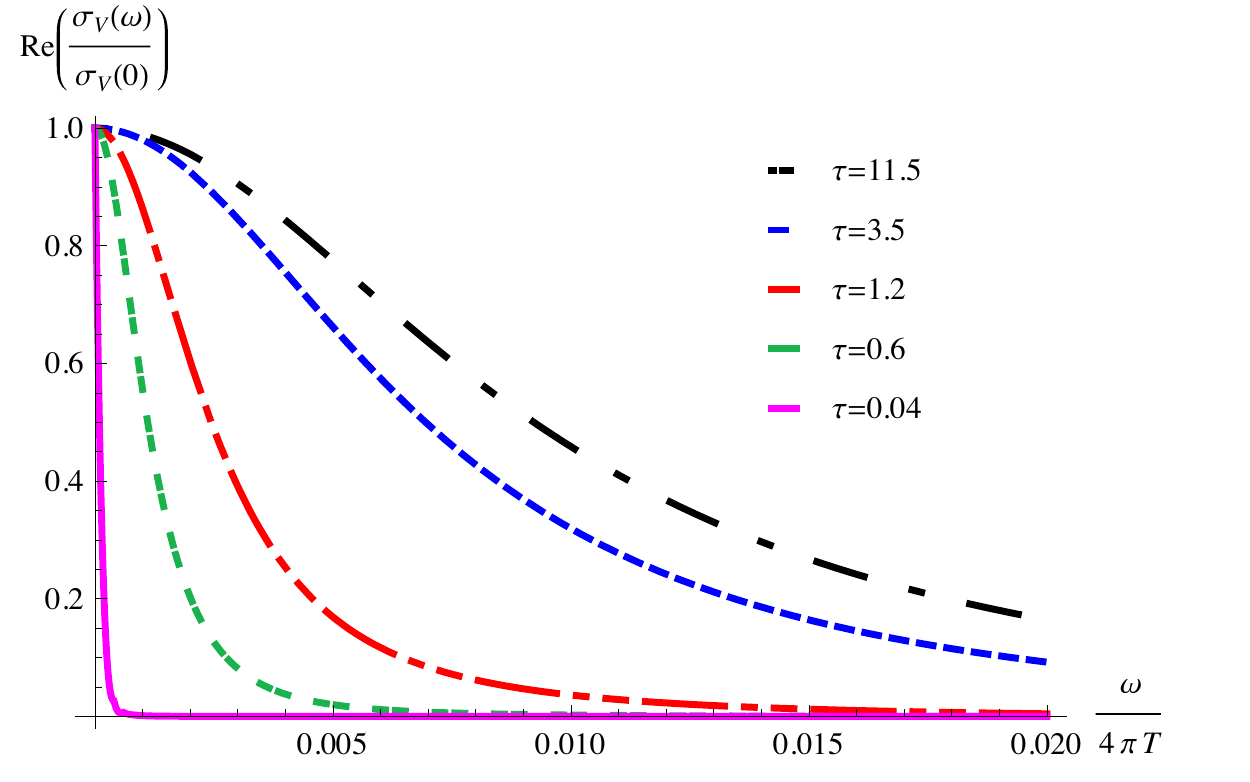}\hfill \includegraphics[scale=0.5]{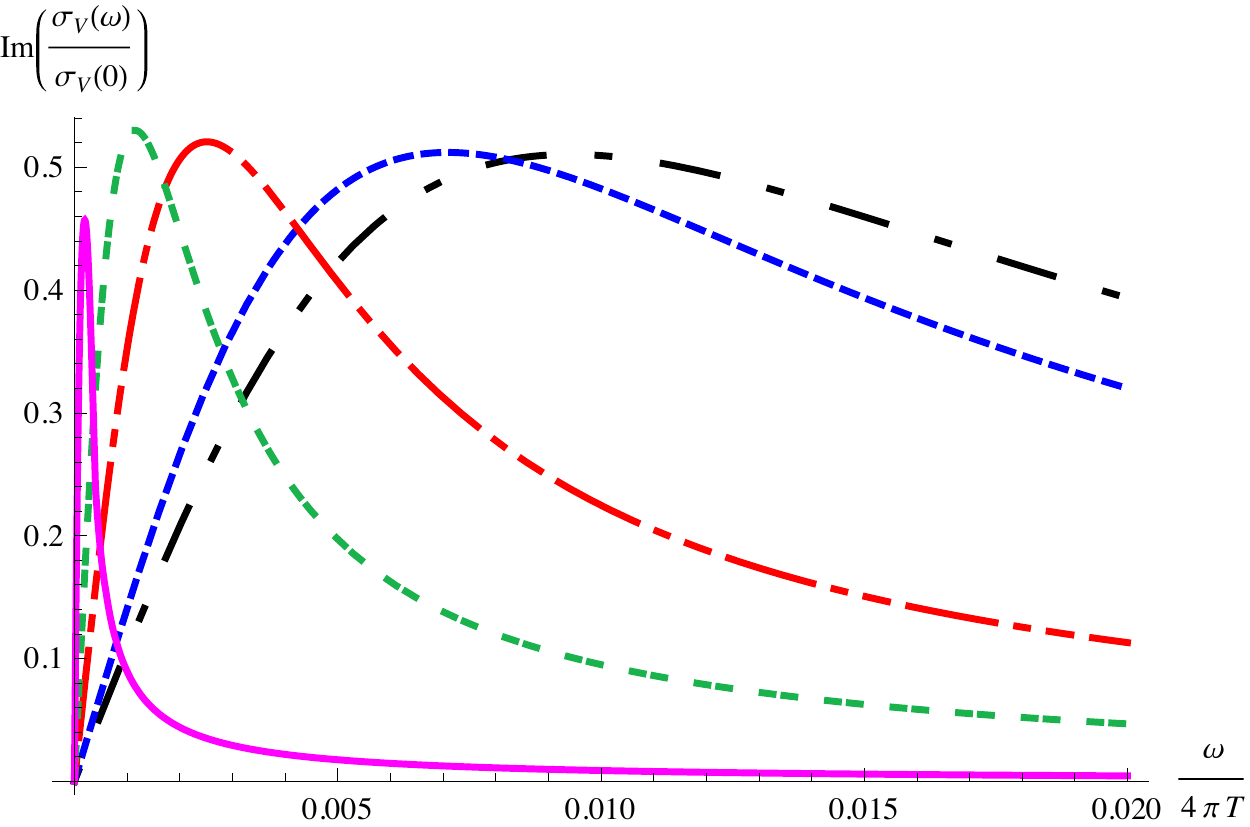} \\
\includegraphics[scale=0.5]{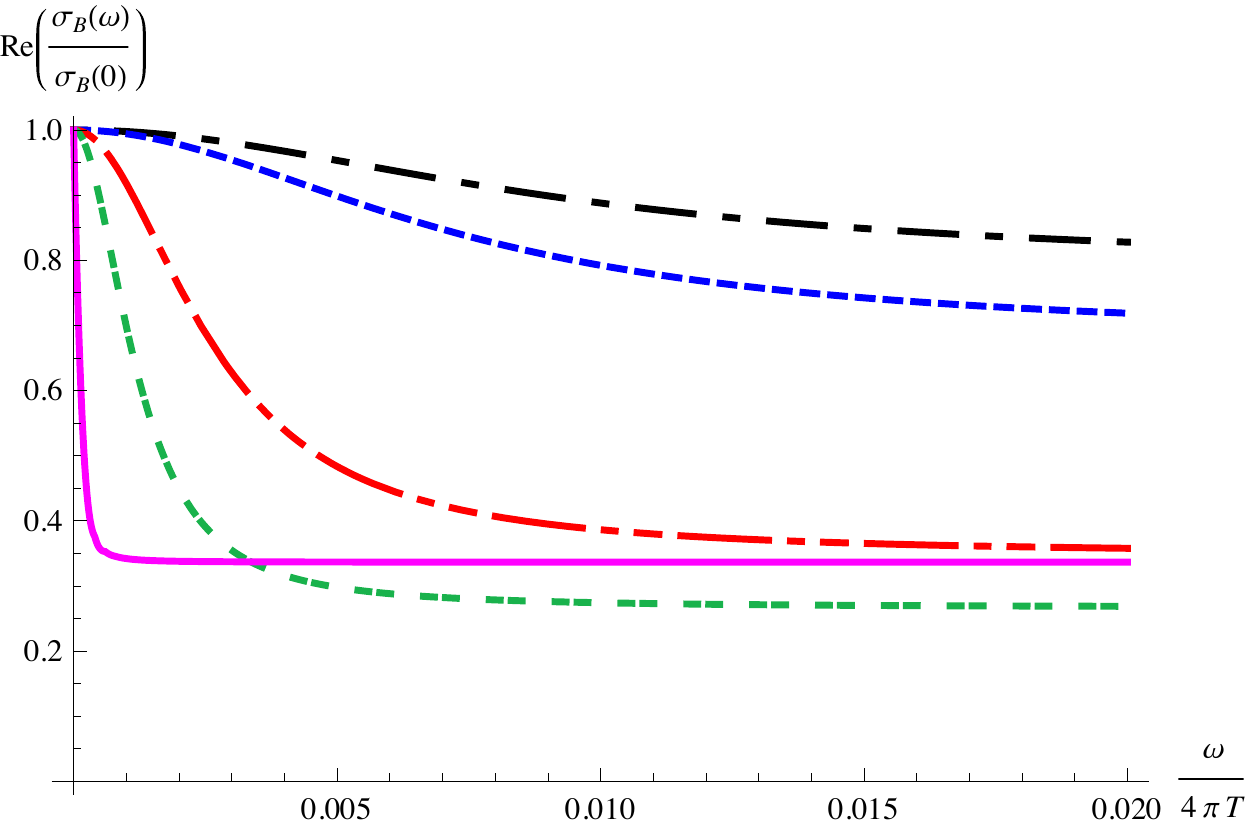} \hfill \includegraphics[scale=0.5]{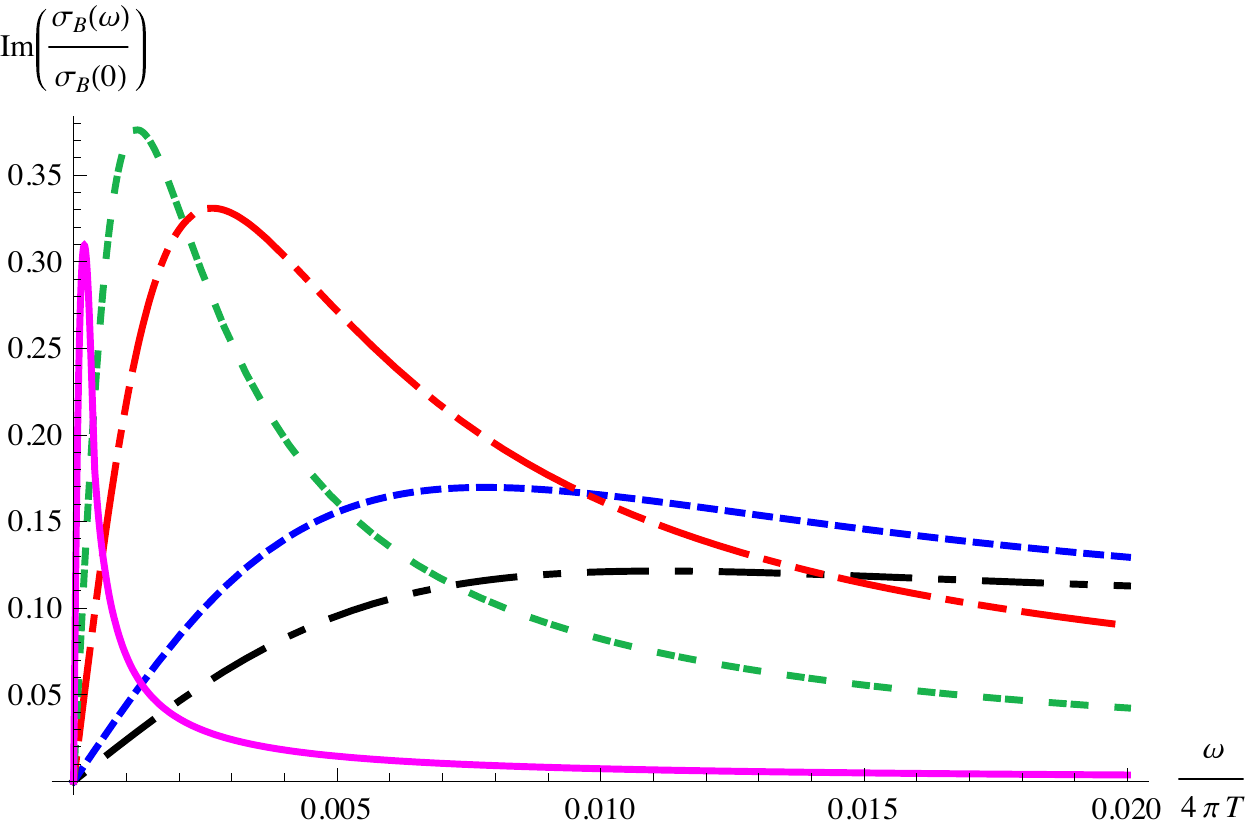} 
\end{center}
\caption{\label{plots2}Chiral vortical (up) and magnetic (bottom) conductivities as function of the frequency close to $\omega=0$. Real (left) and imaginary (right) part of the normalized conductivity for different values of the dimensionless temperature.}
\end{figure}

In Figure \ref{plots1} is illustrated the behavior of the vortical and magnetic conductivities as a function of frequency for two very different values of the dimensionless temperature $\tau=2\pi r_{+}T/\mu$. Both of them go to their corresponding zero frequency analytic result in the $\omega\to0$ limit. The frequency dependent chiral magnetic conductivity was also computed in \cite{Yee:2009vw}, though in that case the possible contributions coming from metric fluctuations were neglected. Our result for $\sigma_B(\omega)$ agrees pretty well with the result found in that work in the case of high temperature when the metrics fluctuations can be neglected and $\lambda=0$ (see Appendix \ref{ap:freqdep}), but it develops a dip close to $\omega=0$ when temperature is decreased (see Figure \ref{plots2}), due to the energy flow effect. For small temperatures, the chiral magnetic conductivity drops to $\sim1/3$ of its zero frequency value as soon as we move to finite frequency. The presence of the gravitational anomaly slightly introduce a dip close to $\omega=0$ also in the case of high temperature. In Appendix \ref{ap:freqdep} we can see how the kappa contribution is dominant for high and slow temperature in the chiral magnetic conductivity.
%, as it happens at weak coupling \cite{Kharzeev:2009pj}\footnote{It should be noted that the decoupling limit of holography is somewhat unnatural from the weak coupling field theoretical point of view. It corresponds to a field theory without energy momentum tensor in which space time symmetries such as translations or rotations are not realized in the operator algebra of the field theory observables. A Lagrangian formulation of a weakly coupled field theory does of course contain an energy momentum tensor. Therefore the weak coupling results do not easily compare to the ones obtained in the holographic decoupling limit.}. 
The behaviour of $\sigma_V$ is slightly different: the damping is much faster and the imaginary part remains small compared with the zero frequency value. But unlike to the chiral magnetic conductivity, at high temperature the dominant contribution to the conductivity $\sigma_V$ comes from the $\lambda$-term and at slow enough temperature the pure gauge anomaly is dominant (see also Appedinx \ref{ap:freqdep}).

\begin{figure}[h!]
\begin{center}
\includegraphics[scale=0.5]{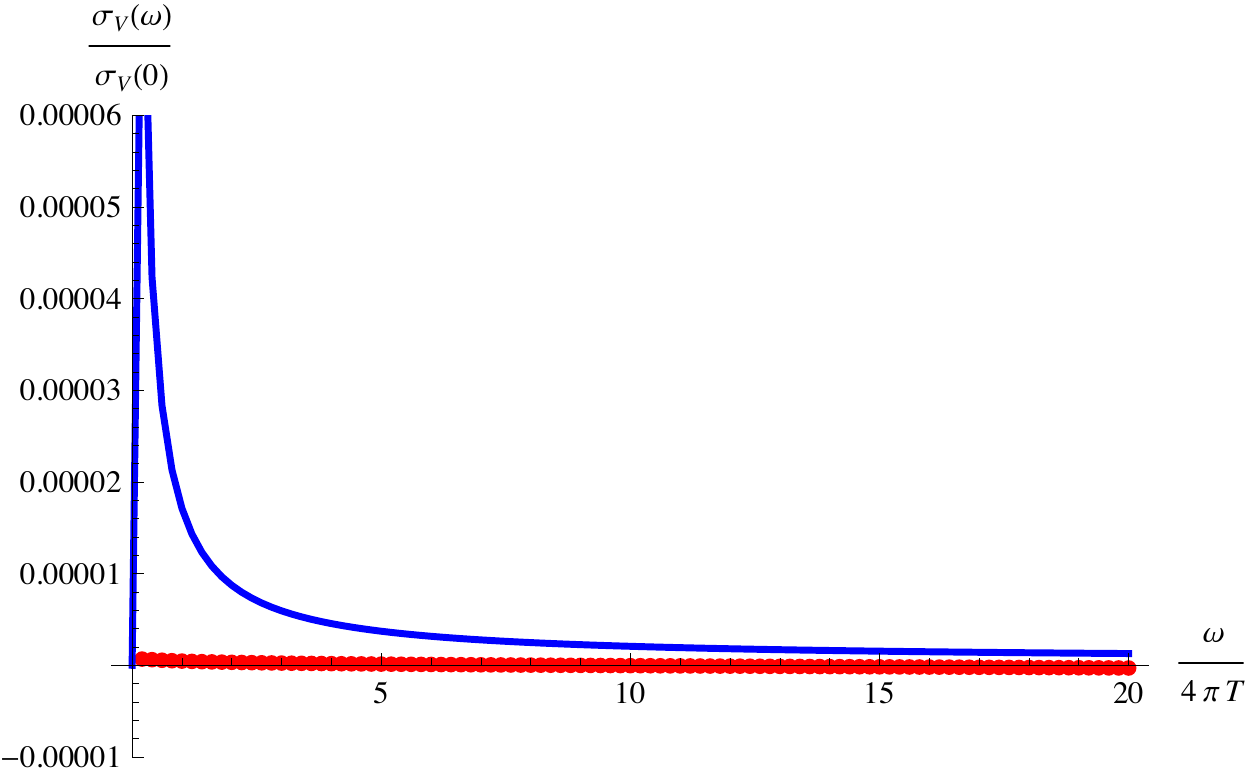}\hfill \includegraphics[scale=0.5]{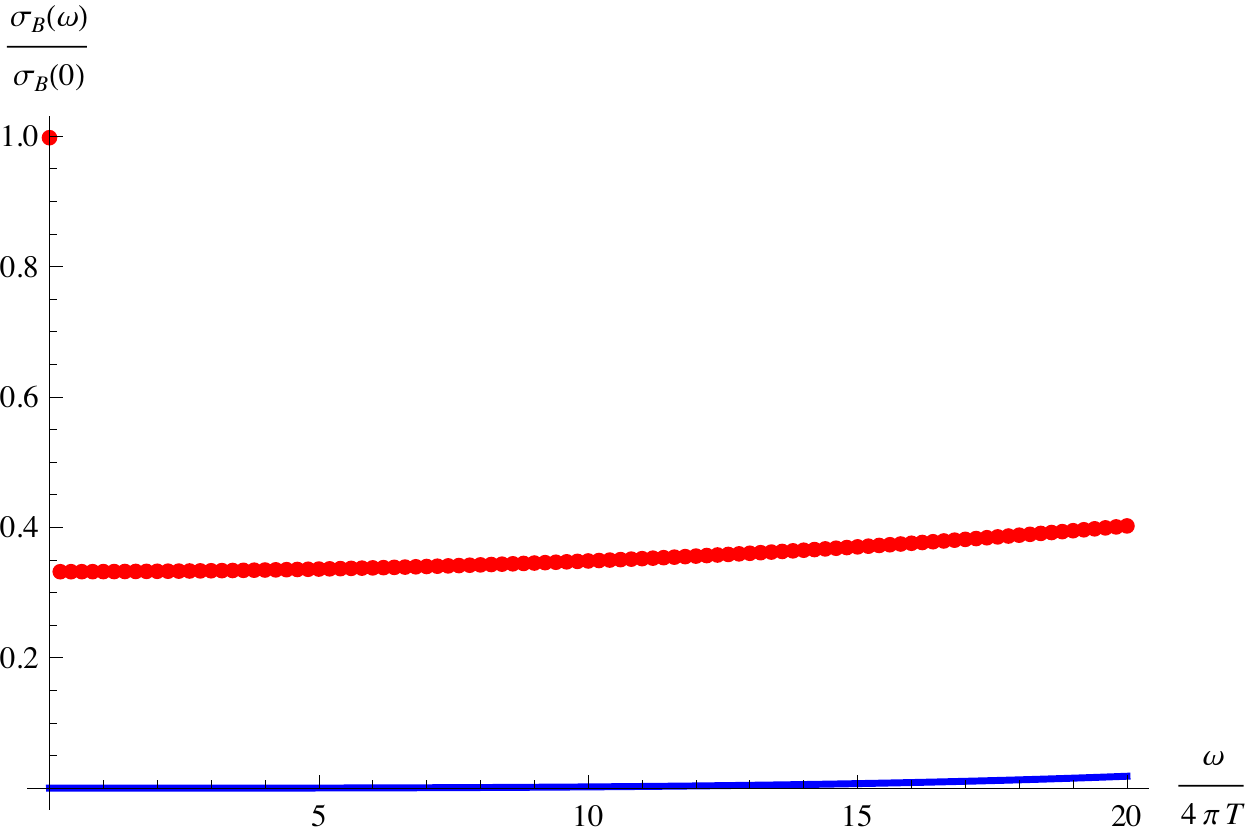}
\end{center}
\caption{\label{plots3}Chiral vortical (left) and magnetic (right) conductivities as function of the frequency at $\tau=0.008$.   Red doted points represent real part and thick blue line the imaginary conductivity.  The real part of $\sigma_V$ at $\omega=0$ is outside the range of the plot.}
\end{figure}

In Figure \ref{plots2} we made a zoom to smaller frequencies in order to see the structure of the dip on $\sigma_B$ and the faster damping on $\sigma_V$. In Figure \ref{plots3} we show the conductivities for very small temperature. From this plots we can infer that at zero temperature the conductivities behave like $\sigma_B = \alpha\sigma_B^0\left(1+\frac{1-\alpha}{\alpha}\delta(\omega)\right)$ and $\sigma_V=\sigma_V^0\delta(\omega)$ with $\alpha$ a constant value of order $1/3$.

% % %%%%%%%%%%%%%%%%%%%%%%%%%%%%%%%%%%%%%%%%%%%%%%%%%%%%%%%%%%%%%%%%%%
  \chapter{Second Order Transport}
  \label{fluid_grav}
  The fluid/gravity correspondence is a very powerful tool to understand the hydrodynamic regime of quantum field theories with holographic dual. This technique has contributed to the understanding of the positivity of the entropy production using techniques of black hole thermodynamics \cite{Bhattacharyya:2008ks,Loganayagam:2008is,Chapman:2012my}. It is also very useful for the computation of transport coefficients. In this chapter we will use this duality introduced in \cite{Bhattacharyya:2008jc} and based on the AdS/CFT correspondence to compute all the second order transport coefficients of the model introduced in the previous section \cite{Fluid:grav}.

\section{Fluid/Gravity Computation}
\label{sec:fluid_gravity}

As we saw above the system of bulk equations of motion~(\ref{eq:Gbulk}) and (\ref{eq:Abulk}) admits an AdS Reissner-Nordstr\"om black-brane solution of the form \footnote{We have set L=1}
\begin{eqnarray}
\label{eq:reiss_nords}
ds^2 &=& -r^2f(r)dt^2+\frac{dr^2}{r^2f(r)}+r^2dx^idx^i\,,\\
A &=&  \phi(r)dt  \,,
\end{eqnarray}
with $f(r)=1-m/r^4+q^2/r^6$ and $\phi(r)=-\sqrt{3}q/r^2$. The real and positive zeros of $f(r)$ are
\begin{eqnarray}
r_+ &=& \frac{\pi T}{2}\left(1 + \sqrt{1 + \frac{2}{3\pi^2}\bar{\mu}^2}\right)\,, \\
r_-^2 &=& \frac{1}{2}r_+^2\left( -1 + \sqrt{9 - \frac{8}{\frac{1}{2}\left( 1 + \sqrt{1 + \frac{2}{3\pi^2}\bar{\mu}^2} \right)}}\right)\,,
\end{eqnarray}
where $r_+$ is the outer horizon and $r_-$ the inner one. The mass of the black hole can be written in terms of hydro variables as
\begin{equation}
m = \frac{\pi^4T^4}{2^4}\left(1 + \sqrt{1 + \frac{2}{3\pi^2}\bar{\mu}^2}\right)^3\left(-1 + 3\sqrt{1 + \frac{2}{3\pi^2}\bar{\mu}^2}\right).
\end{equation}

The boosted version of this blackhole in Eddington-Finkelstein coordinates looks like
\begin{eqnarray}
ds^2 &=& -r^2f(r)u_\mu u_\nu dx^\mu dx^\nu + r^2P_{\mu\nu}dx^\mu dx^\nu-2u_\mu dx^\mu dr\,, \label{eq:ds2Eddington_Finkelstein}\\
A &=& -\phi(r)u_\mu  dx^\mu\,, \label{eq:A0}
\end{eqnarray}
with the normalization condition $u_\mu u^\mu=-1$. The fluid/gravity approach tells us that we have to promote all the parameters to slow varying functions of the space time coordinates, and include corrections to the metric in order to make it a solution of the equations of motion again, for a good review see \cite{Rangamani:2009xk}. 

In order to follow the fluid/gravity techniques \cite{Bhattacharyya:2008jc,Haack:2008cp,Banerjee:2008th,Erdmenger:2008rm}  we will use a Weyl invariant formalism \cite{Bhattacharyya:2008dl} in which the extraction of the transport coefficients is direct.
We start with the ansatz
\begin{eqnarray}\nonumber
ds^2 &=& -2W_1(\rho)u_\mu dx^\mu \left(dr^2+r\mathcal{A}_\nu dx^\nu\right) + \label{eq:weyl:ansatz}r^2\left[W_2(\rho)\eta_{\mu\nu}+W_3(\rho)u_\mu u_\nu + 2\frac{W_{4\sigma}(\rho)}{r_+}  P^\sigma_{\mu}u_\nu \right. \\
&& \left. +\frac{W_{5\mu\nu}(\rho)}{r_+^2}\right] dx^\mu dx^\nu \,,\\
A &=& \left(a^{(b)}_\nu+ a_\mu(\rho)P^\mu_\nu  + r_+ c(\rho) u_\nu\right)dx^\nu \,,
\end{eqnarray}
where $r_+$ is an unknown function $r_+(x)$ and coincides with the radius of the outer horizon of the black hole (Eq. (\ref{eq:reiss_nords})) when the $x^\mu$ dependence is gone. Notice that $\eta_{\mu\nu}$ is the Minkowski metric and $P^\mu_\nu$ is the projector defined in Eq. (\ref{eq:projector}) built up with it, so we will look for metric solutions with flat boundary. The $r$-coordinate has Weyl weight $+1$, in consequence $r_+$ has the same property. By construction the $W$'s are Weyl invariant, then they will depend on~$r$ only in a Weyl invariant way, i.e, $W(r)\equiv W(\rho)$ with $\rho=r/r_+$.  $W_{5\mu\nu}(r)$ obeys the traceless and transversality conditions $W_{5\mu}^\mu(r) =0$, $u^\mu W_{5\mu\nu}(r)=0$. All these scalars, vectors and tensors will be understood in term of a derivative expansion in the transverse coordinates, i.e. $F(\rho)=F^{(0)}(\rho)+ \varepsilon F^{(1)}(\rho)+\varepsilon^2 F^{(2)}(\rho) +O(\varepsilon^3)$ for a generic function $F$, with $\varepsilon$ a parameter counting the number of boundary space-time derivatives.\footnote{This function $F(\rho)$ is basically a Taylor expansion around the point $x_0^\mu$} This solution leads to the current and energy momentum tensor after using the AdS/CFT dictionary (see Eqs.(\ref{eq:Jcov}) and (\ref{eq:Tcov}))
\begin{eqnarray}
\label{Jreno} J_\mu  &=& \frac{1}{8\pi G}\lim_{\epsilon_*\to 0} \left(r^3_+c^{(\bar{2,}\epsilon_*)}u_\mu+ r_+^2 a_\mu^{(\bar{2},\epsilon_*)} + J^{ct}_\mu\right) \\
\label{Treno}\nonumber T_{\mu\nu}  &=& \frac{ 1}{16\pi G}\lim_{\epsilon_*\to 0}\left( r_+^4(W_2+W_3)^{(\bar{4},\epsilon_*)}( 4u_\mu u_\nu +\eta_{\mu\nu} ) \right.\\
 &&\left.\qquad\qquad \qquad+ 4r_+^2W_{5\mu\nu}^{(\bar{4},\epsilon_*)} + 8r_+^3W^{(\bar{4},\epsilon_*)}_{4\sigma} P^\sigma_{(\mu} u_{\nu)} + T^{ct}_{\mu\nu}\right) ,
\end{eqnarray}
where $F^{(\bar{n},\epsilon_*)}$ denotes the coefficient of the term $(\rho^{-1}-\epsilon_*)^{n}$ in an expansion around the regularized boundary, and $1/\epsilon_*$ is the UV cut-off. The counterterms in the current $J^{ct}_\mu$  and energy momentum tensor $T^{ct}_{\mu\nu}$ are needed to make the expressions finite, and they follow from the counterterm of the action~(\ref{eq:Sct}). They write
\begin{eqnarray}
J^{ct}_\mu &=&  \frac{1}{2}\log\epsilon_*\left[ (2\omega_\nu B^\nu - \mathcal{D}_\nu E^\nu)u_\mu +\mathcal{J}^{(7)}_\mu - \mathcal{J}^{(10)}_\mu - \mathcal{J}^{(5)}_\mu + \mathcal{J}^{(6)}_\mu\right]\,, \\
T^{ct}_{\mu\nu}  &=&  \log\epsilon_* \Big[ -\frac{1}{6}(B_\alpha B^\alpha +E_\alpha E^\alpha)P_{\mu\nu} -\frac{1}{2} ( E_\alpha E^\alpha + B_\alpha B^\alpha )u_\mu u_\nu  +  \mathcal{T}^{(9)}_{\mu\nu} +  \mathcal{T}^{(10)}_{\mu\nu} \nonumber \\
&&\qquad - (\mathcal{J}^{(9)}_\mu u_\nu + \mathcal{J}^{(9)}_\nu u_\mu )   \Big] \,. 
\end{eqnarray}

We are considering a flat background metric, and so the divergences appear only through terms involving electromagnetic fields, in addition to the cosmological constant contribution which was already taken into account in Eq. (\ref{Treno}).

The functions at zeroth order in the derivative expansion correspond to the boosted charged blackhole, i.e.~\footnote{Following the notation in~\cite{Erdmenger:2008rm}, barred superscripts $(\bar{n})$ should not be confused with superscripts~$(n)$, where the latter refers to the order in the hydrodynamical expansion.}
\begin{eqnarray}
c^{(0)}(\rho)&=&- \frac{\phi(\rho)}{r_+} \,, \\
W^{(0)}_1(\rho)&=& 1 = W^{(0)}_2(\rho) \,, \\
W^{(0)}_3(\rho)&=&1-f(\rho) \,, \\
W^{(0)}_{4\mu}(\rho)&=& 0 = W^{(0)}_{5\mu\nu}(\rho) \,, \\
a^{(0)}_\mu(\rho)&=&0 \,.
\end{eqnarray}
Then the charge current and energy momentum tensor at this order read~\cite{Erdmenger:2008rm,Banerjee:2008th},
\begin{equation}
 J^{(0)}_\mu  = \frac{\sqrt{3}q}{8\pi G} u_\mu \,, \qquad T^{(0)}_{\mu\nu}  = \frac{m}{16\pi G} \left( 4 u_\mu u_\nu +\eta_{\mu\nu} \right)  \,.
\end{equation}
From this we obtain the equilibrium pressure and charge density $p=\frac{m}{16\pi G}$ and $n=\frac{\sqrt{3}q}{8\pi G}$. For computational reasons it is convenient to define a Weyl invariant charge  $Q=q/r_+^3$ and mass $M=m/r_+^4=1+Q^2$. In terms of these redefined parameters, the black hole temperature and chemical potential read
\begin{eqnarray}
T&=&\frac{r_+}{2\pi}(2-Q^2) \,, \\
\mu &=& \sqrt{3}r_+ Q \,.
\end{eqnarray}
We also define the interior horizon in the $\rho$-coordinate $\rho_2\equiv r_-/r_+$.

\subsection{Einstein-Maxwell equations of motion and Ward identities}

Inserting this ansatz into the Einstein-Maxwell system of equations we find a set of $(2\times\mathbf{1}+2\times\mathbf{3}+\mathbf{5})$ differential equations and $(2\times\mathbf{1}+\mathbf{3})$ constraints relating the allowed $Q(x^\mu)$, $r_+(x^\mu)$, $u^\nu(x^\mu)$ and $a^{(b)}_\nu(x^\mu)$ \cite{Banerjee:2008th}.~\footnote{$\mathbf{1}$, $\mathbf{3}$ and $\mathbf{5}$ denote the $SO(3)$ scalars, vectors and tensors in which the fields are decomposed.} We need to solve the e.o.m. about a a certain point $x_0\mu$ that we choose to be $x_0^\mu=0$. At such point we sit in a frame in which $u^\mu=(1,0,0,0)$ and $a^{(b)}_\nu(0)=0$.

The scalar sector is obtained from the $rr$, $rv$ and $vv$ components of the Einstein equations, and the $r$ and $v$ components of the  Maxwell equations. One finds two constraints 
\begin{eqnarray}
\label{eq:djexp}M^{(n)}_v + r^2f(r)M^{(n)}_r &=& 0\quad \Rightarrow\quad \left(\mathcal{D}_\mu J^{\mu}=c_1F\wedge F\right) ^{(n-1)} \,, \\
\label{eq:dtexp}E^{(n)}_{vv} + r^2f(r) E^{(n)}_{rv} &=& 0 \quad \Rightarrow\quad \left(\mathcal{D}_\mu T^\mu_v=F_{v\alpha}J^\alpha\right)^{(n-1)} \,,
\end{eqnarray}
which, as indicated, correspond to the current and energy momentum non-conservation relations at order~$n-1$.\footnote{Notice that there are no curvature terms in (\ref{eq:djexp}) and (\ref{eq:dtexp}) because we are working with a flat boundary.} The combinations $E_{rr}=0$, $E_{rv} + r^2f(r) E_{rr} = 0$ and $M_r=0$ leads respectively to the set of differential equations
\begin{eqnarray}
\label{ES1} 3 W_1^{(n)}\,'(\rho) - \frac{3}{2}\rho^{-1}\left(\rho^2\partial_\rho W^{(n)}_{2}(\rho)\right)' &=&S^{(n)}(\rho) \,, \\
\label{ES2}  \left(\rho^4 W_3^{(n)}\right)' + 8\rho^3W_1^{(n)} - \sqrt{2}Q  c^{(n)}\,' +(1-4\rho^4) W^{(n)}_2\,' -4\rho^3 W_2^{(n)} &=&K^{(n)}(\rho) \,,\\
\label{MS}  \left(\rho^3\partial_\rho c^{(n)} \right)' - 2\sqrt{3}Q  W^{(n)}_1\,' +3\sqrt{3}Q W_2^{(n)}\,' &=&C^{(n)}(\rho) \,.
\end{eqnarray}
At this stage there is still some gauge freedom in the metric. There are  three (metric) scalar fields~$(W_1\,, W_2$ and $ W_3 )$ but the Einstein's equations give two differential equations, (\ref{ES1}) and (\ref{ES2}). We choose the gauge $W_2(\rho)=1$ in which the system partially decouples and can be solved as
\begin{eqnarray}
W_1^{(n,\epsilon)}(\rho) &=& -\frac{1}{3}\int_\rho^{\frac{1}{\epsilon}} \mathrm{d}x  \,\mathbb S^{(n)}(x)\,,\\
W_3^{(n,\epsilon)}(\rho) &=&  \frac{C_0}{\rho^4} -  \frac{1}{\rho^4}\int_\rho^{\frac{1}{\epsilon}} \mathrm{d}x   \left(\mathbb K^{(n)}(x) - 8 x^3 W_1^{(n,\epsilon)}(x) + \frac{2Q}{\sqrt{3}}\partial_x c^{(n,\epsilon)}(x) \right)\,,\\
c^{(n,\epsilon)}(\rho) &=& c_0\frac{(1-\epsilon^2\rho^2)}{\rho^2} - \int_\rho^{\frac{1}{\epsilon}}\mathrm{d}x \, x^{-3}\int_1^x \mathrm{d}y \left(\mathbb C^{(n)}(y)+\frac{2Q}{\sqrt{3}}\mathbb S^{(n)}(y)\right)  \,. 
\end{eqnarray} 
These solutions have been constructed by requiring Dirichlet boundary conditions at the cut-off surface and demanding regularity at the interior of the bulk. The remaining integration constant $c_0$ is associated to the freedom of choosing frame in the hydrodynamic set up.

In a similar way, the vector sector is constructed with the components of the equations of motion $E_{ri}$, $E_{vi}$ and $M_i$. They lead to a constraint equation,
\begin{equation}
E^{(n)}_{vi} + r^2f(r) E^{(n)}_{ri} = 0\quad \Rightarrow\quad \left(\mathcal{D}_\mu T^\mu_i=F_{i\alpha}J^\alpha\right)^{(n-1)} \,,
\end{equation}
implying the energy conservation equation, and the two dynamical equations
\begin{eqnarray}
\partial_\rho\left(\rho^5\partial_\rho W^{(n)}_{4i} + 2\sqrt{3}Q a^{(n)}_i(\rho)\right) &=&\mathbb J^{(n)}_{i}(\rho) \,, \\
\partial_\rho\left(\rho^3f(\rho)\partial_\rho a^{(n)}_{i}(\rho) +2\sqrt{3}Q\partial_\rho W^{(n)}_{4i}(\rho)\right)&=&\mathbb  A^{(n)}_{i}(\rho) \,, 
\end{eqnarray}
corresponding to $E_{ri} = 0$ and $M_i = 0$ respectively. The general solution of this system in the Landau frame has been found in \cite{Erdmenger:2008rm}.  It is straightforward to generalize the solution to the case in which electromagnetic sources are included. In this case new divergences arise that need to be regulated with the cut-off $1/\epsilon_*$, and then to be substracted with the corresponding counterterms~(\ref{eq:Sct}). The result is
\begin{equation}
\label{a2} a_\nu^{(\bar{2},\epsilon_*)} = \frac{1}{2}\left(\frac{1}{\epsilon_*}\mathbb A_\nu\left(1/\epsilon_*\right) - \int_1^{\frac{1}{\epsilon_*}}\mathrm{d}x \, \mathbb A_{\nu}(x)\right) - \frac{\sqrt{3}Q}{M}C^{(\epsilon_*)}_\nu - \frac{\sqrt{3}Q}{4M}D^{(\epsilon_*)}_\nu,
\end{equation}
where the integration constant $C_\nu^{(\epsilon_*)}$ is determined by fixing the Landau frame, and $D_\nu^{(\epsilon_*)}$ by demanding regularity at the outer horizon. These constants write
\begin{eqnarray}
4C^{(\epsilon_*)}_\nu &=& - \sum_{m=0}^2\frac{(-1)^m\partial^m_\rho \mathbb J_\nu(1/\epsilon_*)}{\epsilon_*^{m+1}(m+1)!} + \int_1^{\frac{1}{\epsilon_*}} \mathrm{d}x \, \mathbb J_\nu(x) - \mathcal J_\nu^{(9)}\log\epsilon_*  \,, \label{eq:4Cnu}\\
D^{(\epsilon_*)}_\nu &=&  - \sqrt{3}Q\int_1^{\frac{1}{\epsilon_*}} \mathrm{d}x \, \frac{\mathbb A_\nu(x)}{x^2} - M\int_1^{\frac{1}{\epsilon_*}} \mathrm{d}x \,  \frac{\mathbb J_\nu(x)}{x^4} - Q^2\int_1^{\frac{1}{\epsilon_*}} \mathrm{d}x \, \frac{\mathbb J_\nu(x)}{x^6}\,. \label{eq:Dnu}
\end{eqnarray}

Finally the tensor equations are the combination $E_{ij}-\frac{1}{3}\delta_{ij}\tr(E_{kl})=0$, which leads to the dynamical equation
\begin{equation}
\partial_\rho\left(\rho^5f(\rho)\partial_\rho W^{(n)}_{5ij}(\rho)\right)=\mathbb P^{(n)}_{ij}(\rho) \,.
\end{equation}
The solution of this equation that satisfies the Dirichlet boundary  and regularity conditions writes
\begin{equation}
W^{(n,\epsilon_*)}_{5\mu\nu}(\rho) = -\int_\rho^{\frac{1}{\epsilon_*}} \mathrm{d}x \, \frac{\int_1^x \mathrm{d}y \, \mathbb P^{(n)}_{\mu\nu}(y)}{x^5f(x)}\,.
\end{equation}
After doing an asymptotic expansion of this solution around the regularized boundary surface, one can extract the relevant quantity to get the energy momentum tensor, cf.~(\ref{Treno}),
\begin{equation}\label{eq:W45}
4W^{(\bar{4},\epsilon_*)}_{5\mu\nu} = - \sum_{m=0}^2\frac{(-1)^m\partial^m_\rho \mathbb P_{\mu\nu}(1/\epsilon_*)}{\epsilon_*^{m+1}(m+1)!} - \int_1^{\frac{1}{\epsilon_*}} \mathrm{d}x \,  \mathbb P_{\mu\nu}(x) \,.
\end{equation}

Note that the form of the homogeneous part in the dynamical equations in the scalar, vector and tensor sectors is the same at any order in the derivative expansion. Each order~$n$ is then characterized by the specific form of the sources. In the next two sections we will compute the sources, and integrate them according to the formulae presented above to get the transport coefficients at first and second order.

\section{First Order Transport Coefficients}
\label{sec:first_order}

The technology presented in Sec.~\ref{sec:fluid_gravity} can be used to construct the solutions of the system at any order in a derivative expansion. As it has been already explained, the solution at zeroth order trivially leads to the charged blackhole with constant parameters~(\ref{eq:ds2Eddington_Finkelstein})-(\ref{eq:A0}). In this section we will solve the system up to first order. The transport coefficients at this order have been obtained previously in the literature using different methods in field theory and holography. In particular, they have been computed within the fluid/gravity approach, but not including external electric fields in this formalism, see eg.~\cite{Erdmenger:2008rm,Banerjee:2008th,Chapman:2012my,Son:2009tf}.

\subsection{Scalar sector}
In the scalar sector, the first order sources look like
\begin{equation}
S^{(1)}(\rho) = K^{(1)}(\rho) =C^{(1)}(\rho) = 0 \,.
\end{equation} 
This very simple situation leads to the solution
\begin{eqnarray}
W_1^{(1,\epsilon)}(\rho) &=& 0 \,, \\
c^{(1,\epsilon)}(\rho) &=& c_0\frac{(1-\epsilon^2\rho^2)}{\rho^2}  \,, \\
W_3^{(1,\epsilon)}(\rho) &=&  \frac{C_0}{\rho^4} + \frac{2Qc_0}{\sqrt{3}}\frac{(1-\epsilon^2\rho^2)}{\rho^6}\,.
\end{eqnarray}
The integration constants $c_0$ and $C_0$ can be fixed to zero because they just redefine the charge and mass of the black hole respectively.
\subsection{Vector and tensor sector}
The first order sources  are given by
\begin{eqnarray}
\mathbb J^{(1)}_{\mu} &=&  -\lambda\frac{96}{\rho^3}\left(\frac{5 Q^2  }{ \rho ^2} - M\right)\frac{B_\mu}{r_+} - \sqrt{3} Q\lambda\left(\frac{1008 Q^2  }{\rho ^7}-\frac{320 M }{\rho ^5}\right) \omega_\mu \,, \\
\nonumber\mathbb A^{(1)}_{\mu} &=& - \frac{\sqrt{3} \pi  T}{M r_+ \rho ^2}P_\mu^{\nu}\mathcal{D}_\nu Q  - \left(1+\frac{9 Q^2}{2 M \rho ^2}\right)\frac{E_\mu}{r_+} - \frac{16\sqrt{3}\kappa Q}{\rho^3}  \frac{B_\mu}{r_+}  - \frac{48\kappa Q^2}{\rho^5} \omega_\mu \\
&&-\frac{48 \lambda  \left(15 Q^4-16 M Q^2 \rho ^2+4 M^2 \rho ^4\right)}{\rho ^{11}} \omega_\mu  \,, \\
\mathbb P^{(1)}_{\mu\nu} &=& -6r_+\rho^2\sigma_{\mu\nu}  \,,
\end{eqnarray}
where $\mathcal{D}_\mu$ is the Weyl covariant derivative and $\mathcal{D}_\alpha Q =\frac{2\pi T^2}{\sqrt{3}r_+^2(1+M)}\mathcal{D}_\alpha\bar{\mu}$. Using equations (\ref{Jreno}),  (\ref{Treno}), (\ref{a2}), (\ref{eq:4Cnu}) and (\ref{eq:Dnu}) it is straightforward to find the first order transport coefficients,
\begin{eqnarray}
\eta &=& \frac{r_+^3}{16\pi G} \,, \quad \sigma =\frac{\pi  r_+^7 T^2}{16 G m^2} \,,\,\,\\
\xi_B &=& -\frac{\sqrt{3} q \left(m+3 r_+^4\right) \kappa }{8 G m \pi  r_+^2}+\frac{\sqrt{3} \pi  q T^2 \lambda }{G m} \,, \,\,\,  \xi_V = -\frac{3 q^2 \kappa }{4 G m \pi }+\frac{2 \pi  \left(2 q^2-r_+^6\right) T^2 \lambda }{G m r_+^2} \,.\,\,
\end{eqnarray}
Chiral magnetic $\xi_B$ and vortical $\xi_V$ conductivities have been computed at first order in holography within the Kubo Formulae formalism in~\cite{Landsteiner:2011iq}. Here we reproduce the same result within the fluid/gravity approach.~\footnote{The gauge gravitational anomaly contribution to the vortical conductivity was also computed recently within the Fluid/Gravity setup in~\cite{Chapman:2012my}.}

Note that to compute the first order transport coefficients one needs only the terms $a_\mu^{(\bar{2},\epsilon_*)}$ and $W^{(\bar{4},\epsilon_*)}$ in the near boundary expansion. However, in order to go to the next order in the derivative expansion, we need to know the exact solutions, which can be written in terms of the sources as
\begin{eqnarray}\label{eq:W4}
W_{4\mu}^{(1)}(\rho) &=& F_1 [\rho]P_\mu^\nu\mathcal{D}_\nu Q(x) + F_2[\rho]\omega_\mu(x) + F_3[\rho]\frac{E_\mu(x)}{r_+} +  F_4[\rho]\frac{B_\mu(x)}{r_+} \,,\\
W_{5\mu\nu}^{(1)}(\rho) &=&  F_5[\rho]\, r_+ \, \sigma_{\mu\nu}(x) \,, \\
\label{eq:amu} a_\mu^{(1)}(\rho) &=&  F_6[\rho]P_\mu^\nu\mathcal{D}_\nu Q(x) +  F_7[\rho]\omega_\mu(x) +  F_8[\rho]\frac{E_\mu(x)}{r_+} +  F_9[\rho]\frac{B_\mu(x)}{r_+} \,,
\end{eqnarray}
 we show in Appendix~\ref{ap:Ffunctions} the expressions for the $F$'s functions. $F_5$ writes
\begin{eqnarray}
\nonumber F_5[\rho] &=& -\frac{2 \log{[1+\rho ]}}{-1+M}-\frac{\left(1+\rho_2+\rho_2^2\right) \log{[\rho -\rho_2]}}{(1+\rho_2) \left(1+2 \rho_2^2\right)}+\frac{2 \left(1+\rho_2^3\right) \log{[\rho +\rho_2]}}{-2-2 \rho_2^2+4 \rho_2^4} \\
 &&+\frac{\log{\left[1+\rho ^2+\rho_2^2\right]}}{2+5 \rho_2^2+2 \rho_2^4}+\frac{2 \left(1+\rho_2^2\right)^{3/2}}{2+5 \rho_2^2+2 \rho_2^4}  \,\mathrm{ArcCot}\left[\frac{\rho }{\sqrt{1+\rho_2^2}}\right] \,.
\end{eqnarray}

\section{Second Order Transport Coefficients}
\label{sec:second_order}

The second order coefficients are much more computationally demanding than the first order ones. The parameter $c^{(\bar{2},\epsilon_*)}$ in~(\ref{Jreno}) can always be chosen to be zero, as it just redefines the charge. On the other hand, because we are working with a conformal fluid in the Landau frame, there is no contribution from the scalar sector to the energy momentum tensor and $(W_2 + W_3)^{(\bar{4},\epsilon_*)}$ is set to zero. We have checked that this is in fact what happens by using the sources for the scalar sector. So, we will focus in this section on the vector and tensor contributions.

\subsection{Vector sector}

The second order sources in the vector sector are shown in the Appendix \ref{ap:Vec_sour}, again using these expressions and the Eqs. (\ref{a2}), (\ref{eq:4Cnu}), (\ref{eq:Dnu}) and (\ref{Jreno}) we can extract the second order transport coefficients. We show first the new non anomalous coefficients
\begin{eqnarray}
\xi_5 &=& \xi_{5,0}(\rho_2) \,, \label{eq:xi5} \\
\xi_6 &=&   \xi_{6,0}(\rho_2) +\frac{3 \left(3+M^2\right)Q^2 \kappa ^2}{4\pi  G M^3 } + \kappa\lambda \xi_{6,\kappa\lambda}(\rho_2) + \lambda^2 \xi_{6,\lambda^2}(\rho_2)  \,, \\
\xi_7 &=& \xi_{7,0}(\rho_2) \,, \\
\xi_8 &=& -\frac{(9 + 12 M + 7 M^2) \pi Q T^3}{128 \sqrt{3} G M^4 (1 + M) r_+^3}   + \kappa^2 \xi_{8,\kappa^2}(\rho_2)+ \kappa\lambda \xi_{8,\kappa\lambda}(\rho_2) + \lambda^2 \xi_{8,\lambda^2}(\rho_2)  \,, \\
\nonumber\xi_9 &=& \frac{Q \left(88+480Q^2M +169Q^6\right)}{512\sqrt{3}\pi  G M^4  r_+} + \frac{1}{r_+}\left( \kappa^2 \xi_{9,\kappa^2}(\rho_2)(\rho_2)\quad \right.\\
&&+\left. \kappa\lambda \xi_{9,\kappa\lambda}(\rho_2)(\rho_2) + \lambda^2 \xi_{9,\lambda^2}(\rho_2)(\rho_2)  \right) \,, \\
\xi_{10} &=& \frac{\left(4+7 Q^2\right)}{64\pi G M  } + \kappa^2 \xi_{10,\kappa^2}(\rho_2) + \kappa\lambda \xi_{10,\kappa\lambda}(\rho_2)+ \lambda^2 \xi_{10,\lambda^2}(\rho_2)   \,. \label{eq:xi10}
\end{eqnarray}
These coefficients had not been computed previously in the literature. The rest of the non anomalous coefficients were obtained in the past without the gravitational anomaly. In this work we have found the $\lambda-$corrected results, which write
\begin{eqnarray}
\xi_1 &=& \frac{\pi  T^3}{8GM^3 (M+1) r_+^2 } \left( Q^2  + \frac{ M^2}{\left(1+2 \rho_2^2\right)} \log{\left[\frac{2+\rho_2^2}{1-\rho_2^2}\right]} \right) \,, \label{eq:xi1} \\
\nonumber \xi_2 &=&  \frac{(3+M) (M (3+M)-6) T^2}{128 G M^3 (M+1) r_+} + \frac{3 \pi  Q^2 T^3 \kappa ^2}{G M^3 (M+1) r_+^2} + r_+\left( \kappa\lambda \xi_{2,\kappa\lambda}(\rho_2)+\lambda^2\xi_{2,\lambda^2}(\rho_2) \right) \,, \\
&&\\
\xi_3 &=& \frac{3 \sqrt{3} Q^3 r_+}{64 \pi G M^2 } \,, \\
\xi_4 &=& \frac{3 \sqrt{3} Q^3 r_+ \kappa ^2}{2\pi  G M^2 } + r_+\kappa\lambda \xi_{4,\kappa\lambda}(\rho_2)+ r_+\lambda^2 \xi_{4,\lambda^2}(\rho_2)  \label{eq:xi4}\,.
\end{eqnarray}
In the anomalous sector, the new coefficients (not computed previously) write
\begin{eqnarray}
\tilde{\xi}_2 &=& \frac{3 \sqrt{3} Q^3 \left(6+M\right) \kappa }{16\pi G M^2   \left(1+2 \rho_2^2\right)^2}+\frac{\sqrt{3} \pi  Q T^2 \kappa  \log{\left[\frac{2+\rho_2^2}{1-\rho_2^2}\right]}}{2G M r_+^2 \left(1+2\rho_2^2\right)^3} + \lambda\tilde \xi_{2,\lambda}(\rho_2) \,, \label{eq:xitilde2} \\
\tilde{\xi}_3 &=& \frac{\sqrt{3}\pi Q T^2}{8r_+^2 M^3G }\left(  Q^2  \kappa +\frac{8  \pi ^2  T^2 \lambda }{ r_+^2}\right) \,, \label{eq:xitilde3}\\
\tilde{\xi}_4 &=& \kappa \tilde \xi_{4,\kappa}(\rho_2) + \lambda\tilde \xi_{4,\lambda}(\rho_2) \,, \\
\tilde{\xi}_5 &=& \kappa \tilde \xi_{5,\kappa}(\rho_2) + \lambda\tilde \xi_{5,\lambda}(\rho_2)\,,  \label{eq:xitilde5}
\end{eqnarray}
while the already known coefficient with the new $\lambda$ contribution writes
\begin{eqnarray}
\tilde{\xi}_1 &=& \frac{3Q^2 r_+ \kappa }{4\pi G M^2  } + \lambda r_+\tilde \xi_{1,\lambda}(\rho_2) \,.  \label{eq:xitilde1}
\end{eqnarray}
The $\xi_{i,(0,\kappa^2,\kappa\lambda,\lambda^2)}(\rho_2)$ and $\tilde\xi_{i,(\kappa,\lambda)}(\rho_2)$ functions are defined in Appendix~\ref{ap:fsecond_order}. These coefficients enter in the constitutive relation for the current through~(\ref{eq:fullconstiJ}) and (\ref{eq:2consti}).

\subsection{Tensor sector}
The second order sources in the tensor sector are shown in Appendix~\ref{ap:Ten_sour}, and again we can extract the transport coefficients at this order after pluging these expressions into Eqs.~(\ref{Treno}) and (\ref{eq:W45}). Due to the length of the expressions, some of them will be shown exactly and the rest are expressed in terms of some functions $\Lambda_{i,(\kappa^2,\kappa\lambda,\lambda^2)}(\rho_2)$ and $\tilde \Lambda_{i,(\kappa,\lambda)}(\rho_2)$ which are presented in the Appendix~\ref{ap:fsecond_order}. Again we split our results in  those non anomalous coefficients which are new,
\begin{eqnarray}
\Lambda_7 &=& -\frac{\sqrt{3} (-3+5 M) Q r_+}{64\pi G M^2  } \,, \label{eq:Lambda7}\\
\Lambda_8 &=& r_+\Lambda_{8,0}(\rho_2) \,, \\
\Lambda_9 &=& \Lambda_{9,0}(\rho_2) \,, \\
\Lambda_{10} &=& \frac{11}{96 \pi G} + \kappa^2 \Lambda_{10,\kappa^2}(\rho_2) + \kappa\lambda \Lambda_{10,\kappa\lambda}(\rho_2) + \lambda^2 \Lambda_{10,\lambda^2}(\rho_2) \,, \\
\Lambda_{11} &=& -\frac{8 \sqrt{3} Q r_+ \kappa  \lambda }{\pi G  } \,, \\
\Lambda_{12} &=& -\frac{\sqrt{3} Q r_+}{16\pi G  } + \frac{3 \sqrt{3} Q^3 r_+ \kappa ^2}{\pi G M } + \kappa\lambda r_+\Lambda_{12,\kappa\lambda}(\rho_2) + \lambda^2r_+ \Lambda_{12,\lambda^2}(\rho_2) \,, \label{eq:lambda12}
\end{eqnarray}
and the rest of the non anomalous ones
\begin{eqnarray}
\Lambda_1 &=& \frac{r_+^2 }{16\pi G  }\left(2+\frac{M }{\sqrt{4 M-3}}\log\left[\frac{3-\sqrt{4 M-3}}{3+\sqrt{4 M-3}}\right]\right) \,, \label{eq:lambda1} \\
\Lambda_2 &=& \frac{r_+^2}{8 \pi G  } \,, \\
\Lambda_3 &=& \frac{r_+^2}{8 G \pi } \left( \frac{ M }{2(1+2 \rho_2^2)}\log{\left[\frac{2+\rho_2^2}{1-\rho_2^2}\right] + 192 Q^2 \kappa  \lambda -\frac{384 (3 M-5) \pi  T \lambda ^2}{r_+}} \right) \,,  \\
\nonumber \Lambda_4 &=& -\frac{Q^2 r_+^2}{16\pi  G } + \frac{3 Q^4 r_+^2 \kappa ^2}{\pi G M} + \frac{18 Q^2 \left(5+Q^2 \left(9 Q^2-16\right)\right) r_+^2 \kappa  \lambda }{5\pi G M  } + \lambda^2r_+^2\Lambda_{4,\lambda^2}(\rho_2) \,, \\
&&\\
\Lambda_5 &=& -\frac{ \pi  QT^3}{16 \sqrt{3} G M^2 (M+1) r_+} \,, \\
\Lambda_6 &=& r_+^2\Lambda_{6,0}(\rho_2) \,. \label{eq:Lambda6} 
\end{eqnarray}
For the anomalous coefficients we get the new ones
\begin{eqnarray}
\tilde\Lambda_4 &=& -\frac{3 Q^2 r_+\kappa }{8\pi G M  }-\frac{ \pi  T^2 \lambda }{G M r_+} \,, \label{eq:Lambdatilde4} \\
\tilde\Lambda_5 &=& \kappa r_+\tilde\Lambda_{5,\kappa}(\rho_2) + \lambda r_+\tilde\Lambda_{5,\lambda}(\rho_2) \,, \\
\tilde\Lambda_6 &=& \kappa \tilde\Lambda_{6,\kappa}(\rho_2) + \lambda\tilde\Lambda_{6,\lambda}(\rho_2) \,, \\
\tilde\Lambda_7 &=& -\frac{2 r_+\lambda }{G \pi } \,, \\
\tilde\Lambda_8 &=& \frac{3 Q^2 \left(Q^2-1\right) r_+ \kappa }{4\pi G M^2  } + \lambda r_+\tilde \Lambda_{8,\lambda}(\rho_2) \,, \label{eq:lambdatilde8}
\end{eqnarray}
and the rest of the anomalous ones
\begin{eqnarray}
\tilde\Lambda_1 &=& -\frac{\sqrt{3} r_+^2  Q^3 \kappa }{4\pi G M  }+\frac{ \sqrt{3}Q r_+(3r_+ + ( Q^2-4)\pi T)}{\pi G M  } \lambda \,, \label{eq:lambdatilde1} \\
\tilde\Lambda_2 &=& \lambda r_+^2 \tilde \Lambda_{2,\lambda}(\rho_2) \,, \\
\tilde\Lambda_3 &=& \frac{2 T^2 \lambda }{G (M+1)} \,. \label{eq:Lambdatilde3}
\end{eqnarray}
These coefficients enter in the constitutive relation for the energy-momentum tensor through (\ref{eq:fullconstiT}) and (\ref{eq:2consti}).

The transport coefficients $\Lambda_1,\ldots,\Lambda_6$, $\tilde{\Lambda}_1,\ldots,\tilde{\Lambda}_3$ and $\xi_1\ldots\xi_4$, $\tilde{\xi}_1$ have been computed in the past in \cite{Erdmenger:2008rm,Banerjee:2008th} without gravitational anomaly. It is interesting to remark that $\Lambda_1, \Lambda_2, \Lambda_5,$ $ \Lambda_6, \Lambda_7, \Lambda_8, \Lambda_9, \xi_1, \xi_3, \xi_5$ and $\xi_7$ do not receive $\lambda-$corrections, actually these coefficients do not depend on $\kappa$ either. It is also remarkable that $\tilde{\Lambda}_2$ and $\tilde{\Lambda}_3$ in the presence of gravitational anomaly are not vanishing. The rest of the transport coefficients we have computed are new.

\subsection{Discussion of second order results}
\label{sec:second_order_discussion}

It would be interesting to compare our results with the predictions done in \cite{Kharzeev:2011ds}. Basically the authors tried to fix the anomalous second order transport coefficients using a generalized version of the method developed by Son \& Surowka \cite{Son:2009tf}. The only issue is that they didn't consider the mixed gauge-gravitational anomaly and neglected all the integration constants as the previous authors. Nowadays we know that at least at first order  these integration constants might be related to the anomalous parameter~$\lambda$. The authors presented a set of algebraic and differential constraints. The algebraic ones are
\begin{eqnarray}
\label{eq:Klambda1}\tilde\Lambda_1 &=&\frac{4\eta }{n}\left(\xi_V - TD_B\right)\,,\\
\label{eq:Klambda4}\tilde\Lambda_4 &=&\frac{2\eta }{n}\left(\xi_B - \bar\kappa\mu\right)\,,\\
\label{eq:Kxi3}\tilde\xi_3 &=&\frac{2\sigma}{n}\left(\xi_V - TD_B\right)\,,\\
\label{eq:Kxi5}\tilde\xi_5 &=& 0\,,
\end{eqnarray}
where $D_B=\frac{\kappa T}{4\pi G  }\bar\mu^2$ is the coefficient multiplying the magnetic field in the entropy current computed in \cite{Son:2009tf,Neiman:2010zi,Kharzeev:2011ds} with only pure gauge anomaly, and $\bar{\kappa}=\frac{\kappa}{2\pi G}$ is the anomalous parameter used by the authors of \cite{Kharzeev:2011ds}. Eqs. (\ref{eq:Klambda1}) and (\ref{eq:Kxi3}) are satisfied by our solutions (\ref{eq:lambdatilde1}) and (\ref{eq:xitilde3}) as long as one fixes the anomaly parameter $\lambda$ to zero. However Eq.~(\ref{eq:Klambda4}) is satisfied  with the gravitational anomaly switched on. So far these constraints are satisfied except Eq. (\ref{eq:Kxi5}), as $\tilde{\xi}_5$ is not vanishing in our model even though we fix the anomalous parameter to vanish.

To check the value we get for $\tilde{\xi}_5$~(\ref{eq:xitilde5}), we may proceed by using the Kubo formula formalism. The Kubo formulae for $\tilde{\xi}_5$ will relate this coefficient to a two point function at second order in a frequency and momentum expansion. Actually it will appear in the same correlator as the chiral magnetic conductivity. To do so we can switch on a gauge field in the $y$ direction $A_y=A_y(t,z)$. In such a situation the Fourier transformed source $\mathcal{J}^{(5)}_\mu$ reduces to
\begin{equation}
\mathcal{J}^{(5)x} = \omega k_z A_y \,,
\end{equation}
so that using the constitutive relation we can read the two point function
\begin{equation}
\label{eq:jjsecorder}\left\langle \mathcal J^x \mathcal J^y \right\rangle = -i\xi_B k_z +  \tilde \xi_5 \omega k_z \,,
\end{equation}

The presence of $\tilde{\xi}_5$ must be captured by  a model in the probe limit. The very simple model of the Chapter \ref{cmecse} is enough to be used as a consistency check. In order to proceed, we will work only with an axial gauge field switched on, so we just need to solve the equations (\ref{eq:probe_axial}) and follow the standard holographic procedure described in the previous chapters in order to obtain the retarded correlator, the final result is
{\small\begin{equation}
\langle \mathcal J^a \mathcal J^b\rangle=\left(
\begin{array}{cc}
 -\frac{i \omega }{g_A}+\omega^2 \frac{ \log 2}{2 g_A}-72  g_A  (k\kappa  \mu) ^2 (-1+\log 4) & -12 i k  \kappa   \mu -12 k \omega  \kappa  \mu \log 2 \\
 12 i k   \kappa   \mu +12 k \omega  \kappa  \mu \log 2 & -\frac{i \omega }{g_A}+\omega^2 \frac{ \log 2}{2 g_A}-72(k\kappa  \mu) ^2  g_A   (-1+\log 4)
\end{array}
\right).
\end{equation}}

In order to relate the Fluid/Gravity model with the one in Chapter \ref{cmecse} we have to do the redefinitions
\begin{eqnarray}\nonumber
\kappa &\rightarrow & 24\pi G \kappa \qquad , \qquad r_+  \rightarrow  1 \\
\nonumber g_A^2 &\rightarrow & 16\pi G \qquad \textrm{and} \qquad T \rightarrow \frac{1}{\pi}.
\end{eqnarray}
Finally we will compare from (\ref{eq:xitilde5}) the $\kappa-$part because the correlator was computed in the probed limit where the backreaction of the gauge field on the background black hole is neglected and in consequence the effect of the gravitational anomaly is subleading. After doing a linear expansion for $\bar{\mu}\ll 1$ we get
\begin{equation}
\tilde{\xi}_5= 12\kappa  \mu  \log 2  - \frac{\lambda  \mu }{\pi G r_+} \left(1 + 2 \log 2\right) +  \mathcal{O}(\bar{\mu}^3)\,.
\end{equation}
This is a non trivial check of the non vanishing of $\tilde\xi_5$. In order to understand the discrepancy between this result and the prediction done by the authors of \cite{Kharzeev:2011ds}, we can analyze the properties under time reversal of the source associated to $\tilde \xi_5$, which reads in the constitutive relations as
\begin{equation}
J^\mu =  \tilde{\xi}_5\epsilon^{\mu\nu\rho\lambda}u_\nu\mathcal{D}_\rho E_\lambda + \ldots  \,.
\end{equation}
This equation in the local rest frame $u^\mu = (1,0,0,0)$ looks like
\begin{equation}
\vec{J}=\tilde \xi_5 \nabla\times \vec{E} + \ldots = - \tilde \xi_5 \frac{\partial \vec{B}}{\partial t}  + \ldots \,.
\end{equation}
The electric field and the operator $\nabla\times$ are even under time reversal while the current is odd, in consequence the conductivity $\tilde\xi_5$ is $\mathcal{T}-$odd. The fact that this transport coefficient is $\mathcal{T}-$odd tells us that such a source might contribute to the entropy production. For this reason demanding a non contribution to the production of entropy might not be well motivated. The situation would be similar as demanding a vanishing contribution from the usual electric conductivity. One can see also the odd property of $\tilde{\xi}_5$ from Eq. (\ref{eq:jjsecorder}), as $\left\langle \mathcal J^x \mathcal J^y \right\rangle$ is $\mathcal{T}-$even and inverting the time is the same as changing $\omega\to - \omega$. 

We have noticed that the anomalous coefficients associated to sources constructed with the second derivative of the fields can be naturally factorized as
\begin{eqnarray}
\tilde{\Lambda}_1 &=& -2\eta\tilde{l}_\omega  \,, \\
\tilde{\Lambda}_4 &=& -2\eta\tilde{l}_B \,, \\
\tilde{\xi}_5 &=& \sigma\tilde{l}_E \,.
\end{eqnarray}
These expressions make their dissipative nature clear, and they suggest the existence of anomalous relaxation lengths in analogy to the relaxation time $\tau_\pi$. These new $\mathcal{T}-$even quantities write
\begin{eqnarray}
\tilde{l}_\omega &=& \frac{2\pi}{ G  p}\left( \frac{\kappa  \mu ^3}{48\pi^2   } +64\mu\lambda\left( 3 r_+^2 - 2\mu^2 - \frac{\pi  T \mu ^2}{r_+}\right) \right) \,, \\
\tilde{l}_B &=& \frac{1}{2\pi G p}\left( \frac{\kappa  \mu ^2}{8} +  \pi^2 T^2 \lambda  \right) \,,\\
\label{eq:le}\tilde{l}_E &=&  -\frac{8 \bar\mu }{\pi^2 T}(\kappa  \log 2 - 2 \lambda  (1+2\log 2))+  \mathcal{O}(\bar{\mu}^3)\,.
\end{eqnarray}

A last interesting observation comes from the result on the dispersion relation of shear waves in \cite{Kharzeev:2011ds}, where they have found that
\begin{equation}
\label{eq:shear_disp}\omega \approx -i\frac{\eta}{4p}k^2 \mp i C k^3+\ldots \,,
\end{equation}
with $C = -\tilde{\Lambda}_1/(8p)$. It would be interesting to generalize the computation of \cite{Sahoo:2009yq} to the case including the mixed gauge-gravitational anomaly to verify whether the result for $C$ is
\begin{equation}
C = \frac{\eta}{4p} \tilde l_\omega \,.
\end{equation}
%\begin{equation}
%C= \frac{\sqrt{3} r_+^6  Q^3  }{2 M^2  }\kappa-\frac{2 \sqrt{3}Q r_+^5(3r_+ + ( Q^2-4)\pi T)}{ M^2  } \lambda\,.
%\end{equation}  
% % %%%%%%%%%%%%%%%%%%%%%%%%%%%%%%%%%%%%%%%%%%%%%%%%%%%%%%%%%%%%%%%%%%
   \chapter{Conclusions and Outlook}
   \label{conclusions}
   In the presence of external sources for the energy momentum tensor and the currents, the anomaly is responsible for a non conservation of the latter. This is conveniently expressed through~\cite{AlvarezGaume:1983ig}
\begin{equation}
 D_\mu J_a^\mu= \epsilon^{\mu\nu\rho\lambda}\left( \frac{d_{abc}}{32\pi^2} 
F^b_{\mu\nu} F^c_{\rho\lambda} + \frac{b_a}{768\pi^2} 
R^\alpha\,_{\beta\mu\nu}
R^\beta\,_{\alpha\rho\lambda}\right) \,,
\end{equation}
where the axial and mixed gauge-gravitational anomaly coefficients, $d_{abc}$ and $b_a$, are given by~(\ref{eq:chiralcoeff}) and (\ref{eq:gravcoeff}) respectively.

We have discussed in Chap.~\ref{chemical} the subtleties of introducing a chemical potential for anomalous charges.  One possible way is by deforming the Hamiltonian according to $H \rightarrow H - \mu Q$, a second, usually equivalent way is by imposing boundary conditions $\phi(t-i\beta) = \pm e^{\mu\beta}\phi(t)$ on the fields along the imaginary time direction \cite{Landsman:1986uw,Evans:1995yz}. These methods are equivalent as long as $Q$ is a non-anomalous charge.    We have argued why the second method could be physically favoured to be the right formalism. Similarly, in holography we can introduce the chemical potential either through a boundary value of the temporal component of the gauge field or through the potential difference between boundary and horizon. Thus, for non-anomalous symmetries, the boundary value of the temporal gauge field can be identified with the chemical potential. Due to the exact gauge invariance of the action, a constant boundary value never enters in correlation functions. In the presence of a Chern-Simons term, however, the gauge symmetry is partially lost and even a constant boundary gauge field becomes observable. This can be seen explicitly from the three-point functions (\ref{eq:jjj5}) and (\ref{eq:threepts6}). Therefore, we should set the axial vector field to zero after having used it as a source for axial current. By defining the corresponding chemical potential as the potential difference between the horizon of the AdS black hole and the holographic boundary we are able to do so. However, the prize we have to pay is a space time with a special topology at the horizon, this puncture in the bulk behave like a flux and allow us to have the twisted boundary conditions for the fields and makes this approach the holographic dual of the formalism B (see table \ref{tab:formalisms}).

We also derived the Kubo formulae that allow the calculation of all the transport coefficients at first order in the hydrodynamic expansion.

Then we computed two- and three-point functions of currents in Chap. \ref{cmecse} at finite density using holographic methods for a simple holographic model incorporating the axial anomaly of the standard model. We were able to reproduce the known weak-coupling results  concerning the chiral magnetic effect and also found a new type of ``conductivity'' in the axial sector alone, $\sigma_{55}$. Although it can not be probed by switching on external fields, as a two-point function it is as well defined as $\sigma_{CME}$. It would be interesting to find a way of also relating this anomalous conductivity to experimentally accessible observables.

Previous calculations of anomalous conductivities have been able to reproduce the weak-coupling result for $\sigma_{\rm axial}\propto\mu$ but not $\sigma_{CME}\propto \mu_5$ unless the contributions from
the Chern-Simons term to the chiral currents were dropped. In our calculation we have used the complete expressions for the currents, but of key importance was a clear distinction between the physical state
variable, the chemical potential, and the external background field. The latter we viewed exclusively as a source that couples to an operator, whereas the chemical potential should correspond in the most elementary way to the cost of energy for adding a unit of charge to the system.

We have confirmed our intuition on the way of introducing anomalous chemical potentials with the computation of three point functions in vacuum and getting the same result as in the holographic model.

Having the Kubo formulae we have computed the magnetic and vortical conductivity at weak coupling and we find contributions that are proportional to the anomaly coefficients (\ref{eq:chiralcoeff}) and (\ref{eq:gravcoeff}). Therefore the non-vanishing value has to be attributed to the presence of chiral and gravitational anomalies.

This result agrees with the known results from AdS/CFT~\cite{Amado:2011zx} up to one important difference: the holographic calculation did not show a contribution proportional to $\tr(T_A)$. This not surprising since only a holographic gauge Chern-Simons term was included. Holographic modelling of the gravitational anomaly called however also for inclusion of a mixed gauge-gravitational Chern-Simons term of the form $A\wedge R\wedge R$.

We find  a non-vanishing vortical conductivity proportional to $\sim T^2$ even in an uncharged fluid.   In ~\cite{Neiman:2010zi} similar terms in the vortical conductivities have been argued for as undetermined integration constant without any relation to the gravitational anomaly. The~$T^2$ behavior had appeared already previously in neutrino physics~\cite{Vilenkin:1979ui}.

In order to perform the analysis at strong coupling via AdS/CFT methods, we have defined in Chap.~\ref{strongcve} a holographic model implementing both type of anomalies via gauge and mixed gauge-gravitational Chern-Simons terms. We have computed the anomalous magnetic and vortical conductivities from a charged black hole background and have found a non-vanishing vortical conductivity proportional to $\sim T^2$. These terms are characteristic for the contribution of the gravitational anomaly and they even appear in an
uncharged fluid.  Very recently  a generalization of the results ~(\ref{eq:sigb})-(\ref{eq:sigve}) to any even space-time dimension as a polynomial in~$\mu$ and $T$~\cite{Loganayagam:2012pz} has been proposed. Finally, the
consequences of this anomaly in hydrodynamics have been studied using a group theoretic approach, which seems to suggest that their effects could be present even at $T=0$~\cite{Nair:2011mk}.

To have a consistent hydrodynamics description of the anomalous holographic plasma we went to second order in the derivative expansion using the {\it Fluid/Gravity correspondence}, and we computed all the transport coefficients. Except the ones associated to the presence of curvature in the fluid background. Within the most important results we have gotten that gravitational anomaly has a non trivial contribution to most of the transport coefficients, even thought to the non anomalous one. In the anomalous side we have found a dissipative conductivity
\begin{equation}
\vec J = \tilde{\xi}_5\nabla\times\vec E \,,
\end{equation}
that was though to be vanish for any anomalous model \cite{Kharzeev:2011ds}, but we have checked our result also using Kubo formulae as a consistency check. The existence of an anomalous dissipative conductivity results counter-intuitive and need to be understood better. We also have found the expression for $\tilde{\lambda}_1$ in presence of the mixed gauge-gravitational anomaly, which in principle determine the coefficient in front of $k^3$ in the chiral dispersion relation of shear waves \cite{Kharzeev:2011ds}.

We have defined the $\mathcal T-$even quantities $\tilde{l}_E$, $\tilde{l}_B$ and $\tilde{l}_\omega$  in analogy with the definition of $\tau_\pi$.  A generalization of the Israel and Stewart theory to hydrodynamics with anomalies would give us a better  physical intuition on these parameters. In particular the authors of \cite{Kharzeev:2011ds} have noticed that the chiral coefficient $C$ (\ref{eq:shear_disp}) in the dispersion relation of shear waves is related to $\tilde{\Lambda}_1$, and in consequence to $\tilde{l}_\omega$,
\begin{equation}
\omega \approx -i\frac{\eta}{4p}k^2\left(1\pm \tilde{l}_\omega k\right)\,.
\end{equation}

There are important phenomenological consequencies of the present study to heavy ion physics. In~\cite{KerenZur:2010zw} enhanced production of high spin hadrons (especially $\Omega^-$ baryons) perpendicular to the reaction plane in heavy ion collisions has been proposed as an observational signature for the chiral separation effect. Three sources of chiral separation have been identified: the anomaly in vacuum, the magnetic and the vortical conductivities of the axial current $J_A^\mu$. Of these the contribution of the vortical effect was judged to be subleading by a relative factor of $10^{-4}$. The~$T^2$ term in (\ref{eq:sigv}) leads however to a significant enhancement. If we take $\mu$ to be the baryon chemical potential $\mu\approx 10$ MeV, neglect $\mu_A$ as in \cite{KerenZur:2010zw} and take a typical RHIC temperature of $T=350$ MeV, we see that the temperature enhances the axial chiral vortical conductivity by a factor of the order of $10^4$. We expect the enhancement at the LHC to be even higher due to the higher temperature. 

Beyond applications to heavy ion collisions leading to charge and chiral separation effects \cite{Kharzeev:2010gr} it is tempting to speculate that the new terms in the chiral vortical conductivity might play a role in the early universe.  Indeed it has been suggested before that the gravitational anomaly might give rise to Lepton number
generation, e.g. in \cite{Alexander:2004us}. The lepton number separation due to the gravitational anomaly could contribute to generate regions with non-vanishing lepton number.

% %%%%%%%%%%%%%%%%%%%%%%%%%%%%%%%%%%%
   \chapter{Resumen y Perspectivas}
   \label{conclusionsE}
   En presencia de fuentes externas para el tensor de energ\'ia momento y las corrientes, las anomal\'ias son responsables de la no conservaci\'on de dicha corriente. Esto se expresa de la siguiente forma~\cite{AlvarezGaume:1983ig}
\begin{equation}
 D_\mu J_a^\mu= \epsilon^{\mu\nu\rho\lambda}\left( \frac{d_{abc}}{32\pi^2} 
F^b_{\mu\nu} F^c_{\rho\lambda} + \frac{b_a}{768\pi^2} 
R^\alpha\,_{\beta\mu\nu}
R^\beta\,_{\alpha\rho\lambda}\right) \,,
\end{equation}
donde los coeficientes axial y el mixto gauge-gravitacional, $d_{abc}$ y $b_a$, están dados por ~(\ref{eq:chiralcoeff}) y (\ref{eq:gravcoeff}) respectivamente.

En el capítulo~\ref{chemical} hemos discutido las sutilezas de introducir un potencial químico para cargas anómalas. Una posibilidad es deformando el Hamiltoniano según $H \rightarrow H - \mu Q$, una segunda forma usualmente equivalente es imponiendo las condiciones de frontera $\phi(t-i\beta) = \pm e^{\mu\beta}\phi(t)$ sobre los campos a lo largo del eje imaginario del tiempo \cite{Landsman:1986uw,Evans:1995yz}. Ambos m\'etodos son equivalentes siempre y cuando $Q$ no sea una carga an\'omala. Hemos dado argumentos f\'isicos de por qu\'e el segundo m\'etodo deber\'ia ser el apropiado. Similarmente, en holograf\'ia hemos introducido los potenciales qu\'imicos a trav\'es  del valor que toma en la frontera el campo gauge o como la diferencia de potencial entre el horizonte de eventos y la frontera del espacio tiempo. Para simetr\'ias no an\'omalas el potencial qu\'imico puede ser idetificado con el valor del campo gauge en la frontera. Debido a la  invariancia gauge de la acci\'on los correladores nunca dependerán del valor en la frontera del campo. Sin embargo, en presencia de un t\'ermino de Chern-Simons la simetr\'ia gauge se pierde parcialmente e inclusive configuraciones constantes en la frontera contribuirán a los correladores y se volverán observables. Esto se ve explicitamente de las funciones a tres puntos (\ref{eq:jjj5}) y (\ref{eq:threepts6}).  Por lo tanto, deber\'iamos poner a cero el campo axial luego de haberlo usado como una fuente para calcular la funci\'on de partici\'on. Al definir el potencial qu\'imico como la diferencia de potencial entre el horizonte y la frontera, podemos darle la interpretaci\'on de energ\'ia necesaria para introducir una unidad de carga en el sistema.

En esta tesis tambien derivamos la formulas de Kubo que permiten el c\'alculo de todos los coficientes de transporte a primer orden en la expansi\'on hidrodin\'amica.

Luego calculamos funciones a dos y tres puntos en el cap\'itulo \ref{cmecse} utilizando métodos hologr\'aficos para un modelo simple que incorpora la anomal\'ia axial del modelo standard. En este modelo fuimos capaces de reproducir el resultado conocido del r\'egimen de acoplamiento debil para la conductividad quiral magn\'etica.

Los c\'alculos previos de conductividades an\'omalas fueron capaces de reproducir el resultado de acoplamiento debil para  $\sigma_{\rm axial}\propto\mu$ pero no para  $\sigma_{CME}\propto \mu_5$ al menos que la contribuci\'on del t\'ermino de Chern-Simons se eliminara. En nuestro c\'alculo hemos utilizado la corriente consistente completa. De gran importancia fue la distinci\'on entre potencial qu\'imico y el valor asintot\'otico del campo gauge.

Nuestra intuici\'on sobre como introducir el potencial qu\'imico fue confirmada por el c\'alculo de funciones a tres puntos a acoplamiento debil y temperatura cero. Este resultado es una verificaci\'on no trivial de que nuestro razonamiento es correcto.

Teniendo conocimiento de las formulas de Kubo las utilizamos para calcular  las conductividades por campos magn\'eticos y por vorticidad en un r\'egimen de acoplamiento debil y obtuvimos resultados proporcionales a los coeficientes an\'omalos (\ref{eq:chiralcoeff}) y (\ref{eq:gravcoeff}). Por lo tanto estas conductividades son distintas de cero si y solo la teoría presenta anomal\'ias.

Este resultado coincide con el c\'alculo en AdS/CFT de~\cite{Amado:2011zx} salvo por una diferencia importante: el c\'alculo hologr\'afico no mostr\'o la contribuci\'on proporcional a $\tr(T_A)$. Pero esto no es una sorpresa porque hologr\'aficamente solo la anomal\'ia gauge se reproduc\'ia en este modelo. Para incluir la anomal\'ia gravitacional es necesario incluir en la acci\'on de gravedad un t\'ermino con la forma  $A\wedge R\wedge R$.

Para realizar un an\'alisis completo a acoplamiento fuerte de estas conductividades definimos un modelo en el contexto de AdS/CFT en el cap\'itulo~\ref{strongcve} en el cual implementamos ambas anomal\'ias. En este modelo calculamos todas las conductividades usando formulas de Kubo y encontramos en la conductividad por vorticidad el t\'ermino proporcional a  $\sim T^2$. Este t\'ermino est\'a presente si y solo si la anomal\'ia gravitacional est\'a en la teor\'ia al igual que en el caso de acoplamiento debil.

Para finalizar y tener una descripci\'on hidrodin\'amica consistente de este tipo de sistemas utilizamos la correspondencia fluido/gravedad que permite calcular los coeffiientes de transporte a segundo orden en la expansi\'on derivativa. Utilizando esta t\'ecnica fuimos capaces de calcular todos los coeficientes de transporte excepto los asociados a la presencia de curvatura en el medio. Dentro de los resultados mas importantes tenemos que inclusive a este orden la anomal\'ia gravitacional presenta contribuciones no triviales. En particular descubrimos una nueva conductividad an\'omala disipativa asociada a la variaci\'on en el tiempo del campo magn\'etico
\begin{equation}
\vec J = -\tilde{\xi}_5\frac{\partial\vec B}{\partial t} \,.
\end{equation}
También encontramos la contribuci\'on de $\lambda$ a $\tilde{\lambda}_1$, este coeficiente de transporte en princio determina  la contribuci\'on  $k^3$ en la relación de dispersi\'on quiral de las conocidas {\it shear waves} \cite{Kharzeev:2011ds}.

Este trabajo tiene consecuencias fenomenol\'ogicas importantes para el estudio de la f\'isica de colisiones de iones pesados. En ~\cite{KerenZur:2010zw} se predijo la producción de hadrones de alto esp\'in (especialmente bariones $\Omega^-$). Esta producci\'on es una consecuencia del efecto de separaci\'on de quiralidad. La anomalía gravitacional contribuye importantemente en este efecto porque su aporte va como $T^2$. Así que a temperaturas obtenidas en el LHC quizás sería posible observar este efecto.

Mas allá de la aplicaci\'on a la f\'isica de colisiones de iones pesados el efecto por separaci\'on de quiralidad \cite{Kharzeev:2010gr} es un candidato para especular sobre la generaci\'on de n\'umero lept\'onico en el universo temprano.

% %%%%%%%%%%%%%%%%%%%%%%%%%%%%%%%%%%%
 \begin{appendices}
 \begin{appendix}
 
 \chapter{Evaluation of triangle diagram}
  \label{app:triangle}
  We wish to compute the integral corresponding to the triangle diagram in Fig.~\ref{fig:triangle},
\begin{eqnarray}
\Gamma^{\mu\nu\rho}(p,q) &=& (-1) (i e)^2 (i g) (i)^3 \int \frac{d^d l}{(2\pi)^d} \mathrm{tr}\left( 
\gamma_5 
 \frac{\slashed{l}-\slashed{p}}{(l-p)^2}
\gamma^\mu
\frac{ \slashed{l}}{l^2} \gamma^\nu \frac{\slashed{l}+\slashed{q}}{(l+q)^2} \gamma^\rho  \right) \nonumber\\
 & &+(\mu \leftrightarrow \nu , p \leftrightarrow q).
\end{eqnarray}
Using Feynman parametrization the integral can be written as
\begin{eqnarray}
\Gamma^{\mu\nu\rho}(p,q) &=&  I_{\alpha \beta \gamma} \left[ \mathrm{tr} \left(
\gamma_5\gamma^\alpha \gamma^\mu \gamma^\beta \gamma^\nu \gamma^\gamma \gamma^\rho \right)
- \mathrm{tr} \left(
\gamma_5\gamma^\gamma \gamma^\nu \gamma^\beta \gamma^\mu \gamma^\alpha \gamma^\rho \right) \right], \\
I_{\alpha\beta\gamma} &=& -2 \int_0^1 dx dy\, \Theta(1-x-y) \int  \frac{d^d l}{(2\pi)^d}  \frac{N_{\alpha\beta\gamma}}{(l^2+D)^3},
\end{eqnarray}
where
\begin{eqnarray}
D &=& x(1-x) p^2 + 2 x y p\cdot q + y(1-y) q^2, \\
r_\mu &=& x p_\mu - y q_\mu, \\
N_{\alpha\beta\gamma} &=& (r-p)_\alpha r_\beta (r+q)_\gamma + \frac{l^2}{d}\Big[ \delta_{\alpha\beta}(r+q)_\gamma +   \delta_{\alpha\gamma}\,r_\beta + \delta_{\beta\gamma}(r-p)_\alpha \Big].
\end{eqnarray}
Here we have already taken into account that with both dimensional and cutoff regularizations, the integral with odd powers of $l$ in the numerator of the integrand vanishes, and the remaining tensor structure is dictated by the rotational symmetry of a momentum shell at fixed $|l|$.

Using
\begin{eqnarray}
\int_0^\Lambda \frac{l^3 dl}{(l^2+D)^3} &=& \frac{1}{4D} + \mathcal{O}\left(\frac{1}{\Lambda^2}\right), \\
\int_0^\Lambda \frac{l^5 dl}{(l^2+D)^3} &=&
\frac 1 2 \left[ \log\left(\frac{\Lambda^2}{D}\right) - \frac 3 2 \right]  + \mathcal{O}\left(\frac{1}{\Lambda^2}\right),
\end{eqnarray}
in the cutoff regularization ($d=4$), and
\begin{eqnarray} 
\left(\frac{\mathrm{e}^{\gamma_E} \mubar^2}{4\pi}\right)^\epsilon\,\int\,\frac{d^{4-2\epsilon}l}{(2\pi)^{4-2\epsilon}}\frac{1}{(l^2+D)^3} & = & \frac{\Gamma(\epsilon)}{16\pi^2}\left(\mathrm{e}^{\gamma_E} \mubar^2 \right)^\epsilon \frac{\epsilon}{2}\frac{1}{D^{1+\epsilon}}, \\
\left(\frac{\mathrm{e}^{\gamma_E} \mubar^2}{4\pi}\right)^\epsilon\,\int\,\frac{d^{4-2\epsilon}l}{(2\pi)^{4-2\epsilon}}\frac{l^2}{(l^2+D)^3} & = & \frac{\Gamma(\epsilon)}{16\pi^2}\left(\mathrm{e}^{\gamma_E} \mubar^2 \right)^\epsilon \left(1-\frac{\epsilon}{2}\right)\frac{1}{D^\epsilon},
\end{eqnarray}
in the dimensional regularization ($d=4-2\epsilon$) with $\ms$ scheme, we find
\begin{eqnarray}
\Gamma^{\mu\nu\rho}_\mathrm{reg}(p,q) &=& \frac{i}{2 \pi^2} \int_0^1\, dx\, dz\,\Theta(1-x-z) \Big[\,
(A^\mathrm{reg} p_\alpha + B^\mathrm{reg} q_\alpha) \epsilon^{\alpha \mu\nu\rho}+  \\
\nonumber & & 
 ( C_1^\mathrm{reg} p^\mu + D_1^\mathrm{reg} q^\mu) p_\alpha q_\beta \epsilon^{\alpha\beta\nu\rho} +
( C_2^\mathrm{reg} p^\nu + D_2^\mathrm{reg} q^\nu) p_\alpha q_\beta \epsilon^{\alpha\beta\mu\rho} \Big],
\end{eqnarray}
with $\mathrm{reg}\in\{\mathrm{CO},\mathrm{DR}\}$. The coefficients are given by
\begin{eqnarray}\label{eq:coeffs_begin}
A^\mathrm{CO} &=&  \frac{ (x-1)r^2 + y q^2}{D} + \left[ \log\left(\frac{\Lambda^2}{D}\right) - \frac 3 2 \right] (3x-1),\\
B^\mathrm{CO} &=& \frac{ (1-y)r^2 - x p^2}{D} + \left[ \log\left(\frac{\Lambda^2}{D}\right) - \frac 3 2 \right] (1-3y), \\
C_1^\mathrm{CO} & =& \frac{2 x(x-1)}{D},  \\
C_2^\mathrm{CO} &=& \frac{2 x y}{D},  \\
D_1^\mathrm{CO} &=& -\frac{2 x y}{D}, \\
D_2^\mathrm{CO} &=& \frac{2 y (1-y)}{D},
 \end{eqnarray}
in the cutoff regularization, and
\begin{eqnarray}
A^\mathrm{DR} & = & \left[\frac{(x-1)(r^2-D) + yq^2}{D}\epsilon+(3x-1) \right]\frac{\Gamma(\epsilon)}{D^{\epsilon}}\left(\mathrm{e}^{\gamma_E} \mubar^2 \right)^\epsilon, \\
B^\mathrm{DR} & = & \left[\frac{(1-y)(r^2-D) - xp^2}{D}\epsilon + (1-3y) \right]\frac{\Gamma(\epsilon)}{D^{\epsilon}}\left(\mathrm{e}^{\gamma_E} \mubar^2 \right)^\epsilon, \\
C_1^\mathrm{DR} & = & \frac{2\epsilon x(x-1)}{D^{1+\epsilon}} \Gamma(\epsilon)\left(\mathrm{e}^{\gamma_E} \mubar^2 \right)^\epsilon, \\
C_2^\mathrm{DR} & = & \frac{2\epsilon xy}{D^{1+\epsilon}}\Gamma(\epsilon)\left(\mathrm{e}^{\gamma_E} \mubar^2 \right)^\epsilon, \\
D_1^\mathrm{DR} & = & -\frac{2\epsilon xy}{D^{1+\epsilon}}\Gamma(\epsilon)\left(\mathrm{e}^{\gamma_E} \mubar^2 \right)^\epsilon, \\ \label{eq:coeffs_end}
D_2^\mathrm{DR} & = & -\frac{2\epsilon y(y-1)}{D^{1+\epsilon}} \Gamma(\epsilon)\left(\mathrm{e}^{\gamma_E} \mubar^2 \right)^\epsilon,
\end{eqnarray}
in the dimensional regularization.
  % %%%%%%%%%%%%%%%%%%%%%%%%%%%%%%%%%%%
 \chapter{Codazzi form of equations of motion}
  \label{app:codazziEq}
    \section{Codazzi form of Equations of Motion}
\label{appendixeq}
We project the equations of motion (\ref{eq:Gbulk}) and (\ref{eq:Abulk}) into the boundary surface and the orthogonal direction and rewrite them in terms of quantities at the regulated boundary. Doing so we get a set of two dynamical equations 
\begin{eqnarray}
\label{eq:Ab2}  0&=& \dot{E}^i + K E^i +  D_j \hat{F}^{ji} 
- 4\epsilon^{ijkl} \bigg( \kappa E_j \hat{F}_{kl} 
+ 4\lambda \dot{K}^s_j D_l K_{sk} + 2\lambda \hat{R}^s\,_{tkl} D_s K^t_j  \nonumber \\
&&\qquad\qquad\qquad\qquad\qquad\qquad\qquad\qquad + 4\lambda K_{ks} K^t_l D_t K^s_j  + 4\lambda K_{st} K^t_j D_l K^s_k \bigg) \,,
\end{eqnarray}
\begin{eqnarray}
\label{eq:Gb3} 0&=&\dot{K}^i_j + K K^i_j - \hat{R}^i_j 
+ \frac{1}{2}E^i E_j + \frac{1}{2} \hat{F}^{im}\hat{F}_{jm} - \frac{\delta^i_j}{(d-1)} \bigg( 2\Lambda + \frac{1}{2} E^m E_m + \frac{1}{4} \hat{F}^{lm}\hat{F}_{lm} \bigg) \nonumber \\
&&+2\lambda \bigg[
-2 \epsilon^{(iklm}\partial_r\left(\hat{F}_{kl}\dot{K}_{mj)}\right)
+2 \epsilon^{[iklm}\partial_r\left(\hat{F}_{kl}K_{ms}K^s_{j]}\right)
+2 \epsilon^{iklm}\hat{F}_{kl}K_{js}\left(\dot{K}^s_m + K^s_t K^t_m\right) \nonumber \\
&&\quad-\epsilon^{klmn}\hat{F}_{kl}\left( K^{(i}_s \hat{R}^s\,_{j)mn} 
+2 K^{(i}_m \dot{K}_{nj)} - 2 K^i_s K^s_m K_{nj}\right)  +4 \epsilon^{(iklm}\partial_r\left(E_k D_m K_{j)l} \right) \nonumber \\
&&\quad+2 \epsilon^{(iklm}D_s\left(\hat{F}_{kl}\left( D_ {j)} K^s_m - D^s K_{j)m} \right) \right) +4 \epsilon^{iklm} E_k K_{js} D_l K^s_m \nonumber \\
&&\quad-4 \epsilon^{klmn} E_k K^{(i}_l D_n K_{mj)} +2\epsilon^{(iklm} D_s\left( E_k (\hat{R}^s\,_{j)lm} - 2 K^s_l K_{j)m}) \right)  
\bigg] \,,
\end{eqnarray}
and three constraints
\begin{eqnarray}
 \label{eq:Gb1} 0&=&K^2 - K_{ij} K^{ij} - \hat{R} - 2 \Lambda - \frac{1}{2} E_i E^i + \frac{1}{4} \hat{F}_{ij}  \hat{F}^{ij} \nonumber \\
&&\qquad\qquad+ 8 \lambda \epsilon^{ijkl} \bigg( D_m ( \hat{F}_{ij} D_k K^m_l ) 
+  \hat{F}_{ij} K_{km} \dot{K}^m_l   + 2 E_i K_{jt} D_l K^t_k \bigg)  \,, \\
\label{eq:Gb2} 0&=&D_j K^{ji} - D^i K + \frac{1}{2} E_j \hat{F}^{ji} +  2 \lambda \epsilon^{klmi}  D_j\left[2E_k D_l K^j_m + \hat F_{kl}\left(\dot K^j_m + K^j_s K^s_m\right)\right] \nonumber\\
&&+\lambda\epsilon^{klmn}\left\{2\hat F_{kl}K^i_j D_mK^j_n + D_j\left[ F_{kl}(\hat R^{ij}\,_{nm}+2K^i_nK^j_m)\right] \right.\nonumber\\
&&\left. +2E_k K^j_m \hat R^i\,_{jnl} +2 \hat F_{kl} K^j_m(D^i K_{nj}-D_j K^i_n)
+2\partial_r(\hat F_{kl}D_nK^i_m)\right\}   \,,  \\
 \label{eq:Ab1}0&=&D_i E^i - \epsilon^{ijkl} \bigg( \kappa \hat{F}_{ij}\hat{F}_{kl} 
+ \lambda  \hat{R}^s\,_{tij} \hat{R}^t\,_{skl}
+ 4 \lambda K_{is} K^t_j\hat{R}^s\,_{tkl} 
+ 8 \lambda D_i K_{sj} D_l K^s_k \bigg)  \,,
\end{eqnarray}
 with the notation
\begin{equation}
X^{(i}\,_{j)} := \frac{1}{2} (X^i\,_j + X_j\,^i) \,, \qquad X^{[i}\,_{j]} := \frac{1}{2} (X^i\,_j - X_j\,^i) \,. \label{eq:defXij}
\end{equation}
We take Eq.~(\ref{eq:defXij}) as a definition, and it should be applied also when $X$ includes derivatives on~$r$, for instance $X^{(i} \dot{K}_{lj)} = \frac{1}{2}(X^i \dot{K}_{lj} + X_j \dot{K}^i_l)$.  

 % %%%%%%%%%%%%%%%%%%%%%%%%%%%%%%%%%%%
 \chapter{Technical details on holographic renormalization}
  \label{app:renorma}
  \section{Technical details on Holographic Renormalization}
\label{sec:app_holo_renorm}

The renormalization procedure follows from an expansion of the four dimensional quantities in eigenfunctions of the dilatation operator

\begin{equation}
\delta_D = 2 \int d^4x \gamma_{ij} \frac{\delta}{\delta \gamma_{ij}} \,. 
\end{equation}
This expansion reads
\begin{eqnarray}
K^i_j &=& K_{(0)}\,^i_j + K_{(2)}\,^i_j + K_{(4)}\,^i_j + \tilde{K}_{(4)}\,^i_j \log e^{-2r} + \cdots \,, \label{eq:Kijexp} \\
A_i &=& A_{(0)}\,_i + A_{(2)}\,_i + \tilde{A}_{(2)}\,_i \log e^{-2r} + \cdots \,, \label{eq:Aiexp}
\end{eqnarray}
where
\begin{eqnarray}
&&\delta_D K_{(0)}\,^i_j = 0 \,, \qquad \delta_D K_{(2)}\,^i_j = -2 K_{(2)}\,^i_j \,, \nonumber \\
&& \delta_D K_{(4)}\,^i_j = -4 K_{(4)}\,^i_j - 2 \tilde{K}_{(4)}\,^i_j \,, \qquad \delta_D \tilde{K}_{(4)}\,^i_j = -4 \tilde{K}_{(4)}\,^i_j \,, \nonumber \\
&& \delta_D A_{(0)}\,_i = 0 \,, \qquad  \delta_D A_{(2)}\,_i  = -2 A_{(2)}\,_i - 2 \tilde{A}_{(2)}\,_i \,, \nonumber \\
&& \delta_D \tilde{A}_{(2)}\,_i = -2 \tilde{A}_{(2)}\,_i \,.
 \end{eqnarray}
Given the above expansion of the fields one has to solve the equations of motion in its Codazzi form, order by order in a recursive way. To do so one needs to identify the leading order in dilatation eigenvalues at which each term contributes. One has
\begin{eqnarray}
\gamma_{ij} &\sim& {\cal O}(-2)\,, \qquad \gamma^{ij} \sim {\cal O}(2)\,, \qquad\quad\; E_i \sim {\cal O}(2)\,, \quad\;\;\; \hat{F}_{ij} \sim {\cal O}(0)\,, \nonumber \\
\sqrt{-\gamma} &\sim& {\cal O}(-4)\,, \qquad  K^i_j \sim {\cal O}(0)\,, \qquad \hat R^i\,_{jkl} \sim {\cal O}(0)\,,  \qquad \nabla_i \sim {\cal O}(0) \,.
\end{eqnarray}
Note that for convenience of notation we define ${\cal O}(n)$ if the leading eigenvalue of the dilatation operator is $-n$.  In practice, in the renormalization procedure one needs to use the equations of motion Eqs.~(\ref{eq:Gb3}) and (\ref{eq:Gb1}) up to ${\cal O}(2)$ and ${\cal O}(4) + {\cal O}(\tilde{4})$ respectively. Up to ${\cal O}(0)$ they write
\begin{eqnarray}
&&0=K_{(0)}^2 - K_{(0)}\,^i_j K_{(0)}\,^j_i - 2\Lambda   \,,  \label{eq:Gb1order0}\\
&&0=\dot{K}_{(0)}\,^i_j + K_{(0)} K_{(0)}\,^i_j - \frac{2\Lambda}{(d-1)}\delta^i_j \,. \label{eq:Gb3order0}
\end{eqnarray}
Order ${\cal O}(2)$ writes
\begin{eqnarray}
&&0=2K_{(0)}K_{(2)} - 2 K_{(0)}\,^i_j K_{(2)}\,^j_i - \hat{R} \,, \label{eq:Gb1order2}  \\
&&0=\dot{K}^i_j|_{(2)} + K_{(0)} K_{(2)}\,^i_j + K_{(2)} K_{(0)}\,^i_j -\hat{R}^i_j \,, \label{eq:Gb3order2}
\end{eqnarray}
and finally orders ${\cal O}(4)$ and ${\cal O}(\tilde{4})$ for Eq.~(\ref{eq:Gb1}) write respectively
\begin{eqnarray}
&&0=2K_{(0)}K_{(4)} + K^2_{(2)} - 2 K_{(0)} \,^i_j K_{(4)}\,^j_i - K_{(2)}\,^i_j K_{(2)}\,^j_i + \frac{1}{4} \hat{F}_{(0)}\,_{ij} \hat{F}_{(0)}\,^{ij}  \,, \label{eq:Gb1order4}\\
&&0=2 \left( K_{(0)} \tilde{K}_{(4)} - K_{(0)}\,^i_j \tilde{K}_{(4)}\,^j_i \right)\log e^{-2r}  \,. \label{eq:Gb1order4tilde}
\end{eqnarray}
The derivative on $r$ can be computed by using 
\begin{equation}
\frac{d}{dr} = \int d^4x \dot{\gamma}_{km} \frac{\delta}{\delta \gamma_{km}} = 2 \int d^4x K^l_m \gamma_{lk} \frac{\delta}{\delta \gamma_{km}}  \,. \label{eq:dr}
\end{equation}
By inserting in this equation the expansion of $K^i_j$ given by Eq.~(\ref{eq:Kijexp}), one gets $d/dr \simeq \delta_D$ at the lowest order. Taking into account this, the computation of $K_{(0)}\,^i_j$ is trivial if one considers the definition of $K_{ij}$, i.e.
\begin{equation}
K_{(0)} \,_{ij} = \frac{1}{2} \dot\gamma_{ij} \big|_{(0)} = \frac{1}{2}\delta_D \gamma_{ij} = \gamma_{ij} \,. \label{eq:K0comp}
\end{equation}
Then the result up to ${\cal O}(0)$ is
\begin{eqnarray}
K_{(0)}\,^i_j = \delta^i_j \,, \qquad K_{(0)} = d \,. \label{eq:K0app}
\end{eqnarray}
Inserting this result into Eq.~(\ref{eq:Gb1order0}) or (\ref{eq:Gb3order0}) one arrives at the well known cosmological constant
\begin{equation}
\Lambda = \frac{d(d-1)}{2} \,.
\end{equation}
We have used in Eq.~(\ref{eq:Gb3order0}) that $\dot{K}_{(0)}\,^i_j = \delta_D K_{(0)}\,^i_j = 0$. The result for $K_{(2)}$ follows immediately from Eqs.~(\ref{eq:Gb1order2}) and (\ref{eq:K0app}),
\begin{equation}
K_{(2)} := P = \frac{\hat{R}}{2(d-1)} \,. \label{eq:P}
\end{equation}
In order to proceed with the computation of $K_{(2)}\,^i_j$ from Eq.~(\ref{eq:Gb3order2}), we should evaluate first $\dot{K}^i_j|_{(2)}$. Using the definition of $d/dr$ given by Eq.~(\ref{eq:dr}), it writes
\begin{eqnarray}
\dot{K}^i_j|_{(2)} &=& 2 \int d^4x K_{(0)}\,^l_m \gamma_{lk} \frac{\delta}{\delta\gamma_{km}} K_{(2)}\,^i_j + 2 \int d^4x K_{(2)}\,^l_m \gamma_{lk} \frac{\delta}{\delta\gamma_{km}} K_{(0)}\,^i_j \nonumber \\
&=& 2 \int d^4x \gamma_{km} \frac{\delta}{\delta\gamma_{km}} K_{(2)}\,^i_j = \delta_D K_{(2)}\,^i_j = -2 K_{(2)}\,^i_j \,. \label{eq:K2pijcomp}
\end{eqnarray}
Because $K_{(0)}\,^i_j$ is the Kronecker's delta, the second term after the first equality is zero, while the first one becomes the dilatation operator acting over $K_{(2)}\,^i_j$. Then one gets from Eq.~(\ref{eq:Gb3order2}) the result
\begin{equation}
K_{(2)}\,^i_j := P^i_j = \frac{1}{(d-2)} \left[ \hat{R}^i_j - P \delta^i_j \right] \,. \label{eq:Pij} 
\end{equation}
Note that the trace of $K_{(2)}\,^i_j$ agrees with Eq.~(\ref{eq:P}). Using all the results above it is straightforward to solve for orders ${\cal O}(4)$ and ${\cal O}(\tilde{4})$. From Eqs.~(\ref{eq:Gb1order4}) and (\ref{eq:Gb1order4tilde}) one gets respectively
\begin{eqnarray}
K_{(4)} &=& \frac{1}{2(d-1)} \bigg[ P^i_j P^j_i - P^2 - \frac{1}{4} \hat{F}_{(0)}\,_{ij} \hat{F}_{(0)}\,^{ij}  \bigg] \,, \label{eq:K4}\\
\tilde{K}_{(4)} &=& 0 \,. \label{eq:K4tilde}
\end{eqnarray}
 
In order to compute the counterterm for the on-shell action, besides the equations of motion an additional equation is needed. Following Ref.~\cite{Papadimitriou:2004ap}, one can introduce a covariant variable $\theta$ and write the on-shell action as
\begin{equation}
S_{on-shell} = \frac{1}{8\pi G} \int_\partial d^4x \sqrt{-h} (K - \theta) \,. \label{eq:Sonshelltheta}
\end{equation}
Then computing $\dot{S}_{on-shell}$ from Eq.~(\ref{eq:Sonshelltheta}), and comparing it with the result obtained by using Eqs.~(\ref{eq:Sb1})-(\ref{eq:Sb3}), one gets the following equation
\begin{eqnarray}
0 &=& \dot{\theta} + K\theta -\frac{1}{(d-1)}\left( 2\Lambda + \frac{1}{2}E_i E^i + \frac{1}{4}\hat{F}_{ij}\hat{F}^{ij} \right) 
-\frac{2}{3}\kappa \epsilon^{ijkl} A_i E_j \hat{F}_{kl} \nonumber \\
&&-\frac{12\lambda}{(d-1)}\epsilon^{ijkl}\bigg[ 
 A_i\hat{R}^n\,_{mkl}D_n K^m_j 
+E_i K_{jm} D_k K^m_l
+\frac{1}{2}\hat{F}_{ik} K_{jm} \dot{K}^m_l  
\bigg]  \,. \label{eq:theta}
\end{eqnarray}
The variable $\theta$ admits also an expansion in eigenfunctions of $\delta_D$ of the form
\begin{equation}
\theta = \theta_{(0)} + \theta_{(2)} + \theta_{(4)} + \tilde\theta_{(4)}\log e^{-2r} + \cdots \,, \label{eq:thetaexpansion}
\end{equation}
where
\begin{eqnarray}
&&\delta_D \theta_{(0)} = 0 \,, \qquad \delta_D \theta_{(2)} = -2 \theta_{(2)} \,, \nonumber \\
&& \delta_D \theta_{(4)} = -4 \theta_{(4)} - 2 \tilde\theta_{(4)} \,, 
\qquad \delta_D \tilde\theta_{(4)} = -4 \tilde\theta_{(4)} \,.
\end{eqnarray}
Inserting expansion~(\ref{eq:thetaexpansion}) into Eq.~(\ref{eq:theta}), one gets the following identities
\begin{eqnarray}
&&0 = \dot{\theta}_{(0)} + K_{(0)}\theta_{(0)} - \frac{2\Lambda}{(d-1)} \,, \label{eq:thetaorder0} \\
&&0=\dot{\theta}|_{(2)} + K_{(2)}\theta_{(0)} + K_{(0)} \theta_{(2)} \,,  \label{eq:thetaorder2}\\
&&0=\dot{\theta}|_{(4)} + K_{(4)}\theta_{(0)} + K_{(2)} \theta_{(2)} + K_{(0)}\theta_{(4)}  - \frac{1}{4(d-1)}  \hat{F}_{(0)}\,_{ij} \hat{F}_{(0)}\,^{ij} \,, \label{eq:thetaorder4}\\
&&0=\dot{\theta}|_{(\tilde{4})}  + \left( \theta_{(0)} \tilde{K}_{(4)} + K_{(0)} \tilde{\theta}_{(4)} \right) \log e^{-2r} \,, \label{eq:thetaorder4tilde}
\end{eqnarray}
corresponding to orders ${\cal O}(0)$, ${\cal O}(2)$, ${\cal O}(4)$ and ${\cal O}(\tilde{4})$ respectively. Following the same procedure as shown in Eqs.~(\ref{eq:K0comp}) and (\ref{eq:K2pijcomp}), one gets
\begin{equation}
\dot{\theta}_{(0)} = 0 \,, \qquad \dot{\theta}_{(2)} = \delta_D \theta_{(2)} = -2\theta_{(2)} \,.
\end{equation}
At this point one can solve Eqs.~(\ref{eq:thetaorder0}) and (\ref{eq:thetaorder2}) to get
\begin{eqnarray}
\theta_{(0)} = 1 \,, \qquad \theta_{(2)} = \frac{P}{(2-d)} \,.  \label{eq:theta02}
\end{eqnarray}
Higher orders are a little bit more involved. Using the definition of
$d/dr$, then $\dot{\theta}|_{(4)}$ writes
\begin{eqnarray}
\dot{\theta}|_{(4)} &=& 2 \int d^4x K_{(0)}\,^l_m \gamma_{lk} \frac{\delta}{\delta\gamma_{km}} \theta_{(4)} 
+ 2 \int d^4x K_{(4)}\,^l_m \gamma_{lk} \frac{\delta}{\delta\gamma_{km}} \theta_{(0)} 
+2 \int d^4x K_{(2)}\,^l_m \gamma_{lk} \frac{\delta}{\delta\gamma_{km}} \theta_{(2)} \nonumber \\
&=& \delta_D \theta_{(4)} 
+ \frac{2}{(2-d)} \int d^4x P_{km} \frac{\delta}{\delta\gamma_{km}} P   \,. \label{eq:theta4pcomp}
\end{eqnarray}
Note that the second term after the first equality vanishes, while the first one writes in terms of $\delta_D$. To evaluate the last term at the r.h.s. of eq.~(\ref{eq:theta4pcomp}) we use
\begin{equation}
\delta \hat{R} = -\hat{R}^{km}\delta \gamma_{km} + D^k D^m \delta \gamma_{km} - \gamma^{km} D_l D^l \delta \gamma_{km} \,.
\end{equation}
After a straightforward computation, one gets
\begin{equation}
\dot{\theta}|_{(4)} = -4 \theta_{(4)} - 2\tilde\theta_{(4)} + \frac{1}{(d-1)(d-2)} \bigg[(d-2)P^i_j P^j_i + P^2 + D_i( D^i P  - D^j P^i_j) \bigg] \,. \label{eq:theta4dot}
\end{equation}
Inserting Eq.~(\ref{eq:theta4dot}) into Eq.~(\ref{eq:thetaorder4}) one can solve the latter, and the result is~\footnote{This result for $\tilde\theta_{(4)}$ includes a total derivative term which has not been computed in Ref.~\cite{Papadimitriou:2004ap}. To compute $\tilde\theta_{(4)}$, in this reference the authors derive the elegant relation $\tilde\theta_{(4)} = \frac{(d-1)}{2}K_{(4)} + \tilde{K}_{(4)}$. This identity is however valid modulo total derivative terms.}
\begin{equation}
\tilde\theta_{(4)} = \frac{1}{4} \bigg[ P^i_j P^j_i - P^2 - \frac{1}{4}  \hat{F}_{(0)}\,_{ij} \hat{F}_{(0)}\,^{ij} + \frac{1}{3} D_i\left( D^i P - D^j P^i_j \right)  \bigg] \,. \label{eq:theta4tilde}
\end{equation}
The computation of $\dot{\theta}|_{(\tilde{4})}$ follows in a similar way, and one gets $\dot{\theta}|_{(\tilde{4})} =  -4 \tilde\theta_{(4)} \log e^{-2r}$. By inserting it into Eq.~(\ref{eq:thetaorder4tilde}), this equation is trivially fulfilled.

The counterterm of the action can be read out from Eq.~(\ref{eq:Sonshelltheta}) by using $K$ and~$\theta$ computed up to order ${\cal O}(\tilde{4})$, i.e.
\begin{equation}
S_{ct} = -S_{on-shell} = -\frac{1}{8\pi G} \int_\partial d^4x \sqrt{-h} \bigg[ (K_{(0)} - \theta_{(0)}) +  (K_{(2)} - \theta_{(2)})  + (\tilde{K}_{(4)} - \tilde{\theta}_{(4)}) \log e^{-2r} \bigg]  \,.
\end{equation}
From this equation and Eqs.~(\ref{eq:K0app}), (\ref{eq:P}), (\ref{eq:K4tilde}), (\ref{eq:theta02}) and (\ref{eq:theta4tilde}), one finally gets
\begin{eqnarray}
S_{ct} &=& - \frac{(d-1)}{8\pi G} \int_\partial d^4x \sqrt{-h} \bigg[
1 + \frac{1}{(d-2)}P \nonumber \\
&&\qquad\qquad- \frac{1}{4(d-1)} \left( P^i_j P^j_i - P^2 -  \frac{1}{4} \hat{F}_{(0)}\,_{ij} \hat{F}_{(0)}\,^{ij} \right)\log e^{-2r} \bigg] \,.
\end{eqnarray}
The last term in Eq.~(\ref{eq:theta4tilde}) is a total derivative, and so it doesn't contribute to the action. As a remarkable fact we find that there is no contribution in the counterterm coming from the gauge-gravitational Chern-Simons term. This is because this term only contributes at higher orders. Indeed as explained above, in the renormalization procedure we use  Eqs.~(\ref{eq:Gb1}) and (\ref{eq:theta}) up to orders ${\cal O}(0)$, ${\cal O}(2)$, ${\cal O}(4)$ and ${\cal O}(\tilde{4})$, and Eq.~(\ref{eq:Gb3}) up to orders ${\cal O}(0)$ and ${\cal O}(2)$. We have explicitly checked that the $\lambda$ dependence starts contributing at ${\cal O}(6)$ in all these three equations.~\footnote{Note that $\dot{K}^i_j$ and $\dot{\theta}$ induce terms proportional to $\lambda$. Up to order ${\cal O}(4) + {\cal O}(\tilde{4})$ these operators write $\dot{K}^i_j|_{(4)+(\tilde{4})} = -4 K_{(4)}\,^i_j + \dots$, and $\dot{\theta}|_{(4)+(\tilde{4})} = -4 \theta_{(4)} + \dots$, where the dots indicate extra terms which are $\lambda$-independent. The only $\lambda$-dependence could appear in $K_{(4)}\,^i_j$ and $\theta_{(4)}$, but these contributions are precisely cancelled by other terms in Eqs.~(\ref{eq:Gb3}) and (\ref{eq:theta}) respectively, so that these equations become $\lambda$-dependent only at ${\cal O}(6)$ and higher.} This means that the gauge-gravitational Chern-Simons term does not induce new divergences, and so the renormalization is not modified by it.

% %%%%%%%%%%%%%%%%%%%%%%%%%%%%%%%%%%%
  \chapter{Equation of motion for shear sector}
   \label{app:sheareq}
    
The perturbative solutions of the system \erf{htx1} and \erf{bx1} up to first order in momentum are 
These are the complete linearized set of six dynamical equations of motion
\begin{eqnarray}
\label{eq_As}\nonumber 0&=& B_\alpha'' ( u )+\frac { f' ( u ) }{f(u)}B_\alpha'( u )+ \frac {b^2  }{uf( u )^2 }\left(w^2-  f( u) 
{k}^{2} \right)B_\alpha( u) -  \frac{ h^{\alpha'}_t ( u
 )}{f(u)} \\
 &&+ik\epsilon_{\alpha\beta}\left(\ \frac{3}{u f(u)} \bar\lambda \left( \frac{2}{3a} (f(u)-1)+ u^3 \right) h_t^{\beta'}(u)+\bar \kappa \frac{B_\beta(u)}{f(u)} \right) \,,\\
\label{eq_Hts}\nonumber 0&=&  h^{\alpha''}_t(u) - \frac{h^{\alpha'}_t(u)}{u} -\frac {b^2}{
uf( u )}\left(k^2 h^\alpha_t(u)+ h^\alpha_y \left( u \right) wk \right)
- 3a u B'_\alpha(u) \\
\nonumber &&i\bar\lambda k \epsilon_{\alpha\beta}\left[\left(24a u^3-6(1-f(u))\right)\frac {B_\beta(u)}{u}+(9a u^3-6(1-f(u)))B'_\beta(u)\right.\\
 &&\left.+2 u (uh^{\beta'}_t ( u ) )'  -\frac {  2u  b^2}{  f ( u) }\left( h_y^\beta( u ) w k+h_t^\beta( u ) k^2\right)\right] \,,\\
\label{eq_Hys}\nonumber 0&=& h^{\alpha''}_y(u)+ \frac{\left(f/u\right)'}{f/u}h^{\alpha'}_y(u)+\frac{b^2}{u f(u)^2}\left(w^2h_y^\alpha(u)+wk h_t^\alpha(u)\right)+2uik\bar{\lambda}\epsilon_{\alpha\beta}\left[u  h^{\beta''}_y(u)\right.\\
 && \left.+\left(9f(u)-6+3a u^3\right)\frac{h^{\beta'}_y(u)  }{f(u)}+\frac{ b^2 }{f(u)^2}\left(wk h_t^\beta(u)+w^2 h_y^\beta(u)\right)\right] \,,
\end{eqnarray}

and two constraints for the fluctuations at $w ,k\neq 0$ 
\begin{eqnarray}
\label{constraints}\nonumber 0&=&  w \left(h^{\alpha'}_t(u)-3a u B_\alpha(u)\right)+f(u)kh^{\alpha'}_y(u)+i k\bar\lambda\epsilon_{\alpha\beta}\left[2u^2 \left(w h^{\beta'}_t+ f( u )  k h^{\beta'}_y (u)\right)\right.\\
 && \left. +\left(9a u^3-6(1-f(u))\right)B_\beta(u)\right] \,.
\end{eqnarray}
% %%%%%%%%%%%%%%%%%%%%%%%%%%%%%%%%%%%
   \chapter{Analytic solutions in the hydrodynamic regime}
    \label{ap:anSol}
     
We write in this appendix the solutions for the system
(\ref{htx1})-(\ref{bx1}). These functions depend explicitly on the
boundary sources $\tilde H^\alpha$ and $\tilde B_\alpha$, and the
anomalous parameters $\bar \kappa,\bar\lambda$. Switching off $\bar
\lambda$ we get the same system obtained in \cite{Amado:2011zx}
{\small
\begin{eqnarray}
\nonumber h^\alpha_t(u) &=& \tilde H^\alpha f(u) - \frac{i k \bar\kappa\epsilon_{\alpha\beta} (u-1)a }{2 (1+4 a)^{3/2}}\left[
\left. (1+4 a)^{3/2} u^2 \tilde H^\beta\right.\right. + \\
\nonumber&&\left. 3 \left(\sqrt{1+4 a} u (2 a u-1)+2 \left(1+u-a u^2\right) \text{ArcCoth}\left[\frac{2+u}{\sqrt{1+4 a} u}\right]\right)\tilde B_\beta\right] +
\\
\nonumber&&k i\bar\lambda \epsilon_{\alpha\beta}(u-1) \left[\tilde B_\beta  \left(-\frac{3 i (u+1)(1+a) \pi }{2  a }+\frac{3   (1+a (5+a)) u}{ (1+4 a)}+\frac{  \left(5+21 a+2 a^3\right) u^2}{ (1+4 a)}\right.\right.\\
\nonumber&&\left.\left.+\frac{3}{2}i  (1+a) \pi  u^2-6   a u^3
%%\right.\right.\\
%% &&\left.\left.
-\frac{3i f(u) (1+a (7+2 a (7+a))) \text{ArcCoth}\left[\frac{2+u}{\sqrt{1+4 a} u}\right]}{(u-1)(-1-4 a)^{3/2}  a}+\right.\right.\\
 &&\left.\left.-\frac{3 f(u) (1+a) }{2a(u-1)} \text{Log}\left[-1-u+a u^2\right]\right)+\right.\\
\nonumber&&\left.\tilde H^\beta  \left(-\frac{2i (u+1) (1+a)^2 \pi }{ a^2 }+\frac{2   (1+a) (2+a (7+2 a)) u}{ a (1+4 a) }+\right.\right.\\
\nonumber&&\left.\left. +\frac{  (4+a (25+a (39+a (-5+4 a)))) u^2}{ a (1+4 a) }+\frac{2i  (1+a)^2 \pi  u^2}{ a }+\right.\right.\\
\nonumber&&\left.\left.   u^3(1-5 a -6au)+\right.\right.\\
\nonumber&&\left.\left. -\frac{4 i f(u) (1+a) (1+2 a) (1+a (5+a)) \text{ArcCoth}\left[\frac{2+u}{\sqrt{1+4 a} u}\right]}{(u-1)(-1-4 a)^{3/2}  a^2}+\right.\right.\\
\nonumber&&\left.\left.-\frac{2  f(u) (1+a)^2 \text{Log}\left[-1-u+a u^2\right]}{ (u-1)a^2}\right)\right]\,,
\end{eqnarray}
\begin{eqnarray}
\nonumber B_\alpha(u)&=& \tilde B_\alpha+\tilde H^\alpha u -i\frac{k\bar\kappa\epsilon_{\alpha\beta}}{2 (1+4 a)^{3/2}} \left(\tilde H^\beta u\left(  1+4 a   \right)^{3/2}+\right.\\
\nonumber&&\left.\tilde B_\beta \left(6 a \sqrt{1+4 a}  u-2 (-2+a (-2+3 u)) \text{ArcCoth}\left[\frac{2+u}{\sqrt{1+4 a} u}\right]\right)\right)\\
\nonumber&&+i k \bar\lambda \epsilon_{\alpha\beta}\left[\tilde B_\beta \left(-\frac{i (1+a)^2  \pi }{ a^2 }+\frac{2   (1+a) (1+a (5+a))  u}{ a (1+4 a )}+\frac{3 i (1+a)  \pi  u}{2  a }\right.\right.\\
\nonumber&&\left.\left. -3 u^2-\frac{i (1+a (7+2 a (7+a))) (-2+a (-2+3 u)) \text{ArcCoth}\left[\frac{2+u}{\sqrt{1+4 a} u}\right]}{(-1-4 a)^{3/2}  a^2}\right.\right.\\
&&\left.\left.-\frac{ (1+a) (-2+a (-2+3 u)) \text{Log}\left[-1-u+a u^2\right]}{2 a^2}\right)\right.\\
\nonumber && \left.+\tilde H^\beta \left(-\frac{4 i (1+a)^3  \pi }{3  a^3 }+\frac{  (8+a (48+a (84+a (29+12 a))))  u}{3  a^2 (1+4 a )}+\right.\right.\\
\nonumber&&\left.\left.+\frac{2 i (1+a)^2  \pi  u}{ a^2 }-\frac{2   (1+a)  u^2}{ a }-3 u^3\right.\right.\\
\nonumber&&\left.\left.-\frac{4 i (1+a) (1+2 a) (1+a (5+a)) (-2+a (-2+3 u)) \text{ArcCoth}\left[\frac{2+u}{\sqrt{1+4 a} u}\right]}{3 (-1-4 a)^{3/2}  a^3}+\right.\right.\\
\nonumber&&\left.\left.-\frac{2 (1+a)^2 (-2+a (-2+3 u)) \text{Log}\left[-1-u+a u^2\right]}{3  a^3}\right)\right]\,.
\end{eqnarray}
}

% %%%%%%%%%%%%%%%%%%%%%%%%%%%%%%%%%%%
   \chapter{Fluid velocity dependence}
    \label{ap:fluiddep}
     We have seen that in the hydrodynamic regime the velocity of the fluid in the Landau frame is determined modulo a $P$-odd term $v_m\sim O(k)$ that is an arbitrary function of the sources. In this appendix we show the independence of the transport coefficients on this arbitrary function, even if the correlators are velocity dependent, and also that these arbitrariness disappears once we correctly impose the physical boundary conditions on the bulk fields. For simplicity we will do this analysis in the case when the mixed-gravitational anomaly vanish.

In order to do so, we are going to solve the system at $\omega=0$, first order in $k$ and for arbitrary value of $v_m$. Again, the system reduces to:
 \bea
 0&=&  h''^i_t(u)-\frac{h'^i_t(u)}{u}-3\, a\, u\, B'_i(u)\,,\\ 
0&=&  B_i(u)+\frac{f'(u)}{f(u)}\,B'_i(u)-i\,\epsilon_{ij}\,\bar\kappa\,\frac{k B_j(u)}{f(u)}-\frac{h'^i_t(u)}{f(u)}\,,
\eea
where $h^i_t(u) = h^{(0)}_i(u)+p \,h^{(1)}_i(u) $ and $B_i(u) = B^{(0)}_i(u)+p \,B^{(1)}_i(u)$. After imposing regularity at the horizon we find the following solutions:
\bea
B^{(0)}_i(u) &=& \tilde{B}_i + A_i\, u\,,\\
B^{(1)}_i(u) &=& C_i u-\frac{2 i (1+a)^2 \bar\kappa \epsilon_{ij} A_j u}{(2-a) (1+4 a) b}-\frac{i\,\bar\kappa\, \epsilon_{ij} \tilde{B}_j}{(2-a)(1+4a)^{3/2}b}\left(9a (1+a) u \sqrt{1+4 a}\right.\\
&&\hspace{-1.8cm}\left. +(2-a)^2 \left( 2 (1+a) \text{ArcCoth}\left[\sqrt{1+4 a}\right]+(2+a(2-3u)) \text{ArcTanh}\left[\frac{-1+2 a u}{\sqrt{1+4 a}}\right]\right)\right)\nn
\eea
\bea
h^{(0)}_i(u) &=& \tilde{h}_i + A_i\,(f(u)-1)\,,\\
h^{(1)}_i(u) &=&  C_i (f(u)-1) - \frac{i\, a\, \bar\kappa (4 (1+a)^2 u -27 a)\,\epsilon_{ij} A_j\, u^2}{2 (2-a) (1+4a) b} \\
&&\hspace{-1.8cm}+\frac{3\,i\,a\,\bar\kappa\, \epsilon_{ij} \tilde{B}_j}{2(2-a)(1+4a)^{3/2}b} \left(((2+a(16+5a))u-6 a (1+a) u^2-(2-a)^2) \sqrt{1+4 a} u\right. \nn\\
&&\hspace{-1.8cm}\left.+ 2 (2-a)^2 \left(\text{ArcCoth}\left[\sqrt{1+4 a}\right]+f(u) \text{ArcTanh}\left[\frac{-1+2 a u}{\sqrt{1+4 a}}\right]\right)\right)\,.\nn
\eea
As we know, this is not enough to solve the boundary value problem since both of the two independent solutions for the metric fluctuations satisfy the regularity condition. However, we can 
use the constitutive relations to try to fix the arbitrariness. In the hydrodynamic description, the stress-energy tensor is given by
\be
T^{ti}=(\epsilon +P) v^i -P \tilde{h}^{ti}\,,
\ee
where the velocity is order $p$. Using the holographic dictionary, we can identify the coefficient of the non-normalizable mode of the asymptotic behavior of a bulk field with the source of the dual operator and the coefficient 
of the normalizable one with its expectation value. Therefore, we can write the metric fluctuation close to the boundary as
\be
h_t^i(u)\sim \tilde{h}_t^i + T_t^i u^2\,,
\ee
so using the hydrodynamic result, we can do the identification order by order in momentum, in such a way that the 
velocity piece of the energy tensor fixes the horizon value of $h^{(1)}_i$. Doing so, the asymptotic behavior of each order becomes
\bea
h^{(0)}_i (u) &\sim& \tilde{h}_t^i (1-P u^2)\,,\nn\\
p h^{(1)}_i (u) &\sim& -(\epsilon +P) v_i u^2\,.
\eea 
We can proceed to construct the matrix of correlators for arbitrary value of the velocity as explained in Chapter \ref{strongcve}. Now, all the correlators pick contributions proportional to the 
velocity. In a compact way, the retarded propagators read
\bea
G_{i,j}&=&-\frac{r_{\rm H}}{\pi G L}\left(\frac{i\sqrt{3a} \,k\, (4+a) \kappa}{8 (1+a)}\epsilon_{ij}-\frac{r_{\rm H}}{2 L^2} \frac{\partial v_i}{\partial \tilde{B}_j}\right)-\frac{i k \beta \kappa}{6 \pi G} \epsilon_{ij}\,,\\
G_{i,tj}&=&-\frac{r_{\rm H}^2}{\pi G L^2}\left(\frac{3\, i\, a\, k\, \kappa}{4 (1+a)} \epsilon_{ij} -\frac{\sqrt{3a} r_{\rm H}}{4 L} \delta_{ij}-\frac{\sqrt{3a}\,r_{\rm H}\,}{2 L^2}\frac{\partial v_i}{\partial \tilde{h}_j}\right)\,,\\
G_{ti,j}&=&\frac{r_{\rm H}^3}{\pi G L^4} \frac{(1+a)}{\sqrt{3a}}\frac{\partial v_i}{\partial \tilde{B}_j}\,,\\
G_{ti,tj}&=&\frac{r_{\rm H}^4}{\pi G L^5 } \left(\frac{(1+a)}{16} \delta_{ij} + (1+a)\frac{\partial v_i}{\partial \tilde{h}_j}\right)\,,
\eea
where $i,j=x,z$. It is straightforward to prove that applying definitions \erf{eq:cvc} and \erf{eq:cmc}
for the chiral vortical and magnetic conductivities, the result is independent of the velocity and coincides with \erf{eq:xiv} and \erf{eq:xib} as expected. Setting the velocities 
to zero, the correlators coincide with those presented in \cite{Matsuo:2009xn}.

If we now impose the correct zero frequency `infalling' condition to the fields $h_t^i$, i. e. vanishing at the horizon, the velocities are not arbitrary anymore, but 
are given in terms of the boundary sources,
\be
v_i=-\frac{i a \bar\kappa \epsilon_{ij}(2 \tilde{h}_j + 3 \tilde{B}_j) k}{16(1+a)}\,.
\ee
Of course, substituting them in the Green functions given above, the antisymmetric correlation matrix spanned by \erf{eq:giti} -- \erf{eq:gtiti} is recovered.

     % %%%%%%%%%%%%%%%%%%%%%%%%%%%%%%%%%%%
   \chapter{Comparing the $\lambda$ and $\kappa$ contribution in the frequency dependence}
    \label{ap:freqdep}
     \begin{figure}[h!]
\begin{center}
\includegraphics[scale=0.58]{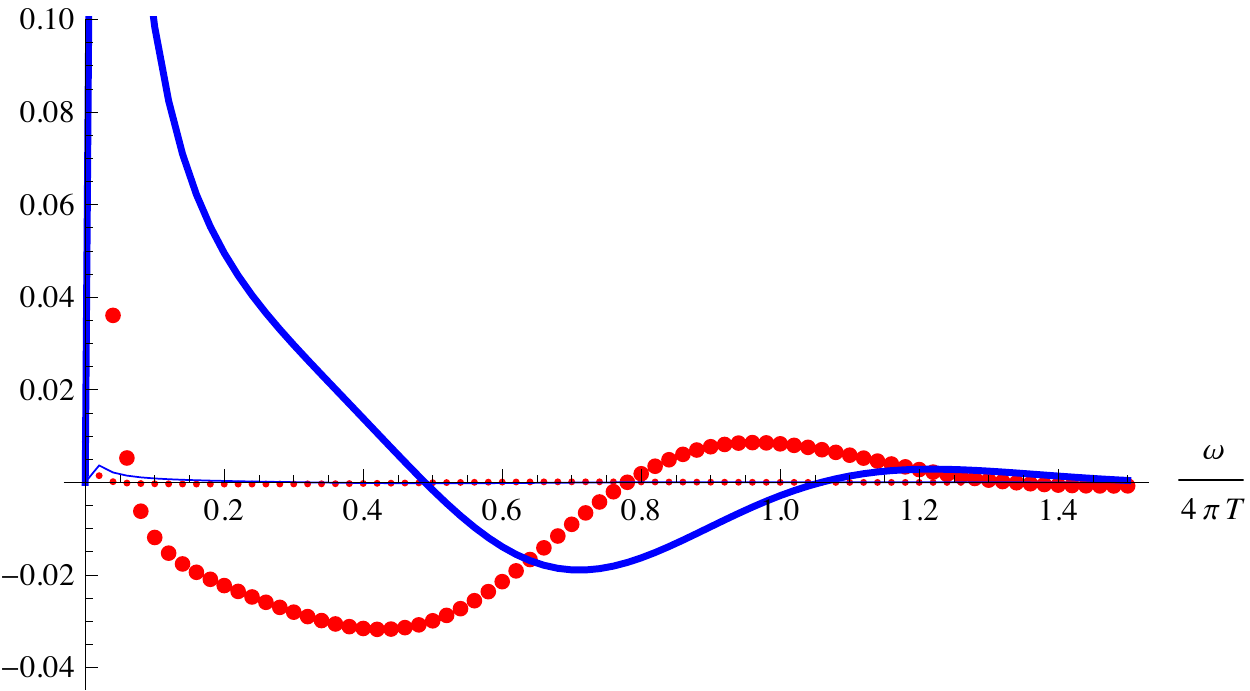}\hfill \includegraphics[scale=0.58]{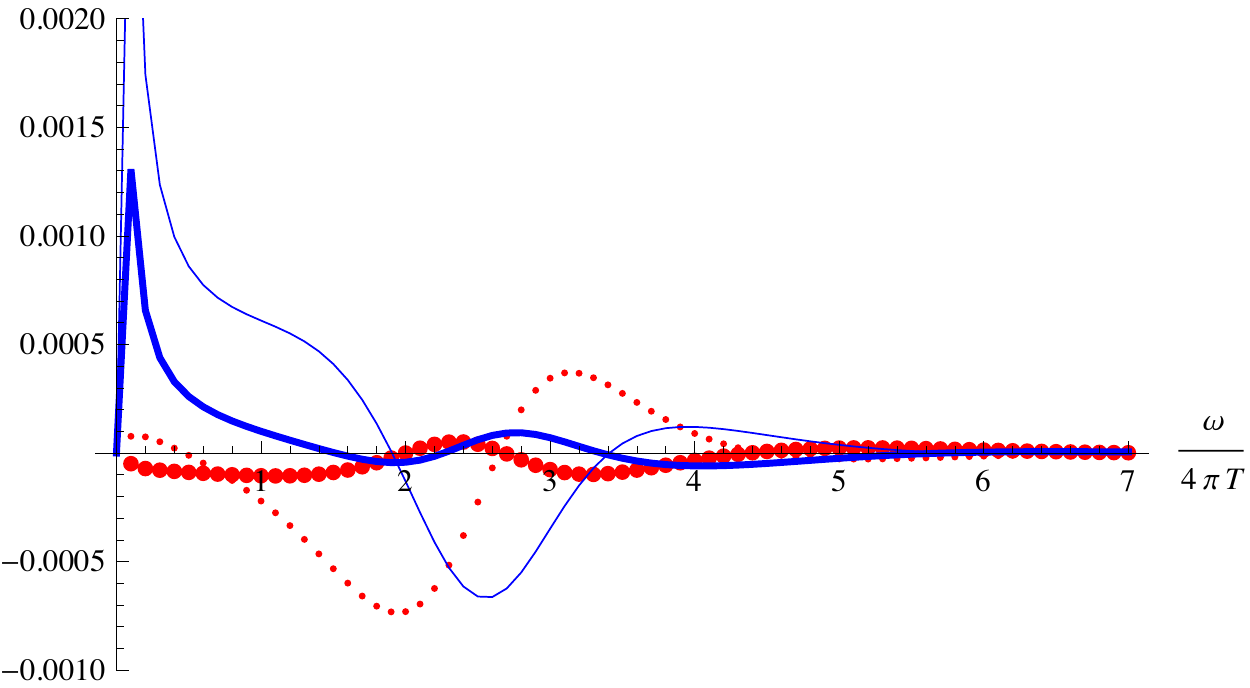} \\
\vspace{0.23cm}
\includegraphics[scale=0.58]{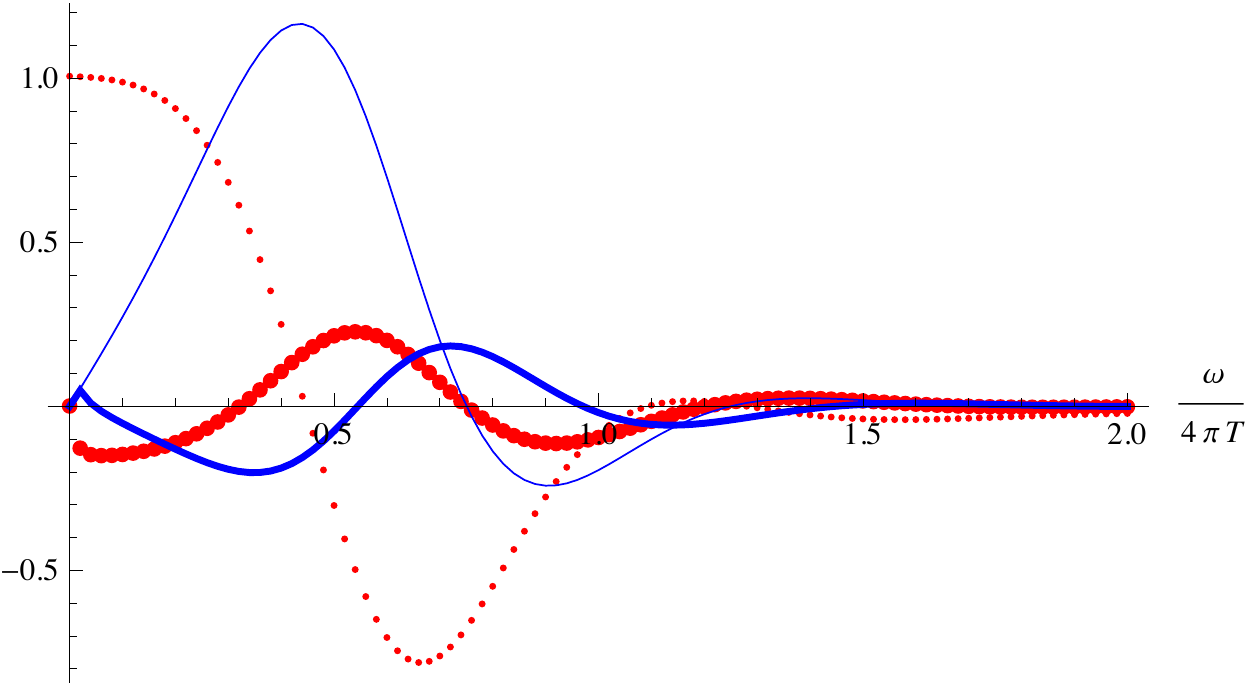} \hfill\includegraphics[scale=0.58]{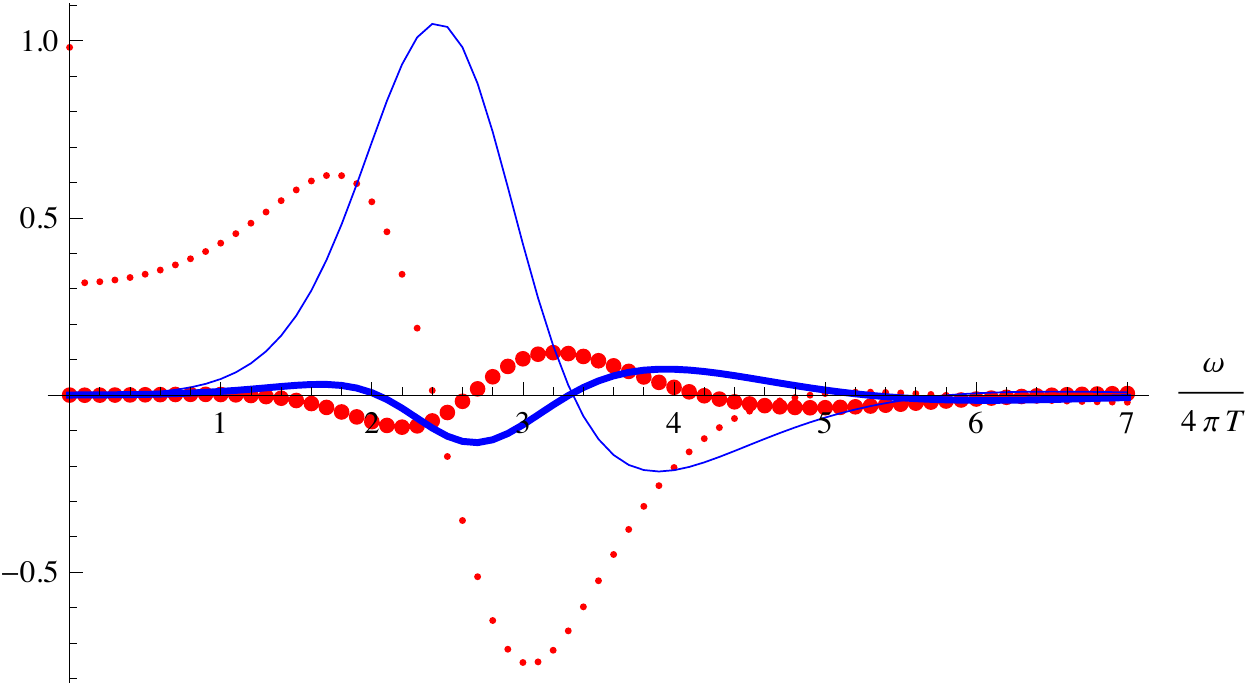} 
\end{center}
\caption{\label{plots4}Chiral vortical (up) and magnetic (bottom) conductivities as function of the frequency at $\tau=36.5$ (left) and $\tau=0.24$ (right). Red doted points represent real part and blue line the imaginary conductivity. Small dots and thin lines represent the conductivities with $\lambda=0$ and the thick case shows the $\kappa=0$ regime. All cases are normalized to zero frequency conductivities with both anomalous parameters switched on.}
\end{figure}

\begin{figure}[h!]
\begin{center}
\includegraphics[scale=0.58]{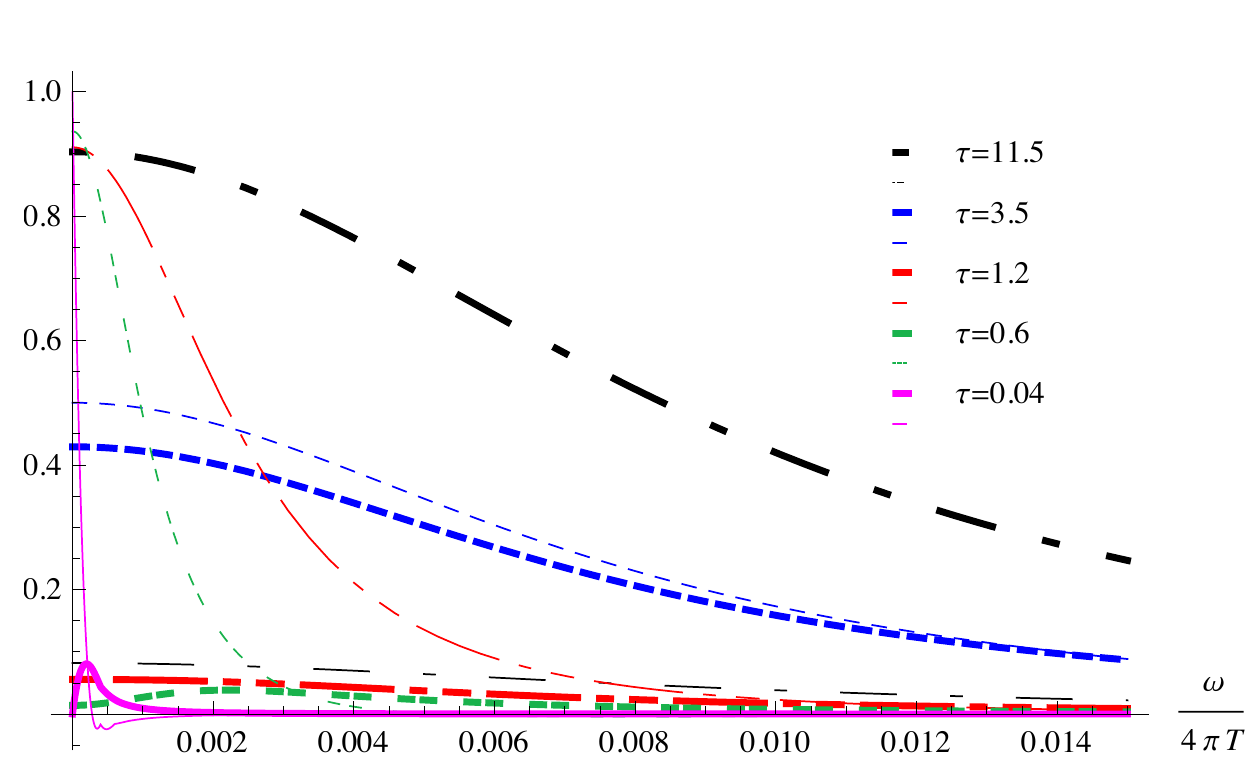}\hfill \includegraphics[scale=0.58]{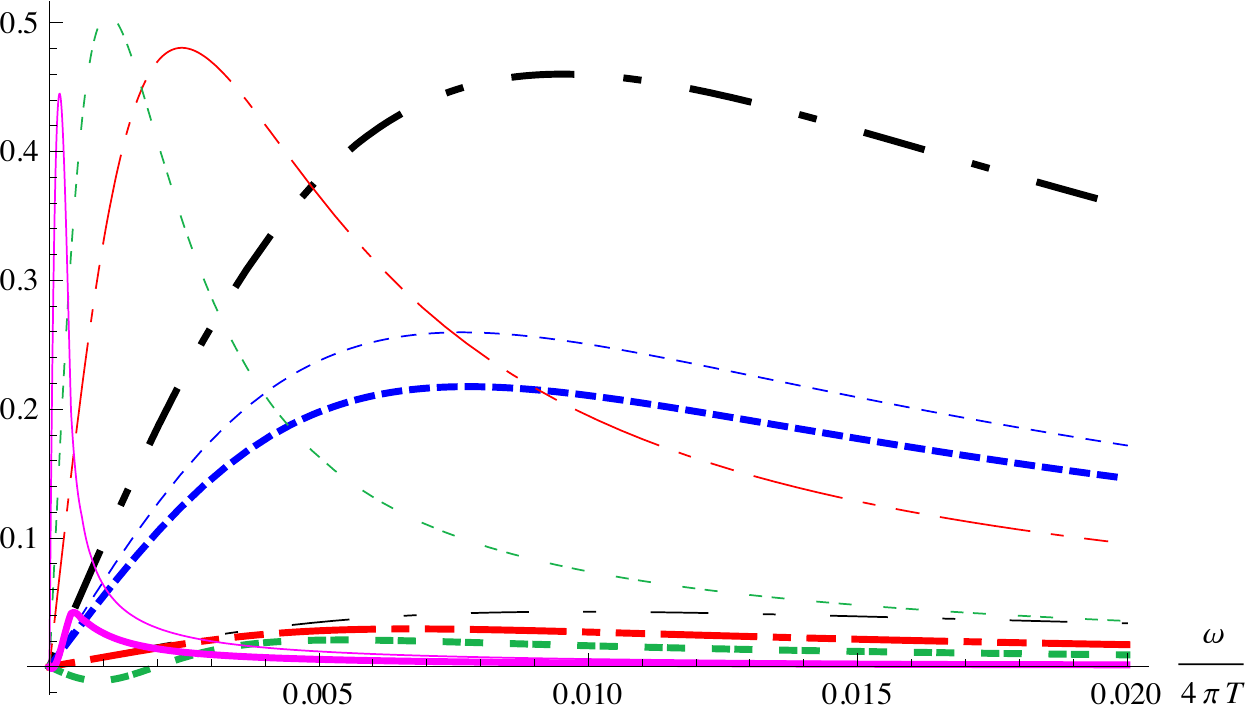} \\
\includegraphics[scale=0.58]{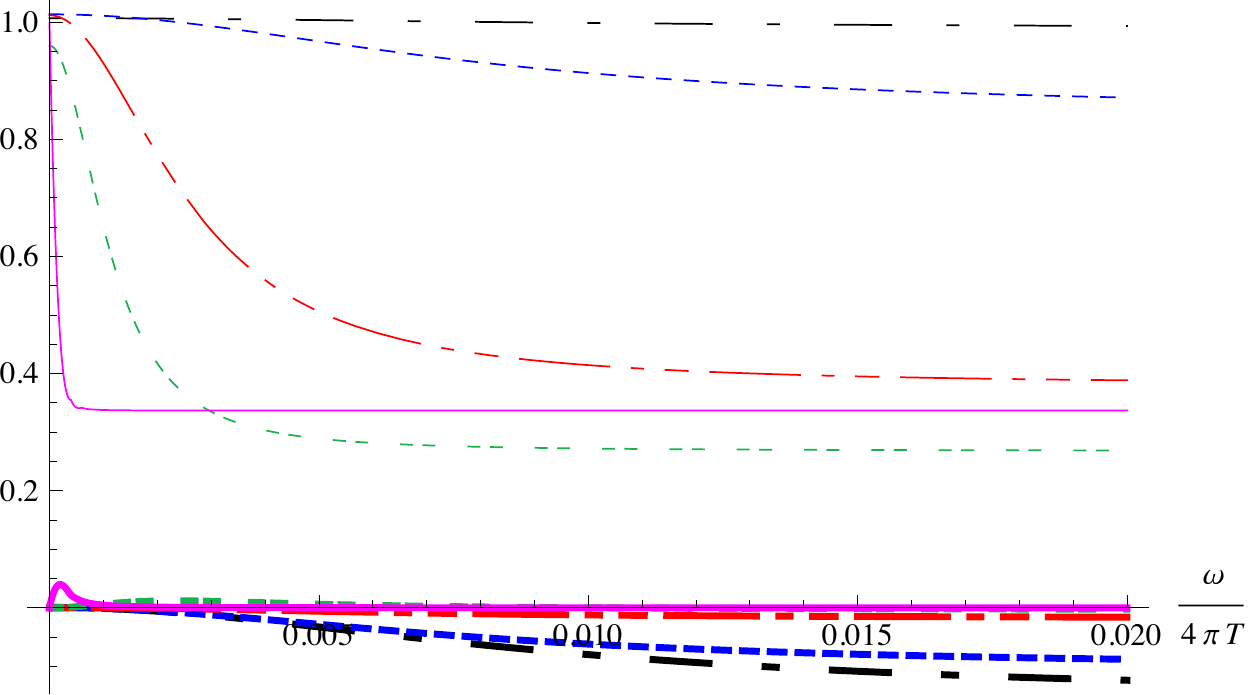} \hfill \includegraphics[scale=0.58]{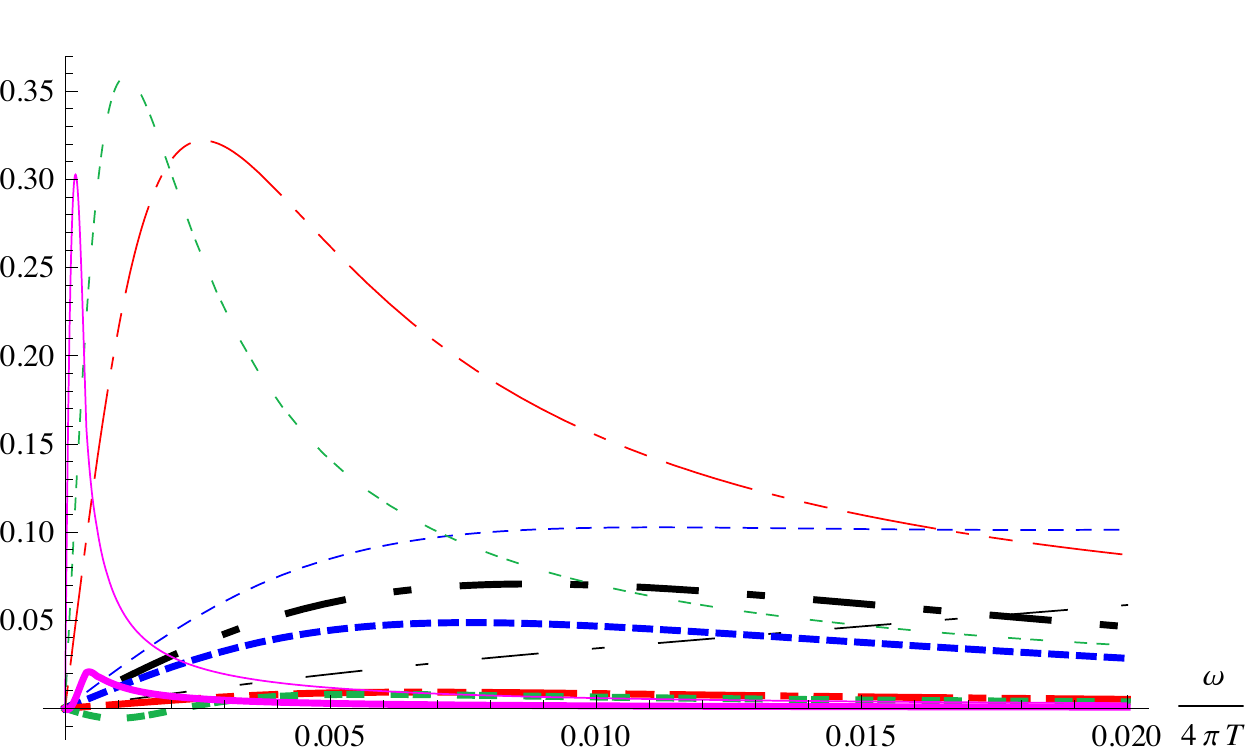} 
\end{center}
\caption{\label{plots5}Chiral vortical (up) and magnetic (bottom) conductivities as function of the frequency close to $\omega=0$. Real (left) and imaginary (right) part of the normalized conductivity for different values of the dimensionless temperature.}
\end{figure}

% %%%%%%%%%%%%%%%%%%%%%%%%%%%%%%%%%%%
   \chapter{First order solutions in the Fluid/Gravity approach}
    \label{ap:Ffunctions}
     Here we show the exact form of the $F_i[\rho]$ functions defined in Eqs. (\ref{eq:W4}) and (\ref{eq:amu})
{\scriptsize
\begin{eqnarray}
\nonumber F_1[\rho] &=& -Q\Bigg(\frac{\left(9 Q^6 \rho^2+2 M^2 \rho^3 (-4 M+3 (1+M) \rho )-27 Q^4 \left(1+M-\rho^2\right)+6 Q^2 \rho  \left(4 M^2+3 \rho -6 M \rho^4\right)\right) \rho_2 \left(2+5 \rho_2^2 +2 \rho_2^4\right)}{8 M^2 \rho^6 \rho_2 \left(-1+\rho_2^2\right) \left(2+\rho_2^2\right)^2 \left(1+2\rho_2^2 \right)^3} \\
\nonumber && +\frac{-6 \pi  Q \left(-1+\rho^2\right) \left(\rho^2-\rho_2^2\right) \left(-1+\rho_2^2\right)^2 \left(1+\rho ^2+\rho_2^2\right) \left(2+5 \rho_2^2+7 \rho_2^4+5 \rho_2^6+2 \rho_2^8\right)}{8 M^2 \rho ^6 \rho_2 \left(-1+\rho_2^2\right) \left(2+\rho_2^2\right)^2 \left(1+2 \rho_2^2\right)^3}  \Bigg)\\
\nonumber && -\frac{3 \left(Q^2-M \rho ^2+\rho ^6\right) \rho_2 \left(-2-3 \rho_2^2+3 \rho_2^6+2 \rho_2^8\right) \text{ArcTan}\left[\frac{\rho }{\sqrt{1+\rho_2^2}}\right]}{2 M \rho ^6 \left(2+\rho_2^2\right)^2 \left(1+2 \rho_2^2\right)^3} + \frac{3 (1+M) Q \left(Q^2-M \rho ^2+\rho ^6\right) \text{Log}[1+\rho ]}{2 M \left(-2+Q^2\right)^2 \rho ^6}\\
\nonumber && -\frac{3 Q \left(Q^2-M \rho ^2+\rho ^6\right) (-1+\rho_2) \rho_2 \left(2+\rho_2^2\right) \text{Log}[\rho -\rho_2]}{4 \rho ^6 (1+\rho_2)^2 (1+(-1+\rho_2) \rho_2) \left(1+2 \rho_2^2\right)^3} -\frac{3 Q \left(Q^2-M \rho ^2+\rho ^6\right) \rho_2 (1+\rho_2) \left(2+\rho_2^2\right) \text{Log}[\rho +\rho_2]}{4 \rho ^6 (-1+\rho_2)^2 \left(1+\rho_2+\rho_2^2\right) \left(1+2 \rho_2^2\right)^3} \\
 && + \frac{3 Q \left(Q^2-M \rho ^2+\rho ^6\right) \left(-2-\rho_2^2+M \rho_2^4\right) \text{Log}\left[1+\rho ^2+\rho_2^2\right]}{4 M \rho ^6 \left(2+\rho_2^2\right)^2 \left(1+2 \rho_2^2\right)^3}
\end{eqnarray}
}

{\scriptsize
\begin{eqnarray}
\nonumber F_2[\rho] &=& -\frac{2 \sqrt{3} \kappa  \rho_2^6 \left(1+3 \rho_2^2 + 2 \rho_2^4\right)^2}{Q \rho ^6 \left(1+2 \rho_2^2\right)^2 \left(Q^2+\rho_2^6\right)} + \frac{2 \sqrt{3} \lambda  \left(12 M \rho_2^6 \left(1+3 \rho_2^2+2 \rho_2^4\right)^2-2 Q^2 \rho ^2 \rho_2^2 \left(1+3 \rho_2^2+3 \rho_2^4+2 \rho_2^6\right)^2\right.}{Q \rho ^{10} \left(1+2 \rho_2^2\right)^2 \left(Q^2+\rho_2^6\right)}\\
\nonumber && +\frac{+6 \rho ^6 \left(Q^2+\rho_2^6\right)^2 \left(1+3 \rho_2^2+4 \rho_2^4+2 \rho_2^6\right)+4 M^2 \rho ^8 \rho_2^2 \left(2+7 \rho_2^2+9 \rho_2^4+4 \rho_2^6+2 \rho_2^8\right)}{Q \rho ^{10} \left(1+2 \rho_2^2\right)^2 \left(Q^2+\rho_2^6\right)} \\
\nonumber && +\frac{\left.-2 \rho ^4 \rho_2^2 \left(4+Q^2 \left(28+Q^2 \left(60+\left(43+34 Q^2\right) \rho_2^2 \left(1+\rho_2^2\right)\right)\right)\right)\right)}{Q \rho ^{10} \left(1+2 \rho_2^2\right)^2 \left(Q^2+\rho_2^6\right)} -\frac{32 \sqrt{3} M^2 \lambda  \left(Q^2-M \rho ^2+\rho ^6\right) \text{Log}[\rho ]}{Q^3 \rho ^6} \\
\nonumber && + \frac{8 \sqrt{3} Q \lambda  \left(Q^2-M \rho ^2+\rho ^6\right) \left(2+12 \rho_2^2+27 \rho_2^4+35 \rho_2^6+27 \rho_2^8+12 \rho_2^{10}+2 \rho_2^{12}\right) \text{Log}[\rho -\rho_2]}{\rho ^6 \rho_2^4 \left(1+2 \rho_2^2\right)^3}\\
\nonumber && + \frac{8 \sqrt{3} Q \lambda  \left(Q^2-M \rho ^2+\rho ^6\right) \left(2+12 \rho_2^2+27 \rho_2^4+35 \rho_2^6+27 \rho_2^8+12 \rho_2^{10}+2 \rho_2^{12}\right) \text{Log}[\rho +\rho_2]}{\rho ^6 \rho_2^4 \left(1+2 \rho_2^2\right)^3} \\
&& -\frac{8 \sqrt{3} \lambda  \left(Q^2-M \rho ^2+\rho ^6\right) \rho_2 \left(-1-3 \rho_2^2-6 \rho_2^4-7 \rho_2^6-3 \rho_2^8+2 \rho_2^{12}\right) \text{Log}\left[1+\rho ^2+\rho_2^2\right]}{\rho ^6 \left(1+\rho_2^2\right)^{3/2} \left(1+2 \rho_2^2\right)^3}  
\end{eqnarray}
}

{\scriptsize
\begin{eqnarray}
\nonumber F_3[\rho] &=& \frac{\sqrt{3} \left(9 Q^5 \left(2+\rho_2^2\right)+2 M Q \rho ^4 \left(2+\rho_2^2\right) \left(-M+2 \rho  \left(1+2 \rho_2^2\right)^2\right)-\left(\rho_2+\rho_2^3\right) \left(3 M Q \rho ^2 \rho_2 \left(2+\rho_2^2\right)\right.\right.}{8 M^2 \rho ^6 \left(2+\rho_2^2\right) \left(1+2 \rho_2^2\right)^2} \\
\nonumber && + \frac{\left.\left.+2 \pi  \left(-1+\rho ^2\right) \left(\rho ^2-\rho_2^2\right) \left(1+\rho ^2+\rho_2^2\right) \left(1+4 \rho_2^2+6 \rho_2^4+5 \rho_2^6+2 \rho_2^8\right)\right)\right)}{8 M^2 \rho ^6 \left(2+\rho_2^2\right) \left(1+2 \rho_2^2\right)^2}\\
\nonumber && \frac{\sqrt{3} \left(Q^2-M \rho ^2+\rho ^6\right) \rho_2 \left(1+\rho_2^2\right)^2 \text{ArcTan}\left[\frac{\rho }{\sqrt{1+\rho_2^2}}\right]}{2 M \rho ^6 \left(2+5 \rho_2^2+2 \rho_2^4\right)} + \frac{\sqrt{3} Q \left(Q^2-M \rho ^2+\rho ^6\right) \text{Log}[1+\rho ]}{2 \rho ^6 \left(-2-\rho_2^2+2 \rho_2^6+\rho_2^8\right)} \\
\nonumber && + \frac{\sqrt{3} Q \left(Q^2-M \rho ^2+\rho ^6\right) \rho_2^2 (4+\rho_2 (-1+4 \rho_2)) \text{Log}[\rho -\rho_2]}{4 \rho ^6 \left(1+2 \rho_2^2\right)^3 \left(1+\rho_2^3\right)} -\frac{\sqrt{3} Q \left(Q^2-M \rho ^2+\rho ^6\right) \rho_2^2 \left(4+\rho_2+4 \rho_2^2\right) \text{Log}[\rho +\rho_2]}{4 \rho ^6 \left(1+2 \rho_2^2\right)^3 \left(-1+\rho_2^3\right)} \\
 && +\frac{\sqrt{3} Q^3 \left(Q^2-M \rho ^2+\rho ^6\right) \left(-1+\rho_2^2\right)^2 \text{Log}\left[1+\rho ^2+\rho_2^2\right]}{4 \rho ^6 \rho_2^2 \left(1+2 \rho_2^2\right)^3 \left(2+3 \rho_2^2+3 \rho_2^4+\rho_2^6\right)}
\end{eqnarray}
}

{\scriptsize
\begin{eqnarray}
\nonumber F_4[\rho] &=& \frac{\kappa  \left(-9 Q^4 \rho ^2+3 M Q^2 \rho ^4-6 M Q^2 \rho ^6\right)}{M \rho ^8 \left(1+2 \rho_2^2\right)^2} + \frac{\lambda  \left(-120 Q^2 \rho ^2-180 Q^4 \rho ^2+18 M Q^4 \rho ^4+72 M Q^2 \rho ^6\right.}{M \rho ^8 \left(1+2 \rho_2^2\right)^2}\\
\nonumber && +\frac{\left.+2 Q^6 \rho ^2 \left(-67+6 \rho ^4\right)+4 \left(-5 \rho ^2+3 \rho ^6+6 \left(\rho_2+2 \rho_2^3\right)^2 \left(1+2 \rho_2^2+2 \rho_2^4+\rho_2^6\right)\right)\right)}{M \rho ^8 \left(1+2 \rho_2^2\right)^2}\\
\nonumber && -\frac{24 M \lambda  \left(Q^2-M \rho ^2+\rho ^6\right) \text{Log}[\rho ]}{Q^2 \rho ^6} -\frac{6 Q^2 \kappa  f[\rho] \text{Log}[\rho -\rho_2]}{\left(1+2 \rho_2^2\right)^3}  +\frac{12 \lambda  f[\rho] \left(1+\rho_2^2\right) \left(1+5 \rho_2^2+9 \rho_2^4+5 \rho_2^6+\rho_2^8\right) \text{Log}[\rho -\rho_2]}{\rho_2^2 \left(1+2 \rho_2^2\right)^3}\\
\nonumber && -\frac{6 \kappa  f[\rho]  Q^2 \text{Log}[\rho +\rho_2]}{\left(1+2 \rho_2^2\right)^3}+\frac{12 \lambda  f[\rho] Q^2\left(1+5 \rho_2^2+9 \rho_2^4+5 \rho_2^6+\rho_2^8\right) \text{Log}[\rho +\rho_2]}{\rho_2^4 \left(1+2 \rho_2^2\right)^3} \\
&& \frac{6 Q^2 \kappa  f[\rho] \text{Log}\left[1+\rho ^2+\rho_2^2\right]}{\left(1+2 \rho_2^2\right)^3}-\frac{12 \lambda  f[\rho] \rho_2^2 \left(1+2 \rho_2^2-\rho_2^6+\rho_2^8\right) \text{Log}\left[1+\rho ^2+\rho_2^2\right]}{\left(1+\rho_2^2\right) \left(1+2 \rho_2^2\right)^3}
\end{eqnarray}
}
{\scriptsize
\begin{eqnarray}
\nonumber F_6[\rho] &=&\frac{\sqrt{3} Q \left(3 Q \rho_2 \left(2+5 \rho_2^2+2 \rho_2^4\right) \left(-8 M^2 \rho +9 (1+M) \rho_2^2+9 (1+M) \rho_2^4\right)\right.}{8 M^2 \rho ^2 \rho_2 \left(-1+\rho_2^2\right) \left(2+\rho_2^2\right)^2 \left(1+2 \rho_2^2\right)^3}\\
\nonumber && + \frac{\left.+2 \pi  \left(3 Q^2-2 M \rho ^2\right) \left(-1+\rho_2^2\right)^2 \left(2+5 \rho_2^2+7 \rho_2^4+5 \rho_2^6+2 \rho_2^8\right)\right)}{8 M^2 \rho ^2 \rho_2 \left(-1+\rho_2^2\right) \left(2+\rho_2^2\right)^2 \left(1+2 \rho_2^2\right)^3} \\
\nonumber && + \frac{\sqrt{3} \left(-3 Q^2+2 M \rho ^2\right) \left(-2-3 \rho_2^2+3 \rho_2^6+2 \rho_2^8\right) \text{ArcTan}\left[\frac{\rho }{\sqrt{1+\rho_2^2}}\right]}{2 M \rho ^2 \sqrt{1+\rho_2^2} \left(2+\rho_2^2\right)^2 \left(1+2 \rho_2^2\right)^3} + \frac{\sqrt{3} (1+M) \left(3 Q^2-2 M \rho ^2\right) \text{Log}[1+\rho ]}{2 M \left(-2+Q^2\right)^2 \rho ^2} \\
\nonumber && +\frac{\sqrt{3} \left(-3 Q^2+2 M \rho ^2\right) (-1+\rho_2) \rho_2 \left(2+\rho_2^2\right) \text{Log}[\rho -\rho_2]}{4 \rho ^2 (1+\rho_2)^2 (1+(-1+\rho_2) \rho_2) \left(1+2 \rho_2^2\right)^3} + \frac{\sqrt{3} \left(-3 Q^2+2 M \rho ^2\right) \rho_2 (1+\rho_2) \left(2+\rho_2^2\right) \text{Log}[\rho +\rho_2]}{4 \rho ^2 (-1+\rho_2)^2 \left(1+\rho_2+\rho_2^2\right) \left(1+2 \rho_2^2\right)^3} \\
 && +\frac{\sqrt{3} \left(3 Q^2-2 M \rho ^2\right) \left(-2-\rho_2^2+\rho_2^4+\rho_2^6+\rho_2^8\right) \text{Log}\left[1+\rho ^2+\rho_2^2\right]}{4 M \rho ^2 \left(2+\rho_2^2\right)^2 \left(1+2 \rho_2^2\right)^3}
\end{eqnarray}
}
{\scriptsize
\begin{eqnarray}
\nonumber F_7[\rho] &=& -\frac{6 \kappa  \rho_2^4 \left(1+3 \rho_2^2+2 \rho_2^4\right)^2}{M \rho ^2 \left(1+\rho_2^2\right) \left(\rho_2+2 \rho_2^3\right)^2} + \frac{\lambda  \left(-4 Q^4 \left(120+77 Q^2+86 Q^4\right) \rho ^4+4 \left(-8 \rho ^4+9 M \rho_2^4 \left(1+3 \rho_2^2+2 \rho_2^4\right)^2\right)\right.}{M \rho ^6 \left(1+\rho_2^2\right) \left(\rho_2+2 \rho_2^3\right)^2}\\
\nonumber && +\frac{\left.+8 Q^2 \rho ^2 \left(-28 \rho ^2+3 \left(1+3 \rho_2^2+3 \rho_2^4+2 \rho_2^6\right)^2\right)\right)}{M \rho ^6 \left(1+\rho_2^2\right) \left(\rho_2+2 \rho_2^3\right)^2} + \frac{32 M^2 \lambda  \left(-3 Q^2+2 M \rho ^2\right) \text{Log}[\rho ]}{Q^4 \rho ^2} \\
\nonumber && + \frac{8 \lambda  \left(3 Q^2-2 M \rho ^2\right) \left(2+12 \rho_2^2+27 \rho_2^4+35 \rho_2^6+27 \rho_2^8+12 \rho_2^{10}+2 \rho_2^{12}\right) \text{Log}[\rho -\rho_2]}{\rho ^2 \rho_2^4 \left(1+2 \rho_2^2\right)^3} \\
\nonumber && + \frac{8 \lambda  \left(3 Q^2-2 M \rho ^2\right) \left(2+12 \rho_2^2+27 \rho_2^4+35 \rho_2^6+27 \rho_2^8+12 \rho_2^{10}+2 \rho_2^{12}\right) \text{Log}[\rho +\rho_2]}{\rho ^2 \rho_2^4 \left(1+2 \rho_2^2\right)^3} \\
&& + \frac{8 \lambda  \left(-3 Q^2+2 M \rho ^2\right) \left(-1-3 \rho_2^2-6 \rho_2^4-7 \rho_2^6-3 \rho_2^8+2 \rho_2^{12}\right) \text{Log}\left[1+\rho ^2+\rho_2^2\right]}{\rho ^2 \left(1+\rho_2^2\right)^2 \left(1+2 \rho_2^2\right)^3}
\end{eqnarray}
}
{\scriptsize
\begin{eqnarray}
\nonumber F_8[\rho] &=& \frac{\left(1+\rho_2^2\right) \left(27 \rho_2^5 \left(1+\rho_2^2\right) \left(2+\rho_2^2\right)+2 \pi  Q \left(-3 Q^2+2 M \rho ^2\right) \left(1+3 \rho_2^2+3 \rho_2^4+2 \rho_2^6\right)\right)}{8 M^2 \rho ^2 \rho_2 \left(2+\rho_2^2\right) \left(1+2 \rho_2^2\right)^2} \\
\nonumber  && + \frac{Q^3 \left(3 Q^2-2 M \rho ^2\right) \text{ArcTan}\left[\frac{\rho }{\sqrt{1+\rho_2^2}}\right]}{2 \rho ^2 \rho_2^3 \left(2+7 \rho_2^2+9 \rho_2^4+7 \rho_2^6+2 \rho_2^8\right)} + \frac{\left(3 Q^2-2 M \rho ^2\right) \text{Log}[1+\rho ]}{2 \rho ^2 \left(-2-\rho_2^2+2 \rho_2^6+\rho_2^8\right)} \\
\nonumber && +\frac{\left(3 Q^2-2 M \rho ^2\right) \rho_2^2 (4+\rho_2 (-1+4 \rho_2)) \text{Log}[\rho -\rho_2]}{4 \rho ^2 \left(1+2 \rho_2^2\right)^3 \left(1+\rho_2^3\right)} + \frac{\left(-3 Q^2+2 M \rho ^2\right) \rho_2^2 \left(4+\rho_2+4 \rho_2^2\right) \text{Log}[\rho +\rho_2]}{4 \rho ^2 \left(1+2 \rho_2^2\right)^3 \left(-1+\rho_2^3\right)} \\
 && \frac{\left(3 Q^2-2 M \rho ^2\right) \left(-1+\rho_2^2\right)^2 \left(1+\rho_2^2\right) \text{Log}\left[1+\rho ^2+\rho_2^2\right]}{4 M \rho ^2 \left(2+\rho_2^2\right) \left(1+2 \rho_2^2\right)^3}
\end{eqnarray}
}
{\scriptsize
\begin{eqnarray}
\nonumber F_9[\rho] &=& -\frac{9 \sqrt{3} Q^3 \kappa  \rho_2^2}{\rho ^2 \left(1+2 \rho_2^2\right)^2 \left(Q^2+\rho_2^6\right)} + \frac{2 \sqrt{3} \lambda  \rho_2^2 \left(-\left(4+24 Q^2+36 Q^4+43 Q^6\right) \rho ^2+6 \left(\rho_2+2 \rho_2^3\right)^2 \left(1+2 \rho_2^2+2 \rho_2^4+\rho_2^6\right)\right)}{Q \rho ^4 \left(1+2 \rho_2^2\right)^2 \left(Q^2+\rho_2^6\right)} \\
\nonumber &&  -\frac{2 \sqrt{3} Q \kappa  \left(3 Q^2-2 M \rho ^2\right) \text{Log}[\rho -\rho_2]}{\rho ^2 \left(1+2 \rho_2^2\right)^3} +\frac{4 \sqrt{3} Q \lambda  \left(3 Q^2-2 M \rho ^2\right) \left(1+5 \rho_2^2+9 \rho_2^4+5 \rho_2^6+\rho_2^8\right) \text{Log}[\rho -\rho_2]}{\rho ^2 \rho_2^4 \left(1+2 \rho_2^2\right)^3} \\
\nonumber &&  -\frac{2 \sqrt{3} Q \kappa  \left(3 Q^2-2 M \rho ^2\right) \text{Log}[\rho +\rho_2]}{\rho ^2 \left(1+2 \rho_2^2\right)^3}+\frac{4 \sqrt{3} Q \lambda  \left(3 Q^2-2 M \rho ^2\right) \left(1+5 \rho_2^2+9 \rho_2^4+5 \rho_2^6+\rho_2^8\right) \text{Log}[\rho +\rho_2]}{\rho ^2 \rho_2^4 \left(1+2 \rho_2^2\right)^3} \\
\nonumber && +\frac{2 \sqrt{3} \kappa  \left(3 Q^2-2 M \rho ^2\right) \rho_2 \sqrt{1+\rho_2^2} \text{Log}\left[1+\rho ^2+\rho_2^2\right]}{\rho ^2 \left(1+2 \rho_2^2\right)^3}-\frac{4 \sqrt{3} \lambda  \left(3 Q^2-2 M \rho ^2\right) \rho_2 \left(1+2 \rho_2^2-\rho_2^6+\rho_2^8\right) \text{Log}\left[1+\rho ^2+\rho_2^2\right]}{\rho ^2 \left(1+\rho_2^2\right)^{3/2} \left(1+2 \rho_2^2\right)^3}\\
&&+\frac{8 \sqrt{3} M \lambda  \left(-3 Q^2+2 M \rho ^2\right) \text{Log}[\rho ]}{\rho ^2 \rho_2^3 \left(1+\rho_2^2\right)^{3/2}}
\end{eqnarray}
}

% %%%%%%%%%%%%%%%%%%%%%%%%%%%%%%%%%%%
  \chapter{Second order sources}
    \label{ap:SOsources}
     \section{Second Order Vector Sources}
\label{ap:Vec_sour}
In this appendix we show the vector sources splitted in terms of the anomalous and non anomalous one, the tildes refer to the anomalous sector
\begin{equation}
\mathbb{J}_{\mu} = \sum_{a=1}^{10}r_a^{(E)} \mathcal{J}^{(a)}_{\mu} + \sum_{a=1}^{5}\tilde r_a^{(E)} \mathcal{\tilde J}^{(a)}_{\mu}\qquad , \qquad \mathbb{A}_{\mu} = \sum_{a=1}^{10}r_a^{(M)} \mathcal{J}^{(a)}_{\mu} + \sum_{a=1}^{5}\tilde r_a^{(M)} \mathcal{\tilde J}^{(a)}_{\mu} \,,
\end{equation}
\subsection{Non-anomalous sources}
{\scriptsize
\begin{eqnarray}
\nonumber r_+^3 r_1^{(E)} &=& \frac{2 \pi  Q T^2 (-1+\rho )^2 \left(-\rho ^2 (1+\rho  (2+3 \rho ))+Q^2 (3+2 \rho  (3+2 \rho  (2+\rho )))\right)}{\sqrt{3}  M (1+M) \rho ^8 f[\rho ]^2} +\frac{2 \pi  T^2 \rho ^3 \partial_QF_5'[\rho ]}{\sqrt{3}  (1+M)} -\frac{4 \pi  T^2 (-1+\rho ) \left(1+\rho +\rho ^2\right) F_1'[\rho ]}{\sqrt{3}  (1+M) f[\rho ]} \\
 &&  \frac{4 \pi  T^2 (-1+\rho )^2 \left(Q^2 (6+\rho  (12+7 \rho  (2+\rho )))-\rho ^2 (4+\rho  (8+3 \rho  (4+\rho  (3+\rho  (2+\rho )))))\right) F_1[\rho ]}{\sqrt{3}  (1+M) \rho ^7 f[\rho ]^2} \,, \\
 \nonumber r_+^3 r_2^{(E)} &=& -\frac{64 \sqrt{3} \pi  Q T^2 \lambda ^2 \left(-9 Q^4 \left(242+131 Q^2\right)+M Q^2 \left(2654+1603 Q^2\right) \rho ^2-4 M^2 \left(146+127 Q^2\right) \rho ^4\right)}{ M (1+M) \rho ^{17} f[\rho ]}\\
 \nonumber && -\frac{64 \sqrt{3} \pi  Q T^2 \lambda ^2 \left(3 \left(16-498 Q^2-225 Q^4+16 Q^6\right) \rho ^6+220 \left(2+3 Q^2+Q^4\right) \rho ^8\right)}{ M (1+M) \rho ^{17} f[\rho ]}-\frac{2 \sqrt{3} \pi  T^2 \rho ^2 F_1[\rho ]}{ (1+M)}  -\frac{4 \pi  T^2 \rho ^3  F_1'[\rho ]}{\sqrt{3}  (1+M)}\\
\nonumber  && -\frac{768 \sqrt{3} \pi  Q^3 T^2 \kappa  \lambda  \left(-3+\rho ^2+Q^2 \left(-2+\rho ^2\right)\right)}{ M (1+M) \rho ^{11} f[\rho ]}+ \frac{64 \sqrt{3} \pi  T^2 \lambda  \left(\rho ^2+Q^2 \left(-5+\rho ^2\right)\right) \partial_Q  F_7[\rho ]}{ (1+M) \rho ^5}  -\frac{16 \pi  Q T^2 \lambda  \partial_QF_2'[\rho ]}{(1+M)}\\
 && + \frac{48 \pi  T^2 \lambda  \left(2 \rho ^2-2 \rho ^6+Q^4 \left(-5+3 \rho ^2\right)-Q^2 \left(8-5 \rho ^2+\rho ^6\right)\right) F_2'[\rho ]}{ M (1+M) \rho ^6 f[\rho ]} + \frac{2 \pi  Q T^2 F_6'[\rho ]}{\ (1+M) \rho }-\frac{16 \pi  Q T^2 \lambda  \rho  \partial_QF_2''[\rho ]}{ (1+M)}\\
\nonumber && + \frac{8 \sqrt{3} \pi  Q T^2 \lambda  \left(-3 Q^2 \left(2+Q^2\right)+M \left(18+7 Q^2\right) \rho ^2-3 \left(14+9 Q^2\right) \rho ^6+8 M \rho ^8\right) F_7'[\rho ]}{ M (1+M) \rho ^{10} f[\rho ]}-\frac{16 \sqrt{3} \pi  T^2 \lambda  \left(-5 Q^2+2 M \rho ^2\right) \partial_QF_7'[\rho ]}{ (1+M) \rho ^4}, 
\end{eqnarray}
\begin{eqnarray}
r_+ r_3^{(E)} &=& \rho^3 F_5'(\rho) \,, \\
\nonumber r_+ r_4^{(E)} &=& -\frac{24192 Q^4 \lambda ^2-7680 M Q^2 \lambda ^2 \rho ^2+\rho ^{12}}{\rho ^{11}} +\frac{96 \lambda  \left(\rho ^2+Q^2 \left(-5+\rho ^2\right)\right) F_7[\rho ]}{ \rho ^5} -\frac{24 \lambda  \left(-7 Q^2+2 M \rho ^2\right) F_7'[\rho ]}{ \rho ^4} + 32 \sqrt{3} Q \lambda  F_2'[\rho ] \,, \\
&&\\
\nonumber r_+^2 r_5^{(E)} &=& \frac{\sqrt{3} Q (-1+\rho )^2 \left(\rho ^2 (1+\rho  (2+3 \rho ))-Q^2 (3+2 \rho  (3+2 \rho  (2+\rho )))\right)}{\ M \rho ^8 f[\rho ]^2} -\frac{2 (-1+\rho ) \left(1+\rho +\rho ^2\right) F_3'[\rho ]}{ f[\rho ]}\\
 \nonumber &&-\frac{2 (-1+\rho ) \left(1+\rho +\rho ^2\right) F_3'[\rho ]}{ f[\rho ]}  + \frac{2Q^2 (-1+\rho )^2  (6+\rho  (12+7 \rho  (2+\rho ))) F_3[\rho ]}{ \rho ^7 f[\rho ]^2}\\
 &&- \frac{2\rho ^2 (-1+\rho )^2  (4+\rho  (8+3 \rho  (4+\rho  (3+\rho  (2+\rho ))))) F_3[\rho ]}{ \rho ^7 f[\rho ]^2} \,, \\
\nonumber r_+^2 r_6^{(E)} &=& -\frac{96 \sqrt{3} Q \lambda  \left(-999 Q^6 \lambda +1813 M Q^4 \lambda  \rho ^2-978 M Q^2 \lambda  \rho ^4+2 M Q^2 (4 \kappa +341 \lambda ) \rho ^8-120 M^2 \lambda  \rho ^{10}\right)}{ M \rho ^{17} f[\rho ]}- 2 \rho ^3 F_3'[\rho ] + \frac{\sqrt{3} Q F_8'[\rho ]}{ \rho }\\
\nonumber && -\frac{96 \sqrt{3} Q \lambda  \left(\left(152 \lambda +456 Q^2 \lambda +152 Q^6 \lambda -3 Q^4 (4 \kappa +121 \lambda )\right) \rho ^6\right)}{ M \rho ^{17} f[\rho ]} -\frac{24 \lambda  \left(9 Q^4-8 M Q^2 \rho ^2+M^2 \rho ^4+3 Q^2 \rho ^6-M \rho ^8\right) F_2'[\rho ]}{ M \rho ^6 f[\rho ]}\\
&& + \frac{4 \sqrt{3} Q \lambda  \left(9 Q^4-19 M Q^2 \rho ^2+6 M^2 \rho ^4+45 Q^2 \rho ^6-22 M \rho ^8\right) F_7'[\rho ]}{ M (-1+\rho ) \rho ^4 (1+\rho ) \left(Q^2-\rho ^2 \left(1+\rho ^2\right)\right)} -3 \rho ^2 F_3[\rho ]\,, \\
r_7^{(E)} &=& 0 \,, \\
\nonumber r_+^4 r_8^{(E)} &=& -\frac{384 \pi  T^2 \lambda  \left(5 Q^4 \left(14+Q^2\right) \lambda -4 M Q^2 \left(22+Q^2\right) \lambda  \rho ^2-M^2 \left(-18+Q^2\right) \lambda  \rho ^4\right)}{ M (1+M) \rho ^{15} f[\rho ]}-\frac{32 \sqrt{3} \pi  Q T^2 \lambda  \left(\rho ^2+Q^2 \left(-5+\rho ^2\right)\right) F_9[\rho ]}{ M (1+M) \rho ^5}\\
\nonumber && -\frac{384 \pi  T^2 \lambda  \left(Q^2 \left(\left(6+4 Q^2\right) \kappa +5 \left(14+Q^2\right) \lambda \right) \rho ^6+M \left(-18 \lambda +Q^2 (-2 \kappa +\lambda )\right) \rho ^8\right)}{ M (1+M) \rho ^{15} f[\rho ]}\\
\nonumber &&  -\frac{16 \pi  T^2 \lambda  \left(-Q^2 \left(24+13 Q^2\right)+M \left(6+7 Q^2\right) \rho ^2-\left(6+Q^2\right) \rho ^6\right) F_4'[\rho]}{ M (1+M) \rho ^6 f[\rho ]}- \frac{64 \sqrt{3} \pi  T^2 \lambda  \left(\rho ^2+Q^2 \left(-5+\rho ^2\right)\right) \partial_Q F_9[\rho ]}{ (1+M) \rho ^5}\\
\nonumber && -\frac{64 \pi  Q T^2 \lambda  \partial_QF_4'[\rho]}{ (1+M)} -\frac{2 \pi  T^2 \rho  F_6'[\rho ]}{\sqrt{3}  (1+M)} + \frac{16 \sqrt{3} \pi  T^2 \lambda  \left(-7 Q^2+2 M \rho ^2\right) \partial_Q F_9'[\rho ]}{ (1+M) \rho ^4}\\
\nonumber && -\frac{16 \sqrt{3} \pi  Q T^2 \lambda  \left(7 \rho ^2+\rho ^4-19 \rho ^6+3 \rho ^8+Q^4 \left(4-3 \rho ^2+\rho ^4\right)\right) F_9'[\rho ]}{ M (1+M) \rho ^{10} f[\rho ]} -\frac{16 \sqrt{3} \pi  Q^3 T^2 \lambda  \left(-1+4 \rho ^2+2 \rho ^4-8 \rho ^6+3 \rho ^8\right) F_9'[\rho ]}{ M (1+M) \rho ^{10} f[\rho ]} \,, \\
\nonumber r_+^3 r_{9}^{(E)} &=& -\frac{48 \sqrt{3} Q \lambda  \left(\rho ^2+Q^2 \left(-5+\rho ^2\right)\right) F_9[\rho ]}{ M \rho ^5}  + \rho  F_8'[\rho ] -\frac{24 \lambda  \left(\rho ^4-\rho ^8-Q^2 \rho ^2 \left(-5+\rho ^2\right) \left(-2+\rho ^4\right)+Q^4 \left(11-10 \rho ^2+\rho ^4\right)\right) F_4'[\rho ]}{ M \rho ^6 f[\rho ]} \\
 && -\frac{8 \sqrt{3} Q \lambda  \left(15 Q^4+2 M \rho ^4 \left(3 M-7 \rho ^4\right)+Q^2 \left(-23 M \rho ^2+33 \rho ^6\right)\right) F_9'[\rho ]}{ M \rho ^{10} f[\rho ]}\\
 \nonumber r_+^2 r_{10}^{(E)} &=& \frac{8\sqrt{3} \lambda ^2 \left(-480  Q^3+96  Q \rho ^2+96  Q^3 \rho ^2\right)}{ \rho ^9}+\frac{96 \lambda  \left(\rho ^6+Q^2 \rho ^4 \left(-5+\rho ^2\right)\right) F_9[\rho ]}{ \rho ^9}+32 \sqrt{3} Q \lambda  F_4'[\rho ] \\
&& +\frac{8 \lambda  \left(21 Q^2 \rho ^5-6 \rho ^7-6 Q^2 \rho ^7\right) F_9'[\rho ]}{ \rho ^9} \,,
\end{eqnarray}
}
{\scriptsize
\begin{eqnarray}
\nonumber r_+^3 r_1^{(M)} &=& \frac{\pi  T^2 (-1+\rho ) \left(8 \rho ^2 \left(1+\rho +\rho ^2\right)+Q^4 \left(3-\rho  \left(-3+4 \rho  (1+2 \rho )^2\right)\right)\right)}{2  M^2 (1+M) \rho ^8 f[\rho ]} \\
\nonumber && + \frac{4 \pi  T^2 F_6'[\rho ]}{\sqrt{3}  (1+M) \rho ^2}  -\frac{8 \pi  Q T^2 (-1+\rho ) \left(1+\rho +\rho ^2\right) F_1[\rho ]}{ (1+M) \rho ^5 f[\rho ]} -\frac{3 \pi  Q T^2 \rho ^3 f[\rho ] \partial_QF_5'[\rho ]}{2  M (1+M)}  \\
&& + \frac{\pi  T^2Q^2 (-1+\rho )  \left(12+\rho  \left(12+\rho  \left(13+\rho +\rho ^2+9 \rho ^3\right)\right)\right)}{2  M^2 (1+M) \rho ^8 f[\rho ]} \,,\\
 \nonumber r_+^3 r_2^{(M)} &=&  \frac{\pi  T^2 \left(Q^6 \left(-3+4 \rho ^2\right)+8 \left(\rho ^4+\rho ^6\right)+Q^4 \left(6+\rho ^2 \left(-1+\rho  \left(-3-\rho +8 \rho ^3\right)\right)\right)\right)}{4  M^2 (1+M) \rho ^5 \left(Q^2-\rho ^2 \left(1+\rho ^2\right)\right)}-\frac{4 \pi  Q T^2 F_1[\rho ]}{ (1+M) \rho ^2} -\frac{32 \pi  Q T^2 \kappa  \partial_QF_7[\rho ]}{ (1+M) \rho ^3}\\
\nonumber  && +\frac{\pi  T^2 Q^2 \rho ^2 (14+\rho  (12+\rho  (2+\rho  (9+\rho  (-16+9 \rho ))))))}{4  M^2 (1+M) \rho ^5 \left(Q^2-\rho ^2 \left(1+\rho ^2\right)\right)}  + \frac{\pi  Q T^2 \left(3 Q^2-7 M \rho ^2+3 \rho ^6\right) F_1'[\rho ]}{2  M (1+M) \rho ^3}-\frac{3 \sqrt{3} \pi  Q^2 T^2 f[\rho ] F_6'[\rho ]}{2  M (1+M) \rho }\\
&& +\frac{24 \sqrt{3} \pi  Q T^2 \lambda  \left(-4-3 Q^2+2 M \rho ^2\right) F_2'[\rho ]}{ M (1+M) \rho ^4} -\frac{16 \sqrt{3} \pi  T^2 \lambda  \left(-5 Q^2+2 M \rho ^2\right) \partial_QF_2'[\rho ]}{ (1+M) \rho ^4} -\frac{3 \pi  Q T^2 \rho ^3 f[\rho ] \partial_QF_5'[\rho ]}{4  M (1+M)} \\
\nonumber && + \frac{8 \pi  T^2 \left(-6 Q^2 \left(4+3 Q^2\right) \lambda +12 M Q^2 \lambda  \rho ^2-\left(2+Q^2\right) \kappa  \rho ^6\right) F_7'[\rho ]}{ M (1+M) \rho ^8} +\frac{384 \pi  Q^2 T^2 \lambda ^2 \left(-4-3 Q^2+2 M \rho ^2\right) \left(-63 Q^2+20 M \rho ^2\right)}{ M (1+M) \rho ^{15}}\,, \\
r_+ r_3^{(M)} &=& -\frac{3 \sqrt{3} Q}{2  M \rho ^2} \,, \\
 r_+ r_4^{(M)} &=& -\frac{\sqrt{3} Q \left(\rho ^{12}-3840 \lambda ^2 \rho ^4 \left(-1+\rho ^4\right)\right)}{ \rho ^{15}}  -\frac{48 \sqrt{3}  Q \lambda  \left(-5 Q^2+2 M \rho ^2\right)F_7[1]}{ \rho ^9}  + \frac{16 \sqrt{3} Q \left(-15 Q^2 \lambda +6 M \lambda  \rho ^2-\kappa  \rho ^6\right) F_7[\rho ]}{ \rho ^9}+\\
\nonumber && \frac{24  \lambda  \left(5 Q^2-2 M \rho ^2\right)F_2'[1]}{ \rho ^9}-\frac{\sqrt{3} Q \left(384 Q^2 \lambda ^2 \rho ^2 \left(-46+20 \rho ^2+25 \rho ^4+\rho ^6\right)+192 Q^4 \lambda ^2 \left(105-92 \rho ^2+20 \rho ^4-55 \rho ^6+22 \rho ^8\right)\right)}{ \rho ^{15}} \,, \\
\nonumber r_+^2 r_5^{(M)} &=& -\frac{(-1+\rho ) \left(8 \rho ^4 \left(1+\rho +\rho ^2\right)+Q^4 \left(-27+\rho  \left(-27+4 \rho  \left(1+\rho +\rho ^2\right) \left(-9+2 \rho ^2\right)\right)\right)\right)}{4  M^2 \rho ^8 f[\rho ]}-F_8[\rho ] -\frac{2 \left(-1+\rho ^3\right) F_8'[\rho ]}{ \rho ^2} \\
 && -\frac{(-1+\rho ) \left(Q^2 \rho ^2 (-9+\rho  (-9+\rho  (7+\rho  (43+16 \rho ))))\right)}{4  M^2 \rho ^8 f[\rho ]} -\frac{4 \sqrt{3} Q (-1+\rho ) \left(1+\rho +\rho ^2\right) F_3[\rho ]}{ \rho ^5 f[\rho ]} \,,
 \end{eqnarray}
 }
 {\scriptsize
 \begin{eqnarray}
\nonumber r_+^2 r_6^{(M)} &=& \frac{3 Q^2 \left(512 M \lambda ^2 \left(-3 Q^2+2 M \rho ^2\right) \left(-63 Q^2+20 M \rho ^2\right)+9 Q^2 \rho ^{10}-6 M \rho ^{12}-9 \rho ^{13}\right)}{8 \ M^2 \rho ^{15}} -\frac{3 \sqrt{3}  QF_3'[1]}{8  M \rho ^2} -\frac{9 Q^2F_8[1] }{4  M \rho ^2}\\
\nonumber &&  -\frac{2 \sqrt{3} Q F_3[\rho ]}{ \rho ^2} -\frac{1}{2 }F_8[\rho ] + \frac{12 \sqrt{3} Q \lambda  \left(-3 Q^2+2 M \rho ^2\right) F_2'[\rho ]}{ M \rho ^4}+ \frac{\sqrt{3} Q \left(3 Q^2-7 M \rho ^2+3 \rho ^6\right) F_3'[\rho ]}{4  M \rho ^3} \\
&&+ \frac{4 \left(-3 Q^2+2 M \rho ^2\right) \left(6 Q^2 \lambda +\kappa  \rho ^6\right) F_7'[\rho ]}{ M \rho ^8} + \frac{\left(-4 \rho ^8+9 Q^4 \left(-1+\rho ^2\right)+Q^2 \rho ^2 \left(9-9 \rho ^4-4 \rho ^6\right)\right) F_8'[\rho ]}{4  M \rho ^7}\, ,\\
r_7^{(M)} &=& -\frac{3 \sqrt{3}  QF_3'[1]}{4 \rho ^2}-\frac{9 Q^2F_8[1]}{2 M \rho ^2} - F_8[\rho ] - 2 \rho  F_8'[\rho ]\, ,\\
\nonumber r_+^4 r_8^{(M)} &=& \frac{32 \pi  Q T^2 \kappa  \partial_Q F_9[\rho ]}{ (1+M) \rho ^3} + \frac{2 \pi  T^2 \rho  F_1'[\rho ]}{\sqrt{3}  (1+M)} -\frac{32 \sqrt{3} \pi  Q T^2 \lambda  \left(-3+\rho ^2+Q^2 \left(-1+\rho ^2\right)\right) F_4'[\rho ]}{ M (1+M) \rho ^4}+\frac{16 \sqrt{3} \pi  T^2 \lambda  \left(-5 Q^2+2 M \rho ^2\right) \partial_QF_4'[\rho ]}{ (1+M) \rho ^4} \\
 && + \frac{3 \pi  Q T^2 \left(Q^2-M \rho ^2+\rho ^6\right) F_6'[\rho ]}{2  M (1+M) \rho ^5} +\frac{8 \pi  T^2 \left(6 Q^2 \left(4+3 Q^2\right) \lambda -12 M Q^2 \lambda  \rho ^2+\left(2+Q^2\right) \kappa  \rho ^6\right) F_9'[\rho ]}{ M (1+M) \rho ^8}\, ,\\
\nonumber r_+^3 r_{9}^{(M)} &=& \frac{3 \sqrt{3} Q \left(1024 M \lambda ^2 \left(15 Q^4-13 M Q^2 \rho ^2+2 M^2 \rho ^4\right)+9 Q^2 \rho ^{10}-2 M \rho ^{12}\right)}{8  M^2 \rho ^{13}} - \rho  F_3'[\rho ]  -\frac{\left(9 Q^2+2 M \rho ^2\right) F_3[\rho ]}{2  M \rho ^2} + \frac{24 Q^2 \kappa  F_9[\rho ]}{ M \rho ^3}\\
&&   + \frac{48 \sqrt{3} Q \lambda  \left(\rho ^2+Q^2 \left(-2+\rho ^2\right)\right) F_4'[\rho ]}{ M \rho ^4} -\frac{3 \sqrt{3} Q \rho  f[\rho ] F_8'[\rho ]}{4  M} + \frac{4 \left(-3 Q^2+2 M \rho ^2\right) \left(6 Q^2 \lambda +\kappa  \rho ^6\right) F_9'[\rho ]}{ M \rho ^8} \, ,\\
r_+^2 r_{10}^{(M)} && -\frac{1}{ \rho }-\frac{16 \sqrt{3} Q \kappa  F_9[\rho ]}{ \rho ^3}-\frac{24 \lambda  \left(2 \rho ^2+Q^2 \left(-5+2 \rho ^2\right)\right) F_4'[\rho ]}{ \rho ^4}\,,
\end{eqnarray}
}
\subsection{Anomalous sources}
{\scriptsize
\begin{eqnarray}
\nonumber r_+\tilde r_1^{(E)} &=& \frac{16 \sqrt{3} Q \lambda  \left(2 \rho ^2 (-3+\rho  (-3+\rho  (-3+2 \rho )))+Q^2 (17+\rho  (17+\rho  (11-10 \rho  (1+\rho ))))\right)}{\rho ^5 (1+\rho ) \left(-Q^2+\rho ^2+\rho ^4\right)}-\frac{16 \sqrt{3} Q \lambda  \left(-63 Q^2+20 M \rho ^2\right) F_5[\rho]}{ \rho^7} \\
\nonumber && -\frac{2 \rho ^5 \left(-Q^2 (6+\rho  (12+7 \rho  (2+\rho )))+\rho ^2 (4+\rho  (8+3 \rho  (4+\rho  (3+\rho  (2+\rho )))))\right) F_2[\rho ]}{(1+\rho )^2 \left(-Q^2+\rho ^2+\rho ^4\right)^2} -\frac{2 (-1+\rho ) \left(1+\rho +\rho ^2\right) F_2'[\rho ]}{\ f[\rho ]} \,, \\
&&\\
\nonumber r_+^2 \tilde r_2^{(E)} &=& -\frac{48 \lambda  \left(\rho ^2+\rho ^3+\rho ^4-\rho ^5+Q^2 (-4+\rho  (-4+\rho  (-3+2 \rho  (1+\rho ))))\right)}{ \rho ^3 (1+\rho ) \left(-Q^2+\rho ^2+\rho ^4\right)}-\frac{96 \lambda  \left(\rho ^2+Q^2 \left(-5+\rho ^2\right)\right) F_5[\rho ]}{ \rho ^5}\\
\nonumber && -\frac{2 \rho ^5 \left(-Q^2 (6+\rho  (12+7 \rho  (2+\rho )))+\rho ^2 (4+\rho  (8+3 \rho  (4+\rho  (3+\rho  (2+\rho )))))\right) F_4[\rho ]}{ (1+\rho )^2 \left(-Q^2+\rho ^2+\rho ^4\right)^2}  -\frac{2 (-1+\rho ) \left(1+\rho +\rho ^2\right) F_4'[\rho ]}{ f[\rho ]} \,, \\
&&\\
r_+^2\tilde r_3^{(E)} &=& \frac{-120 Q^2 \lambda +\frac{\sqrt{3} Q F_9'[\rho ]}{ \rho }+48 \lambda  \rho ^2+48 Q^2 \lambda  \rho ^2}{ \rho ^6}-3 \rho ^2 F_4[\rho ]-2 \rho ^3 F_4'[\rho ]-\rho  F_7'[\rho ] \,, \\
\nonumber r_+^4 \tilde r_4^{(E)} &=& -\frac{8 \sqrt{3} \pi  Q T^2 \lambda  \left(2 \rho ^2 \left(-5+\rho ^2\right)+Q^2 \left(-21+7 \rho ^2+2 \rho ^4\right)\right)}{ M (1+M) \rho ^8 f[\rho ]} -\frac{16 \sqrt{3} \pi  T^2 \lambda  \left(9 Q^4-8 M Q^2 \rho ^2+M^2 \rho ^4+3 Q^2 \rho ^6-M \rho ^8\right) F_1'[\rho ]}{ M (1+M) \rho ^6 f[\rho ]}\\
\nonumber && -\frac{64 \sqrt{3} \pi  T^2 \lambda  \left(\rho ^2+Q^2 \left(-5+\rho ^2\right)\right) \partial_Q F_8[\rho ]}{ (1+M) \rho ^5} -\frac{32 \sqrt{3} \pi  Q T^2 \lambda  \left(\rho ^2+Q^2 \left(-5+\rho ^2\right)\right) F_8[\rho ]}{ M (1+M) \rho ^5}-\frac{64 \pi  Q T^2 \lambda  \partial_Q F_3'[\rho ]}{ (1+M)} \\
\nonumber &&+ \frac{16 \pi  T^2 \lambda  \left(Q^4 \left(13-7 \rho ^2\right)+6 \rho ^2 \left(-1+\rho ^4\right)+Q^2 \left(24-13 \rho ^2+\rho ^6\right)\right) F_3'[\rho ]}{ M (1+M) \rho ^6 f[\rho ]}+ \frac{16 \sqrt{3} \pi  T^2 \lambda  \left(-7 Q^2+2 M \rho ^2\right) \partial_Q F_8'[\rho ]}{ (1+M) \rho ^4}\\
\nonumber  &&  -\frac{8 \pi  Q T^2 \lambda  \left(9 Q^4-19 M Q^2 \rho ^2+6 M^2 \rho ^4+45 Q^2 \rho ^6-22 M \rho ^8\right) F_6'[\rho ]}{ M (1+M) \rho ^{10} f[\rho ]}-\frac{16 \sqrt{3} \pi  Q^3 T^2 \lambda   \left(-1+4 \rho ^2+2 \rho ^4-8 \rho ^6+3 \rho ^8\right) F_8'[\rho ]}{ M (1+M) \rho ^{10} f[\rho ]}\\
&& -\frac{16 \sqrt{3} \pi  Q T^2 \lambda  \left(7 \rho ^2+\rho ^4-19 \rho ^6+3 \rho ^8+Q^4 \left(4-3 \rho ^2+\rho ^4\right)\right) F_8'[\rho ]}{ M (1+M) \rho ^{10} f[\rho ]}\,, \\
\nonumber r_+^2\tilde r_5^{(E)} &=& 32 \sqrt{3} Q \lambda  F_3'[\rho ] + \frac{96 \lambda  \left(\rho ^2+Q^2 \left(-5+\rho ^2\right)\right) F_8[\rho ]}{ \rho ^5}-\frac{24 \lambda  \left(2 \rho ^2+Q^2 \left(-7+2 \rho ^2\right)\right) F_8'[\rho ]}{ \rho ^4} \,, \\
\end{eqnarray}
}

{\scriptsize
\begin{eqnarray}
\nonumber r_+\tilde r_1^{(M)} &=& \frac{12 \lambda  \left(-63 Q^4 \rho +8 M^2 \rho ^2+2 M Q^2 \left(-8+15 \rho ^3\right)\right)}{ M \rho ^9} -\frac{4 \sqrt{3} Q (-1+\rho ) \left(1+\rho +\rho ^2\right) F_2[\rho ]}{ \rho ^5 f[\rho ]}-\frac{3 \sqrt{3} Q \rho ^3 F_2'[\rho ]}{2 M} + \frac{2 \left(1-2 \rho ^3\right) F_7'[\rho ]}{\ \rho ^2}\\
 && + \frac{48 \left(15 Q^4 \lambda +4 M^2 \lambda  \rho ^4+Q^2 \left(-16 M \lambda  \rho ^2+\kappa  \rho ^6\right)\right) F_5[\rho ]}{ \rho ^{11}} -\frac{\left(9 Q^2+2 M \rho ^2\right) F_7[\rho ]}{ M \rho ^2} \,,  \\
\nonumber r_+^2 \tilde r_2^{(M)} &=& -\frac{4 \sqrt{3} Q (-1+\rho ) \left(1+\rho +\rho ^2\right) F_4[\rho ]}{ \rho ^5 f[\rho ]} + \frac{16 \sqrt{3} Q \kappa  F_5[\rho ]}{ \rho ^3} -\frac{3 \sqrt{3} Q \rho ^3 F_4'[\rho ]}{2  M}+ \frac{2 \left(1-2 \rho ^3\right) F_9'[\rho ]}{ \rho ^2}\\
 && \frac{36 \sqrt{3} Q \lambda  \left(-5 Q^2+2 M \rho ^2\right)-\left(9 Q^2 \rho ^4+2 M \rho ^6\right) F_9[\rho ]}{ M \rho ^6}  \,, \\
\nonumber r_+^2\tilde r_3^{(M)} &=& -\frac{9 Q^2F_2'[1] }{4  M^2 \rho ^2} -\frac{9 \sqrt{3}  Q^3F_7[1]}{2 M^2 \rho ^2} + \frac{3 \sqrt{3} F_4'[1] Q}{2  M \rho ^2} +  \frac{9 Q^2F_9[1]}{ M \rho ^2}+ \left(1+\frac{9 Q^2}{2M \rho ^2}\right) F_2[\rho ] -\frac{2 \sqrt{3} Q F_4[\rho ]}{ \rho ^2} + \frac{9 \sqrt{3} Q^3 F_7[\rho ]}{4  M^2 \rho ^2}\\
\nonumber && \frac{9 \sqrt{3} Q \lambda  \left(10 MQ^2 \left(Q^2-2 \rho ^6\right)+\rho ^4 \left(4+12 Q^2+33 Q^4+4 Q^6-2 \left(4-12 Q^2+5 Q^4\right) \rho ^6\right)\right)}{ M^2 \rho ^{12}}\\
\nonumber && \frac{9 \sqrt{3} Q \lambda  M^2 \left(-14 Q^2 \rho ^2+4 \rho ^8\right)}{ M^2 \rho ^{12}}  -\frac{9 Q^2 F_9[\rho ]}{ M \rho ^2} + \frac{\rho  \left(8 M^2+9 Q^2 \rho ^2\right) F_2'[\rho ]}{8  M^2} -\frac{\sqrt{3} Q \left(-3 Q^2+7 M \rho ^2+3 \rho ^6\right) F_4'[\rho ]}{4 M \rho ^3}\\
&& \frac{3 \sqrt{3} Q \left(Q^2-M \rho ^2+\rho ^6\right) F_7'[\rho ]}{4  M \rho ^5} -\frac{9 Q^2 f[\rho ] F_9'[\rho ]}{4  M \rho }  \,, \\
\nonumber r_+^4 \tilde r_4^{(M)} &=&  -\frac{18 \pi ^2 Q^2 T^3 \lambda  \left(-21 Q^2+10 M \rho ^2\right)}{r_+ M^3 (1+M) \rho ^8} + \frac{\pi ^2 T^3 \left(9 Q^2+2 M \rho ^2\right) F_7[\rho ]}{2 r_+ M^3 (1+M) \rho ^2} + \frac{16 \pi  Q^2 T^2 \kappa  F_8[\rho ]}{ M (1+M) \rho ^3}+ \frac{32 \pi  Q T^2 \kappa  \partial_Q F_8[\rho ]}{ (1+M) \rho ^3} \\
\nonumber && + \frac{24 \pi  Q T^2 \lambda  \left(-3 Q^2+2 M \rho ^2\right) F_1'[\rho ]}{ M (1+M) \rho ^4} + \frac{3 \sqrt{3} \pi ^2 Q T^3 \rho ^3 F_2'[\rho ]}{4 r_+ M^3 (1+M)}-\frac{32 \sqrt{3} \pi  Q T^2 \lambda  \left(-3+\rho^2+Q^2 \left(-1+\rho^2\right)\right) F_3'[\rho]}{ (1+M) \rho ^4}  \\
\nonumber && + \frac{16 \sqrt{3} \pi  T^2 \lambda  \left(-5 Q^2+2 M \rho^2\right) \partial_Q F_3'[\rho ]}{ (1+M) \rho ^4}+ \frac{8 \pi  T^2 \left(-3 Q^2+2 M \rho ^2\right) \left(6 Q^2 \lambda +\kappa  \rho ^6\right) F_6'[\rho]}{\sqrt{3}  M (1+M) \rho ^8} + \frac{2 \pi ^2 T^3 \rho  F_7'[\rho]}{r_+ M^2 (1+M)} \\
&& + \frac{8 \pi  T^2 \left(6 Q^2 \left(4+3 Q^2\right) \lambda -12 M Q^2 \lambda  \rho ^2+\left(2+Q^2\right) \kappa  \rho ^6\right) F_8'[\rho]}{M (1+M) \rho^8} \,,\\
\nonumber  r_+^2\tilde r_5^{(M)} &=&  -\frac{9 Q^2F_2'[1] }{8 M^2 \rho ^2} -\frac{9 \sqrt{3} Q^3F_7[1] }{4 M^2 \rho ^2} + \frac{3 \sqrt{3} QF_4'[1]}{4 M \rho ^2} + \frac{9 Q^2F_9[1] }{2  M \rho ^2} -\frac{9 \sqrt{3} Q \left(4-12 Q^2+5 Q^4\right) \lambda }{ M^2 \rho ^2}\\
 && -\frac{\sqrt{3} Q F_7[\rho ]}{2  M} -\frac{16 \sqrt{3} Q \kappa  F_8[\rho ]}{ \rho ^3} + F_9[\rho ] -\frac{24 \lambda  \left(-5 Q^2+2 M \rho ^2\right) F_3'[\rho ]}{ \rho ^4}  -\frac{\sqrt{3} Q \rho  F_7'[\rho ]}{M} + 2 \rho  F_9'[\rho ] \,.
\end{eqnarray}
}

      \section{Tensorial Second Order Sources}
\label{ap:Ten_sour}
In this appendix we show tensor sources splitted in terms of the anomalous and non anomalous, the tildes refer to the anomalous sector
\begin{equation}
\mathbb P_{\mu\nu} = \sum_{a=1}^{12}P_a \mathcal{T}^{(a)}_{\mu\nu} + \sum_{a=1}^{8}\tilde P_a \mathcal{\tilde T}^{(a)}_{\mu\nu} \,,
\end{equation}
\subsection{Non-anomalous sources}
{\scriptsize
\begin{eqnarray}
P_1 &=& 2 \rho -3 \rho ^2 F_5[\rho ]-2 \rho ^3 F_5'[\rho ] \,, \\
P_2 &=& 4 \rho -6 \rho ^2 F_5[\rho ]+\frac{\left(Q^2-M \rho ^2+\rho ^6\right) F_5'[\rho ]^2}{\rho } \,, \\
\nonumber P_3 &=& \frac{768\lambda Q^4 \left(\rho ^2 \left(-5 M \lambda -3 (\kappa -29 \lambda ) \rho ^4\right)+2 Q^2 \lambda \left(9+4 \rho ^6\right)\right)}{\rho ^{11} \left(Q^2-M \rho ^2+\rho ^6\right)} + 3 \rho ^2 F_5[\rho ] + 2 \rho ^3 F_5'[\rho ]+ \frac{32 \sqrt{3} Q \lambda  \left(Q^2-2 \rho ^6\right) F_2'[\rho ]}{Q^2-M \rho ^2+\rho ^6} \\
 \nonumber && +\frac{2 Q^2  \left(9216 \lambda ^2 (\rho ^2 - M^2) +768 M (\kappa -10 \lambda ) \lambda  \rho ^4+\rho ^8\right)+\rho ^2 \left(3072 \lambda ^2-M \rho ^8+\rho ^{12}\right)}{\rho ^{7} \left(Q^2-M \rho ^2+\rho ^6\right)}   \\
 &&+\frac{32 \lambda  \left(-3 Q^2+2 M \rho ^2\right) \left(Q^2-2 \rho ^6\right) F_7'[\rho ]}{\rho ^4 \left(Q^2-M \rho ^2+\rho ^6\right)} \,,\\
  P_4 &=&  \lambda ^2 \left(Q^4 \left(\frac{644352 M}{\rho ^{15}}-\frac{396288}{\rho ^{11}}\right)+Q^6 \left(-\frac{549504}{\rho ^{17}}-\frac{3072}{\rho ^{11}}\right) -\frac{3072}{\rho ^{11}}\right) +\frac{2 Q^2}{\rho ^5}+\frac{M}{\rho ^3} -\rho ^5 F_2'[\rho ]^2\\
\nonumber &&+\rho  \lambda^2 Q^2 \left(-\frac{130560 M^2}{\rho ^{13}}-\frac{9216}{\rho ^{11}}+\frac{107520 M}{\rho ^9}\right)+ f[\rho]\left(192 \sqrt{3} Q \lambda  F_2'[\rho ] -\rho ^3  F_7'[\rho ]^2\right)  +\frac{16 \lambda  \left(27 Q^2+4 M \rho ^2\right) f[\rho ] F_7'[\rho ]}{\rho ^4}  \,, \\
r_+^2 P_5 &=& -\frac{4 \pi  T^2}{\sqrt{3}  (1+M)}\left(\rho^3F_1[\rho ]\right)' \,, \\
 \nonumber P_6 &=& -\left(1+\frac{ \pi T}{ r_+(1+M)}\right)\frac{8\pi  Q T^3\left(\rho^3F_1[\rho ]\right)'}{3 r_+^3 (1+M)^2} -\frac{3 Q \rho ^2 F_1[\rho ]}{M} -2  \partial_Q\left(\rho^3F_1[\rho ]\right)' -\rho ^5 F_1'[\rho ]^2  +\frac{\sqrt{3} \left(2+Q^2\right) F_6'[\rho ]}{M \rho } -\rho ^3 f[\rho ] F_6'[\rho ]^2 \,, \\
 &&\\
r_+ P_7 &=& -2 \left(\rho^3F_3[\rho ]\right)'  \,, \\
 r_+^3 P_8 &=& \frac{6 \pi  Q T^2 \rho ^2 F_1[\rho ]}{r_+^2 M (1+M)} -\frac{4 \sqrt{3} \pi  Q T^2 \rho ^2 F_3[\rho ]}{ M (1+M)} -\frac{4 \sqrt{3} \pi  T^2 \rho ^2 \partial_QF_3[\rho ]}{ (1+M)} -\frac{2 \pi  Q T^2 \rho ^3 F_3'[\rho ]}{\sqrt{3}  M (1+M)}\\
\nonumber && -\frac{4 \pi  T^2 \rho ^5 F_1'[\rho ] F_3'[\rho ]}{\sqrt{3}  (1+M)} -\frac{4 \pi  T^2 \rho ^3 \partial_QF_3'[\rho ]}{\sqrt{3}  (1+M)} -\frac{4 \pi  T^2 \rho ^3 f[\rho ] F_6'[\rho ] F_8'[\rho ]}{\sqrt{3}  (1+M)} + \frac{2 \pi  \left(2+Q^2\right) T^2 F_8'[\rho ]}{ M (1+M) \rho } \,, \\
 r_+^2 P_9 &=& \frac{6 \sqrt{3} Q \rho ^2 F_3[\rho ]}{ M} + \frac{\sqrt{3} Q \rho ^3 F_3'[\rho ]}{ M} -\rho ^5 F_3'[\rho ]^2 + \frac{\left(3 Q^2-2 M \rho ^2\right) F_8'[\rho ]}{ M \rho } - \rho ^3 f[\rho ] F_8'[\rho ]^2 \,, \\
 \nonumber r_+^2 P_{10} && \frac{\lambda ^2 \left(-57600 Q^4+59904 M Q^2 \rho ^2-8448 M^2 \rho ^4-34560 Q^2 \rho ^6+5376 M \rho ^8\right)}{ \rho ^{13}} +\frac{1}{ \rho }\\
 && -\frac{768 Q^2 \kappa  \lambda }{ \rho ^7} + 48 \lambda  \rho ^2 f[\rho ] F_4'[\rho ] - \rho ^5 F_4'[\rho ]^2 + \frac{112 \sqrt{3} Q \lambda  f[\rho ] F_9'[\rho ]}{\rho ^2} - \rho ^3 f[\rho ] F_9'[\rho ]^2 \,, \\
 \nonumber r_+ P_{11} &=& -\frac{256 \sqrt{3} Q \lambda  \left(30 Q^4 \lambda -36 M Q^2 \lambda  \rho ^2+6 M^2 \lambda  \rho ^4-3 Q^2 (\kappa -10 \lambda ) \rho ^6+2 M (\kappa -3 \lambda ) \rho ^8\right)}{\rho ^{15} f[\rho ]} \\
 && -\frac{32 \sqrt{3} Q \lambda  \left(Q^2-2 \rho ^6\right) F_4'[\rho ]}{\rho ^6 f[\rho ]} -\frac{32 \lambda  \left(-3 Q^2+2 M \rho ^2\right) \left(Q^2-2 \rho ^6\right) F_9'[\rho ]}{\text{b0} \rho ^{10} f[\rho ]} \,, \\
  \nonumber r_+ P_{12} &=& -\frac{2 \sqrt{3} Q \lambda ^2 \left(112704 Q^4-127296 M Q^2 \rho ^2+24320 M^2 \rho ^4+71232 Q^2 \rho ^6-16128 M \rho ^8\right)}{ \rho ^{15}} -\frac{2 \sqrt{3} Q \lambda  \left(576 Q^2 \kappa  \rho ^6+128 M \kappa  \rho ^8\right)}{\rho ^{15}}\\
\nonumber  &&  + \frac{2 \sqrt{3} Q}{ \rho ^3} + 48 \lambda  \rho ^2 f[\rho ] F_2'[\rho ] + \frac{2 \left(96 \sqrt{3} Q \lambda  \left(Q^2-M \rho ^2+\rho ^6\right)-\rho ^{11} F_2'[\rho ]\right) F_4'[\rho ]}{ \rho ^6} + \frac{112 \sqrt{3} Q \lambda  f[\rho ] F_7'[\rho ]}{ \rho ^2} \\
&& +  \frac{16 \lambda  \left(27 Q^2+4 M \rho ^2\right) f[\rho ] F_9'[\rho ]}{ \rho ^4} - 2 \rho ^3 f[\rho ] F_7'[\rho ] F_9'[\rho ] \,, 
\end{eqnarray}
}
\subsection{Anomalous sources}
{\scriptsize
\begin{eqnarray}
\tilde P_1 &=& \frac{64 \sqrt{3} Q \lambda }{\rho ^5} - 2 \left(\rho ^3 F_2[\rho ]\right)' \,,  \\
\nonumber r_+^2 \tilde P_2 &=& \frac{32 \pi  T^2 \lambda  \left(-\pi  T \left(51 Q^2+4 M \rho ^2\right)+12 r_+ M Q \left(F_1'[1]+2 \sqrt{3} F_6[1] Q\right) \rho ^3 f[\rho ]\right)}{r_+ M (1+M) \rho ^8} -\frac{2 \sqrt{3} \pi  Q T^2 \rho ^2 F_2[\rho ]}{ M\left(1+M\right)} -\frac{4 \sqrt{3} \pi  T^2 \rho ^2 \partial_QF_2[\rho ]}{(1+M)} \\
 && -\frac{768 \sqrt{3} \pi  Q^2 T^2 \lambda  f[\rho ] F_6[\rho ]}{ (1+M) \rho ^5} -\frac{2 \pi  T^2 \left(2 F_1'[1] M+4 \sqrt{3} F_6[1] M Q+5 Q \rho ^3\right) F_2'[\rho ]}{\sqrt{3}  M\left(1+M\right)} + \frac{8 \pi  Q T^2 F_6[\rho ] F_2'[\rho ]}{ (1+M)}\\
\nonumber&& -\frac{4 \pi  T^2 \rho ^3 \partial_Q F_2'[\rho ]}{\sqrt{3}  (1+M)} + \frac{32 \pi  T^2 \lambda  \left(27 Q^2+4 M \rho ^2\right) f[\rho ] F_6'[\rho ]}{\sqrt{3}  (1+M) \rho ^4} + \frac{4 \pi ^2 T^3 F_7'[\rho ]}{r_+ M\left(1+M\right) \rho }-\frac{4 \pi  T^2 \rho ^3 f[\rho ] F_6'[\rho ] F_7'[\rho ]}{\sqrt{3}  (1+M)} -\frac{2 \pi  Q T^2 \rho ^4 F_2''[\rho ]}{\sqrt{3} M \left(1+M\right)} \,, \\
\nonumber r_+^2\tilde P_3 &=& \frac{16 \pi  Q^6 T^2 \lambda  \left(-1+\rho ^2 \left(3-5 \rho +7 \rho ^3-4 \rho ^5\right)\right)}{ M (1+M) \rho ^{17} f[\rho ]^2} +\frac{64 \pi  T^2 \lambda  \left(2+\rho ^4 \left(-3+\rho ^3 \left(-5+5 \rho +\rho ^4\right)\right)\right)}{ M (1+M) \rho ^{13} f[\rho ]^2}\\
\nonumber &&+\frac{16 \pi  Q^4 T^2 \lambda  \left(-6+\rho ^2 \left(13+\rho  \left(-6+\rho  \left(8+\rho -33 \rho ^2-8 \rho ^3+14 \rho ^4+39 \rho ^5-22 \rho ^7\right)\right)\right)\right)}{ M (1+M) \rho ^{17} f[\rho ]^2}\\
\nonumber &&+\frac{32 \pi  Q^2 T^2 \lambda  \left(5+\rho ^2 \left(8+\rho  \left(-3+\rho  \left(-23+\rho  \left(-2+\rho +29 \rho ^2-21 \rho ^4+5 \rho ^5+\rho ^8\right)\right)\right)\right)\right)}{ M (1+M) \rho ^{15} f[\rho ]^2}\\
\nonumber &&+ \frac{128 \sqrt{3} \pi  T^2 \lambda  Q^2\left(Q^2-2  \rho ^6\right) F_6[\rho ]}{ (1+M) \rho ^{11} f[\rho ]} + \frac{32 \pi  Q T^2 \lambda  \left(-5 Q^2+3 M \rho ^2+\rho ^6\right) \partial_QF_5'[\rho ]}{ (1+M) \rho ^6}\\
\nonumber &&+ \frac{8 \pi  T^2 \lambda  \left(4 \rho ^6+3 Q^4 \left(-5+4 \rho ^2\right)+2 Q^2 \left(-5+6 \rho ^2+\rho ^6\right)\right) F_5'[\rho ]}{ M (1+M) \rho ^6}+ \frac{64 \pi  T^2 \lambda  \left(3 Q^2-2 M \rho ^2\right) \left(Q^2-2 \rho^6\right) F_6'[\rho ]}{\sqrt{3}  (1+M) \rho ^{10} f[\rho ]} \\
&&-\frac{64 F_1'[1] \pi  T^2 \lambda  Q\left(Q^2-2  \rho ^6\right)}{ (1+M) \rho ^{11} f[\rho ]} -\frac{128 \sqrt{3} F_6[1] \pi  T^2 \lambda Q^2 \left(Q^2-2  \rho ^6\right)}{ (1+M) \rho ^{11} f[\rho ]} + \frac{32 \pi  Q T^2 \lambda  \rho  f[\rho ] \partial_Q F_5''[\rho ]}{(1+M)} \,, \\
 r_+ \tilde P_4 &=& - 2 \left(\rho ^3 F_4[\rho ]\right)' \,, \\
\nonumber r_+^3\tilde P_5 &=& \frac{16 \sqrt{3} \pi  Q T^2 \lambda  \left(-18-11 Q^2+4 M \rho ^2\right)}{ M (1+M) \rho ^6} -\frac{4 \sqrt{3} \pi   T^2 \rho ^2(Q F_4[\rho ] + M\partial_QF_4[\rho ]}{ M\left(1+M\right)} + \frac{32 \sqrt{3} \pi  T^2 \lambda  \rho ^2 f[\rho ] F_1'[\rho ]}{(1+M)} -\frac{2 \pi  Q T^2 \rho ^3 F_4'[\rho ]}{\sqrt{3} M\left(1+M\right)} \\
&&-\frac{4 \pi  T^2 \rho ^5 F_1'[\rho ] F_4'[\rho ]}{\sqrt{3} (1+M)} -\frac{4 \pi  T^2 \rho ^3 \partial_QF_4'[\rho ]}{\sqrt{3}  (1+M)} +\frac{224 \pi  Q T^2 \lambda  f[\rho ] F_6'[\rho ]}{ (1+M) \rho ^2} + \frac{2 \pi  \left(2+Q^2\right) T^2 F_9'[\rho ]}{M \left(1+M\right) \rho } -\frac{4 \pi  T^2 \rho ^3 f[\rho ] F_6'[\rho ] F_9'[\rho ]}{\sqrt{3}  (1+M)} \,, \\
\nonumber r_+^2\tilde P_6 &=& \frac{56 \sqrt{3} Q \lambda  \left(-3 Q^2+2 M \rho ^2\right)}{ M \rho ^6} + \frac{6 \sqrt{3} Q \rho ^2 F_4[\rho ]}{M} + 48 \lambda  \rho ^2 f[\rho ] F_3'[\rho ] + \rho ^3 \left(\frac{\sqrt{3} Q}{M}-2 \rho ^2 F_3'[\rho ]\right) F_4'[\rho ]\\
&& +\frac{112 \sqrt{3} Q \lambda  f[\rho ] F_8'[\rho ]}{\rho ^2} + \frac{\left(3 Q^2-2 M \rho ^2\right) F_9'[\rho ]}{ M \rho } -2 \rho ^3 f[\rho ] F_8'[\rho ] F_9'[\rho ]  \,,\\
\nonumber r_+\tilde P_7 &=& -\frac{32 \sqrt{3} Q \lambda  \left(Q^2-2 \rho ^6\right) F_3'[\rho ]}{ \rho ^6 f[\rho ]} -\frac{12 \lambda  \left(4 \rho ^8+Q^4 \left(-5+2 \rho ^2\right)+2 Q^2 \left(\rho ^2-\rho ^6+2 \rho ^8\right)\right) F_5'[\rho ]}{ M \rho ^6}\\
\nonumber &&  -\frac{32 \lambda  \left(-3 Q^2+2 M \rho ^2\right) \left(Q^2-2 \rho ^6\right) F_8'[\rho ]}{ \rho ^{10} f[\rho ]} + \frac{12 \lambda  \left(-4 \rho ^8+Q^4 \left(-5+4 \rho ^2\right)+4 Q^2 \rho ^2 \left(1+\rho ^4-\rho ^6\right)\right) F_5''[\rho ]}{ M \rho ^5}\\
&& -\frac{4 \lambda  \left(-3 Q^2+2 M \rho ^2\right) \left(-18 Q^2 \rho ^2-4 M \rho ^4+\left(Q^2-M \rho ^2+\rho ^6\right)^2 F_5^{(3)}[\rho ]\right)}{ M \rho ^{10} f[\rho ]} \,, \\
\nonumber r_+ \tilde P_8 &=&\frac{3 \sqrt{3} Q \rho ^2 F_2[\rho ]}{ M}-2 \left(-96 \sqrt{3} Q \lambda  f[\rho ]+\rho ^5 F_2'[\rho ]\right) F_3'[\rho ]  \\
&&-\frac{\left(-8 \lambda  \left(27 Q^2+4 M \rho ^2\right)+\rho ^7 F_7'[\rho ]\right) \left(-3 Q^2+2 M \rho ^2+2 M \rho ^4 f[\rho ] F_8'[\rho ]\right)}{ M \rho ^8} \,,
\end{eqnarray}
}

% %%%%%%%%%%%%%%%%%%%%%%%%%%%%%%%%%%%
 \chapter{Transport coefficients at second order}
    \label{ap:fsecond_order}
     In this appendix we will write the expressions for transport coefficients up to second order.
\section{Vector sector}

The solutions for the non anomalous coefficients $\xi_1 \,, \dots \,, \xi_{10}$  as written in (\ref{eq:xi5})-(\ref{eq:xi4}), are given in terms of functions $\xi_{i,(0,\kappa^2,\kappa\lambda,\lambda^2)}$  whose expressions are
{\footnotesize
\begin{eqnarray}
\xi_{2,\kappa\lambda}(\rho_2)  &=&  \frac{2 \pi T^3}{5 G M^3 (M + 1) Q^4 (1 + 2 \rho_2^2)^4r_+^3}  \Bigg( \bigg( 2430 - 14121 M + 32625 M^2 - 36279 M^3 + 17151 M^4  \nonumber \\
&&+ 286 M^5 -2648 M^6 + 656 M^7 \bigg) + \frac{10 M^3 (1 + 2 \rho_2^2)^4}{Q^2} (9 + 6 M - 7 M^2 + 2 M^3)  \log[1-\rho_2^2] \nonumber \\
&&+ \frac{10 M^2 \rho_2^6 (2 + \rho_2^2)}{Q^2 (1 + 2 \rho_2^2)} \Big( 9(1-\rho_2^2)  + 3 M (5 + 13 \rho_2^2)  - 6 M^2 (13 - 7 \rho_2^2) + 8 M^3 (17 - 14 \rho_2^2) \nonumber \\
&&- 32 M^4 (4 - \rho_2^2) +   48 M^5 \Big)  \log\left[\frac{2 + \rho_2^2}{1-\rho_2^2}\right]
\Bigg)   \,,
\end{eqnarray}
\begin{eqnarray}
\xi_{2,\lambda^2}(\rho_2) &=&   \frac{\pi T^3}{210 G M^3 (M+1) \rho_2^8 (1 + 2 \rho_2^2)^4 r_+^3} 
\Bigg(
\frac{1}{ M^2 + Q^2 + \rho_2^2(2M-1)} \Big(-1837080 + 15089571 M \nonumber \\
&& - 54047817 M^2 + 109739475 M^3 - 136610865 M^4 + 102222345 M^5 - 35693475 M^6 \nonumber \\
&&- 7547847 M^7 + 12206741 M^8 - 3911944 M^9 + 427856 M^{10} \Big) 
+\frac{1680(1-\rho_2^2)}{\rho_2^2 (1 + 2 \rho_2^2)} \Big( 81(1+2\rho_2^2) \nonumber \\
&& - M (351 + 783 \rho_2^2) + 
 M^2 (540 + 1674 \rho_2^2)    + M^3 (225 - 1719 \rho_2^2) - M^4 (1497 - 672 \rho_2^2) \nonumber \\
&&+ M^5 (1332 - 12 \rho_2^2) - M^6 (340 - 32 \rho_2^2)  +  32 M^7  \Big) \log[1-\rho_2^2]
- \frac{1680 M^2 \rho_2^8(2+\rho_2^2)}{(1 + \rho_2^2)^5(1 + 2\rho_2^2)} \nonumber \\
&&\times\Big(  81 (1 + 2 \rho_2^2)-  27 M (16 + 29 \rho_2^2) + 54 M^2 (21 + 31 \rho_2^2) - 9 M^3 (216 + 191 \rho_2^2) \nonumber \\
&&+ 3 M^4 (723 + 224 \rho_2^2) - 12 M^5 (112 + \rho_2^2) + 4 M^6 (93 + 8 \rho_2^2)  - 32 M^7 \Big) \log\left[2 + \rho_2^2\right]
\Bigg)  \,,
\end{eqnarray}
\begin{eqnarray}
\xi_{4,\kappa\lambda}(\rho_2)  &=& -\frac{ 2 \sqrt{3} }{5 \pi G M^2 Q^3 (1 + 2 \rho_2^2)^3} \Bigg( (1+2\rho_2^2) \big( 405 - 1944 M + 3708 M^2 - 3612 M^3 + 1903 M^4 - 507 M^5 \nonumber \\
&&+ 57 M^6 \big) 
+\frac{5 M^3 (M^2-3) (1 + 2 \rho_2^2)^2 }{Q^2 \rho_2^8 }\bigg(2 + \rho_2^2 - 
    5M + M^2 (3 - 2 \rho_2^2) \bigg)\log[1-\rho_2^2] \nonumber \\
&&+  \frac{5 M^3 \rho_2^6 (2 + \rho_2^2)^3}{Q^2} \bigg(  3 \rho_2^2  - 3M (1 + 2 \rho_2^2)  + 2M^2 \bigg)    \log\left[\frac{2 + \rho_2^2}{1-\rho_2^2}\right]
\Bigg)   \,, 
\end{eqnarray}
\begin{eqnarray}
\xi_{4,\lambda^2}(\rho_2) &=& \frac{1}{280\sqrt{3}\pi G Q^7 M^2 (1 + 2\rho_2^2)^2} \Bigg( 
\big( 1224720 - 8421651 M + 24887169 M^2 - 41179203 M^3 \nonumber \\
&&+ 41512749 M^4 - 25998933 M^5 + 10023123 M^6 - 2364497 M^7 + 330895 M^8 - 21092 M^9 \big) \nonumber \\
&&- \frac{3360 M^4 (1+2\rho_2^2)^2}{Q^2}\big(9 - 60 M + 105 M^2 - 70 M^3 + 20 M^4 - 2 M^5 \big)  \log[1-\rho_2^2] \nonumber \\
&&+ \frac{3360 M^4 \rho_2^8 (2 + \rho_2^2)^3 ( M \rho_2^2 - 3Q^2) }{Q^2 (1 + 2 \rho_2^2)} ( -3\rho_2^2  + 3 M (1 + 2 \rho_2^2) - 2 M^2 ) \log\left[\frac{2 + \rho_2^2}{1-\rho_2^2}\right]
\Bigg)  \,,
\end{eqnarray}
\begin{eqnarray}
\xi_{5,0}(\rho_2) &=& \frac{3Q^2}{256 \pi G M^3(1 + 2 \rho_2^2)^2}  \Bigg(
(5 M-3 )(11 M - 15) + \frac{8 M^3 (1 + 2 \rho_2^2)^2}{3 Q^2} \log[1-\rho_2^2] \nonumber \\
&&-  \frac{2 M^3 \rho_2^2 (2 + \rho_2^2)^2 }{3 Q^2  (Q^2 + \rho_2^6) (1 + 2 \rho_2^2) }\big(9 \rho_2^2 + M (1 - 16 \rho_2^2)\big) \log\left[\frac{2 + \rho_2^2}{1-\rho_2^2}\right]
\Bigg)  \,,  
\end{eqnarray}
\begin{eqnarray}
\xi_{6,0}(\rho_2)  &=&  -\frac{1}{512 \pi G (1 + Q^2)^3 (1 + 2 \rho_2^2)^2} \Bigg(
\big( 243 - 513 M + 189 M^2 - 39 M^3 + 152 M^4 \big)  \nonumber \\
\nonumber &&+ 8 M^3 (1 + 2 \rho_2^2)^2  \log[1-\rho_2^2]
- \frac{2M^2}{(1 + 2 \rho_2^2)} \big( 3\rho_2^2 + M(4 + \rho_2^2) \big) \big(15 (1 + \rho_2^2) - 16M \big)   \log\left[\frac{2 + \rho_2^2}{1-\rho_2^2}\right]
\Bigg)   \,,
\end{eqnarray}
\begin{eqnarray}
\xi_{6,\kappa\lambda}(\rho_2)  &=& \frac{1}{10 \pi G M^3 Q^2 (1 + 2 \rho_2^2)^2}    
\Bigg( \big( 2430 - 9477 M + 13824 M^2 - 9576 M^3 + 3444 M^4 - 591 M^5 - 14 M^6
\big) \nonumber \\
&&+\frac{20 M^3 (1 + 2 \rho_2^2)^2 }{Q^2} (18 - 27 M + 12 M^2 - M^3) \log[1-\rho_2^2]
+\frac{10 M^2 \rho_2^4 (2 + \rho_2^2)^3}{Q^2 (1 + 2 \rho_2^2)} \times \nonumber \\
&&\big(3 \rho_2^2 - 3 M (1 + 2 \rho_2^2) + 2 M^2 \big) \big(3 + 6 \rho_2^2 - M (4 + 5 \rho_2^2) \big)  \log\left[\frac{2 + \rho_2^2}{1-\rho_2^2}\right]
\Bigg)  \,, 
\end{eqnarray}
\begin{eqnarray}
\xi_{6,\lambda^2}(\rho_2) &=&  - \frac{1}{280 \pi G M^3 Q^6 (1 + 2 \rho_2^2)^2} \Bigg(
\Big( 612360 - 3485349 M + 7477191 M^2 - 6516837 M^3 - 444309 M^4 \nonumber \\
&&+ 5355933 M^5 - 4409403 M^6 + 1747097 M^7 - 359455 M^8 + 29492 M^9 \Big) 
+ \frac{3360 M^3 (1 + 2 \rho_2^2)^2 }{Q^2} \times \nonumber \\
&&\big(81 - 315 M + 453 M^2 - 303 M^3 + 101 M^4 - 16 M^5 + M^6\big) \log[1-\rho_2^2]
+\frac{1680 M^2 \rho_2^{10} (2 + \rho_2^2)^3}{Q^2 (1 + 2 \rho_2^2)}\times \nonumber \\
&&  \Big(3 \rho_2^2 -3 M (1 + 2 \rho_2^2) + 2 M^2 \Big) \Big(  9 (1 + 2 \rho_2^2) - 12 M (1 + 2 \rho_2^2)  + 3 M^2 (1 + \rho_2^2)  - M^3 \Big)   \log\left[\frac{2 + \rho_2^2}{1-\rho_2^2}\right]
\Bigg)
 \,, \nonumber\\
 &&
 \end{eqnarray}
\begin{eqnarray}
\xi_{7,0}(\rho_2) &=&  -\frac{1}{256 \pi G M^3} \frac{1}{(1+2\rho_2^2)^3} \Bigg( (32 + 224 Q^2 + 426 Q^4 + 443 Q^6 + 128 Q^8) (1 + 2 \rho_2^2)  \nonumber \\
&&+ M^2 \big(90  - 168 M + 82 M^2 + 4 M (1 + 2 \rho_2^2)^3 \big) \log\left[\frac{2 + \rho_2^2}{1-\rho_2^2}\right]  \Bigg)\,, 
\end{eqnarray}
\begin{eqnarray}
\xi_{8,\kappa^2}(\rho_2) &=& -\frac{\sqrt{3} \pi Q T^3}{2 G M^3 (1 + M) (1 + 2 \rho_2^2)^5r_+^3} \Bigg( (18 -69M + 63M^2 - 28M^3) (1 + 2 \rho_2^2)  \nonumber \\
&&+2 M (3 - M)(3 - M  + 4 M^2) \log\left[\frac{2 + \rho_2^2}{1-\rho_2^2}\right]    \Bigg) \,, 
\end{eqnarray}
\begin{eqnarray}
\xi_{8,\kappa\lambda}(\rho_2)  &=&  \frac{\sqrt{3}\pi T^3}{  G M^3 (1 + M) Q^5 (1 + 2\rho_2^3)^5r_+ ^3}  \Bigg( Q^2 \Big(216 - 1233 M + 2691 M^2 - 2823 M^3 + 1379 M^4 \nonumber \\
&&- 202 M^5 - 24 M^6\Big)(1 + 2 \rho_2^2)  +2 (3 - 4 M)^2 M^3 (1 + 2M - M^2 ) (1 + 2 \rho_2^2) \log[1-\rho_2^2] \nonumber \\
&&- M\Big(216  - 1062 M + 2094 M^2 - 1950 M^3 + 634 M^4 + 264 M^5 - 
 244 M^6 + 48 M^7 \nonumber \\
&&- 18 M^2 \rho_2^2 + 12 M^3 \rho_2^2 + 
 82 M^4 \rho_2^2 - 112 M^5 \rho_2^2 + 32 M^6 \rho_2^2\Big)  \log\left[\frac{2 + \rho_2^2}{1-\rho_2^2}\right]  \Bigg)\,,  
 \end{eqnarray}
\begin{eqnarray}
\xi_{8,\lambda^2}(\rho_2)  &=&  \frac{\pi T^3}{35 \sqrt{3} G M^3 (1 + M) Q^7(1 + 2 \rho_2^2)^4r_+^3} \Bigg( \Big(-102060 + 846855 M - 2955555 M^2 + 5599602 M^3 \nonumber \\
&&- 5989080 M^4 + 
 3085503 M^5 + 60033 M^6 - 797012 M^7 + 279586 M^8 - 32072 M^9\Big) \nonumber \\
&&+ \frac{420M^2}{Q^2(1+2\rho_2^2)}(3 - 4 M)^2 (18 - 93 M + 207 M^2 - 191 M^3 + 54 M^4 - 6 M^5 + M^6) \log[1-\rho_2^2] \nonumber \\
&&+  \frac{420M}{Q^2(1+2\rho_2^2)} \Big( 243 + 162 M (-11 + \rho_2^2)  - 27 M^2 (-207 + 47 \rho_2^2)
- M^3 (9243 - 4383 \rho_2^2) \nonumber \\
&&- 
 15 M^4 (-497 + 545 \rho_2^2)  - M^5 (372 - 8382 \rho_2^2) - 2M^6 (2243 + 2203 \rho_2^2) + 
 9 M^7 (404 + 113 \rho_2^2) \nonumber \\
&&- M^8 (1219 + 120 \rho_2^2)   + 
 M^9 (191 + 16 \rho_2^2)  - 12 M^{10}  \Big)\log\left[\frac{2 + \rho_2^2}{1-\rho_2^2}\right]   \Bigg) \,, 
\end{eqnarray}
\begin{eqnarray}
\xi_{9,\kappa^2}(\rho_2) &=&  \frac{-\sqrt{3}Q}{4 \pi G M^3 (1 + 2 \rho_2^2)^2} \Bigg(
9 (3 - 4 M + M^3) + 3 (3 - 4 M) M^2 \log[1-\rho_2^2]  \nonumber \\
&&-\frac{M}{2 (1 + 2 \rho_2^2)}\bigg( 54 -9 M (10 + 2 \rho_2^2) + 24 M^2 (2 + \rho_2^2)  + 4 M^3 )\bigg)  \log\left[\frac{2 + \rho_2^2}{1-\rho_2^2}\right]
 \Bigg) \,, 
 \end{eqnarray}
\begin{eqnarray}
\xi_{9,\kappa\lambda}(\rho_2) &=& \frac{2 \sqrt{3}}{M^3 (-3 + 4 M) Q}  \Bigg(
\big(648 - 2079 M + 2475 M^2 - 1365 M^3 + 357 M^4 - 52 M^5 \big) \nonumber \\
&&+ \frac{M(-3 + 4 M)}{Q^2 (1 + 2 \rho_2^2)^3} \Big(2 (1 - \rho_2^2)^3 (1 + \rho_2^2) \big(-36 + 129 M - 
      129 M^2 + 28 M^3 \nonumber \\
&&+ \rho_2^2(36 -39M - 12 M^2 + 8 M^3) \big) + 
  2M \big(324 -1269M + 1935 M^2 - 1443 M^3 \nonumber \\
&&+ 537 M^4 - 92 M^5 + 
         8 M^6 + \rho_2^2M(-3 + 4 M) (-18 + 9 M + M^3) \big) \Big)   \log[1-\rho_2^2] \nonumber \\
&&+ \frac{2M}{Q^2 (1 + 2 \rho_2^2)^3)} \Big(-972 + 
   5103M - 10881 M^2 + 12069 M^3 - 7383 M^4 + 2424 M^5 \nonumber \\
&&- 392 M^6 + 32 M^7 + \rho_2^2 M(3 - 4 M)^2 (-18 + 9 M + M^3) \Big)  \log\left[\frac{2 + \rho_2^2}{1-\rho_2^2}\right] \Bigg)  \,, 
\end{eqnarray}
\begin{eqnarray}
\xi_{9,\lambda^2}(\rho_2)  &=& \frac{ \sqrt{3} \rho_2^2}{140\pi G M^3 Q^7 (1 + 2 \rho_2^2)^3}  \Bigg(  (-1 + 2 M + \rho_2^2) \Big(102060 - 496935 M + 963855 M^2 \nonumber \\
&&- 954012 M^3 + 518949 M^4 - 159036 M^5 + 27094 M^6 - 2411 M^7 + 16 M^8 \Big)  \nonumber \\
&&+\frac{420 M^2 (1 + 2 \rho_2^2) }{\rho_2^2} \Big(243 - 1269 M + 2673 M^2 - 2919 M^3 + 1779 M^4 - 610 M^5 \nonumber \\
&&+ 110 M^6 - 8 M^7\Big)  \log[1-\rho_2^2] + \nonumber \\
&&+ \frac{420 M}{\rho_2^2} \Big(243 - 243 M (6 - \rho_2^2)  + 27 M^2 (134 - 47 \rho_2^2) - 9 M^3 (529 - 297 \rho_2^2)  \nonumber \\
&&+ M^4 (3513 - 2919 \rho_2^2)  - 3 M^5 (456 - 593 \rho_2^2) + 
   M^6 (181 - 610 \rho_2^2) + 10 M^7 (5 + 11 \rho_2^2) \nonumber \\
&& -4 M^8 (5 + 2 \rho_2^2)    + 2 M^9  \Big)  \log\left[\frac{2 + \rho_2^2}{1-\rho_2^2}\right]   \Bigg)  \,,
\end{eqnarray}
\begin{eqnarray}
\xi_{10,\kappa^2}(\rho_2)  &=&  -\frac{9 \rho_2^2 }{4 \pi G M^2  (1 + 2 \rho_2^2)^3} \Bigg( 
 (3 - 4 M + 3 M^2) (-1 + 2 M + \rho_2^2) 
 -\frac{2 M^2 (1 + 2 \rho_2^2)^3}{3 \rho_2^2}  \log[1-\rho_2^2] \nonumber \\
&&-\frac{2}{3} M^2 \Big(3 + 6 \rho_2^2 + M (5 + \rho_2^2) \Big)  \log\left[\frac{2 + \rho_2^2}{1-\rho_2^2}\right]
\Bigg) \,, 
\end{eqnarray}
\begin{eqnarray}
\xi_{10,\kappa\lambda}(\rho_2)  &=& \frac{3}{2 \pi G M^2 (1 + 2 \rho_2^2)^2} \Bigg(
\big(-108 + 333 M - 372 M^2 + 163 M^3 - 16 M^4\big)  \\
\nonumber&&+6 (3 - 4 M) M^2   \log[1-\rho_2^2]
+\frac{6 M^2 (6 - 5 M - M^2 + (3 - 4 M) \rho_2^2)}{1 + 2 \rho_2^2}   \log\left[\frac{2 + \rho_2^2}{1-\rho_2^2}\right]
 \Bigg)  \,, 
 \end{eqnarray}
\begin{eqnarray}
\xi_{10,\lambda^2}(\rho_2) &=&  \frac{4 Q^{10}}{35 M^2(1 + 2 \rho_2^2)^2} \Bigg(
\Big( 51030 - 308205 M + 782505 M^2 - 1075848 M^3 + 866631 M^4 \nonumber \\
&&-421824 M^5 + 126626 M^6 - 22369 M^7 + 1874 M^8 \Big)
+  \frac{420 M^3 (1 - \rho_2^2)^3 (1 + \rho_2^2)^4}{\rho_2^2(1 + 
  2 \rho_2^2)} \nonumber \\
&&\times\big(9 - 3 M (7 + \rho_2^2)  + 6 M^2 (2 + \rho_2^2)  + M^3 \big) \log[1-\rho_2^2]
+ \frac{420 M^3 \rho_2^{10}}{Q^2 (1 + 2 \rho_2^2)} \big(27 + 54 \rho_2^2  \\
&&- 27 M (1 + 4 \rho_2^2) + 3 M^2 (-7 + 16 \rho_2^2)  + M^3 (6 - 12 \rho_2^2) -M^4 (1 - \rho_2^2)   \big)   \log\left[2 + \rho_2^2\right]
\Bigg) \,, \nonumber
\end{eqnarray}
}
and for the anomalous coefficients $\tilde{\xi}_1 \,, \dots \,, \tilde{\xi}_5$, (\ref{eq:xitilde2})-(\ref{eq:xitilde1}), one has functions $\tilde \xi_{i,(\kappa,\lambda)}$ that write
{\footnotesize
\begin{eqnarray}
\tilde \xi_{1,\lambda}(\rho_2) &=& \frac{2 \pi T^2}{G M^2 Q^2 (1 + 2 \rho_2^2)^2) r_+^2 } \Bigg(
\Big( 9 - 24 M + 14 M^2 + 3 M^3 \Big)
+ \frac{2 M^3}{Q^2} (1 + 2 \rho_2^2) \log[1-\rho_2^2] \nonumber \\
&&-\frac{M^2 \rho_2^2}{(1 + \rho_2^2)(1 + 2\rho_2^2)^2} \big( 3 \rho_2^2 - 
    3 M (1 + 2 \rho_2^2) + 2 M^2 \big)  \log\left[\frac{2 + \rho_2^2}{1-\rho_2^2}\right]
\Bigg) \,,
\end{eqnarray}
\begin{eqnarray}
\tilde \xi_{2,\lambda}(\rho_2)  &=&  \frac{\sqrt{3}\pi T^2}{2 G M^2 Q (1 + 2 \rho_2^2)^2 r_+^2} \Bigg( \big(-6 + 7 M + M^2\big)
+ \frac{2 M^2}{Q^2} (1 + 2 \rho_2^2)^2 \log[1-\rho_2^2] \nonumber \\
&&-\frac{2 M \rho_2^4}{Q^2 (1 + 2 \rho_2^2)} (M^2 - 3 Q^2 \rho_2^2) \log\left[\frac{2 + \rho_2^2}{1-\rho_2^2}\right]
\Bigg)  \,, 
\end{eqnarray}
\begin{eqnarray}
\tilde \xi_{4,\kappa}(\rho_2) &=& \frac{3 \pi Q^2T^3 }{8 G M^4 (M + 1) (1 + 2 \rho_2^2)^4 r_+^3} \Bigg( 
(5 M-3) (15 - 14 M +  4 M^2) \nonumber \\
&&-\frac{2 M^2(M-3)}{3 Q^2 (1 + 2 \rho_2^2)} (3 - 16 M + 12 M^2)  \log\left[\frac{2 + \rho_2^2}{1-\rho_2^2}\right]
\Bigg)  \,, 
\end{eqnarray}
\begin{eqnarray}
\tilde \xi_{4,\lambda}(\rho_2)  &=& \frac{T^3}{4GM^4(M + 1) Q^2 (1 + 2 \rho_2^2)^4r_+^3 } \Bigg( 
\Big( 1215 - 6075 M + 11880 M^2 - 11367 M^3 + 5355 M^4 \nonumber \\
&&- 1096 M^5 + 100 M^6 \Big)
-\frac{2 \pi M^3(M -3) (2M + 1) (1 + 2 \rho_2^2)^4}{Q^2} \log[1-\rho_2^2]  \\
&&+\frac{2 \pi (M-3) M^3 \rho_2^4}{Q^2 (1 + 2 \rho_2^2)} \big( 9 (1 + \rho_2^2) + 3 M (1 + 6 \rho_2^2)  - 14 M^2 (1 + 2 \rho_2^2) + 4 M^3 \big) \log\left[\frac{2 + \rho_2^2}{1-\rho_2^2}\right]
\Bigg)\,,  \nonumber
\end{eqnarray}
\begin{eqnarray}
\tilde \xi_{5,\kappa}(\rho_2) &=&  \frac{-3 \sqrt{3} \pi Q^3 T^2}{4 G M^3 (1 + 2 \rho_2^2)^2r_+^2} \Bigg(
1 + \frac{2 M^2}{3 Q^2 (1 + 2 \rho_2^2)}  \log\left[\frac{2 + \rho_2^2}{1-\rho_2^2}\right]
\Bigg) \,,  \label{eq:xi5tkappa}
\end{eqnarray}
\begin{eqnarray}
\tilde \xi_{5,\lambda}(\rho_2)  &=& \frac{\sqrt{3} \pi T^2 }{2 G M^3 Q (1 + 2 \rho_2^2)^2 r_+^2} \Bigg(
\big(27 - 72 M + 54 M^2 - 7 M^3\big) \nonumber \\
&&+\frac{2 M^3 (4 M-3) (Q^2 - \rho_2^2)}{Q^2 \rho_2^4} \log[1-\rho_2^2]  
- \frac{2 M^3 \rho_2^2 ( M\rho_2^2 -3 Q^2 )}{Q^2 (1 + 2 \rho_2^2)} \log\left[\frac{2 + \rho_2^2}{1-\rho_2^2}\right]
\Bigg) \,. 
\end{eqnarray}
}

\section{Tensor sector}

In this sector the non anomalous coefficients $\Lambda_1 \,, \dots \,, \Lambda_{12}$  written in (\ref{eq:Lambda7})-(\ref{eq:Lambda6}) and the anomalous ones $\tilde{\Lambda}_1 \,, \dots \,, \tilde{\Lambda}_8$ (\ref{eq:Lambdatilde4})-(\ref{eq:Lambdatilde3}), are given in terms of functions $\Lambda_{i,(0,\kappa^2,\kappa\lambda,\lambda^2)}$ and $\Lambda_{i,(\kappa,\lambda)}$ respectively.  The expressions for these functions are in general very complicated, and we present here the result as an expansion at footnotesize $\rho_2$ up to order ${\cal O}(\rho_2^6)$, which is equivalent to order ${\cal O}(\bar{\mu}^6)$. For the non anomalous coefficients we get
{\footnotesize
\begin{eqnarray}
\Lambda_{4,\lambda^2}(\rho_2) &=&  \frac{64}{15\pi G} \Big( (-4 + 15 \log{2}) - \frac{1}{16} (557 + 840 \log{2}) \rho_2^2 - 
 \frac{3}{112} (2789 - 9660 \log{2}) \rho_2^4  + {\cal O}(\rho_2^6) \Big) \,, \nonumber\\
 &&
 \end{eqnarray}
\begin{eqnarray}
\Lambda_{6,0}(\rho_2)  &=& \frac{ 1}{384 \pi ^3G} \Big(  2(47-66 \log 2)\rho_2^2 -4 (1-3 \log 2)- 3(89-101 \log 2)\rho_2^4 + {\cal O}(\rho_2^6)\Big)  \,, 
\end{eqnarray}
\begin{eqnarray}
\Lambda_{8,0}(\rho_2) &=&  -\frac{1}{128\pi^2 G} \Big(4 + 2(1 - 14\log{2}) \rho_2^2 - (137 - 224 \log{2}) \rho_2^4 + {\cal O}(\rho_2^6) \Big) \,, 
\end{eqnarray}
\begin{eqnarray}
\Lambda_{9,0}(\rho_2)  &=& \frac{1}{768\pi G} \Big( 8 (11 + 3 \log{2}) + 12 (7 - 8\log{2}) \rho_2^2 - 3 (91 - 226 \log{2}) \rho_2^4 + {\cal O}(\rho_2^6)  \Big) \,, 
\end{eqnarray}
\begin{eqnarray}
\Lambda_{10,\kappa^2}(\rho_2) &=& \frac{1}{2\pi G} \Big( -12 (1 - 2 \log{2}) \rho_2^2 + 3 (25 - 36 \log{2}) \rho_2^4 + {\cal O}(\rho_2^6)  \Big) \,, \\
\Lambda_{10,\kappa\lambda}(\rho_2)  &=& \frac{1}{2\pi G}  \Big( 8 (5 - 12 \log{2}) \rho_2^2 - 3 (91 - 144 \log{2}) \rho_2^4   + {\cal O}(\rho_2^6) \Big)\,, 
\end{eqnarray}
\begin{eqnarray}
\Lambda_{10,\lambda^2}(\rho_2) &=& \frac{1}{5\pi G} \Big( -90 + 4 (29 + 60 \log{2}) \rho_2^2 + (617 - 1080 \log{2}) \rho_2^4   + {\cal O}(\rho_2^6)  \Big) \,, 
\end{eqnarray}
\begin{eqnarray}
\Lambda_{12,\kappa\lambda}(\rho_2)  &=&  \frac{2}{5\sqrt{3}\pi G} \Big( 
20(-5 + 12 \log{2}) \rho_2 + 5(71 - 192 \log{2}) \rho_2^3 - (2713 -3930\log{2})\rho_2^5 + {\cal O}(\rho_2^7) \Big) \,, 
\end{eqnarray}
\begin{eqnarray}
\Lambda_{12,\lambda^2}(\rho_2) &=& \frac{2}{35\sqrt{3} \pi G} \Big(
-56 (29 + 60 \log{2}) \rho_2 + 840 (1 + 16 \log{2}) \rho_2^3  + (37403 -51660  \log{2}) \rho_2^5 + {\cal O}(\rho_2^7)
\Big) \,, \nonumber\\
\end{eqnarray}
}
and for the anomalous coefficients
{\footnotesize
\begin{eqnarray}
\tilde \Lambda_{2,\lambda}(\rho_2) &=& \frac{1}{12\pi^2 G} \Big(-
24 (1 - \log{2}) + 4 (26 - 45 \log{2}) \rho_2^2 - (527 - 822 \log{2}) \rho_2^4 + {\cal O}(\rho_2^6)
 \Big) \,, 
 \end{eqnarray}
\begin{eqnarray}
\tilde \Lambda_{5,\kappa}(\rho_2)  &=& \frac{\sqrt{3}}{32\pi^2 G} \Big(
-8(1 - 2 \log{2}) \rho_2 + 16 (5 - 4 \log{2}) \rho_2^3 - (389 - 566\log{2})\rho_2^5 + {\cal O}(\rho_2^7)
\Big) \,, 
\end{eqnarray}
\begin{eqnarray}
\tilde \Lambda_{5,\lambda}(\rho_2) &=& \frac{\sqrt{3}}{48\pi^2 G} \Big(-
  24(1 + 2 \log{2}) \rho_2 - 8(17 - 48 \log{2}) \rho_2^3 +3( 379 - 550\log{2} )\rho_2^5 + {\cal O}(\rho_2^7)
\Big) \,, 
\end{eqnarray}
\begin{eqnarray}
\tilde \Lambda_{6,\kappa}(\rho_2)  &=& \frac{\sqrt{3}}{16\pi G} \Big( -
8 (\log{2}) \rho_2 - 12(2 - 5 \log{2}) \rho_2^3 +(174 - 257\log{2})\rho_2^5 + {\cal O}(\rho_2^7) \Big) \,, \\
\tilde \Lambda_{6,\lambda}(\rho_2) &=&  \frac{\sqrt{3}}{48\pi G} \Big( 
24 (5 + 2 \log{2}) \rho_2 - 4(23 + 90 \log{2}) \rho_2^3 -  (991-1494\log{2})\rho_2^5 + {\cal O}(\rho_2^7) 
 \Big) \,, 
 \end{eqnarray}
\begin{eqnarray}
\tilde \Lambda_{8,\lambda}(\rho_2)  &=& \frac{1}{12\pi G} \Big( 
-24 \log{2} + 4 (25 + 42 \log{2}) \rho_2^2 + (199 - 750 \log{2}) \rho_2^4 + {\cal O}(\rho_2^6)
\Big) \,.  
\end{eqnarray}
}
Note that in some cases the order ${\cal O}(\rho_2^6)$ vanishes, so that the corresponding expressions are valid up to ${\cal O}(\rho_2^7)$.

% %%%%%%%%%%%%%%%%%%%%%%%%%%%%%%%%%%%
  \cleardoublepage
 \end{appendix}
 \end{appendices}
 \listoftables
 \listoffigures
  \bibliographystyle{utcaps}
  \bibliography{biblio}

\providecommand{\href}[2]{#2}\begingroup\raggedright\begin{thebibliography}{100}

\bibitem{'tHooft:1973jz}
G.~'t~Hooft, ``{A Planar Diagram Theory for Strong Interactions},''
  \href{http://dx.doi.org/10.1016/0550-3213(74)90154-0}{{\em Nucl.Phys.}
  {\bfseries B72} (1974) 461}.

\bibitem{Maldacena:1997re}
J.~M. Maldacena, ``{The large N limit of superconformal field theories and
  supergravity},'' {\em Adv. Theor. Math. Phys.} {\bfseries 2} (1998) 231--252,
  \href{http://arxiv.org/abs/hep-th/9711200}{{\ttfamily arXiv:hep-th/9711200}}.

\bibitem{Kovtun:2004de}
P.~Kovtun, D.~Son, and A.~Starinets, ``{Viscosity in strongly interacting
  quantum field theories from black hole physics},''
  \href{http://dx.doi.org/10.1103/PhysRevLett.94.111601}{{\em Phys.Rev.Lett.}
  {\bfseries 94} (2005) 111601},
  \href{http://arxiv.org/abs/hep-th/0405231}{{\ttfamily arXiv:hep-th/0405231
  [hep-th]}}.

\bibitem{Erdmenger:2011tj}
J.~Erdmenger, P.~Kerner, and H.~Zeller, ``{Transport in Anisotropic
  Superfluids: A Holographic Description},''
  \href{http://dx.doi.org/10.1007/JHEP01(2012)059}{{\em JHEP} {\bfseries 1201}
  (2012) 059}, \href{http://arxiv.org/abs/1110.0007}{{\ttfamily arXiv:1110.0007
  [hep-th]}}.

\bibitem{Landau}
L.~Landau and E.~Lifshitz, {\em Fluid Mechanics (Course on Theoretical Physics
  Volume 6)}.
\newblock Butterworth-Heinemann; 2 edition, 1987.

\bibitem{Muller:1967zza}
I.~Muller, ``{Zum Paradoxon der Warmeleitungstheorie},'' {\em Z. Phys.}
  {\bfseries 198} (1967) 329.

\bibitem{Israel:1976tn}
W.~Israel, ``{Nonstationary Irreversible Thermodynamics: A Causal Relativistic
  Theory},'' {\em Annals Phys.} {\bfseries 100} (1976) 310.

\bibitem{Baier:2007ix}
R.~Baier, P.~Romatschke, D.~T. Son, A.~O. Starinets, and M.~A. Stephanov,
  ``{Relativistic viscous hydrodynamics, conformal invariance, and
  holography},'' \href{http://dx.doi.org/10.1088/1126-6708/2008/04/100}{{\em
  JHEP} {\bfseries 0804} (2008) 100},
  \href{http://arxiv.org/abs/0712.2451}{{\ttfamily arXiv:0712.2451 [hep-th]}}.

\bibitem{Gubser:1996de}
S.~Gubser, I.~R. Klebanov, and A.~Peet, ``{Entropy and temperature of black
  3-branes},'' \href{http://dx.doi.org/10.1103/PhysRevD.54.3915}{{\em
  Phys.Rev.} {\bfseries D54} (1996) 3915--3919},
  \href{http://arxiv.org/abs/hep-th/9602135}{{\ttfamily arXiv:hep-th/9602135
  [hep-th]}}.

\bibitem{Witten:1998zw}
E.~Witten, ``{Anti-de Sitter space, thermal phase transition, and confinement
  in gauge theories},'' {\em Adv. Theor. Math. Phys.} {\bfseries 2} (1998)
  505--532, \href{http://arxiv.org/abs/hep-th/9803131}{{\ttfamily
  arXiv:hep-th/9803131}}.

\bibitem{CasalderreySolana:2011us}
J.~Casalderrey-Solana, H.~Liu, D.~Mateos, K.~Rajagopal, and U.~A. Wiedemann,
  ``{Gauge/String Duality, Hot QCD and Heavy Ion Collisions},''
  \href{http://arxiv.org/abs/1101.0618}{{\ttfamily arXiv:1101.0618 [hep-th]}}.

\bibitem{Son:2004tq}
D.~T. Son and A.~R. Zhitnitsky, ``{Quantum anomalies in dense matter},''
  \href{http://dx.doi.org/10.1103/PhysRevD.70.074018}{{\em Phys. Rev.}
  {\bfseries D70} (2004) 074018},
  \href{http://arxiv.org/abs/hep-ph/0405216}{{\ttfamily arXiv:hep-ph/0405216}}.

\bibitem{Metlitski:2005pr}
M.~A. Metlitski and A.~R. Zhitnitsky, ``{Anomalous axion interactions and
  topological currents in dense matter},''
  \href{http://dx.doi.org/10.1103/PhysRevD.72.045011}{{\em Phys. Rev.}
  {\bfseries D72} (2005) 045011},
  \href{http://arxiv.org/abs/hep-ph/0505072}{{\ttfamily arXiv:hep-ph/0505072}}.

\bibitem{Newman:2005as}
G.~M. Newman and D.~T. Son, ``{Response of strongly-interacting matter to
  magnetic field: Some exact results},''
  \href{http://dx.doi.org/10.1103/PhysRevD.73.045006}{{\em Phys. Rev.}
  {\bfseries D73} (2006) 045006},
  \href{http://arxiv.org/abs/hep-ph/0510049}{{\ttfamily arXiv:hep-ph/0510049}}.

\bibitem{Charbonneau:2009ax}
J.~Charbonneau and A.~Zhitnitsky, ``{Topological Currents in Neutron Stars:
  Kicks, Precession, Toroidal Fields, and Magnetic Helicity},''
  \href{http://arxiv.org/abs/0903.4450}{{\ttfamily arXiv:0903.4450
  [astro-ph.HE]}}.

\bibitem{KerenZur:2010zw}
B.~Keren-Zur and Y.~Oz, ``{Hydrodynamics and the Detection of the QCD Axial
  Anomaly in Heavy Ion Collisions},''
  \href{http://dx.doi.org/10.1007/JHEP06(2010)006}{{\em JHEP} {\bfseries 06}
  (2010) 006}, \href{http://arxiv.org/abs/1002.0804}{{\ttfamily arXiv:1002.0804
  [hep-ph]}}.

\bibitem{Kharzeev:2007jp}
D.~E. Kharzeev, L.~D. McLerran, and H.~J. Warringa, ``{The effects of
  topological charge change in heavy ion collisions: 'Event by event P and CP
  violation'},'' \href{http://dx.doi.org/10.1016/j.nuclphysa.2008.02.298}{{\em
  Nucl. Phys.} {\bfseries A803} (2008) 227--253},
  \href{http://arxiv.org/abs/0711.0950}{{\ttfamily arXiv:0711.0950 [hep-ph]}}.

\bibitem{Fukushima:2008xe}
K.~Fukushima, D.~E. Kharzeev, and H.~J. Warringa, ``{The Chiral Magnetic
  Effect},'' \href{http://dx.doi.org/10.1103/PhysRevD.78.074033}{{\em Phys.
  Rev.} {\bfseries D78} (2008) 074033},
  \href{http://arxiv.org/abs/0808.3382}{{\ttfamily arXiv:0808.3382 [hep-ph]}}.

\bibitem{Kharzeev:2004ey}
D.~Kharzeev, ``{Parity violation in hot QCD: Why it can happen, and how to look
  for it},'' \href{http://dx.doi.org/10.1016/j.physletb.2005.11.075}{{\em Phys.
  Lett.} {\bfseries B633} (2006) 260--264},
  \href{http://arxiv.org/abs/hep-ph/0406125}{{\ttfamily arXiv:hep-ph/0406125}}.

\bibitem{Kharzeev:2007tn}
D.~Kharzeev and A.~Zhitnitsky, ``{Charge separation induced by P-odd bubbles in
  QCD matter},'' \href{http://dx.doi.org/10.1016/j.nuclphysa.2007.10.001}{{\em
  Nucl. Phys.} {\bfseries A797} (2007) 67--79},
  \href{http://arxiv.org/abs/0706.1026}{{\ttfamily arXiv:0706.1026 [hep-ph]}}.

\bibitem{Kharzeev:2009pj}
D.~E. Kharzeev and H.~J. Warringa, ``{Chiral Magnetic conductivity},''
  \href{http://dx.doi.org/10.1103/PhysRevD.80.034028}{{\em Phys. Rev.}
  {\bfseries D80} (2009) 034028},
  \href{http://arxiv.org/abs/0907.5007}{{\ttfamily arXiv:0907.5007 [hep-ph]}}.

\bibitem{Alekseev:1998ds}
A.~Y. Alekseev, V.~V. Cheianov, and J.~Fr{\"o}hlich, ``{Universality of
  transport properties in equilibrium, Goldstone theorem and chiral anomaly},''
  {\em Phys. Rev. Lett.} {\bfseries 81} (1998) 3503--3506,
  \href{http://arxiv.org/abs/cond-mat/9803346}{{\ttfamily
  arXiv:cond-mat/9803346}}.

\bibitem{Abelev:2009uh}
{\bfseries STAR} Collaboration, B.~I. Abelev {\em et~al.}, ``{Azimuthal
  Charged-Particle Correlations and Possible Local Strong Parity Violation},''
  \href{http://dx.doi.org/10.1103/PhysRevLett.103.251601}{{\em Phys. Rev.
  Lett.} {\bfseries 103} (2009) 251601},
  \href{http://arxiv.org/abs/0909.1739}{{\ttfamily arXiv:0909.1739 [nucl-ex]}}.

\bibitem{Voloshin:2008jx}
{\bfseries STAR} Collaboration, S.~A. Voloshin, ``{Probe for the strong parity
  violation effects at RHIC with three particle correlations},''
  \href{http://arxiv.org/abs/0806.0029}{{\ttfamily arXiv:0806.0029 [nucl-ex]}}.

\bibitem{Abelev:2012pa}
{\bfseries ALICE Collaboration} Collaboration, B.~Abelev {\em et~al.},
  ``{Charge separation relative to the reaction plane in Pb-Pb collisions at
  $\sqrt{s_{NN}}= 2.76$ TeV},'' {\em Phys.Rev.Lett.} {\bfseries 110} (2013)
  012301, \href{http://arxiv.org/abs/1207.0900}{{\ttfamily arXiv:1207.0900
  [nucl-ex]}}.

\bibitem{Wang:2009kd}
F.~Wang, ``{Effects of Cluster Particle Correlations on Local Parity Violation
  Observables},'' \href{http://arxiv.org/abs/0911.1482}{{\ttfamily
  arXiv:0911.1482 [nucl-ex]}}.

\bibitem{Asakawa:2010bu}
M.~Asakawa, A.~Majumder, and B.~Muller, ``{Electric Charge Separation in Strong
  Transient Magnetic Fields},''
  \href{http://dx.doi.org/10.1103/PhysRevC.81.064912}{{\em Phys. Rev.}
  {\bfseries C81} (2010) 064912},
  \href{http://arxiv.org/abs/1003.2436}{{\ttfamily arXiv:1003.2436 [hep-ph]}}.

\bibitem{Buividovich:2009wi}
P.~V. Buividovich, M.~N. Chernodub, E.~V. Luschevskaya, and M.~I. Polikarpov,
  ``{Numerical evidence of chiral magnetic effect in lattice gauge theory},''
  \href{http://dx.doi.org/10.1103/PhysRevD.80.054503}{{\em Phys. Rev.}
  {\bfseries D80} (2009) 054503},
  \href{http://arxiv.org/abs/0907.0494}{{\ttfamily arXiv:0907.0494 [hep-lat]}}.

\bibitem{Buividovich:2010tn}
P.~V. Buividovich {\em et~al.}, ``{Magnetic-Field-Induced insulator-conductor
  transition in SU(2) quenched lattice gauge theory},''
  \href{http://arxiv.org/abs/1003.2180}{{\ttfamily arXiv:1003.2180 [hep-lat]}}.

\bibitem{Bhattacharyya:2008jc}
S.~Bhattacharyya, V.~E. Hubeny, S.~Minwalla, and M.~Rangamani, ``{Nonlinear
  Fluid Dynamics from Gravity},''
  \href{http://dx.doi.org/10.1088/1126-6708/2008/02/045}{{\em JHEP} {\bfseries
  0802} (2008) 045}, \href{http://arxiv.org/abs/0712.2456}{{\ttfamily
  arXiv:0712.2456 [hep-th]}}.

\bibitem{Erdmenger:2008rm}
J.~Erdmenger, M.~Haack, M.~Kaminski, and A.~Yarom, ``{Fluid dynamics of
  R-charged black holes},''
  \href{http://dx.doi.org/10.1088/1126-6708/2009/01/055}{{\em JHEP} {\bfseries
  01} (2009) 055}, \href{http://arxiv.org/abs/0809.2488}{{\ttfamily
  arXiv:0809.2488 [hep-th]}}.

\bibitem{Banerjee:2008th}
N.~Banerjee {\em et~al.}, ``{Hydrodynamics from charged black branes},''
  \href{http://dx.doi.org/10.1007/JHEP01(2011)094}{{\em JHEP} {\bfseries 01}
  (2011) 094}, \href{http://arxiv.org/abs/0809.2596}{{\ttfamily arXiv:0809.2596
  [hep-th]}}.

\bibitem{Son:2009tf}
D.~T. Son and P.~Surowka, ``{Hydrodynamics with Triangle Anomalies},''
  \href{http://dx.doi.org/10.1103/PhysRevLett.103.191601}{{\em Phys. Rev.
  Lett.} {\bfseries 103} (2009) 191601},
  \href{http://arxiv.org/abs/0906.5044}{{\ttfamily arXiv:0906.5044 [hep-th]}}.

\bibitem{Neiman:2010zi}
Y.~Neiman and Y.~Oz, ``{Relativistic Hydrodynamics with General Anomalous
  Charges},'' \href{http://dx.doi.org/10.1007/JHEP03(2011)023}{{\em JHEP}
  {\bfseries 03} (2011) 023}, \href{http://arxiv.org/abs/1011.5107}{{\ttfamily
  arXiv:1011.5107 [hep-th]}}.

\bibitem{Kalaydzhyan:2011vx}
T.~Kalaydzhyan and I.~Kirsch, ``{Fluid/gravity model for the chiral magnetic
  effect},'' {\em Phys. Rev. Lett.} {\bfseries 106} (2011) 211601,
  \href{http://arxiv.org/abs/1102.4334}{{\ttfamily arXiv:1102.4334 [hep-th]}}.

\bibitem{Amado:2011zx}
I.~Amado, K.~Landsteiner, and F.~Pena-Benitez, ``{Anomalous transport
  coefficients from Kubo formulas in Holography},''
  \href{http://dx.doi.org/10.1007/JHEP05(2011)081}{{\em JHEP} {\bfseries 1105}
  (2011) 081}, \href{http://arxiv.org/abs/1102.4577}{{\ttfamily arXiv:1102.4577
  [hep-th]}}.

\bibitem{Landsteiner:2011cp}
K.~Landsteiner, E.~Megias, and F.~Pena-Benitez, ``{Gravitational Anomaly and
  Transport},'' {\em Phys. Rev. Lett.} {\bfseries 107} (2011) 021601,
  \href{http://arxiv.org/abs/1103.5006}{{\ttfamily arXiv:1103.5006 [hep-ph]}}.

\bibitem{Vilenkin:1979ui}
A.~Vilenkin, ``{Macrospcopic parity violating effects: neutrino fluxes from
  rotating black holes and in rotating thermal radiation},''
  \href{http://dx.doi.org/10.1103/PhysRevD.20.1807}{{\em Phys.Rev.} {\bfseries
  D20} (1979) 1807--1812}.

\bibitem{Landsteiner:2011iq}
K.~Landsteiner, E.~Megias, L.~Melgar, and F.~Pena-Benitez, ``{Holographic
  Gravitational Anomaly and Chiral Vortical Effect},''
  \href{http://dx.doi.org/10.1007/JHEP09(2011)121}{{\em JHEP} {\bfseries 1109}
  (2011) 121}, \href{http://arxiv.org/abs/1107.0368}{{\ttfamily arXiv:1107.0368
  [hep-th]}}.

\bibitem{Golkar:2012kb}
S.~Golkar and D.~T. Son, ``{Non-Renormalization of the Chiral Vortical Effect
  Coefficient},'' \href{http://arxiv.org/abs/1207.5806}{{\ttfamily
  arXiv:1207.5806 [hep-th]}}.

\bibitem{Hou:2012xg}
D.-F. Hou, H.~Liu, and H.-c. Ren, ``{A Possible Higher Order Correction to the
  Vortical Conductivity in a Gauge Field Plasma},''
  \href{http://dx.doi.org/10.1103/PhysRevD.86.121703}{{\em Phys.Rev.}
  {\bfseries D86} (2012) 121703},
  \href{http://arxiv.org/abs/1210.0969}{{\ttfamily arXiv:1210.0969 [hep-th]}}.

\bibitem{Loganayagam:2012pz}
R.~Loganayagam and P.~Surowka, ``{Anomaly/Transport in an Ideal Weyl gas},''
  \href{http://dx.doi.org/10.1007/JHEP04(2012)097}{{\em JHEP} {\bfseries 1204}
  (2012) 097}, \href{http://arxiv.org/abs/1201.2812}{{\ttfamily arXiv:1201.2812
  [hep-th]}}.

\bibitem{Loganayagam:2011mu}
R.~Loganayagam, ``{Anomaly Induced Transport in Arbitrary Dimensions},''
  \href{http://arxiv.org/abs/1106.0277}{{\ttfamily arXiv:1106.0277 [hep-th]}}.

\bibitem{Jensen:2012jy}
K.~Jensen, ``{Triangle Anomalies, Thermodynamics, and Hydrodynamics},''
  \href{http://dx.doi.org/10.1103/PhysRevD.85.125017}{{\em Phys.Rev.}
  {\bfseries D85} (2012) 125017},
  \href{http://arxiv.org/abs/1203.3599}{{\ttfamily arXiv:1203.3599 [hep-th]}}.

\bibitem{Banerjee:2012iz}
N.~Banerjee, J.~Bhattacharya, S.~Bhattacharyya, S.~Jain, S.~Minwalla, {\em
  et~al.}, ``{Constraints on Fluid Dynamics from Equilibrium Partition
  Functions},'' \href{http://dx.doi.org/10.1007/JHEP09(2012)046}{{\em JHEP}
  {\bfseries 1209} (2012) 046},
  \href{http://arxiv.org/abs/1203.3544}{{\ttfamily arXiv:1203.3544 [hep-th]}}.

\bibitem{Jensen:2012kj}
K.~Jensen, R.~Loganayagam, and A.~Yarom, ``{Thermodynamics, gravitational
  anomalies and cones},'' \href{http://arxiv.org/abs/1207.5824}{{\ttfamily
  arXiv:1207.5824 [hep-th]}}.

\bibitem{Yee:2009vw}
H.-U. Yee, ``{Holographic Chiral Magnetic Conductivity},''
  \href{http://dx.doi.org/10.1088/1126-6708/2009/11/085}{{\em JHEP} {\bfseries
  11} (2009) 085}, \href{http://arxiv.org/abs/0908.4189}{{\ttfamily
  arXiv:0908.4189 [hep-th]}}.

\bibitem{Rebhan:2009vc}
A.~Rebhan, A.~Schmitt, and S.~A. Stricker, ``{Anomalies and the chiral magnetic
  effect in the Sakai- Sugimoto model},''
  \href{http://dx.doi.org/10.1007/JHEP01(2010)026}{{\em JHEP} {\bfseries 01}
  (2010) 026}, \href{http://arxiv.org/abs/0909.4782}{{\ttfamily arXiv:0909.4782
  [hep-th]}}.

\bibitem{Gynther:2010ed}
A.~Gynther, K.~Landsteiner, F.~Pena-Benitez, and A.~Rebhan, ``{Holographic
  Anomalous Conductivities and the Chiral Magnetic Effect},''
  \href{http://dx.doi.org/10.1007/JHEP02(2011)110}{{\em JHEP} {\bfseries 02}
  (2011) 110}, \href{http://arxiv.org/abs/1005.2587}{{\ttfamily arXiv:1005.2587
  [hep-th]}}.

\bibitem{Landsteiner:2011tf}
K.~Landsteiner, E.~Megias, L.~Melgar, and F.~Pena-Benitez, ``{Gravitational
  Anomaly and Hydrodynamics},''
  \href{http://dx.doi.org/10.1088/1742-6596/343/1/012073}{{\em
  J.Phys.Conf.Ser.} {\bfseries 343} (2012) 012073},
  \href{http://arxiv.org/abs/1111.2823}{{\ttfamily arXiv:1111.2823 [hep-th]}}.

\bibitem{Landsteiner:2012kd}
K.~Landsteiner, E.~Megias, and F.~Pena-Benitez, ``{Anomalous Transport from
  Kubo Formulae},'' \href{http://arxiv.org/abs/1207.5808}{{\ttfamily
  arXiv:1207.5808 [hep-th]}}.

\bibitem{Bardeen:1969md}
W.~A. Bardeen, ``{Anomalous Ward identities in spinor field theories},''
  \href{http://dx.doi.org/10.1103/PhysRev.184.1848}{{\em Phys. Rev.} {\bfseries
  184} (1969) 1848--1857}.

\bibitem{Dragon:2012au}
N.~Dragon and F.~Brandt, ``{BRST Symmetry and Cohomology},''
  \href{http://arxiv.org/abs/1205.3293}{{\ttfamily arXiv:1205.3293 [hep-th]}}.

\bibitem{Bertlmann:1996xk}
R.~A. Bertlmann, ``{Anomalies in quantum field theory},''. Oxford, UK:
  Clarendon (1996) 566 p. (International series of monographs on physics: 91).

\bibitem{Delbourgo:1972xb}
R.~Delbourgo and A.~Salam, ``{The gravitational correction to pcac},''
  \href{http://dx.doi.org/10.1016/0370-2693(72)90825-8}{{\em Phys.Lett.}
  {\bfseries B40} (1972) 381--382}.

\bibitem{AlvarezGaume:1983ig}
L.~Alvarez-Gaume and E.~Witten, ``{Gravitational Anomalies},''
  \href{http://dx.doi.org/10.1016/0550-3213(84)90066-X}{{\em Nucl.Phys.}
  {\bfseries B234} (1984) 269}.

\bibitem{Moore:2010jd}
G.~D. Moore and M.~Tassler, ``{The Sphaleron Rate in SU(N) Gauge Theory},''
  \href{http://dx.doi.org/10.1007/JHEP02(2011)105}{{\em JHEP} {\bfseries 1102}
  (2011) 105}, \href{http://arxiv.org/abs/1011.1167}{{\ttfamily arXiv:1011.1167
  [hep-ph]}}.

\bibitem{Landsteiner:2012hc}
K.~Landsteiner, E.~Megias, L.~Melgar, and F.~Pena-Benitez, ``{Gravitational
  anomaly and hydrodynamics in AdS/CFT},''
  \href{http://dx.doi.org/10.1002/prop.201200021}{{\em Fortsch.Phys.}
  {\bfseries 60} (2012) 1064--1070}.

\bibitem{Kovtun:2012rj}
P.~Kovtun, ``{Lectures on hydrodynamic fluctuations in relativistic
  theories},'' \href{http://dx.doi.org/10.1088/1751-8113/45/47/473001}{{\em
  J.Phys.} {\bfseries A45} (2012) 473001},
  \href{http://arxiv.org/abs/1205.5040}{{\ttfamily arXiv:1205.5040 [hep-th]}}.

\bibitem{Kharzeev:2011ds}
D.~E. Kharzeev and H.-U. Yee, ``{Anomalies and time reversal invariance in
  relativistic hydrodynamics: the second order and higher dimensional
  formulations},'' \href{http://arxiv.org/abs/1105.6360}{{\ttfamily
  arXiv:1105.6360 [hep-th]}}.

\bibitem{Mashhoon:2003ax}
B.~Mashhoon, ``{Gravitoelectromagnetism: A Brief review},''
  \href{http://arxiv.org/abs/gr-qc/0311030}{{\ttfamily arXiv:gr-qc/0311030
  [gr-qc]}}.

\bibitem{Landsteiner:2011tg}
K.~Landsteiner, E.~Megias, and F.~Pena-Benitez, ``{Anomalies and Transport
  Coefficients: The Chiral Gravito-Magnetic Effect},''
  \href{http://arxiv.org/abs/1110.3615}{{\ttfamily arXiv:1110.3615 [hep-ph]}}.

\bibitem{Loganayagam:2008is}
R.~Loganayagam, ``{Entropy Current in Conformal Hydrodynamics},'' {\em arXiv}
  {\bfseries 0805} (2008) 087.

\bibitem{Lifschytz:2009si}
G.~Lifschytz and M.~Lippert, ``{Anomalous conductivity in holographic QCD},''
  \href{http://dx.doi.org/10.1103/PhysRevD.80.066005}{{\em Phys. Rev.}
  {\bfseries D80} (2009) 066005},
  \href{http://arxiv.org/abs/0904.4772}{{\ttfamily arXiv:0904.4772 [hep-th]}}.

\bibitem{Hill:2006ei}
C.~T. Hill, ``{Anomalies, Chern-Simons terms and chiral delocalization in extra
  dimensions},'' \href{http://dx.doi.org/10.1103/PhysRevD.73.085001}{{\em Phys.
  Rev.} {\bfseries D73} (2006) 085001},
  \href{http://arxiv.org/abs/hep-th/0601154}{{\ttfamily arXiv:hep-th/0601154}}.

\bibitem{Sakai:2004cn}
T.~Sakai and S.~Sugimoto, ``{Low energy hadron physics in holographic QCD},''
  \href{http://dx.doi.org/10.1143/PTP.113.843}{{\em Prog. Theor. Phys.}
  {\bfseries 113} (2005) 843--882},
  \href{http://arxiv.org/abs/hep-th/0412141}{{\ttfamily arXiv:hep-th/0412141}}.

\bibitem{Sakai:2005yt}
T.~Sakai and S.~Sugimoto, ``{More on a holographic dual of QCD},''
  \href{http://dx.doi.org/10.1143/PTP.114.1083}{{\em Prog. Theor. Phys.}
  {\bfseries 114} (2005) 1083--1118},
  \href{http://arxiv.org/abs/hep-th/0507073}{{\ttfamily arXiv:hep-th/0507073}}.

\bibitem{Gorsky:2010xu}
A.~Gorsky, P.~N. Kopnin, and A.~V. Zayakin, ``{On the Chiral Magnetic Effect in
  Soft-Wall AdS/QCD},''
  \href{http://dx.doi.org/10.1103/PhysRevD.83.014023}{{\em Phys. Rev.}
  {\bfseries D83} (2011) 014023},
  \href{http://arxiv.org/abs/1003.2293}{{\ttfamily arXiv:1003.2293 [hep-ph]}}.

\bibitem{Rubakov:2010qi}
V.~A. Rubakov, ``{On chiral magnetic effect and holography},''
  \href{http://arxiv.org/abs/1005.1888}{{\ttfamily arXiv:1005.1888 [hep-ph]}}.

\bibitem{Son:2002sd}
D.~T. Son and A.~O. Starinets, ``{Minkowski-space correlators in AdS/CFT
  correspondence: Recipe and applications},'' {\em JHEP} {\bfseries 09} (2002)
  042, \href{http://arxiv.org/abs/hep-th/0205051}{{\ttfamily
  arXiv:hep-th/0205051}}.

\bibitem{Herzog:2002pc}
C.~P. Herzog and D.~T. Son, ``{Schwinger-Keldysh propagators from AdS/CFT
  correspondence},'' {\em JHEP} {\bfseries 03} (2003) 046,
  \href{http://arxiv.org/abs/hep-th/0212072}{{\ttfamily arXiv:hep-th/0212072}}.

\bibitem{Kaminski:2009dh}
M.~Kaminski, K.~Landsteiner, J.~Mas, J.~P. Shock, and J.~Tarrio, ``{Holographic
  Operator Mixing and Quasinormal Modes on the Brane},''
  \href{http://dx.doi.org/10.1007/JHEP02(2010)021}{{\em JHEP} {\bfseries 02}
  (2010) 021}, \href{http://arxiv.org/abs/0911.3610}{{\ttfamily arXiv:0911.3610
  [hep-th]}}.

\bibitem{Landsman:1986uw}
N.~Landsman and C.~van Weert, ``{Real and Imaginary Time Field Theory at Finite
  Temperature and Density},''
  \href{http://dx.doi.org/10.1016/0370-1573(87)90121-9}{{\em Phys.Rept.}
  {\bfseries 145} (1987) 141}.

\bibitem{Eguchi:1976db}
T.~Eguchi and P.~G. Freund, ``{Quantum Gravity and World Topology},''
  \href{http://dx.doi.org/10.1103/PhysRevLett.37.1251}{{\em Phys.Rev.Lett.}
  {\bfseries 37} (1976) 1251}.

\bibitem{Gubser:1998bc}
S.~Gubser, I.~R. Klebanov, and A.~M. Polyakov, ``{Gauge theory correlators from
  noncritical string theory},''
  \href{http://dx.doi.org/10.1016/S0370-2693(98)00377-3}{{\em Phys.Lett.}
  {\bfseries B428} (1998) 105--114},
  \href{http://arxiv.org/abs/hep-th/9802109}{{\ttfamily arXiv:hep-th/9802109
  [hep-th]}}.

\bibitem{Witten:1998qj}
E.~Witten, ``{Anti-de Sitter space and holography},'' {\em Adv. Theor. Math.
  Phys.} {\bfseries 2} (1998) 253--291,
  \href{http://arxiv.org/abs/hep-th/9802150}{{\ttfamily arXiv:hep-th/9802150}}.

\bibitem{Aharony:1999ti}
O.~Aharony, S.~S. Gubser, J.~M. Maldacena, H.~Ooguri, and Y.~Oz, ``{Large N
  field theories, string theory and gravity},''
  \href{http://dx.doi.org/10.1016/S0370-1573(99)00083-6}{{\em Phys. Rept.}
  {\bfseries 323} (2000) 183--386},
  \href{http://arxiv.org/abs/hep-th/9905111}{{\ttfamily arXiv:hep-th/9905111}}.

\bibitem{Andreev:2006ct}
O.~Andreev and V.~I. Zakharov, ``{Heavy-quark potentials and AdS/QCD},''
  \href{http://dx.doi.org/10.1103/PhysRevD.74.025023}{{\em Phys. Rev.}
  {\bfseries D74} (2006) 025023},
  \href{http://arxiv.org/abs/hep-ph/0604204}{{\ttfamily arXiv:hep-ph/0604204}}.

\bibitem{Galow:2009kw}
B.~Galow, E.~Megias, J.~Nian, and H.~J. Pirner, ``{Phenomenology of AdS/QCD and
  Its Gravity Dual},''
  \href{http://dx.doi.org/10.1016/j.nuclphysb.2010.03.022}{{\em Nucl. Phys.}
  {\bfseries B834} (2010) 330--362},
  \href{http://arxiv.org/abs/0911.0627}{{\ttfamily arXiv:0911.0627 [hep-ph]}}.

\bibitem{Gursoy:2008za}
U.~Gursoy, E.~Kiritsis, L.~Mazzanti, and F.~Nitti, ``{Holography and
  Thermodynamics of 5D Dilaton-gravity},''
  \href{http://dx.doi.org/10.1088/1126-6708/2009/05/033}{{\em JHEP} {\bfseries
  05} (2009) 033}, \href{http://arxiv.org/abs/0812.0792}{{\ttfamily
  arXiv:0812.0792 [hep-th]}}.

\bibitem{Megias:2010ku}
E.~Megias, H.~J. Pirner, and K.~Veschgini, ``{QCD-Thermodynamics using 5-dim
  Gravity},'' \href{http://dx.doi.org/10.1103/PhysRevD.83.056003}{{\em Phys.
  Rev.} {\bfseries D83} (2011) 056003},
  \href{http://arxiv.org/abs/1009.2953}{{\ttfamily arXiv:1009.2953 [hep-ph]}}.

\bibitem{Veschgini:2010ws}
K.~Veschgini, E.~Megias, and H.~J. Pirner, ``{Trouble Finding the Optimal
  AdS/QCD},'' \href{http://dx.doi.org/10.1016/j.physletb.2011.01.011}{{\em
  Phys. Lett.} {\bfseries B696} (2011) 495--498},
  \href{http://arxiv.org/abs/1009.4639}{{\ttfamily arXiv:1009.4639 [hep-th]}}.

\bibitem{Jackiw:2003pm}
R.~Jackiw and S.~Pi, ``{Chern-Simons modification of general relativity},''
  \href{http://dx.doi.org/10.1103/PhysRevD.68.104012}{{\em Phys.Rev.}
  {\bfseries D68} (2003) 104012},
  \href{http://arxiv.org/abs/gr-qc/0308071}{{\ttfamily arXiv:gr-qc/0308071
  [gr-qc]}}.

\bibitem{Alexander:2009tp}
S.~Alexander and N.~Yunes, ``{Chern-Simons Modified General Relativity},''
  \href{http://dx.doi.org/10.1016/j.physrep.2009.07.002}{{\em Phys.Rept.}
  {\bfseries 480} (2009) 1--55},
  \href{http://arxiv.org/abs/0907.2562}{{\ttfamily arXiv:0907.2562 [hep-th]}}.

\bibitem{Saremi:2011ab}
O.~Saremi and D.~T. Son, ``{Hall viscosity from gauge/gravity duality},''
  \href{http://arxiv.org/abs/1103.4851}{{\ttfamily arXiv:1103.4851 [hep-th]}}.

\bibitem{Delsate:2011qp}
T.~Delsate, V.~Cardoso, and P.~Pani, ``{Anti de Sitter black holes and branes
  in dynamical Chern-Simons gravity: perturbations, stability and the
  hydrodynamic modes},'' \href{http://dx.doi.org/10.1007/JHEP06(2011)055}{{\em
  JHEPA,1106,055.2011} {\bfseries 1106} (2011) 055},
  \href{http://arxiv.org/abs/1103.5756}{{\ttfamily arXiv:1103.5756 [hep-th]}}.

\bibitem{Martelli:2002sp}
D.~Martelli and W.~Mueck, ``{Holographic renormalization and Ward identities
  with the Hamilton-Jacobi method},''
  \href{http://dx.doi.org/10.1016/S0550-3213(03)00060-9}{{\em Nucl.Phys.}
  {\bfseries B654} (2003) 248--276},
  \href{http://arxiv.org/abs/hep-th/0205061}{{\ttfamily arXiv:hep-th/0205061
  [hep-th]}}.

\bibitem{Papadimitriou:2004ap}
I.~Papadimitriou and K.~Skenderis, ``{AdS / CFT correspondence and geometry},''
  \href{http://arxiv.org/abs/hep-th/0404176}{{\ttfamily arXiv:hep-th/0404176
  [hep-th]}}.

\bibitem{Yee:2011yn}
H.-U. Yee and I.~Zahed, ``{Holographic two dimensional QCD and Chern-Simons
  term},'' \href{http://arxiv.org/abs/1103.6286}{{\ttfamily arXiv:1103.6286
  [hep-th]}}.

\bibitem{Kraus:2005zm}
P.~Kraus and F.~Larsen, ``{Holographic gravitational anomalies},''
  \href{http://dx.doi.org/10.1088/1126-6708/2006/01/022}{{\em JHEP} {\bfseries
  0601} (2006) 022}, \href{http://arxiv.org/abs/hep-th/0508218}{{\ttfamily
  arXiv:hep-th/0508218 [hep-th]}}.

\bibitem{Grumiller:2008ie}
D.~Grumiller, R.~B. Mann, and R.~McNees, ``{Dirichlet boundary value problem
  for Chern-Simons modified gravity},''
  \href{http://dx.doi.org/10.1103/PhysRevD.78.081502}{{\em Phys.Rev.}
  {\bfseries D78} (2008) 081502},
  \href{http://arxiv.org/abs/0803.1485}{{\ttfamily arXiv:0803.1485 [gr-qc]}}.

\bibitem{Clark:2010fs}
T.~Clark, S.~Love, and T.~ter Veldhuis, ``{Holographic Currents and
  Chern-Simons Terms},''
  \href{http://dx.doi.org/10.1103/PhysRevD.82.106004}{{\em Phys.Rev.}
  {\bfseries D82} (2010) 106004},
  \href{http://arxiv.org/abs/1006.2400}{{\ttfamily arXiv:1006.2400 [hep-th]}}.

\bibitem{Neiman:2011mj}
Y.~Neiman and Y.~Oz, ``{Anomalies in Superfluids and a Chiral Electric
  Effect},'' \href{http://arxiv.org/abs/1106.3576}{{\ttfamily arXiv:1106.3576
  [hep-th]}}.

\bibitem{Bonora:2011mf}
L.~Bonora, M.~Cvitan, P.~D. Prester, S.~Pallua, and I.~Smolic, ``{Gravitational
  Chern-Simons Lagrangian terms and spherically symmetric spacetimes},''
  \href{http://arxiv.org/abs/1105.4792}{{\ttfamily arXiv:1105.4792 [hep-th]}}.

\bibitem{Bhattacharyya:2008ks}
S.~Bhattacharyya, V.~E. Hubeny, R.~Loganayagam, G.~Mandal, S.~Minwalla,
  T.~Morita, M.~Rangamani, and H.~S. Reall, ``{Local fluid dynamical entropy
  from gravity},'' {\em Journal of High Energy Physics} {\bfseries 2008}
  no.~06, (June, 2008) 055--055.

\bibitem{Chapman:2012my}
S.~Chapman, Y.~Neiman, and Y.~Oz, ``{Fluid/Gravity Correspondence, Local Wald
  Entropy Current and Gravitational Anomaly},''
  \href{http://dx.doi.org/10.1007/JHEP07(2012)128}{{\em JHEP} {\bfseries 1207}
  (2012) 128}, \href{http://arxiv.org/abs/1202.2469}{{\ttfamily arXiv:1202.2469
  [hep-th]}}.

\bibitem{Fluid:grav}
E.~Megias and F.~Pena-Benitez, ``{Holographic Gravitational Anomaly in First
  and Second Order Hydrodynamics},'' {\em To appear} (2013) .

\bibitem{Rangamani:2009xk}
M.~Rangamani, ``{Gravity and Hydrodynamics: Lectures on the fluid-gravity
  correspondence},''
  \href{http://dx.doi.org/10.1088/0264-9381/26/22/224003}{{\em
  Class.Quant.Grav.} {\bfseries 26} (2009) 224003},
  \href{http://arxiv.org/abs/0905.4352}{{\ttfamily arXiv:0905.4352 [hep-th]}}.

\bibitem{Haack:2008cp}
M.~Haack and A.~Yarom, ``{Nonlinear viscous hydrodynamics in various dimensions
  using AdS/CFT},'' \href{http://dx.doi.org/10.1088/1126-6708/2008/10/063}{{\em
  JHEP} {\bfseries 0810} (2008) 063},
  \href{http://arxiv.org/abs/0806.4602}{{\ttfamily arXiv:0806.4602 [hep-th]}}.

\bibitem{Bhattacharyya:2008dl}
S.~Bhattacharyya, R.~Loganayagam, S.~Minwalla, S.~Nampuri, S.~P. Trivedi, and
  S.~R. Wadia, ``{Forced Fluid Dynamics from Gravity},'' {\em arXiv.org}
  {\bfseries hep-th} (May, 2008) .

\bibitem{Sahoo:2009yq}
B.~Sahoo and H.-U. Yee, ``{Holographic chiral shear waves from anomaly},''
  \href{http://dx.doi.org/10.1016/j.physletb.2010.04.076}{{\em Phys.Lett.}
  {\bfseries B689} (2010) 206--212},
  \href{http://arxiv.org/abs/0910.5915}{{\ttfamily arXiv:0910.5915 [hep-th]}}.

\bibitem{Evans:1995yz}
T.~Evans, ``{The Condensed matter limit of relativistic QFT},''
  \href{http://arxiv.org/abs/hep-ph/9510298}{{\ttfamily arXiv:hep-ph/9510298
  [hep-ph]}}.

\bibitem{Nair:2011mk}
V.~Nair, R.~Ray, and S.~Roy, ``{Fluids, Anomalies and the Chiral Magnetic
  Effect: A Group-Theoretic Formulation},''
  \href{http://dx.doi.org/10.1103/PhysRevD.86.025012}{{\em Phys.Rev.}
  {\bfseries D86} (2012) 025012},
  \href{http://arxiv.org/abs/1112.4022}{{\ttfamily arXiv:1112.4022 [hep-th]}}.

\bibitem{Kharzeev:2010gr}
D.~E. Kharzeev and D.~T. Son, ``{Testing the chiral magnetic and chiral
  vortical effects in heavy ion collisions},''
  \href{http://dx.doi.org/10.1103/PhysRevLett.106.062301}{{\em Phys. Rev.
  Lett.} {\bfseries 106} (2011) 062301},
  \href{http://arxiv.org/abs/1010.0038}{{\ttfamily arXiv:1010.0038 [hep-ph]}}.

\bibitem{Alexander:2004us}
S.~H.-S. Alexander, M.~E. Peskin, and M.~M. Sheikh-Jabbari, ``{Leptogenesis
  from gravity waves in models of inflation},''
  \href{http://dx.doi.org/10.1103/PhysRevLett.96.081301}{{\em Phys.Rev.Lett.}
  {\bfseries 96} (2006) 081301},
  \href{http://arxiv.org/abs/hep-th/0403069}{{\ttfamily arXiv:hep-th/0403069
  [hep-th]}}.

\bibitem{Matsuo:2009xn}
Y.~Matsuo, S.-J. Sin, S.~Takeuchi, and T.~Tsukioka, ``{Magnetic conductivity
  and Chern-Simons Term in Holographic Hydrodynamics of Charged AdS Black
  Hole},'' \href{http://dx.doi.org/10.1007/JHEP04(2010)071}{{\em JHEP}
  {\bfseries 1004} (2010) 071},
  \href{http://arxiv.org/abs/0910.3722}{{\ttfamily arXiv:0910.3722 [hep-th]}}.

\end{thebibliography}\endgroup
%%%%%%%%%%%%%%%%%%%%%%%%%%%%%%%%%%%%%%%%%
\begin{titlepage}
\fancyhf{}
\newpage
\mbox{}
\newpage
\mbox{}
\newpage\thispagestyle{empty}
\end{titlepage}
\end{document}